%% file: doctor04.tex
\documentclass[10pt]{book}
\usepackage{amsmath,amssymb,bm,graphicx,color}
\usepackage{cite,mathrsfs}
\date{}
\makeatletter
 
  \@addtoreset{equation}{section}
 \makeatother
 
\newlength{\myheight}
\begin{document}
\begin{flushright}
\vskip 0.6cm
%March, 2010
\end{flushright}

\thispagestyle{empty}

\vskip 1.8cm
\begin{center}
 
  {\LARGE \bf Recent Attempts in the Analysis \\of Black Hole Radiation}

\vskip 0.6cm
  
%{\large （ブラックホール放射の研究における最近の試み）}
  
\vskip 2.3cm
\normalsize

  {\Large Koichiro Umetsu 
  \footnote{E-mail: umetsu@phys.cst.nihon-u.ac.jp}
  }

\vskip 0.5cm

  {\large \it Graduate School of Quantum Science and Technology,\\ Nihon University,
               Tokyo 101-8308, Japan}

\vskip 0.5cm

{\large January, 2010}

%%%%%%%%%%%%%%%%%%%%%%%%%%%%%%%%%%%%%%%%%%%%%%%%               
%\vskip 2cm

%\begin{figure}[h]
%  \begin{center}
%    \includegraphics*[width = 5 cm]{Nlogo.eps}
%  \end{center}
%\end{figure}
%\vskip 1cm
%%%%%%%%%%%%%%%%%%%%%%%%%%%%%%%%%%%%%%%%%%%%%%%%%
\vskip 6cm
%\vskip 7cm

{\large A Dissertation in candidacy for the degree\\
 of Doctor of Philosophy}

\newpage 
\thispagestyle{empty}
~

\newpage
\pagenumbering{roman}
\setcounter{page}{1}
\vspace{0.5cm}

{\bf Abstract}
\end{center}

In this thesis, we first present a brief review of black hole radiation which is commonly called Hawking radiation.
%Hawking radiation is derived by taking into account the quantum effects in the framework of general relativity.
%Hawking radiation is thus one of very interesting phenomena where both of general theory of relativity and quantum theory play a role at the same time.
The existence of Hawking radiation by itself is well established by now because the same result is derived by several different methods.
On the other hand, there remain several aspects of the effect which have yet to be clarified.
We clarify some arguments in previous works on the subject and then attempt to present the more satisfactory derivations of Hawking radiation.
To be specific, we examine the analyses in the two recent derivations of Hawking radiation which are based on anomalies and tunneling;
both of these derivations were initiated by Wilczek and his collaborators.
We then present a simple derivation based on anomalies
by emphasizing a systematic use of covariant currents and covariant anomalies combined with boundary conditions which have clear physical meaning.
We also extend a variant of the tunneling method proposed by Banerjee and Majhi to a Kerr-Newman black hole by using the technique of the dimensional reduction near the horizon.
We directly derive the black body spectrum for a Kerr-Newman black hole
on the basis of the tunneling mechanism.

\baselineskip 7mm

\clearpage

\tableofcontents

\baselineskip 6.22mm

%%%%%%%%%%%%%%%%%%%%%%%%%%%%%%%%%%%%%%%%%%%%%%%%%%%%%%%%%%%%%%%%%%%%%%%%%%%%%%%%%%%%%%%%%%%%%%%%%
%%                                                                                             %%
%%                                        Chapter 1                                            %%
%%                                                                                             %%
%%%%%%%%%%%%%%%%%%%%%%%%%%%%%%%%%%%%%%%%%%%%%%%%%%%%%%%%%%%%%%%%%%%%%%%%%%%%%%%%%%%%%%%%%%%%%%%%%
\newpage
\pagenumbering{arabic}
\setcounter{page}{1}
\chapter{Introduction}

\quad~ General theory of relativity and quantum theory are two fundamental theories in modern physics.
According to the current understanding of physics, it is well known that
all the forces which have been identified in nature can be explained by the electromagnetic force, weak force, strong force and gravity.
The first three of them are described by quantum field theory and the remaining gravity is described by the general theory of relativity.

Furthermore, the idea of a unified field theory, which can describe the four forces by a single theory, was advanced.
Electromagnetic and weak forces were unified by Weinberg-Salam theory and 
a grand unified theory which combines the strong interaction with the electroweak interactions was proposed.
However, gravity has yet to be successfully included in a theory of everything. 
A simple attempt to combine the gravitational interaction with the strong and electroweak interactions runs into fundamental difficulties since the resulting theory is not renormalizable.
This means that physically meaningful observables contain nonremovable infinities. 
The string theory has potentiality which solves these problems. 
However, we have not obtained any solid result in string theory yet and it depends on the future progress.
We have not yet formulated a widely accepted, consistent theory that combines the general theory of relativity with the principle of quantum theory.
In any case, we must study a framework where both of the general theory of relativity and quantum theory are consistently incorporated.

From the above point of view, the black hole radiation which was suggested by Hawking is very interesting.
Although we have not yet confirmed that black holes do really exist, it has been predicted that they exist as a consequence of the general theory of relativity, namely, as special solutions of the basic Einstein equation.
According to the Einstein equation, the space-time is curved by the effects of gravity.
The space-time curved by a very strong gravity can form a closed region from which nothing, not even photons, can escape.
The closed region is the black hole.
Thus black holes cannot classically allow the emission of radiation. 
However, by using quantum field theory in black hole physics, 
a mechanism by which black holes can radiate was proposed by Hawking \cite{haw00,haw01}.
The radiation from the black hole is commonly called the Hawking radiation.
In this sense, 
it can be said that Hawking radiation is one of precious phenomena where both of the general theory of relativity and quantum theory
play a role at the same time.
When any new theory of quantum gravity is constructed, it must be checked if it correctly describes Hawking radiation in the proposed theory.

Hawking's original derivation is very direct and physical \cite{haw01}.
The analysis calculates the Bogoliubov coefficients between the in- and out-states for a body collapsing to form a black hole.
It is well-known that the characteristic spectrum found in the original derivation agrees with the black body spectrum
with a characteristic temperature associated with the black hole if we ignore the back scattering of particles falling into the black hole.
Namely, it was found that a black hole behaves as a black body and the black hole emits radiation.

After Hawking's original derivation, various derivations of Hawking radiation have been suggested.
All of them reproduce the same result that the black hole entropy is described by the surface area of the black hole and 
the temperature of the black hole is described by a surface gravity of the black hole.
Hawking radiation is thus one of the most striking effects which are widely accepted by now.
However, there are several aspects which have not been completely clarified yet.
In particular, although the entropy is interpreted as a count of the number of states in statistical mechanics,
the entropy of a black hole with a finite temperature has not been derived by counting the number of quantum states associated with the black hole.
It is considered that this problem is closely related to the fact that quantum theory of gravity has not been explicitly formulated yet,
and it is very difficult to construct a consistent quantum gravity.
By examining the various derivations of Hawking radiation,
we find that each derivation has both merits and demerits.
In this sense, it may be fair to say that these known derivations of Hawking radiation have not reached an impeccable conclusion yet.

Recently, Robinson and Wilczek suggested a new method of deriving Hawking radiation by the consideration of anomalies \cite{rob01}.
The basic idea of the approach is that the flux of Hawking radiation is determined by anomaly cancellation conditions in the background of a Schwarzschild black hole.
Iso, Umetsu and Wilczek improved the approach by Robinson and Wilczek, and they extended the method to a charged black hole \cite{iso01} and a rotating black hole \cite{iso02}.
The approach of Iso, Umetsu and Wilczek \cite{iso02} is very transparent and interesting.
However, there remain several points to be clarified.
We have presented arguments which clarify the basic idea of the derivation and given a simple derivation by using the Ward Identities and boundary conditions \cite{ume01}.
We would like to explain our simple derivation as comprehensibly as possible in the present thesis.

A straightforward derivation on the basis of the tunneling mechanism was also suggested by Parikh and Wilczek \cite{par01}.
The analysis of tunneling mechanism was mainly confined to the derivation of the temperature of a black hole, and the black body spectrum itself has not been much discussed.
More recently, this problem of the black body spectrum was emphasized by Banerjee and Majhi \cite{maj01}.
They showed how to reproduce the black body spectrum directly, which agrees with Hawking's original result, by using the properties of the tunneling mechanism.
Thus the derivation on the basis of the tunneling mechanism became more satisfactory.
Their result is valid only for black holes with a spherically symmetric geometry.
However, it is known that 4-dimensional black holes have not only a mass and a charge but also angular momentum,
and the geometry of a rotating black hole becomes spherically asymmetric because of its own rotation.
We have recently attempted to extend Banerjee and Majhi's method to a rotating black hole by using a technique valid only near the horizon,
which is called the dimensional reduction \cite{ume02}.
We showed that the result agrees with the previous result.
We explain our method which shows how to directly derive the black body spectrum for a rotating black hole on the basis of the tunneling mechanism in this thesis.
To the best of my knowledge, there is no derivation of the spectrum by using the technique of the dimensional reduction in the tunneling mechanism.
Therefore, we believe that this derivation clarifies some aspects of the tunneling mechanism.

The contents of the present thesis are as follows.
In Chapter 2, we review some properties of black holes. These properties will be useful to understand the contents of the following chapters.
In Chapter 3, we review the original derivation of Hawking radiation by Hawking and briefly explain other representative derivations of Hawking radiation.
It will be argued that there are several aspects to be clarified in the existing derivations of Hawking radiation.
In Chapter 4, we discussed the derivation of Hawking radiation which is based on anomalies.
We clarify some aspects in previous works on this subject and present a simple derivation of Hawking radiation from anomalies.
In Chapter 5, we discussed the derivation of Hawking radiation which is based on quantum tunneling.
We present a generalization of the derivation of Hawking radiation by Banerjee and Majhi on the basis of the tunneling mechanism to a rotating black hole and also give some clarifying comments.
Chapter 6 is devoted to discussion and conclusion.
Some of the technical details are given in appendices.

In this paper, we use the natural system of units 
\begin{align}
c=G=\hbar=1
\label{1unit01}
\end{align} 
unless stated otherwise,
where $c$ is the speed of light in vacuum,
$G$ is the gravitational constant and $\hbar$ is the Planck constant (Dirac's constant).
%%%%%%%%%%%%%%%%%%%%%%%%%%%%%%%%%%%%%%%%%%%%%%%%%%%%%%%%%%%%%%%%%%%%%%%%%%%%%%%%%%%%%%%%%%%%%%%%%
%%                                                                                             %%
%%                                        Chapter 2                                            %%
%%                                                                                             %%
%%%%%%%%%%%%%%%%%%%%%%%%%%%%%%%%%%%%%%%%%%%%%%%%%%%%%%%%%%%%%%%%%%%%%%%%%%%%%%%%%%%%%%%%%%%%%%%%%
\newpage

\chapter{Properties of Black Hole}

\quad~ The existence of black holes has been predicted by the general theory of relativity.
To understand Hawking radiation, we have to know some classical properties of black holes first.
In this chapter we would like to review some properties of black holes.

The contents of this chapter are as follows.
In Section 2.1, we review the general theory of relativity and black holes.
We would also like to mention various types of black holes.
In Section 2.2, we refer to the Penrose diagram and show how to describe it.
In Section 2.3, we discuss how to extract energy from a rotating black hole classically.
In Section 2.4, we would like to discuss the dimensional reduction near the event horizon which is a boundary between our universe and a black hole. By using the technique of the dimensional reduction, we show that a 4-dimensional metric associated with a charged and rotating black hole  effectively becomes a 2-dimensional spherically symmetric metric.
In Section 2.5, we would like to discuss analogies between black hole physics and thermodynamics.
In Section 2.6, we review the argument, which was suggested by Bekenstein, that black holes have entropy.
These introductory discussions will be useful to understand the contents of the following chapters.
%%%%%%%%%%%%%%%%%%%%%%%%%%%%%%%%%%%%%%%%%%%%%%%%%%%%%%%%%%%%%%%%%%%%%%%%%%%%%%%%%%%%%%%%%%%%%%%%%
%%                                                                                             %%
%%                                     Section 2.1                                             %%
%%                                                                                             %%
%%%%%%%%%%%%%%%%%%%%%%%%%%%%%%%%%%%%%%%%%%%%%%%%%%%%%%%%%%%%%%%%%%%%%%%%%%%%%%%%%%%%%%%%%%%%%%%%%
\section{General Theory of Relativity and Black Hole}

\quad~ General theory of relativity is the theory of space-time and gravitation formulated by Einstein in 1915 \cite{ein01}.
The Einstein equation which describes the general theory of relativity is given by \cite{ein02}
\begin{equation}
R_{\mu\nu}-\frac{1}{2}R g_{\mu\nu}=\frac{8\pi G}{c^4} T_{\mu\nu},
\label{2ein01}
\end{equation}
where 
$R_{\mu\nu}$ is the Ricci tensor, 
$R$ is the Ricci scalar, 
$g_{\mu\nu}$ is the metric of space-time, 
$G$ is the gravitational constant, 
$c$ is the speed of light,
and $T_{\mu\nu}$ is the energy-momentum tensor.
These quantities are defined by
\begin{align}
R&\equiv R^\mu_{~\mu} =g^{\mu\nu}R_{\nu\mu}\\
R_{\mu\nu}&\equiv R^{\rho}_{\mu\nu\rho}\\
R^{\rho}_{\mu\nu\sigma}&\equiv \partial_{\nu} \Gamma^{\rho}_{\mu\sigma} - \partial_{\sigma} \Gamma^{\rho}_{\mu\nu}
+\Gamma^{\alpha}_{\mu\sigma}\Gamma^{\rho}_{\alpha\nu}
-\Gamma^{\alpha}_{\mu\nu}\Gamma^{\rho}_{\alpha\sigma}\\
\Gamma^{\rho}_{\mu\nu}&\equiv \frac{1}{2} g^{\rho \alpha} \left( \partial_{\nu} g_{\alpha\mu} + \partial_{\mu} g_{\alpha\nu}
-\partial_{\alpha} g_{\mu\nu}
\right)
\label{2chr01}\\
ds^2&\equiv g_{\mu\nu}dx^\mu dx^\nu,
\end{align}
where $R^{\rho}_{\mu\nu\sigma}$ is the Riemann-Christoffel tensor or the curvature tensor, 
$\Gamma^{\rho}_{\mu\nu}$ is the Christoffel symbol,
and $ds$ is the line element.
The expression on the left-hand side of the equation (\ref{2ein01}) represents the curvature of space-time as determined by the metric, 
and the expression on the right-hand side represents the distribution of matter fields. 
The Einstein equation is then interpreted as a set of equations dictating how the curvature of space-time is related to the distribution of matter and energy in the universe.

It is difficult to solve the general solution for the Einstein equation 
because the Einstein equation is the quadratic nonlinear differential equation.
However, it is known that there are several exact solutions for the Einstein equation.
In 1916, Schwarzshild found an exact solution for the Einstein equation \cite{sch01} which describes the gravitational field outside a black hole which depends only on the mass.

It is considered that black holes are formed as a result of the gravitational collapse of a star with a very large mass.
The original stars which will form the black hole have various physical quantities and properties.
As soon as a black hole is formed by the gravitational collapse, the state of the black hole becomes a stationary state.
It is known that the stationary state is characterized by only three physical parameters, namely, the mass, the angular momentum and the electrical charge.
This means that a black hole does not retain the various information of the original star except for these three parameters.
In other words, a black hole can uniquely be decided by the mass, the angular momentum and the charge.
This consequence is called the black hole uniqueness theorem \cite{isr01, car01, haw02} or the no-hair theorem \cite{ruf01},
and the uniqueness theorem is shown in a 4-dimensional theory if the solutions of the Einstein equation satisfy the four conditions

1. Only electromagnetic field exits.

2. Asymptotically flat.

3. Stationary.

4. No singularity exists on and outside the event horizon.\\
Here the fourth condition is based on the cosmic censorship hypothesis proposed by Penrose \cite{pen01}.

Black holes are divided into four groups by depending on parameters and each has its own name (Tab. 2.1).  
The Schwarzschild black hole depends on only the mass.
The Schwarzschild metric, which describes the space-time outside the Schwarzschild black hole, is given by
\begin{align}
ds^2 = -\left( 1- \frac{2M}{r} \right)dt^2 + \frac{1}{1- \frac{2M}{r}} dr^2 + r^2 d\theta^2 + r^2 \sin^2 \theta d \varphi^2,
\label{2sch01}
\end{align}
where $r$, $\theta$ and $\varphi$ are commonly used variables in polar coordinates, and $M$ is the mass of the black hole.
The Reissner-Nordstr\"om black hole depends on both the mass and the charge.
The Kerr black hole depends on both the mass and the angular momentum.
The Kerr-Newman black hole depends on the mass, the charge and the angular momentum.
The Kerr-Newman metric is given by
\begin{align}
ds^2=&-\frac{\Delta -a^2 \sin^2\theta}{\Sigma}dt^2-\frac{2a\sin^2\theta}{\Sigma}(r^2+a^2-\Delta)dtd\varphi \notag\\
&-\frac{a^2\Delta \sin^2\theta-(r^2+a^2)^2}{\Sigma}\sin^2\theta d\varphi^2+\frac{\Sigma}{\Delta}dr^2+\Sigma d\theta^2,
\label{2kn01}
\end{align}
where $a$ is defined in order to adjust the dimensions by
\begin{align}
a\equiv \frac{L}{M},
\label{2kn02}
\end{align}
and for simplicity, the symbols are respectively defined by
\begin{align}
\Sigma&\equiv r^2+a^2\cos^2 \theta ,
\label{2kn03}\\
\Delta&\equiv r^2-2Mr+a^2+Q^2.
\label{2kn04}
\end{align}
In 4 dimensions, the Kerr-Newman black hole is the most general black hole.
We can thus obtain the Kerr metric by taking $Q=0$ in the metric (\ref{2kn01}), 
and we also obtain the Reissner-Nordstr\"om metric by taking $L=0$, namely, $a=0$.
Of course, by taking both $Q=0$ and $a=0$ in (\ref{2kn01}),
it can be checked that the Kerr-Newman metric (\ref{2kn01}) actually becomes the Schwarzschild metric (\ref{2sch01}).
We also note that the metric (\ref{2kn01}) is asymptotically flat, i.e., it approaches the Minkowski metric 
\begin{align}
ds^2=\eta_{\mu\nu}dx^\mu dx^\nu =-dt^2+dx^2+dy^2+dz^2.
\label{2min01}
\end{align}
which stands for the flat space-time in our universe.

\newpage

%%%%%%%%%%%%%%%%%       Tab. 2.1        %%%%%%%%%%%%%%%%%%%%%
\begin{center}
\vspace{0.6cm}
Tab. 2.1 \quad The types of black holes
\vspace{0.3cm}\\
  \begin{tabular}{|c|c|c|}
    \hline
    \parbox[c][\myheight][c]{0cm}{}
       & Non-rotating ($a=0$) & Rotating ($a\neq 0$)   \\
    \hline
    \parbox[c][\myheight][c]{0cm}{}
     Uncharged ($Q=0$) & Schwarzschild black hole   & Kerr black hole   \\
    \hline
    \parbox[c][\myheight][c]{0cm}{}
     Charged ($Q\neq 0$) & Reissner-Nordstr\"om black hole   & Kerr-Newman black hole   \\
    \hline
  \end{tabular}
\vspace{0.6cm}\\
   \end{center}
%%%%%%%%%%%%%%%%%       Tab. 2.1        %%%%%%%%%%%%%%%%%%%%%

Here we consider the event horizon which is the surface of the black hole.
For the sake of convenience, by using the metric $g_{\mu\nu}$, we describe (\ref{2kn01}) as
\begin{align}
ds^2&\equiv g_{\mu\nu} dx^\mu dx^\nu\\
&=g_{tt}dt^2+g_{rr}dr^2+g_{\theta\theta}d\theta^2 +2g_{t\varphi}dtd\varphi +g_{\varphi\varphi}d\varphi^2,
\label{21met01}
\end{align}
with
\begin{align}
(g_{\mu\nu})=
\left( 
\begin{array}{ccccc}
\displaystyle
-\frac{\Delta -a^2 \sin^2\theta}{\Sigma}&0&0& \displaystyle -\frac{a\sin^2\theta}{\Sigma}(r^2+a^2-\Delta)\\
0& \displaystyle \frac{\Sigma}{\Delta}&0&0\\
0&0&\displaystyle \Sigma&0\\
\displaystyle -\frac{a\sin^2\theta}{\Sigma}(r^2+a^2-\Delta)&0&0& \displaystyle -\frac{a^2\Delta \sin^2\theta-(r^2+a^2)^2}{\Sigma}\sin^2\theta
\end{array}
\right).
\label{21met02}
\end{align}
The black hole is the region that even the light cannot escape from its surface.
The event horizon of the Kerr-Newman black hole thus appears at the point where $g_{rr}=\infty$, i.e., 
\begin{align}
\qquad \qquad \qquad \Delta=0, \qquad ({\rm at~the~horizon}).
\label{21hor01}
\end{align}
From (\ref{21hor01}), the distance from the center of the black hole to the event horizon is given by
\begin{align}
r_\pm = M \pm \sqrt{M^2-a^2-Q^2},
\label{2hor01}
\end{align}
where we assumed $M^2>a^2+Q^2$ since the mass of the black hole is very generally large.
By using (\ref{2hor01}), $\Delta$ in (\ref{2kn04}) can be written as
\begin{align}
\Delta =(r-r_+)(r-r_-).
\label{21hor02}
\end{align}
There are two event horizons in the case of the Kerr-Newman black hole, i.e., $r_+$ and $r_-$, and they are respectively called the outer event horizon and the inner event horizon.
The inner event horizon $r_-$ exists inside the outer event horizon.
We do not care the existence of the inner event horizon since we cannot know information inside the outer horizon.
In what follows, we simply describe the outer event horizon as the horizon.

On the horizon $r=r_+$, the metric (\ref{2kn01}) becomes the intrinsic metric given by
\begin{align}
ds^2=-\frac{a^2\Delta_+ \sin^2\theta-(r_+^2+a^2)^2}{\Sigma_+}\sin^2\theta d\varphi^2+\Sigma_+ d\theta^2,
\end{align}
since both $t$ and $r$ are constant on the horizon, i.e., $dt=dr=0$.
Here we defined $\Delta_+\equiv \Delta(r_+)=0$ and $\Sigma_+\equiv \Sigma(r_+)$.
We thus find that the area of the black hole $A$ is given by
\begin{align}
A=\int \sqrt{g_{\theta\theta}(r_+) g_{\varphi\varphi}(r_+) } d\theta d\varphi=4\pi(r_+^2+a^2).
\label{2a01}
\end{align}
From $\Delta_+=r_+^2 -2Mr_+ + a^2+Q^2=0$, we can also write the black hole area as
\begin{align}
A=4\pi (2 M r_+ -Q^2).
\label{2a02}
\end{align}
By taking the total differentiation of (\ref{2a02}), we obtain
\begin{align}
dM = \frac{\kappa}{8\pi} dA + \Omega_{{\rm H}} dL + \Phi_{{\rm H}} dQ,
\label{2ec01}
\end{align}
where $\kappa$, $\Omega_{{\rm H}}$ and $\Phi_{{\rm H}}$ are respectively the surface gravity, the angular velocity and the electrical potential on the horizon,
which are defined by
\begin{align}
\kappa&\equiv \frac{4\pi (r_+ - M)}{A},
\label{2sg01}\\
\Omega_{{\rm H}}&\equiv \frac{4\pi a}{A},
\label{2av01}\\
\Phi_{{\rm H}}&\equiv \frac{4\pi r_+ Q}{A}.
\label{2ep01}
\end{align}
It is known that the relation (\ref{2ec01}) is the energy conservation law in black hole physics.
It is easy to find that the each term has the dimension of the energy in the natural system of units.

Before closing this section, we would like to state black holes in various dimensions.
It is known that a vacuum solution of the Einstein equation without the cosmological constant in three dimensions ((2+1)-dimensions), 
corresponds to a flat solution and no black hole solution exists.
However, when we consider the Einstein equation with a negative cosmological constant which behaves as attraction,
we can obtain black hole solutions in 3-dimensions.
It is called the BTZ black hole, which was found by Ba$\tilde{{\rm n}}$ados, Teitelboim and Zanelli \cite{btz01,btz02}.
It is known that it is the lowest dimensional black hole.

In dimensions higher than four, there are several black hole solutions 
because the restriction of topology with respect to the horizon is alleviated.
Therefore, a black hole cannot be uniquely decided even if the mass, the angular momentum and the charge are given.
This suggests that the uniqueness theorem is not satisfied.
For example, in five dimensions, there are the Myers-Perry black hole which has two independent rotation parameters \cite{mye01} and 
the black ring \cite{emp01}.
We thus have these two solutions with the same mass and the same angular momenta. 
It is thus known that the black hole uniqueness theorem does not hold in the higher dimensional theory. 

%%%%%%%%%%%%%%%%%%%%%%%%%%%%%%%%%%%%%%%%%%%%%%%%%%%%%%%%%%%%%%%%%%%%%%%%%%%%%%%%%%%%%%%%%%%%%%%%%
%%                                                                                             %%
%%                                     Section 2-2                                             %%
%%                                                                                             %%
%%%%%%%%%%%%%%%%%%%%%%%%%%%%%%%%%%%%%%%%%%%%%%%%%%%%%%%%%%%%%%%%%%%%%%%%%%%%%%%%%%%%%%%%%%%%%%%%%
\section{Penrose Diagram}

\quad~ Penrose diagram is very useful to understand the global structure of black hole space-time.
It was proposed by Penrose in 1964 \cite{pen02}.
In this section, we would like to recall the advantages of using the Penrose diagram.
Then we will show how to describe the Penrose diagram.

For simplicity, we consider the case of the Schwarzschild black hole.
The Schwarzschild metric is given by
\begin{align}
ds^2=-\left( 1- \frac{2M}{r} \right)dt^2 + \frac{1}{1- \frac{2M}{r}} dr^2 + r^2 d\Omega^2,
\label{2sch02}
\end{align}
where $d\Omega^2$ stands for a 2-dimensional unit sphere defined by
\begin{align}
d\Omega^2 \equiv d \theta^2 + \sin ^2 \theta d\varphi^2.
\label{2sch03}
\end{align}
It follows from the expression (\ref{2sch02}) that there are two singularities $r=0$ and $r=2M$
in the Schwarzschild metric. 
A singularity at $r=0$ is the curvature singularity
which cannot be removed while the other at $r=2M$ is a fictitious singularity arising
merely from an improper choice of coordinates. 
We therefore know that the singularity at $r=2M$ can be removed by using appropriate coordinates.

The Penrose diagram is drawn for the Schwarzschild metric as in Fig. 2.1.
The notations ${\mathcal I}^0$, ${\mathcal I}^\pm$ and ${\mathcal J}^{\pm}$ appearing in Fig. 2.1, respectively stand for the following regions
\begin{align}
&{\mathcal I}^0=
\left\{ 
\begin{array}{l}
t~;~{\rm finite}\\
r \to \infty~,
\end{array}\right. \qquad \quad
{\mathcal I}^\pm=\left\{ \begin{array}{l}
t \to \pm\infty\\
r~;~{\rm finite}~,
\end{array}\right.
\label{2region01}\\
&{\mathcal J}^-=\left\{ \begin{array}{l}
t \to -\infty\\
r \to +\infty~,
\end{array}\right.\qquad
{\mathcal J}^+=\left\{ \begin{array}{l}
t \to +\infty\\
r \to +\infty~,
\end{array}\right.
\label{2region02}
\end{align}
and two double lines of ${\mathcal R}$ stand for the curvature singularity of the Schwarzschild metric. 
The heavy lines ${\mathcal H}^+$ and ${\mathcal H}^-$ also stand for
\begin{align}
{\mathcal H}^+=\left\{ \begin{array}{l}
t \to +\infty\\
r = 2M~,
\end{array}\right.\qquad
{\mathcal H}^-=\left\{ \begin{array}{l}
t \to -\infty\\
r = 2M~,
\end{array}\right.
\label{2region03}
\end{align}
and ${\mathcal H}^+$ and ${\mathcal H}^-$ are respectively called the future event horizon and the past event horizon.

%\newpage

%%%%%%%%%%%%%     Fig.2.1       %%%%%%%%%%%%
\begin{center}
\vspace{0.6cm}
\input{pen01}\\
\vspace{0.6cm}
Fig. 2.1 \quad The Penrose diagram for the Schwarzschild solution.
\vspace{0.6cm}
\end{center}
%%%%%%%%%%%%%%%%%%%%%%%%%%%%%%%%%%%%%%%%%%

We can draw the Penrose diagram through some coordinate transformations (see, for example, \cite{tow01}).
As a first step of coordinate transformations, we use the tortoise coordinate defined by \cite{whe01, reg01}
\begin{align}
dr_* \equiv \frac{1}{1-\frac{2M}{r}} dr.
\label{2sch04}
\end{align}
The metric (\ref{2sch02}) is then written by
\begin{align}
ds^2=-\left( 1-\frac{2M}{r} \right)(dt-dr_*)(dt+dr_*) + r^2 d\Omega^2.
\label{2sch05}
\end{align}
By integrating (\ref{2sch04}) over $r$ from 0 to $r$, we obtain
\begin{align}
r_* = r + 2M \ln \left| \frac{r}{2M} -1 \right|.
\label{2sch06}
\end{align}

As the second step we use the Eddington-Finkelstein coordinates defined by \cite{edd01, fin01}
\begin{align}
\left\{
\begin{array}{lll}
\displaystyle v \equiv t+r_* = t + r + 2M \ln \left| \frac{r}{2M} -1 \right|,
\vspace{0.2cm}\\
\displaystyle u \equiv t-r_* = t - r - 2M \ln \left| \frac{r}{2M} -1 \right|,
\end{array}
\right.
\label{2sch07}
\end{align}
where $v$ is called the advanced time and $u$ is called the retarded time.
The metric (\ref{2sch05}) is then written as
\begin{align}
ds^2=-\left( 1-\frac{2M}{r} \right)dvdu + r^2 d\Omega^2.
\label{2sch08}
\end{align}

As the third step we use the Kruskal-Szekeres coordinates \cite{kru01, sze01}.
When $r>2M$, these coordinates are defined by
\begin{align}
\left\{
\begin{array}{lll}
V\equiv \displaystyle \exp\left[\frac{v}{4M} \right],
\vspace{0.2cm}\\
U\equiv \displaystyle -\exp\left[ -\frac{u}{4M}\right],
\end{array}
\right.
\label{2sch09}
\end{align}
and the metric (\ref{2sch08}) is written as
\begin{align}
ds^2=-\frac{32M^3}{r} \exp\left[ -\frac{r}{2M}\right] dV dU +r^2 d\Omega^2, \qquad {\rm when}\quad r>2M.
\label{2sch10}
\end{align}
When $r<2M$, these coordinates are defined by
\begin{align}
\left\{
\begin{array}{lll}
V\equiv \displaystyle \exp\left[\frac{v}{4M} \right],
\vspace{0.2cm}\\
U\equiv \displaystyle \exp\left[ -\frac{u}{4M}\right],
\end{array}
\right.
\label{2sch11}
\end{align}
and the metric (\ref{2sch08}) is then written as
\begin{align}
ds^2=\frac{32M^3}{r} \exp\left[ -\frac{r}{2M}\right] dV dU +r^2 d\Omega^2, \qquad {\rm when}\quad r<2M.
\label{2sch12}
\end{align}

As the fourth step we use the following coordinate transformations defined by
\begin{align}
\left\{
\begin{array}{ccc}
\displaystyle \tilde{V}=\tan^{-1} \left( \frac{V}{4M\sqrt{2M}} \right),
\vspace{0.2cm}\\
\displaystyle \tilde{U}=\tan^{-1} \left( \frac{U}{4M\sqrt{2M}} \right).
\end{array}
\right.
\label{2sch13}
\end{align}
We find that infinities appeared in $V$ or $U$ are converted to finite values such as $\displaystyle \frac{\pi}{2}$ or $-\displaystyle\frac{\pi}{2}$.

As the final step we use the following coordinate transformations defined by
\begin{align}
\left\{
\begin{array}{ccc}
\displaystyle \tilde{T}=\frac{1}{2} \left( \tilde{V} + \tilde{U} \right),
\vspace{0.2cm}\\
\displaystyle \tilde{R}=\frac{1}{2} \left( \tilde{V} - \tilde{U} \right).
\end{array}
\right.
\label{2sch14}
\end{align}
Penrose diagram is drawn by choosing the vertical axis as $\tilde{T}$ and the horizontal axis as $\tilde{R}$.

As an illustration, we draw ${\mathcal I}^+$ and ${\mathcal J}^+$.
First, ${\mathcal I}^+$ is expressed by
\begin{align}
{\mathcal I}^+=\left\{ 
\begin{array}{l}
t \to +\infty\\
r~;~{\rm finite}~.
\end{array}\right.
\label{2region04}
\end{align}
Since $r$ is finite, we need to consider two cases of $r>2M$ and $r<2M$.
When $r>2M$, by substituting (\ref{2region04}) into (\ref{2sch07}), $v$ and $u$ become 
\begin{align}
{\mathcal I}^+=\left\{ 
\begin{array}{l}
v\to +\infty\\
u\to +\infty~.
\end{array}
\right. 
\label{2region05}
\end{align}
While when $r<2M$, by substituting (\ref{2region04}) into (\ref{2sch11}), $v$ and $u$ agree with (\ref{2region05}).
By substituting (\ref{2region05}) into (\ref{2sch09}), $V$ and $U$ become
\begin{align}
{\mathcal I}^+=\left\{ 
\begin{array}{l}
V\to +\infty\\
U\to 0\quad~,
\end{array}
\right. 
\label{2region06}
\end{align}
and by substituting (\ref{2region06}) into (\ref{2sch13}), $\tilde{V}$ and $\tilde{U}$ become
\begin{align}
{\mathcal I}^+=\left\{ 
\begin{array}{l}
\displaystyle \tilde{V}\to +\frac{\pi}{2}\\
\displaystyle \tilde{U}\to 0\quad.
\end{array}
\right. 
\label{2region07}
\end{align}
By substituting (\ref{2region07}) into (\ref{2sch14}), $\tilde{T}$ and $\tilde{R}$ become
\begin{align}
{\mathcal I}^+=\left\{ 
\begin{array}{l}
\displaystyle \tilde{T}\to +\frac{\pi}{4}
\vspace{0.2cm}\\
\displaystyle \tilde{R}\to +\frac{\pi}{4}.
\end{array}
\right. 
\label{2region08}
\end{align}
We thus find that the region ${\mathcal I}^+$ as in (\ref{2region04}) is represented by $\displaystyle (\tilde{R},\tilde{T})=\left( \frac{\pi}{4},\frac{\pi}{4} \right)$
in the Penrose diagram, when $r$ takes finite values except $r=2M$ (Fig. 2.2).

%%%%%%%%%%%%%     Fig.2.2       %%%%%%%%%%%%
\begin{center}
\vspace{0.6cm}
\input{pen02}\\
\vspace{0.6cm}
Fig. 2.2 \quad The region of ${\mathcal I}^+$ in the Penrose diagram.
\vspace{0.6cm}
\end{center}
%%%%%%%%%%%%%%%%%%%%%%%%%%%%%%%%%%%%%%%%%%

Next, we similarly draw ${\mathcal J}^+$.
The region ${\mathcal J}^+$ is expressed by
\begin{align}
{\mathcal J}^+=\left\{ 
\begin{array}{l}
t \to +\infty\\
r \to +\infty\\
u~;~{\rm finite}~.
\end{array}\right.
\label{2region09}
\end{align}
Since $r$ is at infinity, we have only to consider the case of $r>2M$.
By substituting (\ref{2region09}) into (\ref{2sch07}), $v$ and $u$ become 
\begin{align}
{\mathcal J}^+=\left\{ 
\begin{array}{l}
v\to +\infty\\
u~;~{\rm finite}~.
\end{array}
\right. 
\label{2region10}
\end{align}
By substituting (\ref{2region10}) into (\ref{2sch09}), $V$ and $U$ become
\begin{align}
{\mathcal J}^+=\left\{ 
\begin{array}{l}
V\to +\infty\\
U~;~{\rm finite}~,
\end{array}
\right. 
\label{2region11}
\end{align}
and by substituting (\ref{2region11}) into (\ref{2sch13}), $\tilde{V}$ and $\tilde{U}$ become
\begin{align}
{\mathcal J}^+=\left\{ 
\begin{array}{l}
\displaystyle \tilde{V}\to +\frac{\pi}{2}\\
\displaystyle \tilde{U}~;~{\rm finite}~.
\end{array}
\right. 
\label{2region12}
\end{align}
Finally, by substituting (\ref{2region12}) into (\ref{2sch14}), $\tilde{T}$ and $\tilde{R}$ become
\begin{align}
{\mathcal J}^+=\left\{ 
\begin{array}{l}
\displaystyle \tilde{T}=\frac{1}{2} \left( \frac{\pi}{2} + \tilde{U} \right)
\vspace{0.2cm}\\
\displaystyle \tilde{R}=\frac{1}{2} \left( \frac{\pi}{2} - \tilde{U} \right).
\end{array}
\right. 
\label{2region13}
\end{align}
From these two relations (\ref{2region13}), 
we thus find that the region ${\mathcal J}^+$ as in (\ref{2region09}) is represented by the segment of a line
\begin{align}
\tilde{T}=\frac{\pi}{2} -\tilde{R}
\end{align}
in the Penrose diagram (Fig. 2.3).

%%%%%%%%%%%%%     Fig.2.3       %%%%%%%%%%
\begin{center}
\vspace{0.6cm}
\input{pen03}\\
\vspace{0.6cm}
Fig. 2.3 \quad The region of ${\mathcal J}^+$ in the Penrose diagram.
\vspace{0.6cm}
\end{center}
%%%%%%%%%%%%%%%%%%%%%%%%%%%%%%%%%%%%%%%%%%

We can similarly draw other points and segments (Tab. 2.2).
In Tab. 2.2, when a variable is finite and is not uniquely fixed, the name of the variable is retained.
The diagram drawn by using Tab. 2.2, is expressed as in Fig. 2.4.
The regions ${\mathcal R}^+$ and ${\mathcal R}^-$ respectively stand for the following regions
\begin{align}
{\mathcal R}^+=\left\{ \begin{array}{l}
t \to +\infty\\
r = 0\quad ~,
\end{array}\right.\qquad
{\mathcal R}^-=\left\{ \begin{array}{l}
t \to -\infty\\
r = 0\quad ~,
\end{array}\right.
\label{2region14}
\end{align}
and the double line ${\mathcal R}$ combines between ${\mathcal R}^+$ and ${\mathcal R}^-$.
The region ${\mathcal R}$ stands for $r=0$ with finite $t$,
but we cannot uniquely decide the point in the region ${\mathcal R}$.
This means that we do not know how to draw an exact line of the region ${\mathcal R}$.
We therefore drew a double line as the line ${\mathcal R}$.
Also by comparison with Fig. 2.1, there are some missing parts in Fig. 2.4.
We can draw them by defining the other universe where time proceeds reversely by comparison with our universe.
We however skip them because they are not important in the body of the present thesis.

\begin{center}
\vspace{0.6cm}
Tab. 2.2 \quad Coordinate values in each region
\vspace{0.3cm}\\
\begin{tabular}{|c|c|c|c|c|c|}\hline
\parbox[c][\myheight][c]{0cm}{}
Region&$\displaystyle \left(t,r\right)$&$(v, u)$&$(V,U)$&$\left(\tilde{V},\tilde{U}\right)$&$\left(\tilde{T},\tilde{R}\right)$\\ \hline
\parbox[c][\myheight][c]{0cm}{}
$\displaystyle{\mathcal I}^+$&$\left(+\infty,r\right)$&$\left(+\infty,+\infty\right)$&$\left(+\infty,0\right)$
&$\displaystyle\left(+\frac{\pi}{2},0\right)$&$\displaystyle\left(+\frac{\pi}{4},+\frac{\pi}{4}\right)$\\ \hline
\parbox[c][\myheight][c]{0cm}{}
$\displaystyle{\mathcal I}^-$&$\left(-\infty,r\right)$&$\left(-\infty,-\infty\right)$&$\left(0,-\infty\right)$
&$\displaystyle\left(0,-\frac{\pi}{2}\right)$&$\displaystyle\left(-\frac{\pi}{4},+\frac{\pi}{4}\right)$\\ \hline
\parbox[c][\myheight][c]{0cm}{}
${\mathcal I}^0$&$\left(t,+\infty\right)$&$\left(+\infty,-\infty\right)$&$\left(+\infty,-\infty\right)$
&$\displaystyle\left(+\frac{\pi}{2},-\frac{\pi}{2}\right)$&$\displaystyle\left(0,+\frac{\pi}{2}\right)$\\ \hline
\parbox[c][\myheight][c]{0cm}{}
${\mathcal J}^+$&$\left(+\infty,+\infty\right)$&$\left(+\infty,u\right)$&$\left(+\infty,U\right)$
&$\displaystyle\left(\frac{\pi}{2},\tilde{U}\right)$&$\displaystyle \tilde{T}=\frac{\pi}{2}-\tilde{R}$\\ \hline
\parbox[c][\myheight][c]{0cm}{}
${\mathcal J}^-$&$\left(-\infty,+\infty\right)$&$\left(v,-\infty\right)$&$\left(V,-\infty\right)$
&$\displaystyle\left(\tilde{V},-\frac{\pi}{2}\right)$&$\displaystyle \tilde{T}=\tilde{R}-\frac{\pi}{2}$\\ \hline
\parbox[c][\myheight][c]{0cm}{}
${\mathcal H}^+$&$\left(+\infty,2M\right)$&$\left(v,+\infty\right)$&$\left(V,0\right)$
&$\left(\tilde{V},0\right)$&$\tilde{T}=\tilde{R}$\\ \hline
\parbox[c][\myheight][c]{0cm}{}
${\mathcal H}^-$&$\left(-\infty,2M\right)$&$\left(-\infty,u\right)$&$\left(0,U\right)$
&$\left(0,\tilde{U}\right)$&$\tilde{T}=-\tilde{R}$\\ \hline
\parbox[c][\myheight][c]{0cm}{}
${\mathcal R}^+$&$\left(+\infty,0\right)$&$\left(+\infty,+\infty\right)$&$\left(+\infty,0\right)$
&$\displaystyle\left(+\frac{\pi}{2},0\right)$&$\displaystyle\left(+\frac{\pi}{4},+\frac{\pi}{4}\right)$\\ \hline
\parbox[c][\myheight][c]{0cm}{}
${\mathcal R}^-$&$\left(-\infty,0\right)$&$\left(-\infty,-\infty\right)$&$\left(0,\infty\right)$
&$\displaystyle\left(0,+\frac{\pi}{2}\right)$&$\displaystyle\left(+\frac{\pi}{4},-\frac{\pi}{4}\right)$\\ \hline
\end{tabular}
\vspace{0.6cm}
\\
\end{center}
\newpage
%%%%%%%%%%%%%     Fig.2.4       %%%%%%%%%%
\begin{center}
\vspace{0.6cm}
\input{pen04}\\
\vspace{0.6cm}
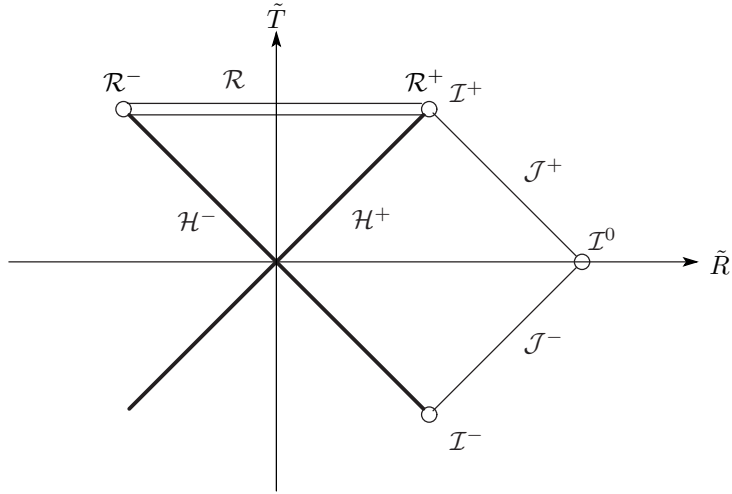
Fig. 2.4 \quad The Penrose diagram corresponding to Tab. 2.1.
\vspace{0.6cm}
\end{center}
%%%%%%%%%%%%%%%%%%%%%%%%%%%%%%%%%%%%%%%%%%

As above, the Penrose diagram can represent infinite time or radial coordinates as points or lines.
In this diagram null geodesics is also represented as lines of $\pm 45^\circ$ to the vertical.
Each point of the diagram represents a 2-dimensional sphere of area $4\pi r^2$.
Namely, angular coordinates $\theta$ and $\varphi$ as in (\ref{2sch02}) are attached to each point of the coordinate.
For this reason, the Penrose diagram is also called the conformal diagram.

The Penrose diagram is divided into four regions by the two diagonal lines ${\mathcal H}^+$ and ${\mathcal H}^-$ (Fig. 2.5).
The region I represents our universe.
The region I\hspace{-.1em}I represents a black hole.
The region I\hspace{-.1em}I\hspace{-.1em}I represents the other universe that time reversely proceeds by comparison with our universe.
The region I\hspace{-.1em}V represents a white hole which is the time reversal of a black hole and ejects matter from the horizon.
For example, we find that null geodesics in the region I 
can arrive at ${\mathcal J}^+$ or the black hole through the horizon ${\mathcal H}^+$
but null geodesics in the region I\hspace{-.1em}I (inside the black hole) 
cannot arrive at our universe through the horizon ${\mathcal H}^+$.

\newpage

%%%%%%%%%%%%%     Fig.2.5       %%%%%%%%%%
\begin{center}
\vspace{0.6cm}
\input{pen05}\\
\vspace{0.6cm}
Fig. 2.5 \quad The Penrose diagram for the Schwarzschild solution.
\vspace{0.6cm}
\end{center}
%%%%%%%%%%%%%%%%%%%%%%%%%%%%%%%%%%%%%%%%%%

Now we consider that a black hole is formed by the gravitational collapse of a star with a heavy mass.
This was comprehensibly discussed by Hawking in the literature \cite{haw01}.
Hence we would faithfully like to present the argument by following Hawking's exposition.
For simplicity, we assume that the gravitational collapse is spherically symmetric.
Such a object starts to collapse at the point ${\mathcal I}^-$.
Since the collapsing object has a mass, the passing is later than light (Fig. 2.6).
In Fig. 2.6, the time-like geodesic with an angle, which is smaller than $45^\circ$, represents the surface of the collapsing object
and the shaded region represents inside the collapsing object.

%%%%%%%%%%%%%     Fig.2.6       %%%%%%%%%%
\begin{center}
\vspace{0.6cm}
\input{pen06}\\
\vspace{0.6cm}
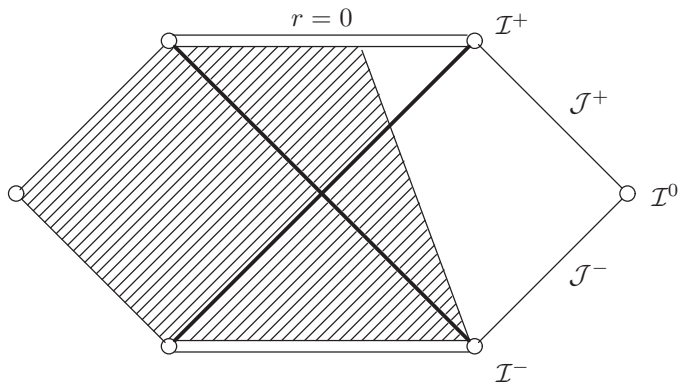
Fig. 2.6 \quad The development of the collapsing object in the Penrose diagram.
\vspace{0.6cm}
\end{center}
%%%%%%%%%%%%%%%%%%%%%%%%%%%%%%%%%%%%%%%%%%

In the case of exactly spherical collapse, the metric is exactly the Schwarzschild metric 
everywhere outside the surface of the collapsing object.
O the other hand, inside the object the metric is completely different. 
Thus, the past event horizon, the past curvature singularity and the other asymptotically flat region do not exist 
and are replaced by a time-like curve representing the origin of polar coordinates.
The appropriate Penrose diagram is shown in Fig. 2.7.
We represented the origin as the vertical dotted line because the metric inside the object might be nonsingular at the origin.

%%%%%%%%%%%%%     Fig.2.7       %%%%%%%%%%
\begin{center}
\vspace{0.6cm}
\input{pen07}\\
\vspace{0.6cm}
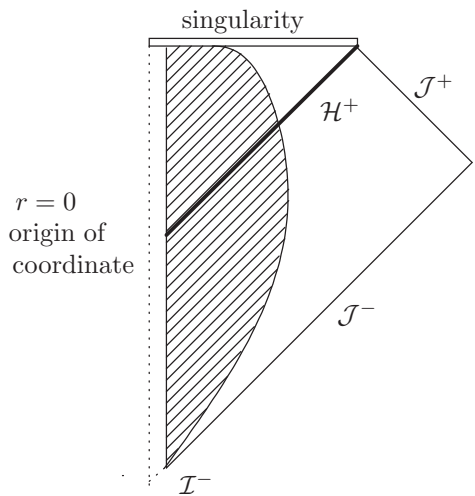
Fig. 2.7 \quad The Penrose diagram of a spherically symmetric \\  \quad collapsing body producing a black hole.
\vspace{0.6cm}
\end{center}
%%%%%%%%%%%%%%%%%%%%%%%%%%%%%%%%%%%%%%%%%%

%%%%%%%%%%%%%%%%%%%%%%%%%%%%%%%%%%%%%%%%%%%%%%%%%%%%%%%%%%%%%%%%%%%%%%%%%%%%%%%%%%%%%%%%%%%%%%%%%
%%                                                                                             %%
%%                                     Section 2-3                                             %%
%%                                                                                             %%
%%%%%%%%%%%%%%%%%%%%%%%%%%%%%%%%%%%%%%%%%%%%%%%%%%%%%%%%%%%%%%%%%%%%%%%%%%%%%%%%%%%%%%%%%%%%%%%%%
\section{Energy Extraction from Rotating Black Holes}

\quad~ By definition, a black hole is a ``region of no escape."
It might thus seem that energy cannot be extracted from a black hole.
However, the mechanism of energy extraction from a rotating black hole was proposed by Penrose \cite{pen03}.
The process is called the Penrose process and the radiance is called the black hole superradiance.
This can be explained in the classical theory.

%%%%%%%%%%%%%%%%%%%%%%%%%%%%%%%%%%%%%%%%%%%%%%%%%%%%%%%%%%%%%%%%%%%%%%%%%%%%%%%%%%%%%%%%%%%%%%%%%
%%                                                                                             %%
%%                                     Section 2-3-1                                           %%
%%                                                                                             %%
%%%%%%%%%%%%%%%%%%%%%%%%%%%%%%%%%%%%%%%%%%%%%%%%%%%%%%%%%%%%%%%%%%%%%%%%%%%%%%%%%%%%%%%%%%%%%%%%%

\subsection{Penrose process}

\quad~ In this subsection, we would like to show how to extract energy from a rotating black hole.
To begin with, we shall present an intuitive explanation.
The metric of a rotating black hole is given by the Kerr metric
\begin{align}
ds^2=&-\frac{\Delta -a^2 \sin^2\theta}{\Sigma}dt^2-\frac{2a\sin^2\theta}{\Sigma}(r^2+a^2-\Delta)dtd\varphi \notag\\
&-\frac{a^2\Delta \sin^2\theta-(r^2+a^2)^2}{\Sigma}\sin^2\theta d\varphi^2+\frac{\Sigma}{\Delta}dr^2+\Sigma d\theta^2.
\end{align}
This form agrees with the Kerr-Newman metric (\ref{2kn01}) 
but the contents of both $\Delta$ and $\Sigma$ are different because of $Q=0$.
The event horizon in the Kerr coordinate system is defined by $g_{rr}=\infty$ except at the curvature singularity.
The surface defined by $g_{tt}=0$ is called the {\it ergosphere}.
The region enclosed by the ergosphere and the event horizon is called the ergoregion (Fig. 2.8).
In the case of the Schwarzschild metric, the ergosphere agrees with the event horizon.

%%%%%%%%%%%%%     Fig.2.8       %%%%%%%%%%
\begin{center}
\vspace{0.6cm}
\input{ergo01}\\
\vspace{0.6cm}
Fig. 2.8 \quad Event horizon and ergosphere
\vspace{0.6cm}
\end{center}
%%%%%%%%%%%%%%%%%%%%%%%%%%%%%%%%%%%%%%%%%%

We consider a process where a particle breaks up into two fragments in the ergoregion.
The energy of the original particle at infinity is represented as $E_0$.
By defining the four dimensional momentum of the particle as $p_0^\mu$, the energy is given by
\begin{align}
E_0=-p^\mu_0 \xi_\mu,
\label{23ene01}
\end{align}
where $\xi^\mu$ is the Killing field defined by
\begin{align}
\xi^\mu \equiv \left( \frac{\partial}{\partial t} \right)^\mu, 
\label{23tki01}
\end{align}
which becomes a time translation asymptotically at infinity and is space-like in the ergoregion.
When the particle enters the ergoregion, we arrange to have it break up into two fragments (Fig. 2.9).
By the local momentum conservation law, we have
\begin{align}
p_0^\mu = p_1^\mu + p_2^\mu,
\label{23pcl01}
\end{align}
where $p_1^\mu$ and $p_2^\mu$ are the four dimensional momenta of the two fragments.
By contracting the equation (\ref{23pcl01}) with $\xi_\mu$, we obtain the local energy conservation law
\begin{align}
E_0=E_1+E_2.
\label{23ecl01}
\end{align}
The energy need not be positive in the ergoregion since $\xi^\mu$ is space-like there.
We can arrange the breakup so that one of the fragments has negative total energy,
\begin{align}
E_1 <0.
\label{23ene02}
\end{align}
The fragment with the negative energy falls into the black hole through the event horizon,
while the other can escape to infinity since it does not pass through the event horizon.
Therefore, we can obtain 
\begin{align}
E_2>E_0.
\label{23ene03}
\end{align}
This means that energy can be classically extracted from a black hole.
The above process is called the Penrose process.

%%%%%%%%%%%%%     Fig.2.9       %%%%%%%%%%
\begin{center}
\vspace{0.6cm}
\input{ergo02}\\
\vspace{0.6cm}
Fig. 2.9 \quad Energy extraction from black hole.
\vspace{0.6cm}
\end{center}
%%%%%%%%%%%%%%%%%%%%%%%%%%%%%%%%%%%%%%%%%%

Of course, all energy cannot be extracted from the black hole by the Penrose process.
The negative energy particle also carry a negative angular momentum, i.e., the angular momentum opposite to that of the black hole.
As a result, the black hole gradually decreases its angular momentum.
When the black hole loses the total angular momentum, it becomes a Schwarzschild black hole.
Since the ergosphere no longer exist in the case of a Schwarzschild black hole, no further energy extraction can occur.

To see the limit on energy extraction, we use the Killing field $\chi^\mu$ defined by
\begin{align}
\chi^\mu\equiv \xi^\mu + \Omega_{{\rm H}} \psi^\mu,
\label{23ki01}
\end{align}
where $\Omega_{{\rm H}}$ is the angular velocity defined by (\ref{2av01}) and $\psi^\mu$ is the axial Killing field defined by
\begin{align}
\psi^\mu\equiv \left( \frac{\partial}{\partial \varphi}\right)^\mu.
\label{23aki01}
\end{align}
The Killing field is tangent to the null geodesic generators of the horizon and is future directed null on the horizon.
Since the Killing field (\ref{23ki01}) is future directed null on the horizon and $p^\mu$ is future-directed timelike or null, we have
\begin{align}
-p^\mu\chi_\mu\geq 0.
\label{23pen01}
\end{align}
By substituting (\ref{23ki01}) into (\ref{23pen01}), we obtain
\begin{align}
-p^\mu(\xi_\mu+\Omega_{{\rm H}} \psi_\mu)
=\omega-m \Omega_{{\rm H}}  \geq 0,
\label{23pen02}
\end{align}
where $\omega$ is the energy of the fragment which enters the black hole and $m=p^\mu \psi_\mu$ is an angular momentum of it.
In a Kerr black hole background, the system is stationary and has the axial symmetry.
We therefore find that both the energy and the angular momentum are conserved and these quantities are
%respectively carried by the corresponding Killing fields (\ref{23tki01}) and (\ref{23aki01}).
identified at asymptotic infinity (the Minkowski space).
The relation (\ref{23pen02}) is also written as
\begin{align}
m \leq \frac{\omega}{\Omega_{{\rm H}}}.
\end{align}
If $\omega$ is negative, $m$ is also negative.
Thus the angular momentum of the black hole is reduced.
The mass and the angular momentum of the black hole are respectively $M+\delta M$ and $L+\delta L$ 
where $\delta M =\omega$ and $\delta L= m$.
Thus we obtain
\begin{align}
\delta L \leq \frac{\delta M}{\Omega_{{\rm H}}}= \frac{2M\left(M^2+\sqrt{M^4-L^2} \right)}{L}\delta M,
\end{align}
where we used the formula for $\Omega_{{\rm H}}$.
This is equivalent to
\begin{align}
\delta\left( \frac{1}{2} \left[ M^2 + \sqrt{M^4-J^2} \right]\right)\geq 0.
\label{23mir01}
\end{align}
Christodoulou defined the irreducible mass $M_{{\rm ir}}$ by \cite{chr01}
\begin{align}
M_{{\rm ir}}^2&\equiv \frac{1}{2} \left[ M^2 + \sqrt{M^4-J^2} \right]\\
&=\frac{1}{2} \left[ M^2+M\sqrt{M^2-a^2} \right].
\label{23mir02}
\end{align}
The irreducible mass can also be written in terms of the black hole area as in (\ref{2a01}), i.e.,
\begin{align}
M_{{\rm ir}}^2=\frac{A}{16\pi}.
\label{23mir03}
\end{align}
By substituting (\ref{23mir03}) into (\ref{23mir01}),
the energy extraction by Penrose process is thus limited by the requirement that
\begin{align}
\delta A \geq 0.
\end{align}
This result agrees with Hawking's black hole area theorem that the black hole area never decreases \cite{haw03}. 

%%%%%%%%%%%%%%%%%%%%%%%%%%%%%%%%%%%%%%%%%%%%%%%%%%%%%%%%%%%%%%%%%%%%%%%%%%%%%%%%%%%%%%%%%%%%%%%%%
%%                                                                                             %%
%%                                     Subsection 2-3-2                                        %%
%%                                                                                             %%
%%%%%%%%%%%%%%%%%%%%%%%%%%%%%%%%%%%%%%%%%%%%%%%%%%%%%%%%%%%%%%%%%%%%%%%%%%%%%%%%%%%%%%%%%%%%%%%%%

\subsection{Superradiance}

\quad~ There is a wave analog of the Penrose process \cite{mis01,pre01}. It is called superradiant scattering or superradiance.
It is known that scalar fields display superradiance.
To find this, we consider the energy current defined by
\begin{align}
J_{\mu}\equiv -T_{\mu\nu}\xi^\nu,
\label{23j01}
\end{align}
where $T_{\mu\nu}$ is an energy-momentum tensor which is a symmetric tensor in this case.
We take the covariant derivative $\nabla^\mu$ of (\ref{23j01})
\begin{align}
\nabla^\mu J_\mu =-(\nabla^\mu T_{\mu\nu})\xi^\nu -T_{\mu\nu} (\nabla^\mu \xi^\nu).
\label{23j02}
\end{align}
By the general coordinate invariance, the energy-momentum tensor satisfies
\begin{align}
\nabla^{\mu}T_{\mu\nu}=0.
\end{align}
We thus find that the relation (\ref{23j02}) becomes
\begin{align}
\nabla^\mu J_\mu&=-T_{\mu\nu} (\nabla^\mu \xi^\nu)\\
&=-\frac{1}{2} T_{\mu\nu} (\nabla^\mu \xi^\nu +\nabla^\nu \xi^\mu)\\
&=0,
\label{23j03}
\end{align}
where we used the facts that the energy-momentum tensor is a symmetric tensor and Killing fields satisfy the Killing equation
\begin{align}
\nabla^\mu \xi^\nu +\nabla^\nu \xi^\mu=0.
\end{align}

If we integrate (\ref{23j03}) over the region ${\cal K}$ of space-time whose boundary consists of two spacelike hypersurfaces
$\Sigma_1$ at $t$ and $\Sigma_2$ at $t+\delta t$ (the constant time slice $\Sigma_2$ is a time translate of $\Sigma_1$ by $\delta t$)
and two timelike hypersurfaces ${\cal H}$ (the event horizon at $r=r_+$) 
and ${\mathscr S}{(\infty)}$ (large sphere at spatial infinity $r\to \infty$),
we can know the presence or absence of the superradiance.
The intuitive figure is shown in Fig. 2.10.
Strictly speaking, this figure is not precise.
The precise figure is shown in Fig. 2.11 by using Penrose diagram.

%%%%%%%%%%%%%     Fig.2.10       %%%%%%%%%%
\begin{center}
\vspace{0.6cm}
\input{gauss01}\\
\vspace{0.6cm}
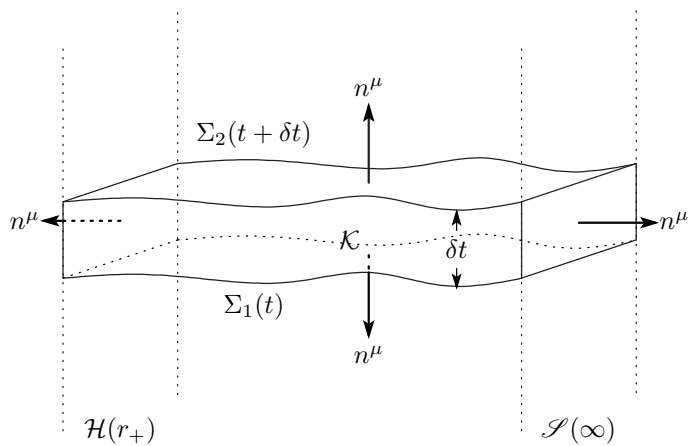
Fig. 2.10 \quad Intuitive figure with respect to Gauss's theorem
\vspace{0.6cm}
\end{center}
%%%%%%%%%%%%%%%%%%%%%%%%%%%%%%%%%%%%%%%%%%

%%%%%%%%%%%%%     Fig.2.11       %%%%%%%%%%
\begin{center}
%\vspace{0.6cm}
\input{gauss02}\\
%\vspace{0.6cm}
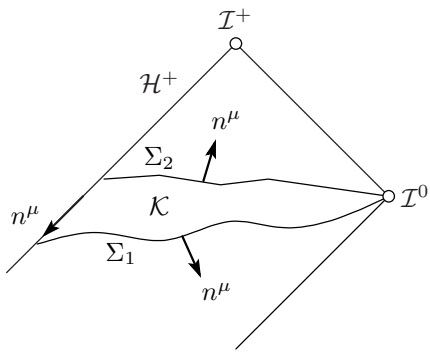
Fig. 2.11 \quad Precise figure using the Penrose diagram
%\vspace{0.6cm}
\end{center}
%%%%%%%%%%%%%%%%%%%%%%%%%%%%%%%%%%%%%%%%%%

By using Gauss's theorem, we obtain
\begin{align}
0&=\int_{{\cal K}}\sqrt{-g} d^4x (\nabla_\mu J^\mu)\\
&=\int_{\partial {\cal K}}d\Sigma_\mu J^\mu\\
&=\int_{\Sigma_1 (t)} n_\mu J^\mu d\Sigma+ \int_{\Sigma_2 (t+\delta t)} n_\mu J^\mu d\Sigma 
+\int_{{\cal H}(r_+)} n_\mu J^\mu d\Sigma + \int_{ {\mathscr S}{(\infty)} } n_\mu J^\mu d\Sigma,
\label{23j06}
\end{align}
where $\partial {\cal K}$ is the boundary of the region ${\cal K}$, $d \Sigma_\mu \equiv n_\mu d\Sigma$ is a 3-dimensional suitable area element and the unit vector $n^\mu$ is outwardly normal to the region ${\cal K}$.
In the last line, the first two terms cancel with each other because the system has the time translation symmetry 
and the two directions of $n_{\mu}$ are opposite to each other.
The third term represents the flow of the net energy current flux into the black hole.
The last term represents the net energy current flux flow out of ${\cal K}$ to infinity.
Thus the relation (\ref{23j06}) becomes
\begin{align}
\int_{ {\mathscr S}{(\infty)} } n_\mu J^\mu d\Sigma =- \int_{{\cal H}(r_+)} n_\mu J^\mu d\Sigma.
\label{23j07}
\end{align}
If the quantity on the right-hand side in (\ref{23j07}) is positive (negative),
this means that the outgoing energy current flux is larger (smaller) than the incident one
and the superradiance is present (absent).

We would like to evaluate the quantity on the right-hand side in (\ref{23j07}).
The vector $n^\mu$ is normal to the event horizon.
The normal vector $n^\mu$ can be written in terms of the Killing field $\chi^\mu$ as
\begin{align}
n^\mu=-\chi^\mu,
\label{23n01}
\end{align}
where $\chi^\mu$ is the Killing field defined by (\ref{23ki01}).
As already stated, one may recall that the Killing field is tangent to the horizon.
One might therefore wonder the appearance of the relation (\ref{23n01}).
This result is known by the fact that the vector which is normal to the horizon is tangent to itself on the horizon (the null hypersurface).
We show a proof in Appendix A.
The sign of (\ref{23n01}) is decided by the direction of $n^\mu$ toward the horizon which is opposite to the future directed Killing field.
We thus obtain 
\begin{align}
\int_{{\cal H}(r_+)} n_\mu J^\mu d\Sigma&=- \int_{{\cal H}(r_+)} \chi_\mu J^\mu d\Sigma\\
&=-\int_{{\cal H}(r_+)} \chi_\mu \left( - T^\mu_{~\nu} \xi^\nu\right) d\Sigma\\
&=\int_{{\cal H}(r_+)} \chi^\mu T_{\mu\nu} \xi^\nu  d\Sigma,
\label{23j08}
\end{align}
where we used the definition (\ref{23j01}).

Here we would like to find a concrete form of the energy-momentum tensor $T_{\mu\nu}$.
For sake of simplicity, we consider the action for a massless scalar field without interactions.
In curved space-time, the action is given by
\begin{align}
S&\equiv \int \sqrt{-g}d^4 x \left[ {\cal L} \right]
\label{23act01}\\
&=\int \sqrt{-g}d^4 x \left[ \frac{1}{2} \nabla_\mu \phi \nabla^\mu \phi \right],
\label{23act02}
\end{align}
where ${\cal L}$ is the Lagrangian density.
%This action is invariant under the general coordinate transformation and the corresponding N\"other current exits.
According to field theory, the energy-momentum tensor $T_{\mu\nu}$ is then defined by
\begin{align}
T_{\mu\nu}&\equiv \frac{\partial {\cal L} }{\partial \left( \nabla^\mu \phi \right)}\nabla_\nu \phi  -g_{\mu\nu} {\cal L}
\label{23emt01}\\
&=\frac{1}{2} (\nabla_\mu \phi) (\nabla_\nu \phi) - \frac{1}{2}g_{\mu\nu}(\nabla_\alpha \phi) (\nabla^\alpha \phi),
\label{23emt02}
\end{align}
where we used the Lagrangian density in (\ref{23act02}).
By substituting (\ref{23emt02}) into (\ref{23j08}), we obtain
\begin{align}
\int_{{\cal H}(r_+)} n_\mu J^\mu d\Sigma
&=\int_{{\cal H}(r_+)}d\Sigma \left[ \frac{1}{2} \left( \chi^\mu \nabla_\mu \phi \right) \left( \xi^\mu \nabla_\mu \phi \right)
-\frac{1}{2} \chi^\mu \xi_\mu \left( \nabla_\alpha \phi \right)\left( \nabla^\alpha \phi \right)
\right]\\
&=\int_{{\cal H}(r_+)}d\Sigma  \left[ \frac{1}{2}\left( \chi^\mu \nabla_\mu \phi \right) \left( \xi^\mu \nabla_\mu \phi \right) \right], 
\label{23j09}
\end{align}
where we used the fact that $\chi^\mu \xi_\mu=0$ on the horizon.
Since we consider the case of a Kerr black hole which is stationary and axisymmetric, 
the scalar field can be written asymptotically as
\begin{align}
\phi (x)=\phi_0(r,\theta) \cos(\omega t - m \varphi).
\end{align}
Also we asymptotically have
\begin{align}
\chi^\mu \nabla_\mu &=\frac{\partial}{\partial t}+\Omega_{{\rm H}} \frac{\partial}{\partial \varphi},\\
\xi^\mu \nabla_\mu &=\frac{\partial}{\partial t}.
\end{align}
We then find that the integrand of (\ref{23j09}) asymptotically becomes
\begin{align}
\frac{1}{2} \left( \chi^\mu \nabla_\mu \phi \right) \left( \xi^\mu \nabla_\mu \phi \right)
=\frac{1}{2} \omega (\omega -m\Omega_{{\rm H}}) {\tilde{\phi}}^2(x)
\end{align}
where we defined $\tilde{\phi} (x)\equiv \phi_0 (r,\theta) \sin (\omega t -m \varphi)$.
This quantity carried by the Killing field is invariant on the horizon.
The relation (\ref{23j09}) is thus given by
\begin{align}
\int_{{\cal H}(r_+)} n_\mu J^\mu d\Sigma=\frac{1}{2} \omega (\omega -m\Omega_{{\rm H}}) \int_{{\cal H}(r_+)}d\Sigma \tilde{\phi}^2(x).
\label{23j10}
\end{align}
We note that $d\Sigma =dA dv$ on the horizon where $A$ is the surface area of the horizon and the retarded time $v$ is an affine parameter on the horizon.
The relation (\ref{23j10}) generally diverges because of the integration with respect to $v$.
We hence evaluate the energy current flux per unit time. 
The time averaged flux becomes
\begin{align}
\int_{ {\mathscr S}{(\infty)} } n_\mu J^\mu dA &=-\int_{{\cal H}(r_+)} n_\mu J^\mu dA\\
&=-\frac{1}{2} \omega (\omega -m\Omega_{{\rm H}}) \left| \tilde{\phi}_0 \right|^2.
\label{23j11}
\end{align}
where we defined $\displaystyle \left| \tilde{\phi}_0 \right|^2\equiv \int_{{\cal H}(r_+)}dA \tilde{\phi}^2(x)$.
The right-hand side of (\ref{23j11}) is positive for $\omega$ in the range
\begin{align}
0< \omega <m\Omega_{{\rm H}}.
\end{align}
Therefore we find that the outgoing energy current flux is larger than the incident one and the superradiance is present for the scalar field.
The above discussion can be similarly performed for fermion fields.
However, it is known that the right-hand side of (\ref{23j07}) alway becomes zero and 
the superradiance is hence absent in the fermionic case \cite{unr01,guv01}.

%%%%%%%%%%%%%%%%%%%%%%%%%%%%%%%%%%%%%%%%%%%%%%%%%%%%%%%%%%%%%%%%%%%%%%%%%%%%%%%%%%%%%%%%%%%%%%%%%
%%                                                                                             %%
%%                                     Section 2-4                                             %%
%%                                                                                             %%
%%%%%%%%%%%%%%%%%%%%%%%%%%%%%%%%%%%%%%%%%%%%%%%%%%%%%%%%%%%%%%%%%%%%%%%%%%%%%%%%%%%%%%%%%%%%%%%%%
\section{Dimensional Reduction near the Horizon}

\quad~ As stated in Section 2.1, the black hole uniqueness theorem is valid only in four dimensions
and the Kerr-Newman solution is the most general solution in the 4-dimensional theory.
The space-time outside the Kerr-Newman black hole is represented by the Kerr-Newman metric 
and its geometry becomes spherically asymmetric because of its own rotation.
It is known that 
the 4-dimensional Kerr-Newman metric effectively becomes
a 2-dimensional spherically symmetric metric by using the technique of the dimensional reduction near the horizon.

The essential
idea is as follows: We consider the action for a scalar field. We can then ignore the
mass, potential and interaction terms in the action because the kinetic term dominates
in the high-energy theory near the horizon. By expanding the scalar field in terms of
the spherical harmonics and using the above properties at horizon, we find that the integrand in
the action dose not depend on angular variables. Thus we find that the 4-dimensional
action with the Kerr-Newman metric effectively becomes a 2-dimensional action with a spherically symmetric metric.

In this section, we would like to discuss the dimensional reduction near the event horizon and 
actually show that the 4-dimensional Kerr-Newman metric effectively becomes
a 2-dimensional spherically symmetric metric by using the technique of the dimensional reduction near the horizon.

For simplicity, we consider the 4-dimensional action for a complex scalar field
\begin{align}
S=\int d^4x \sqrt{-g}g^{\mu\nu}( \partial_\mu +ieV_\mu ) \phi^* ( \partial_\nu -ieV_\nu )\phi +S_{{\rm int}},
\label{KNact01}
\end{align}
where the first term is the kinetic term and the second term $S_{{\rm int}}$ represents the mass, potential and interaction terms.
The gauge field $V_\mu$ associated with the Coulomb potential of the black hole, is given by
\begin{align}
( V_{\mu} )=\left( -\frac{Qr}{r^2+a^2}, 0, 0, 0\right).
\label{gauge01}
\end{align}
By substituting both the Kerr-Newman metric (\ref{2kn01}) and (\ref{gauge01}) to (\ref{KNact01}), we obtain
\begin{align}
S=\int& dt dr d\theta d\varphi \sin \theta \phi^*
\Bigg[  \left( \frac{(r^2+a^2)^2}{\Delta}-a^2\sin^2 \theta \right) \left( \partial_t+\frac{ieQr}{r^2+a^2} \right) ^2\notag\\
&+2ia \left( \frac{r^2+a^2}{\Delta} -1 \right) \left( \partial_t+\frac{ieQr}{r^2+a^2} \right) \hat{L}_z
-\partial_r\Delta \partial_r +\hat{\bm L}^2 -\frac{a^2}{\Delta}\hat{L}^2_z
\Bigg]\phi+S_{{\rm int}},
\label{kerract2}
\end{align}
where we used
\begin{align}
\hat{\bm L}^2&=-\frac{1}{\sin \theta}\partial_\theta \sin \theta \partial_\theta-\frac{1}{\sin^2 \theta}\partial_\varphi^2,\\
\hat{L}_z&=-i\partial_\varphi.
\end{align}
By performing the partial wave decomposition of $\phi$ in terms of the spherical harmonics
\begin{align}
\phi=\sum_{l,m}\phi_{lm}(t,r)Y_{lm}(\theta,\varphi),
\label{kyu1}
\end{align}
we obtain

\begin{align}
S=&\int dt dr d\theta d\varphi \sin\theta \sum_{l',m'}\phi_{l'm'}^* Y_{l'm'}^*
\Bigg[ \frac{(r^2+a^2)^2}{\Delta}\left( \partial_t+\frac{ieQr}{r^2+a^2} \right)^2-a^2\sin^2 \theta \left( \partial_t+\frac{ieQr}{r^2+a^2} \right)^2 \notag\\
&+2ima\frac{r^2+a^2}{\Delta}\left( \partial_t+\frac{ieQr}{r^2+a^2} \right)
-2ima\left( \partial_t+\frac{ieQr}{r^2+a^2} \right)-\partial_r\Delta \partial_r +l(l+1) -\frac{m^2a^2}{\Delta}
\Bigg]\notag\\
&\times \sum_{l,m}\phi_{lm}Y_{lm}+S_{{\rm int}},
\label{kerract4}
\end{align}
where we used eigenvalue equations for $\hat{\bm L}^2$ and $\hat{L}_z$
\begin{align}
\hat{\bm L}^2Y_{lm}&=l(l+1)Y_{lm},
\label{kaku1}\\
\hat{L}_zY_{lm}&=mY_{lm}.
\label{kaku2}
\end{align}
Here $l$ is the azimuthal quantum number and $m$ is the magnetic quantum number.
Now, we transform the radial coordinate $r$ into the tortoise coordinate $r_*$ defined by
\begin{align}
\frac{dr_*}{dr}=\frac{r^2+a^2}{\Delta}\equiv\frac{1}{f(r)}.
\label{fdef}
\end{align}
After this transformation, the action (\ref{kerract4}) is written by
\begin{align}
S=&\int dt dr_* d\theta d\varphi \sin\theta \sum_{l',m'}\phi_{l'm'}^*Y_{l'm'}^* \Bigg[ (r^2+a^2)\left( \partial_t+\frac{ieQr}{r^2+a^2} \right)^2 -f(r)a^2\sin^2\theta \left( \partial_t+\frac{ieQr}{r^2+a^2} \right)^2\notag\\
&
+2ima\left( \partial_t+\frac{ieQr}{r^2+a^2} \right)-F(r)2ima \left( \partial_t+\frac{ieQr}{r^2+a^2} \right)
-\partial_{r_*}(r^2+a^2)\partial_{r_*}\notag\\
&+f(r)l(l+1)-\frac{m^2a^2}{r^2+a^2}
\Bigg]\sum_{l,m}\phi_{lm}Y_{lm}+S_{{\rm int}}.
\label{kerract5}
\end{align}

Here we consider this action in the region near the horizon.
Since $f(r_+)=0$ at $r \to r_+$,  we only retain dominant terms in (\ref{kerract5}).
We thus obtain the effective action near the horizon $S_{({\rm H})}$
\begin{align}
S_{({\rm H})}=&\int dt dr_* d\theta d\varphi \sin\theta \sum_{l',m'}\phi_{l'm'}^*Y_{l'm'}^*
\Bigg[ (r^2+a^2)\left( \partial_t+\frac{ieQr}{r^2+a^2} \right)^2 +2ima\left( \partial_t+\frac{ieQr}{r^2+a^2} \right)\notag\\
&-\partial_{r_*}(r^2+a^2)\partial_{r_*}-\frac{m^2a^2}{r^2+a^2}
\Bigg]\sum_{l,m}\phi_{lm}Y_{lm},
\label{kerract6}
\end{align}
where we ignored $S_{{\rm int}}$ by using $f(r_+)=0$ at $r \to r_+$. 
Because the theory becomes the high-energy theory near the horizon and the kinetic term dominates, 
we can ignore all the terms in $S_{{\rm int}}$.
For example, we consider the case of a mass term.
In this case, a mass term is usually given by  
\begin{align}
\int dx^4 \left( \mu^2 \phi^* \phi \right)&=\int dt dr d\theta d\varphi \sin\theta \left( \mu^2 \phi^* \phi \right)\\
&=\int dt dr_* d\theta d\varphi \sin\theta \left( f(r) \mu^2 \phi^* \phi \right),
\end{align}
where $\mu$ is a mass of the scalar field and we used (\ref{fdef}) in the last line.
We find that the term vanishes by using $f(r_+)=0$ at $r \to r_+$.
The same is equally true of other interaction terms $S_{{\rm int}}$.
After this analysis, we return to the expression written in terms of $r$, and we obtain
\begin{align}
S_{({\rm H})}=-\sum_{l,m}&\int dt dr(r^2+a^2)\phi_{lm}^* \Bigg[ -\frac{r^2+a^2}{\Delta}\left( \partial_t +\frac{ieQr}{r^2+a^2} +\frac{ima}{r^2+a^2}\right)^2+\partial_r \frac{\Delta}{r^2+a^2} \partial_r \Bigg] \phi_{lm},
\label{kerract9}
\end{align}
where we used the orthonormal condition for the spherical harmonics
\begin{align}
\int d\theta d\varphi \sin\theta Y_{l'm'}^*Y_{lm}=\delta_{l',l}\delta_{m',m}.
\end{align}

From (\ref{kerract9}), we find that $\phi_{lm}$ can be considered as a (1+1)-dimensional complex scalar field in the backgrounds of 
the dilaton $\Phi$, metric $g_{\mu\nu}$ and two $U(1)$ gauge fields $V_\mu$, $U_\mu$
\begin{align}
&\Phi=r^2+a^2,\\
&g_{tt}=-f(r),\quad 
g_{rr}=\frac{1}{f(r)},\quad 
g_{rt}=0,
\label{kerr001}
\\
&V_t =-\frac{Qr}{r^2+a^2}, \quad V_r =0,\\
&U_t =-\frac{a}{r^2+a^2}, \quad U_r =0.
\end{align}
There are two $U(1)$ gauge fields:
One is the original gauge field as in (\ref{gauge01})
while the other is the induced gauge field associated with the isometry along the $\varphi$ direction.
The induced $U(1)$ charge of the 2-dimensional field $\phi_{lm}$ is given by $m$.
Then the gauge potential $A_t$ is a sum of these two fields,
\begin{align}
A_t\equiv eV_t+mU_t=-\frac{eQr}{r^2+a^2}-\frac{ma}{r^2+a^2}, \quad A_r=0.
\label{gauge02}
\end{align}
By using the above notations, the action (\ref{kerract9}) is rewritten as
\begin{align}
S_{({\rm H})}=-\sum_{l,m}&\int dt dr \Phi \phi_{lm}^* \Bigg[ g^{tt}\left( \partial_t -iA_t \right)^2+\partial_r g^{rr} \partial_r \Bigg] \phi_{lm},
\label{kerract10}
\end{align}

From (\ref{kerr001}), we find that the 4-dimensional spherically non-symmetric Kerr-Newman metric (\ref{2kn01}) effectively behaves as
a 2-dimensional spherically symmetric metric in the region near the horizon only
\begin{align}
ds^2=-f(r)dt^2+\frac{1}{f(r)}dr^2.
\label{kerr5}
\end{align}

For confirmation, we show how to derive the surface gravity on the horizon of the Kerr-Newman black hole from $f(r)$ defined by (\ref{fdef}).
Actually by calculating the surface gravity, we can obtain
\begin{align}
\kappa_\pm \equiv \frac{1}{2} f'(r)\Bigg |_{r=r_\pm}=\frac{r_\pm-r_\mp}{2(r_\pm^2+a^2)},
\end{align}
where $\{'\}$ represents differentiation with respect to $r$.
This result agrees with the well-known surface gravity on the horizon of the Kerr-Newman black hole as in (\ref{2sg01}). 
%%%%%%%%%%%%%%%%%%%%%%%%%%%%%%%%%%%%%%%%%%%%%%%%%%%%%%%%%%%%%%%%%%%%%%%%%%%%%%%%%%%%%%%%%%%%%%%%%
%%                                                                                             %%
%%                                     Section 2-5                                             %%
%%                                                                                             %%
%%%%%%%%%%%%%%%%%%%%%%%%%%%%%%%%%%%%%%%%%%%%%%%%%%%%%%%%%%%%%%%%%%%%%%%%%%%%%%%%%%%%%%%%%%%%%%%%%
\section{Analogies between Black Hole Physics and Thermodynamics}

\quad~ To understand properties of black holes, it is very useful to understand the black hole physics in the context of generalized thermodynamics.
The main reason is that there are various analogies between the black hole physics and thermodynamics.
It is said that the idea of making use of thermodynamic methods in black hole physics appears to have been first considered by Greif.
He examined the possibility of defining the entropy of a black hole, but lacking many of the recent results in black hole physics,
he did not make a concrete proposal \cite{gre01}.
Afterward, properties of black holes were analyzed by Bekenstein, Bardeen, Carter and Hawking and others,
and analogies between black hole physics and thermodynamics were clarified \cite{bek01,bar01}.
The discussion is as follows.

By definition, a black hole can absorb matter but nothing, not even light, can classically escape from it.
A black hole has a property that as a black hole absorbs matter, the black hole area increases.
For example, we consider that two Schwarzschild black holes with masses $M_1$ and $M_2$ merge 
and then a black hole with a mass $M=M_1+M_2$ is formed (Fig. 2.12).
Before the merger, areas of two black holes are respectively $A_1=16\pi M_1^2$ and $A_2=16 \pi M_2^2$.
An area of the black hole after the merger is $A=16\pi (M_1+M_2)^2$.
Compared between the sum of two black hole areas before the merger and the black hole area after the merger,
we obtain an inequality for black hole areas
\begin{align}
A_1+A_2 \leq A.
\end{align}
This means that a black hole area after the merger is the same as the sum of each black hole area before the merger or is larger than it.
A black hole area never classically decreases,
\begin{align}
\delta A \geq 0,
\label{25al01}
\end{align}
since no black hole radiates matter or splits into any black holes.
This result is known as Hawking's black hole area theorem\cite{haw03}. 

\newpage

%%%%%%%%%%%%%     Fig.2.12       %%%%%%%%%%
\begin{center}
\vspace{0.6cm}
\input{area01}\\
\vspace{0.6cm}
Fig. 2.12 \quad The merger of black holes.
\vspace{0.6cm}
\end{center}
%%%%%%%%%%%%%%%%%%%%%%%%%%%%%%%%%%%%%%%%%%

Here we would like to state the properties of entropy in thermodynamics.
In thermodynamics, entropy represents the degree of concentration of matter and energy in a system.
As a famous example of explaining entropy, we consider that one puts a drop of ink into a glass of water.
At first, a drop of ink localizes in a certain part of the water.
This is a state with low entropy.
As time advances, a drop of ink distributes all over the water and 
the water achieve an even color at some time.
This is that entropy is a state with higher entropy.
The entropy of an isolated system ${\cal S}$ never decreases and rather increases over time, i.e.,
\begin{align}
\delta {\cal S} \geq 0.
\end{align}
This is well-known as the second law of thermodynamics.
Thus, both of the black hole area and entropy tend to increase irreversibly.

As with entropy, the black hole area is closely related to a degradation of energy, in other words, an unavailable energy.
In thermodynamics, an increase of entropy means that a part of energy is unavailable, 
namely, the energy is no longer converted into work.
There is the same relation in black hole physics.
In Section 2.3, we showed that a part of energy can be extracted from a rotating black hole 
such as a Kerr black hole by the Penrose process. 
But all energy cannot be extracted from the black hole.
The Kerr black hole gradually decreases the angular momentum by the Penrose process.
When the black hole loses the total angular momentum, it becomes a Schwarzschild black hole.
By the Hawking's black hole area theorem, the mass of the black hole is then larger than a mass of a Schwarzschild black hole 
obtained by taking $a=0$ for the original Kerr black hole.
This mass is called an {\it irreducible mass}.
In the case of a Kerr-Newman black hole, the irreducible mass $M_{{\rm ir}}$ is given by
\begin{align}
M_{{\rm ir}}=\sqrt{\frac{A}{16\pi}}.
\label{25irm01}
\end{align}
It is regarded as an inactive energy which cannot be converted to work.
The increase of an irreducible mass $M_{{\rm ir}}$, i.e., the increase of a black hole area $A$ 
thus corresponds to a degradation of the black hole energy in the thermodynamic sense.

As found from the above consideration, it is said that properties possessed by a black hole area $A$ are similar to ones possessed by the thermodynamic entropy and the Hawking's black hole area theorem corresponds to the second law of thermodynamics.
Furthermore, by comparing the energy conservation law in the black hole physics with the first law of thermodynamics,
we would like to clarify corresponding physical quantities in these two phenomena.

In general, the first law of thermodynamics is given by
\begin{align}
d {\cal E} = {\cal T}d {\cal S} - d{\cal W},
\label{26ec01}
\end{align}
where ${\cal E}$ is the energy of the system, ${\cal T}$ is the temperature, ${\cal S}$ is the entropy and ${\cal W}$ is the work done by the system,
while as already stated as in (\ref{2ec01}) of Section 2.1, the energy conservation law in black hole physics is given by
\begin{align}
dM = \frac{\kappa}{8\pi} dA + \Omega_{{\rm H}} dL + \Phi_{{\rm H}} dQ,
\label{26ec02}
\end{align}
Now we make comparisons between the relations (\ref{26ec01}) and (\ref{26ec02}).
We make a table of the corresponding relationships between physical quantities of black hole physics and thermodynamics (Tab. 2.3).
The correspondence relationship between the mass $M$ and the energy ${\cal E}$ in the left side of each relation is clear
and it is well-known as the mass-energy equivalence by Einstein.
The second term and the third term in (\ref{26ec02}) stand for work terms done by the rotation and the electromagnetism.
It is considered that they correspond to the work term $-d{\cal W}$ done by the system in thermodynamics.
We shall compare the remaining first term in each relation, i.e., $\displaystyle \frac{\kappa}{8\pi}dA$ and ${\cal T} d {\cal S}$.
By making a black hole area correspond to entropy, we find that the surface gravity corresponds to the temperature in thermodynamics.

\newpage

%%%%%%%%%%%%%%%%% Table 2.3 %%%%%%%%%%%%%%%%%%%
\begin{center}
\vspace{0.6cm}
Tab. 2.3 \quad The corresponding relationships between physical\\ \qquad \qquad \qquad \quad quantities of thermodynamics and black hole physics
\vspace{0.3cm}\\
  \begin{tabular}{|c|c|}
    \hline
      Thermodynamics & Black hole physics   \\
    \hline\parbox[c][\myheight][c]{0cm}{}
      Energy: ${\cal E}$ & Mass: $M$   \\
    \hline\parbox[c][\myheight][c]{0cm}{}
      Temperature: ${\cal T}$ & Surface gravity: $\kappa$   \\
    \hline\parbox[c][\myheight][c]{0cm}{}
      Entropy: ${\cal S}$ & Black hole area: $A$   \\
    \hline
      ~~~Work term done by system:~~~  & ~~~Work terms done by rotation and~~~   \\
     $-d {\cal W}$ & ~electromagnetism: $\Omega_{{\rm H}} dL+\Phi_{{\rm H}} dQ$ \\
    \hline
  \end{tabular}
  \vspace{0.6cm}\\
 \end{center}
%%%%%%%%%%%%%%%%% Table 2.3 %%%%%%%%%%%%%%%%%%%

Here we recall properties of both the surface gravity of a black hole and temperature.
By definition, a surface gravity of the black hole represents the strength of the gravitational field on the event horizon.
As found from (\ref{2sg01}), the surface gravity $\kappa$ is constant over the horizon in the stationary black hole.
In thermal equilibrium, temperature also possesses the same property.
It is well-known as the zeroth law of thermodynamics.

In passing, the third law of thermodynamics states that the temperature of the system cannot achieve the absolute zero temperature
by a physical process. This is also called Nernst's theorem.
It corresponds to the speculation that the surface gravity cannot achieve $\kappa=0$ by a physical process in black hole physics.
A reason for believing it is that if one could reduce it to zero by a finite sequence of operations, then presumably one could carry the process further, thereby creating a naked singularity.

From the above discussion, we find that the relationships between the laws of black hole physics and thermodynamics 
may be more than an analogy (Tab. 2.4, \cite{wal01}).
However, we do not know from only the above discussion that black holes actually have entropy and temperature.
In the next section, we would like to discuss the consideration that black holes have entropy, which is due to Bekenstein.

\newpage

%%%%%%%%%%%%%%%%% Table 2.4 %%%%%%%%%%%%%%%%%%%
\begin{center}
\vspace{0.6cm}
Tab. 2.4 \quad The corresponding relationships between the laws\\\qquad \qquad  of thermodynamics and black hole physics
\vspace{0.3cm}\\
  \begin{tabular}{|c|c|c|}
    \hline\parbox[c][\myheight][c]{0cm}{}
      Law& Thermodynamics & Black hole physics   \\
    \hline
    & & \\
      ~~~Zeroth~~~ & ${\cal T}$ constant throughout body  & $\kappa$ constant over horizon  \\
      & in thermal equilibrium & of stationary black hole\\
    & & \\
    \hline
    & & \\
      First & $d {\cal E} = {\cal T}d {\cal S} - d{\cal W}$ & $\displaystyle dM = \frac{\kappa}{8\pi} dA + \Omega_{{\rm H}} dL + \Phi_{{\rm H}} dQ$   \\
    & & \\
    \hline
    & & \\
      Second &$\delta {\cal S} \geq 0$ ~in any process & $\delta A \geq 0$ ~in any process   \\
    & & \\
    \hline
    & & \\
      Third & ~~~Impossible to achieve ${\cal T}=0$~~~  & ~~~Impossible to achieve $\kappa=0$~~~  \\
    &  by a physical process &  by a physical process \\
    & & \\
    \hline
  \end{tabular}
  \vspace{0.6cm}\\
 \end{center}
%%%%%%%%%%%%%%%%% Table 2.4 %%%%%%%%%%%%%%%%%%%

%%%%%%%%%%%%%%%%%%%%%%%%%%%%%%%%%%%%%%%%%%%%%%%%%%%%%%%%%%%%%%%%%%%%%%%%%%%%%%%%%%%%%%%%%%%%%%%%%
%%                                                                                             %%
%%                                     Section 2-6                                             %%
%%                                                                                             %%
%%%%%%%%%%%%%%%%%%%%%%%%%%%%%%%%%%%%%%%%%%%%%%%%%%%%%%%%%%%%%%%%%%%%%%%%%%%%%%%%%%%%%%%%%%%%%%%%%
\section{Black Holes and Entropy}

\quad~ In 1973, Bekenstein proposed that a black hole has its entropy.
He stated that the black hole entropy is represented by a function of the black hole area from the above analogies. 
If the black hole entropy is related to the black hole area, it has to satisfy the black hole area theorem.
From this, he presumed that the black hole entropy is proportional to the black hole area.

Then, to find the proportionality coefficient, he considered that a particle with the least information falls into a black hole.
When the particle falls into a black hole, the information of the particle is lost.
In other words, it means an increase in the black hole entropy.
He evaluated the proportionality coefficient by the conjecture that 
the black hole entropy equals the minimum area increased by dropping a matter into the black hole.

In this section, we would like to show the derivation of black hole entropy by Bekenstein.
In Subsection 2.6.1, we simply explain the entropy of a particle with the least information in information theory.
In Subsection 2.6.2, we explain that minimum entropy is increased when a particle falls into a black hole.
In Subsection 2.6.3, we evaluate the black hole entropy by using assumptions that these two quantities are equal to each other.

%%%%%%%%%%%%%%%%%%%%%%%%%%%%%%%%%%%%%%%%%%%%%%%%%%%%%%%%%%%%%%%%%%%%%%%%%%%%%%%%%%%%%%%%%%%%%%%%%
%%                                                                                             %%
%%                                     Subsection 2-6-1                                        %%
%%                                                                                             %%
%%%%%%%%%%%%%%%%%%%%%%%%%%%%%%%%%%%%%%%%%%%%%%%%%%%%%%%%%%%%%%%%%%%%%%%%%%%%%%%%%%%%%%%%%%%%%%%%%
\subsection{Entropy in information theory}

\quad~ In physics, entropy represents a degree of concentration of matter and energies.
The entropy is defined by Boltzmann's formula
\begin{align}
{\cal S}=k\ln W,
\label{261bol01}
\end{align}
where $W$ is the number of states and $k$ is Boltzmann's constant.

The connection between entropy and information is well-known \cite{sha01,sha02}.
In information theory, the uncertain information and the missing information of the system are measured by the entropy.
The probability of the $n$-th state in all known states of the system is defined as $P_n$
The entropy of the system is then defined by Shannon's formula
\begin{align}
{\cal S}= -\sum_n P_n \ln P_n.
\label{261sha01}
\end{align}
Here we note that entropy is dimensionless. We will present the discussion of dimensions later.

When a new piece of information is available for the system, we can find that probabilities $P_n$ are provided with some restrictions.
For example, we consider the case of a die. The probabilities are respectively $\frac{1}{6}$ from 1 to 6.
The entropy is then $\ln 6$ from (\ref{261sha01}).
Now if we get new information that ``There are odd numbers (or odd numbers are given)",
then the probability of getting even numbers is zero, i.e., $P_2=P_4=P_6=0$.
The probability of getting odd numbers is $\frac{1}{3}$ and thus the entropy is $\ln 3$.
As found in the above discussion, as we get new information, the entropy locally decreases.
This property is given by Brillouin's identification \cite{bri01}
\begin{align}
\Delta {\cal I} = - \Delta {\cal S}, 
\label{261bri01}
\end{align}
where $\Delta {\cal I}$ stands for the new information (bound information).
This relation means that the bound information corresponds to the decrease of the entropy.

Here we would like to discuss the dimensions of the physical quantities.
Entropy appearing in Boltzmann's formula (\ref{261bol01}) has the dimension of the energy divided by the temperature.
Although it is possible to decide the dimension of information by using it, 
it is a custom in the information theory to treat information as a dimensionless quantity.
Thus we adopt a unit system that both entropy and information is dimensionless.
This means we select to measure temperature by the unit of energy.
Then Boltzmann's constant is also dimensionless.
By adopting the above unit system, the equations (\ref{261sha01}) and (\ref{261bri01}) are satisfied as dimensionless quantities.

The conventional unit of information is the ``bit" which may be defined as the information available when the answer to a yes-or-no question is precisely known, i.e., the entropy is zero.
Of course, the unit is dimensionless. 
According to (\ref{261bri01}), a bit is also numerically equal to the maximum entropy that can be associated with a yes-or-no question, i.e., the entropy when no information whatsoever is available about the answer.
From (\ref{261sha01}), the entropy in the yes-or-no question is written as
\begin{align}
{\cal S}&=-P_{{\rm yes}}\ln {P_{{\rm yes}}}-P_{{\rm no}} \ln {P_{{\rm no}}}\\
&=-P_{{\rm yes}}\ln {P_{{\rm yes}}} -(1-P_{{\rm yes}}) \ln (1-P_{{\rm yes}}).
\end{align}
We thus find that the entropy is the maximum value $\ln 2$ when $\displaystyle P_{{\rm yes}}=P_{{\rm no}}=\frac{1}{2}$
and one bit is equal to $\ln 2$ of information.

Let us now return to our original subject, black hole.
We consider that a particle falls into a black hole.
An amount of information of the particle would depend on how much is known about the internal states of the particle.
The minimum information loss for the particle would be contained in the answer to the question ``Does the particle exist or not?"
Before the particle drops into the black hole, the answer is known to be ``yes".
But after the particle drops into the black hole, we have no information whatever about the answer.
This is because one knows nothing about the physical conditions inside the black hole, and thus one cannot assess the likelihood of the particle continuing to exist or being destroyed.
One must, therefore, admit the loss of one bit of information at the very least.
This means that the entropy is increased by 
\begin{align}
\Delta {\cal S}=\ln 2,
\label{261s01}
\end{align} 
before and after the particle with the tiniest information falls into the black hole.  

%%%%%%%%%%%%%%%%%%%%%%%%%%%%%%%%%%%%%%%%%%%%%%%%%%%%%%%%%%%%%%%%%%%%%%%%%%%%%%%%%%%%%%%%%%%%%%%%%
%%                                                                                             %%
%%                                     Subsection 2-6-2                                        %%
%%                                                                                             %%
%%%%%%%%%%%%%%%%%%%%%%%%%%%%%%%%%%%%%%%%%%%%%%%%%%%%%%%%%%%%%%%%%%%%%%%%%%%%%%%%%%%%%%%%%%%%%%%%%
\subsection{The minimum increase of the black hole area}

\quad~ In this subsection, we calculate the minimum possible increase in the black hole area, which must result when a spherical particle of rest mass $\mu$ and proper radius $b$ is captured by a Kerr-Newman black hole.
Bekenstein used the ``rationalized area" of a black hole $\alpha$ defined by
\begin{align}
\alpha\equiv \frac{A}{4\pi},
\label{262area01}
\end{align}
where $A$ is the black hole area as in (\ref{2a01}).
The first law of black hole physics (\ref{2ec01}) is then written as
\begin{align}
dM = \Theta_{{\rm H}} d \alpha + \Omega_{{\rm H}} dL+ \Phi_{{\rm H}} dQ,
\label{262ene00}
\end{align}
where $\Theta_{{\rm H}}$ is defined by
\begin{align}
\Theta_{{\rm H}} \equiv \frac{r_+ - M}{2\alpha}.
\label{262the01}
\end{align}

There are several ways in which a particle may fall into a black hole. All these bring the increase of the black hole area.
We are interested in the method for inserting the particle which results in the smallest increase.
This method has already been discussed by Christodoulou 
in connection with his introduction of the concept of irreducible mass \cite{chr01,chr02}.
The essence of Christodoulou's method is that if a freely falling point particle is captured by a Kerr-Newman black hole, then the irreducible mass and, consequently, the area of the black hole is left unchanged.
Bekenstein generalized Christodoulou's method to a particle with a proper radius 
and showed the increased area of the black hole is no longer precisely zero when the particle falls into the black hole.

We assume that a freely falling particle is neutral.
The trajectory of the particle follows a geodesic of the Kerr-Newman metric (\ref{2kn01}).
The horizon is located at $r=r_+$ where $r_\pm$ are defined by (\ref{2hor01}).

First integrals for geodesic motion in the Kerr-Newman background have been given by Carter \cite{car02}.
Christodoulou used the first integral
\begin{equation}
E^2[r^4+a^2 (r^2 + 2Mr -Q^2 )] -2E(2Mr-Q^2)ap_{\varphi}-(r^2 -2Mr+Q^2) p_{\varphi} ^2
- (\mu ^2 r^2 + q) \Delta = ( p_r \Delta )^2,
\label{262ene01}
\end{equation}
as a starting point of his analysis. 
We show the derivation in Appendix B.
In (\ref{262ene01}), $E=-p_t$ is the conserved energy, $p_{\varphi}$ is the conserved component of angular momentum in the direction of the axis of symmetry, $q$ is Carter's fourth constant of the motion, $\mu$ is the rest mass of the particle and $p_r$ is its covariant radial momentum.

By following Christodoulou, we solve (\ref{262ene01}) for $E$:
\begin{equation}
E= {\cal B} a p_{\varphi} + \sqrt{ \left( {\cal B}^2 a^2 + \frac{r^2 - 2Mr + Q^2}{{\cal A}} \right)p_{\varphi} ^2
+ \frac{(\mu ^2 r^2 + q) \Delta + (p_r \Delta) ^2}{{\cal A}} },
\label{262ene02}
\end{equation}
where
\begin{align}
{\cal A} &\equiv r^4 + a^2 ( r^2 + 2Mr - Q^2 ),
\label{262A01}\\
{\cal B} &\equiv \frac{(2Mr - Q^2 )}{{\cal A}}.
\label{262B01}
\end{align}
The definitions (\ref{262A01}) and (\ref{262B01}) at the horizon as in (\ref{21hor01}) are written as
\begin{align}
{\cal A}(r=r_+)&\equiv {\cal A}_+ =(r_+^2+a^2)^2,\\
{\cal B}(r=r_+)&\equiv {\cal B}_+ =\frac{1}{r_+^2 +a^2}.
\end{align}
Furthermore, at the horizon, we obtain 
\begin{align}
{\cal B}_+ a =\Omega_{{\rm H}}.
\end{align}
where $\Omega_{{\rm H}}$ is defined by (\ref{2av01}).
The coefficient of $p_\varphi^2$ at the horizon vanishes
\begin{align}
{\cal B}_+^2 a^2 + \frac{r_+^2 - 2Mr_+ + Q^2}{{\cal A}_+}
=\frac{a^2}{(r_+^2+a^2)^2}+\frac{r_+^2 -2Mr_+ +Q^2}{(r_+^2 + a^2)^2}
=\frac{\Delta}{(r_+^2+a^2)^2}=0,
\end{align}
and the coefficient of $\mu^2 r^2+ q$ also vanishes.
However, since $p_r \Delta$ cannot be defined at the horizon because of $p_r=g_{rr} p^r$,
we retain the term as
\begin{align}
p_r&=\frac{\Sigma}{\Delta}p^r\\
\Leftrightarrow \quad p_r \Delta &=(r^2+a^2 \cos^2 \theta)p^r.
\end{align}
If the particle's orbit intersects the horizon, we then have from (\ref{262ene02}) that
\begin{align}
E= \Omega_{{\rm H}} p_{\varphi} + \frac{|p_r \Delta |_+}{\sqrt{{\cal A}_+}}.
\label{262ene03}
\end{align}
As a result of the capture, the mass of the black hole increases by $E$ and its component of the angular momentum in the direction of the symmetry axis increases by $p_{\varphi}$.
By comparing (\ref{2ec01}) with (\ref{262ene03}), the black hole's rationalized area $\alpha$ increases by $\displaystyle \frac{|p _r \Delta |_+}{\Theta_{{\rm H}} \sqrt{{\cal A}_+}}$.
As pointed out by Christodoulou, by taking
\begin{align}
|p_r \Delta |_+=0,
\end{align}
the relation (\ref{262ene03}) becomes
\begin{align}
E=\Omega_{{\rm H}} p_\varphi,
\label{262ene04}
\end{align}
and the increase of the black hole area vanishes.
The above analysis shows that it is possible for a black hole to capture a point particle without increasing its area.

Here, by following Bekenstein's extension, we would like to show how this conclusion is changed 
if the particle has a nonzero proper radius $b$.
The relation (\ref{262ene02}) always describes the motion of the particle's center of mass at the moment of capture.
It should be clear that to generalize Christodoulou's result to the present case one should evaluate (\ref{262ene02}) not at $r=r_+$, 
but $r=r_+ + \delta$, where $\delta$ is determined by
\begin{align}
\int_{r_+} ^{r_+ + \delta} \sqrt{g_{rr} }dr = b.
\end{align}
$r=r_+ +\delta$ is a point a proper distance $b$ outside the horizon.
By using the component $g_{rr}$ as in (\ref{21met02}), we find
\begin{align}
b = 2 \sqrt{ \frac{\delta ( r_+ ^2 + a^2 \cos ^2 \theta )}{r_+ - r_-}},
\label{262rad01}
\end{align}
where we assumed that $r_+ - r_- \gg \delta$.
Expanding the argument of the square root in (\ref{262ene02}) in powers of $\delta$, 
replacing $\delta$ by its value given by (\ref{262rad01}),
and keeping only terms to $O(b)$ we obtain
\begin{align}
E=\Omega_{{\rm H}} p_{\varphi} + \sqrt{\left( \frac{r_+ ^2 - a^2}{r_+ ^2 + a^2} \right) p_{\varphi} ^2 + \mu ^2 r_+ ^2 +q}
\times \frac{1}{2} b \frac{r_+ -r_-}{(r_+ ^2 + a^2)} \times \frac{1}{\sqrt{r_+ ^2 + a^2 \cos ^2 \theta}}.
\label{262ene05}
\end{align}
This relation (\ref{262ene05}) is the generalization to $O(b)$ of Christodoulou's result (\ref{262ene04}).
Carter's kinetic constant $q$ is given by
\begin{align}
q = \cos^2 \theta \left[ a^2(\mu^2 -E^2) + \frac{p_\varphi ^2}{\sin^2 \theta} \right] + p_\theta ^2
\label{262q01}
\end{align}
This constant appeared in the derivation of (\ref{262ene01}) (see Appendix B).
We can obtain a lower bound for it as follows.
From the requirement that the $\theta$ momentum $p_{\theta}$ is real in (\ref{262q01}), we obtain
\begin{align}
q \geq \cos ^2 \theta \left[ a^2 (\mu ^2 - E^2 ) + \frac{p_{\varphi} ^2}{\sin ^2 \theta }\right],
\label{262q02}
\end{align}
where the equality holds when $p_\theta=0$.
If we replace $E$ in (\ref{262q02}) by $\Omega_{{\rm H}}p_{\varphi}$ as in (\ref{262ene04}),
we obtain
\begin{align}
q \geq \cos ^2 \theta \left[ a^2 \mu ^2 + p_{\varphi} ^2 \left( \frac{1}{\sin^2 \theta} - a^2 \Omega_{{\rm H}} ^2 \right) \right].
\end{align}
We know that $\displaystyle \frac{1}{\sin^2 \theta} \geq 1$ 
and it is easily shown that $ \displaystyle a^2 \Omega_{{\rm H}}^2 \leq \frac{1}{4}$
for a Kerr-Newman black hole.
Since the coefficient of $p_{\varphi}^2$ is positive, we can take the constant $q$ as a smaller value
\begin{align}
q \geq a^2 \mu ^2 \cos^2 \theta,
\label{262q03}
\end{align}
when $p_\varphi=0$.
By substituting (\ref{262q03}) into (\ref{262ene05}), we obtain
\begin{align}
E \geq \Omega_{{\rm H}} p_{\varphi} + \frac{1}{2} \mu b \frac{r_+ - r_-}{r_+ ^2 + a^2},
\label{262ene06}
\end{align}
where the equality holds when $p_\varphi=p_\theta=p^r=0$.
This relation is correct to $O(b)$.
The increase in the rationalized area of the black hole, computed by means of (\ref{262ene00}), (\ref{262the01}) and (\ref{262ene06}), is
given by
\begin{align}
\Delta \alpha \geq 2 \mu b.
\label{262area02}
\end{align}
This gives the fundamental lower bound on the increase in the rationalized area of the black hole $\Delta \alpha$,
\begin{align}
(\Delta \alpha)_{{\rm min}} =2 \mu b.
\label{262area03}
\end{align}
We note that it is independent of $M$, $Q$ and $L$.

We can make $(\Delta \alpha)_{{\rm min}}$ smaller by making $b$ smaller.
However, we must remember that $b$ can be no smaller than the particle's Compton wavelength $\displaystyle \frac{\hbar}{\mu}$, or the Schwarzschild radius $2\mu$.
If the Compton wavelength is larger than the Schwarzschild radius $\displaystyle \frac{\hbar}{\mu}\geq 2\mu$, 
namely, the mass of the particle satisfies $\displaystyle \mu \leq \sqrt{\frac{\hbar}{2}}$,
we can make $b$ smaller to $\displaystyle \frac{\hbar}{\mu}$.
If the Schwarzschild radius is larger than the Compton wavelength $\displaystyle \frac{\hbar}{\mu}< 2\mu$, 
namely, the mass of the particle satisfies $\displaystyle \mu > \sqrt{\frac{\hbar}{2}}$,
we can make $b$ smaller to $2\mu$.
The relation (\ref{262area03}) is thus given by $2\hbar$, when $b= \displaystyle \frac{\hbar}{\mu}$, and given by 
$4\mu^2$, when $b\simeq 2\mu$.
Since $4\mu^2 > 2\hbar$, we can determine a lower bound of the rationalized area of a Kerr-Newman black hole as
\begin{align}
(\Delta \alpha)_{{\min}}=2\hbar,
\label{262area04}
\end{align}
when the black hole captures the particle.

%%%%%%%%%%%%%%%%%%%%%%%%%%%%%%%%%%%%%%%%%%%%%%%%%%%%%%%%%%%%%%%%%%%%%%%%%%%%%%%%%%%%%%%%%%%%%%%%%
%%                                                                                             %%
%%                                     Subsubsection 2-6-3                                     %%
%%                                                                                             %%
%%%%%%%%%%%%%%%%%%%%%%%%%%%%%%%%%%%%%%%%%%%%%%%%%%%%%%%%%%%%%%%%%%%%%%%%%%%%%%%%%%%%%%%%%%%%%%%%%
\subsection{Information loss and black hole entropy}

\quad~ In Section 2.5, we already stated that a black hole area is similar to the entropy in thermodynamics.
Although there are clear analogies between them, we do not know how to identify the black hole area as the black hole entropy.
In this subsection, we would like to present the discussion by Bekenstein \cite{bek01}.

To begin with, we consider that a black hole is formed by the gravitational collapse of a very heavy star.
According to the no-hair theorem \cite{ruf01}, 
the stationary state of the black hole is completely characterized by three parameters, i.e., the mass, the angular momentum and the charge.
Thus black holes do not depend on the internal configuration of the collapsed body.
This means that a lot of information is lost by the gravitational collapse.
It is then natural to introduce the concept of black hole entropy as the measure of the inaccessibility of information to an exterior observer.
Furthermore, we consider that the black hole entropy is associated with the black hole area.

Bekenstein assumed that the entropy of a black hole ${\cal S}_{{\rm BH}}$ is some monotonically increasing function of its rationalized area as in (\ref{262area01}):
\begin{align}
{\cal S}_{{\rm BH}}=f(\alpha).
\label{263s01}
\end{align}
The entropy of an evolving thermodynamic system increases due to the gradual loss of information which is a consequence of the washing out of the most of the initial conditions.
Now, as a black hole approaches equilibrium, the effects of the initial conditions are also washed out (the black hole loses its hair).
One would thus expect that the loss of information about initial peculiarities of the black hole will be reflected in a gradual increase in 
${\cal S}_{{\rm BH}}$.
Indeed the relation (\ref{263s01}) predicts just this.

One possible choice for $f$ in (\ref{263s01}),
\begin{align}
f(\alpha) \propto \sqrt{\alpha},
\label{263s02}
\end{align}
is untenable on some reasons.
We consider two black holes which start off very distant from each other.
Since they interact weakly, we can take the total black hole entropy to be the sum of ${\cal S}_{{\rm BH}}$ of each black hole.
The black holes now move closer together and finally merge, and form a black hole which settles down to equilibrium.
In the process no information about the black hole interior can become available.
On the contrary, much information is lost as the final black hole loses its hair.
Thus, we expect the final black hole entropy to exceed the initial one.
By the assumption (\ref{263s02}), this implies that the irreducible mass (\ref{25irm01}) of the
final black hole exceeds the sum of irreducible masses of the initial black holes. 
Now suppose that all three black holes are Schwarzschild ($M=M_{{\rm ir}}$). 
We are then confronted with the prediction that the final black hole mass exceeds the initial one.
But this is nonsensical since the total black hole mass can only decrease due to gravitational radiation losses.
We thus see that the choice as in (\ref{263s02}) is untenable.

The next simplest choice for $f$ is
\begin{align}
f(\alpha)=\gamma \alpha,
\label{263s03}
\end{align}
where $\gamma$ is a constant.
Repetition of the above argument for this new $f$ leads to the conclusion that the final black-hole area must exceed the total 
initial black hole area.
But we know this to be true from Hawking's theorem \cite{haw03}. 
Thus the choice (\ref{263s03}) leads to no contradiction.
Therefore, Bekenstein adopted (\ref{263s03}) for the moment.

Comparison of (\ref{263s02}) and (\ref{263s03}) shows that $\gamma$ must have the units of [length]$^{-2}$. 
But there is no constant with such units in classical general relativity.
And so Bekenstein found only one truly universal constant $\hbar^{-1}$ with the correct units,
where $\hbar$ is the Planck constant.
Until now, although we used the natural system of units (\ref{1unit01}), 
we shall clearly describe the constant $\hbar$ in this subsection.
Thus Bekenstein represented (\ref{263s01}) as
\begin{align}
{\cal S}_{{\rm BH}}=\frac{\eta \alpha}{\hbar},
\label{263s04}
\end{align}
where $\eta$ is a dimensionless constant.
This expression was also proposed by Bekenstein earlier from a different point of view \cite{bek02}.
It is well known that the Planck constant $\hbar$ also appears in the formulas for the entropy in thermodynamics, 
for example, the Sackur-Tetrode equation (see, for example, \cite{thr01}).

To determine the value of $\eta$, Bekenstein considered that a particle falls into a Kerr-Newman black hole.
In Subsection 2.6.1, we showed that the loss of one bit of information before and after the particle with the least information falls into a black hole, i.e., the increased entropy is $\Delta {\cal S}=\ln 2$.
In Subsection 2.6.2, we showed that when a spherical particle with a radius, which is as large as the Compton wavelength, falls into a black hole, the minimum increase of the black hole area is given by (\ref{262area03}).
From (\ref{262area03}), we obtain the increase of black hole entropy given by
\begin{align}
(\Delta {\cal S}_{{\rm BH}})_{{\rm min}}=2\hbar \frac{df(\alpha)}{d\alpha}.
\label{263ent01}
\end{align}
Bekenstein conjectured that this entropy agrees with the loss of one bit of information (\ref{261s01}) (see Fig. 2.13).
We thus obtain
\begin{align}
2\hbar \frac{df(\alpha)}{d\alpha}=\ln 2.
\label{263ent02}
\end{align}
In the left-hand side of (\ref{263ent02}), the limit as in (\ref{262area04}) can be attained only for a particle whose dimension is given by its Compton wavelength.
Only such an ``elementary particle" may be regarded as having no internal structure.
We can thus consider that the loss of information associated with the loss of such a particle should be minimum.
By integrating (\ref{263ent02}) over $\alpha$, we obtain
\begin{align}
f(\alpha)=\left( \frac{1}{2} \ln 2 \right)\frac{\alpha}{\hbar}.
\end{align}
From (\ref{263s01}), we can obtain the black hole entropy
\begin{align}
{\cal S}_{{\rm BH}}=\left(\frac{1}{2} \ln 2 \right)\frac{\alpha}{\hbar}.
\end{align}
This form agrees with that of (\ref{263s04}).

\newpage

%%%%%%%%%%%%%     Fig.2.13       %%%%%%%%%%
\begin{center}
\vspace{0.6cm}
\input{beke01}\\
\vspace{0.6cm}
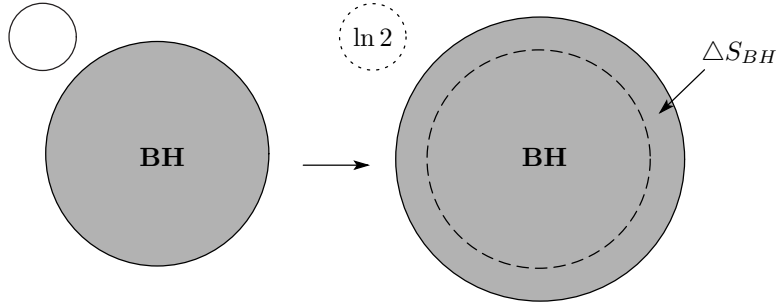
Fig. 2.13 \quad Information loss and increase of entropy
\vspace{0.6cm}
\end{center}
%%%%%%%%%%%%%%%%%%%%%%%%%%%%%%%%%%%%%%%%%%

Bekenstein showed the dependence of the black hole entropy ${\cal S}_{{\rm BH}}$ on the black hole area $\alpha$ from the above discussion,
and the black hole entropy is given by
\begin{align}
{\cal S}_{{\rm BH}}=\frac{\ln 2}{8\pi}\frac{kc^3}{\hbar G}A,
\label{263s05}
\end{align}
where we write the conventional units explicitly.
However, he used some conjectures, and the relation $\eta=\frac{1}{2} \ln 2$ is derived by a certain assumption.
The assumption is that the smallest possible radius of a particle is precisely equal to its Compton wavelength
whereas the actual radius is not so sharply defined.
Furthermore, an amount of information of such a particle might be more than $\ln 2$ 
because the particle has information for the mass and the radius.
According to the current understanding, the black hole entropy is given by
\begin{align}
{\cal S}_{{\rm BH}}=\frac{1}{4}\frac{kc^3}{\hbar G}A.
\label{263s06}
\end{align}
In comparison between (\ref{263s05}) and (\ref{263s06}), the value of $\eta$ is slightly different. 
However, Bekenstein had stated in his paper \cite{bek01} that it would be somewhat pretentious to attempt to calculate the precise value of the constant $\displaystyle \frac{\eta}{\hbar}$ without a full understanding of the quantum reality which underlies a ``classical" black hole.
Surprisingly, he already suggested that the derivation of black hole radiation needs the consideration of quantum theory.

Bekenstein also defined a characteristic temperature for a Kerr-Newman black hole by
\begin{align}
\frac{1}{{\cal T}_{{\rm BH}}}=\left( \frac{\partial {\cal S}_{{\rm BH}}}{\partial M} \right)_{L,Q},
\label{263tem01}
\end{align}
which is an analogue of the thermodynamic relation
\begin{align}
\frac{1}{{\cal T}}=\left( \frac{\partial {\cal S}}{\partial {\cal E}} \right)_{V}.
\label{263tem02}
\end{align}
By using both (\ref{262ene00}) and (\ref{262the01}) in (\ref{263tem01}), we can obtain
\begin{align}
{\cal T}=\frac{2\hbar}{\ln 2}\Theta_{{\rm H}}.
\end{align}
But Bekenstein did not regard this temperature as the temperature of the black hole.
Because if a black hole has a temperature, some radiation from the black hole may appear.
This conflicts with the classical definition.
By definition, a black hole can only absorb matter but cannot radiate matter.
Bekenstein did not suggest that a black hole has a temperature for the above reason.

%%%%%%%%%%%%%%%%%%%%%%%%%%%%%%%%%%%%%%%%%%%%%%%%%%%%%%%%%%%%%%%%%%%%%%%%%%%%%%%%%%%%%%%%%%%%%%%%%
%%                                                                                             %%
%%                                        Section 3                                            %%
%%                                                                                             %%
%%%%%%%%%%%%%%%%%%%%%%%%%%%%%%%%%%%%%%%%%%%%%%%%%%%%%%%%%%%%%%%%%%%%%%%%%%%%%%%%%%%%%%%%%%%%%%%%%
\newpage

\chapter{Black Hole Radiation}

\quad~ In the classical theory, black holes can only absorb matter but cannot radiate matter.
Bekenstein proposed that a black hole has entropy from the point of view of information theory but could not suggest that a black hole has a temperature from his reasoning.
Therefore, the complete correspondence between black hole physics and thermodynamics cannot be obtained.
However, Hawking showed that a black hole continuously radiates its energy by taking quantum effects into account \cite{haw01}.
Furthermore, it was found that a black hole behaves as the a black body with a certain temperature
and continuously performs its radiation.
This is consistent with the classical definition of black holes. 
This black hole radiation is commonly called Hawking radiation.

The intuitive explanation is as follows \cite{schu01}:
According to quantum field theory, 
it is considered that a particle-antiparticle pair is formed by a fluctuation of energy everywhere in our universe.
The antiparticle (negative energy state) formed by the pair creation can exist only for a very short time
since it is unstable in our universe.
The particle-antiparticle pair therefore vanishes by the pair annihilation after a certain short period of time.

Now we consider a pair creation very close to the event horizon outside a black hole.
The pair creation arises in globally curved space-time because general relativity is based on the assumption that the space-time can be made locally flat.
If a particle-antiparticle pair is formed very close to the horizon, 
the antiparticle (negative energy state) can fall into the black hole through the horizon in a certain short period of time.
It is possible to show that the antiparticle can be put into a realizable orbit inside the event horizon.
For an external observer, the black hole decreases its energy by absorbing the negative energy (antiparticle),
while the particle with the positive energy, which is the same amount as the decreased energy of the black hole, can escape to infinity,
since it can stably exist in our universe (Fig. 3.1).
Therefore, we can understand that black holes radiate particles.
This is the mechanism of Hawking radiation.

%%%%%%%%%%%%%     Fig.3.1       %%%%%%%%%%
\begin{center}
\vspace{0.6cm}
\input{haw01}\\
\vspace{0.6cm}
Fig. 3.1 \quad Hawking mechanism
\vspace{0.6cm}
\end{center}
%%%%%%%%%%%%%%%%%%%%%%%%%%%%%%%%%%%%%%%%%%

In this chapter, we would like to present several previous works on Hawking radiation.
For sake of simplicity, we consider the case of a Schwarzschild black hole.
The contents of this chapter are as follows.
In Section 3.1, we review Hawking's original derivation of Hawking radiation.
In Section 3.2, we would like to discuss some representative derivations of Hawking radiation briefly.
%%%%%%%%%%%%%%%%%%%%%%%%%%%%%%%%%%%%%%%%%%%%%%%%%%%%%%%%%%%%%%%%%%%%%%%%%%%%%%%%%%%%%%%%%%%%%%%%%
%%                                                                                             %%
%%                                        Subsection 3.1                                       %%
%%                                                                                             %%
%%%%%%%%%%%%%%%%%%%%%%%%%%%%%%%%%%%%%%%%%%%%%%%%%%%%%%%%%%%%%%%%%%%%%%%%%%%%%%%%%%%%%%%%%%%%%%%%%
\section{Hawking's Original Derivation}

\quad~ Hawking showed that black holes radiate matter by using quantum field theory in black hole physics.
In this section, we would like to show the original derivation of Hawking radiation by Hawking \cite{haw01}.

For sake of simplicity, we consider a free massless scalar field.
In Minkowski space, it satisfies the Klein-Gordon equation
\begin{align}
\eta^{\mu\nu}\partial_{\mu} \partial_{\nu}\phi=0,
\end{align}
where $\phi$ is a massless hermitian scalar field, $\eta^{\mu\nu}$ is the Minkowski metric (\ref{2min01})
and $\partial_\mu$ is the partial derivative defined by
\begin{align}
\partial_\mu \equiv \frac{\partial}{\partial x^\mu}.
\end{align}
The ordinary derivative of $\phi$ is also written as $\phi_{,\mu}$.
We can decompose the field into positive and negative frequency components
\begin{align}
\phi=\sum_i(\varphi_i\bm{a}_i + \varphi^*_i \bm{a}_i^{\dagger}),
\label{31phi01}
\end{align}
where $\{ \varphi_i \}$ are a complete orthonormal family of complex valued solutions of the wave equation
\begin{align}
\eta^{\mu\nu}\partial_\mu \partial_\nu \varphi=0.
\end{align}
It contains only positive frequencies with respect to the usual Minkowski time coordinate.
The operators $\bm{a}_i$ and $\bm{a}_i^\dagger$ are interpreted as the annihilation and creation operators respectively for particles in the $i$-th state.
The vacuum state $|0\rangle$ is defined to be the state from which one cannot annihilate any particle, i.e.,
\begin{align}
\bm{a}_i | 0 \rangle =0, \qquad {\rm for~all~} i.
\end{align}
The orthonormal condition is then given by
\begin{align}
\rho_M(\varphi_i,\varphi_j)\equiv
\frac{1}{2}i \int_V ( \varphi_i \partial_t \varphi^*_{j} - \varphi^*_j \partial_t \varphi_{i})d x^3=\delta_{ij},
\end{align}
where $V$ is a suitable closed space.

We considered quantum field theory in Minkowski space so far.
Here, we would like to extend Minkowski space-time to curved space-time which is produced by the intense gravity of a black hole.
In the curved space-time, the metric changes from the Minkowski metric to the metric of the curved space-time.
Also physical laws must hold in any coordinate system.
The partial derivatives contained in these laws must be replaced by the covariant derivatives in the curved space-time \cite{dir01}.
The covariant derivative is commonly represented by
\begin{align}
\nabla_{\mu}\equiv \phi_{;\mu}.
\label{31cov01}
\end{align}
The covariant derivative for a scalar field $\phi$ is given by
\begin{align}
\nabla_\mu \phi=\partial_\mu \phi,
\label{31cov02}
\end{align}
and the covariant derivative for a vector field $A_{\nu}$ is given by
\begin{align}
\nabla_\mu A_{\nu}=\partial_{\mu} A_\nu + \Gamma^\alpha_{\nu\mu}A_{\alpha},
\label{31cov03}
\end{align}
where $\Gamma^\alpha_{\nu\mu}$ is the Christoffel symbol defined by (\ref{2chr01}).
In curved space-time, the Klein-Gordon equation for a massless hermitian scalar field is thus represented by
\begin{align}
g^{\mu\nu}\nabla_{\mu}\nabla_{\nu}\phi=0.
\label{31kg01}
\end{align}
We can use the relation (\ref{31phi01}) in the flat space.
However, we cannot decompose the field into positive and negative frequency components in curved space.
One can still require that the $\{ f_i \}$ and the $\{ f^*_i \}$ together form a complete the basis for solutions of the wave equations with
\begin{align}
\rho(\varphi_i,\varphi_j)=-\frac{1}{2}i \int_\Sigma
( \varphi_i \nabla_{\mu} \varphi^*_{j} - \varphi^*_j \nabla_{\mu} \varphi_{i} ) d \Sigma^\mu=\delta_{ij},
\label{31nor01}
\end{align}
where $d\Sigma$ stands for an area element and $\Sigma$ is called a Cauchy surface which represents a suitable surface.

Here we recall the Penrose diagram drawn in Fig. 2.7.
In the past null infinity ${\cal J}^-$, the Schwarzschild metric is asymptotically flat (the Minkowski metric) since $r \to \infty$.
We can thus expand the field operator $\phi$ which satisfies the Klein-Gordon equation (\ref{31kg01}) as
\begin{align}
\phi=\sum_i \{ f_i \bm a_i + f^*_i \bm a_i ^\dagger \},
\label{31phi02}
\end{align}
where $\{ f_i \}$ is a family of solutions of the wave equation
\begin{align}
\eta^{\mu\nu}\nabla_\mu \nabla_\nu f_i=0.
\end{align}
In a manner similar to (\ref{31nor01}), they satisfy the orthonormal condition at ${\cal J}^-$
\begin{align}
\rho(f_i, f^*_j)=\frac{1}{2}i \int_\Sigma 
( f_i \nabla_{\mu}f^*_{j} - f^*_j \nabla_\mu f_{i})d \Sigma^\mu=\delta_{ij},
\label{31nor02}
\end{align}
where we note that $\{ f_i \}$ only contain positive frequencies with respect to the canonical affine parameter on ${\cal J}^-$.
It is natural that the operators $\bm a_i$ and $\bm a_i^\dagger$ are respectively regarded as the annihilation and creation operators at ${\cal J}^-$.
The vacuum at ${\cal J}^-$ is thus defined by
\begin{align}
\bm a_i |0_-\rangle=0.
\label{31vac01}
\end{align}
Similarly, in the future null infinity ${\cal J}^+$, the Schwarzschild metric is asymptotically flat since $r \to \infty$.
We can also expand the field operator $\phi$ by
\begin{align}
\phi = \sum_i \left\{ p_i \bm b_i + p^*_i \bm b_i ^\dagger + q_i \bm c_i + q^*_i \bm c_i ^\dagger 
\right\},
\label{31phi03}
\end{align}
where $\{ p_i \}$ are solutions of the wave equation which can escape to ${\cal J}^+$ and
$\{ q_i \}$ are solutions of the wave equation which cannot escape to ${\cal J}^+$ since they are absorbed by the future event horizon ${\cal H}^+$, namely, $\{ p_i \}$ are zero at ${\cal H}^+$ and $\{ q_i \}$ are zero at ${\cal J}^+$.
The operators $\bm b_i$ and $\bm b_i ^\dagger$ respectively stand for the annihilation and creation operators at ${\cal J}^+$, and 
the operators $\bm c_i$ and $\bm c_i ^\dagger$ respectively stand for the annihilation and creation operators at ${\cal H}^+$.
The vacua at ${\cal J}^+$ and ${\cal H}^+$ are thus defined by
\begin{align}
\bm b_i|0_+\rangle=0,\\
\bm c_i|0_{{\mathcal H}^+}\rangle=0,
\end{align}
where we also note that $\{ p_i \}$ contain positive frequencies only with respect to the canonical affine parameter on ${\cal J}^+$.
Although it is not clear whether one should impose some positive frequency condition on $\{ q_i \}$,
we would like to consider particles which start from ${\cal J}^-$, 
pass through the collapsing body and can escape to ${\cal J}^+$.
The choice of the $\{ q_i \}$ does not affect the calculation of the emission of particle to ${\cal J}^+$ 
since the $\{ q_i \}$ are zero at ${\cal J}^+$.
We require that $\{ p_i \}$ and $\{ p^*_i \}$ are a complete orthonormal family which satisfies
\begin{align}
\rho'(p_i,p^*_j)=\frac{1}{2}i \int_{\Sigma'} 
( p_i \nabla_\mu p^*_{j} - p^*_j \nabla_\mu p_{i})d \Sigma'^\mu=\delta_{ij}.
\label{31nor03}
\end{align}

Here we would like to show that the relation (\ref{31nor03}) is satisfied even if one uses $\Sigma$ which appeared in (\ref{31nor02}) instead of $\Sigma'$.
We thus consider $\rho(p_i,p^*_j)-\rho'(p_i,p^*_j)$.
If the stable surface $\Sigma'$ differs from $\Sigma$, $\Sigma'$ can smoothly intersect with $\Sigma$ at certain points 
since $\Sigma'$ is not parallel to $\Sigma$.
We represent the 4-dimensional volume enclosed by these two surfaces as $V$ (Fig. 3.2).
By using the 4-dimensional Gauss theorem, we obtain
\begin{align}
\rho(p_i,p^*_j)-\rho'(p_i,p^*_j)
=\int_V d^4x\sqrt{-g} \nabla^{\mu}(p_i \nabla_{\mu}p^*_{j} - p^*_j \nabla_{\mu}p_{i}),
\label{31nor04}
\end{align}
where $g$ stands for the determinant of the metric $g_{\mu\nu}$ defined by
\begin{align}
g\equiv \det \left( g_{\mu\nu} \right),
\end{align}
namely, $\sqrt{-g}$ stands for the Jacobian with respect to the transformation from $d^4 x$ to $d\Sigma$.
By calculating the integrand of the right-hand side in (\ref{31nor04}), we obtain
\begin{align}
\nabla^{\mu}(p_i\nabla_{\mu}p^*_{j} - p^*_j \nabla_{\mu}p_{i})
=&\nabla^{\mu}p_i \nabla_{\mu}p^*_{j}+p_i\nabla^{\mu}\nabla_{\mu}p^*_{j}
-\nabla^{\mu}p^*_j \nabla_{\mu}p_{i}-p^*_j \nabla^{\mu}\nabla_{\mu}p_{i}\notag\\
=&p_i \nabla^{\mu} \nabla_{\mu}p^*_{j}-p^*_j \nabla^{\mu} \nabla_{\mu}p_{i}.
\label{31nor05}
\end{align}
By using the Klein-Gordon equation (\ref{31kg01}), the relation (\ref{31nor05}) becomes
\begin{align}
\nabla^{\mu}(p_i\nabla_{\mu}p^*_{j} - p^*_j \nabla_{\mu}p_{i})=0.
\end{align}
Since the right-hand side of (\ref{31nor03}) is zero, we obtain
\begin{align}
\rho(p_i,p^*_j)=\rho'(p_i,p^*_j).
\end{align}
This means that $\rho(p_i,p^*_j)$ does not depend on $\Sigma$.
Namely, if the Gauss theorem is satisfied, 
it means that we can freely choose the surface $\Sigma$ in (\ref{31nor01})
\begin{align}
\rho(p_i,p^*_j)=\frac{1}{2}i \int_{\Sigma} 
( p_i \nabla_\mu p^*_{j} - p^*_j \nabla_\mu p_{i})d \Sigma^\mu=\delta_{ij}.
\label{31nor06}
\end{align}
It is known that the above discussion is also valid for a scalar field with a mass \cite{dew01}.

%%%%%%%%%%%%%     Fig.3.2       %%%%%%%%%%
\begin{center}
\vspace{0.6cm}
\input{nor01}\\
\vspace{0.6cm}
Fig. 3.2 \quad The volume $V$ enclosed by two surfaces $\Sigma$ and $\Sigma'$
\vspace{0.6cm}
\end{center}
%%%%%%%%%%%%%%%%%%%%%%%%%%%%%%%%%%%%%%%%%%

In the transitional time between $\{ f_i \}$ and $\{ p_i \}$, a collapsing body will appear.
We do not know the corresponding solutions since we do not know the metric inside this region.
By using the analogy of the tunneling effect, we can represent $\{ p_i \}$ which appear at ${\cal J}^+$ as the linear combinations of $\{ f_i \}$ with $\{ f^*_i \}$ 
\begin{align}
p_i = \sum_j (\alpha _{ij} f_j + \beta _{ij} f^*_j ),
\label{31pi01}
\end{align}
where $\alpha _{ij}$ and $\beta _{ij}$ are proportionality coefficients which stand for the amplitude ratio (Fig. 3.3).
We substitute (\ref{31pi01}) into (\ref{31phi03}).
Since $\{ q_i \}=0$ at ${\cal J}^+$, the relation (\ref{31phi03}) becomes
\begin{align}
\phi=\sum_{i}\left\{ \sum_j(\bm b_j \alpha_{ij} +\bm b_j^\dagger \beta^*_{ij}) f_{i}
+\sum_j(\bm b_j \beta_{ij}+\bm b_j^\dagger\alpha^*_{ij})f^*_{i}\right\}.
\label{31phi04}
\end{align}
By comparison between (\ref{31phi02}) and (\ref{31phi04}), we obtain
\begin{align}
\bm a_i&=\sum_j (\bm b_j \alpha_{ij} +\bm b_j^\dagger \beta^*_{ij}),
\label{31ai01}\\
\bm a_i^\dagger&=\sum_j (\bm b_j \beta_{ij}+\bm b_j^\dagger\alpha^*_{ij}).
\label{31ai02}
\end{align}
We also find that the inverse transformations with respect to $\bm b_j$ and $\bm b_j^\dagger$ are obtained by
\begin{align}
\bm b_i &= \sum_j (\alpha^* _{ij} \bm a_j - \beta^* _{ij} \bm a_j^\dagger ),
\label{31bi01}\\
\bm b_i^\dagger&=\sum_j (\alpha_{ij} \bm a_j^\dagger - \beta_{ij} \bm a_j).
\label{31bi02}
\end{align}
These transformations are called the Bogoliubov transformations.
For details of this calculation, see Appendix C.

%%%%%%%%%%%%%     Fig.3.3       %%%%%%%%%%
\begin{center}
\vspace{0.6cm}
\input{bog01}\\
\vspace{0.6cm}
Fig. 3.3 \quad The relationship between $f_i$ and $p_i$
\vspace{0.6cm}
\end{center}
%%%%%%%%%%%%%%%%%%%%%%%%%%%%%%%%%%%%%%%%%%

We already defined the initial vacuum as in (\ref{31vac01}).
By operating the annihilation operator $\bm b_i$ on the initial vacuum state $|0_-\rangle$, we obtain
\begin{align}
\bm b_i |0_-\rangle &=\sum_j (\alpha^* _{ij} \bm a_j - \beta^* _{ij} \bm a_j^\dagger )|0_-\rangle
\notag\\
&=\sum_j - \beta^* _{ij} \bm a_j^\dagger |0_-\rangle \neq 0.
\end{align}
Since $\beta_{ij}\neq 0$ in general, the initial vacuum state cannot be regarded as a vacuum state for an observer on ${\cal J}^+$.
This means that particles are created.

We would like to find how many particles are created at ${\cal J}^+$ from the initial vacuum $|0_-\rangle$.
By using the number operator $\bm{N}_i$ defined by
\begin{align}
\bm{N}_i \equiv \bm b_i^\dagger \bm b_i,
\end{align}
we find that the vacuum expectation value of the particle number is given by
\begin{align}
N_i & \equiv \langle 0_- | \bm N_i | 0_- \rangle\\
&=\langle 0_- | \bm b_i^\dagger \bm b_i| 0_- \rangle\\
&=\sum_{j,k}\langle 0_- |\beta_{ik} \beta^*_{ij}\bm a_k \bm a_j^\dagger| 0_- \rangle.
\label{31num01}
\end{align}
By using the commutation relation of the creation-annihilation operators given by
\begin{align}
[\bm a_i,\bm a_j^\dagger]=\delta_{ij},
\end{align}
the relation (\ref{31num01}) becomes
\begin{align}
N_i&=\sum_{j,k}\beta_{ik}\beta^*_{ij}\delta_{jk}=\sum_j |\beta_{ij}|^2.
\label{31num02}
\end{align}
This stands for the number of particles which propagate to infinity among the particle pairs created by the vacuum.
To determine the value, we calculate the coefficients $\beta_{ij}$.

By actually solving the Klein-Gordon equation (\ref{31kg01}), we obtain
\begin{align}
f_{\omega' lm}&=\frac{F_{\omega'}(r)}{r\sqrt{2\pi \omega'}}e^{i\omega' v}Y_{lm}(\theta,\varphi),
\label{31fo01}\\
p_{\omega l m}&=\frac{P_{\omega}(r)}{r\sqrt{2\pi \omega}}e^{i\omega u}Y_{lm}(\theta,\varphi).
\label{31po01}
\end{align}
For details of these calculations, see Appendix D.
Here $Y_{lm}(\theta,\varphi)$ is the spherical harmonics, $l$ is an azimuthal quantum number, $m$ is a magnetic quantum number.
The frequencies $\omega$ and $\omega'$ are eigenvalues given by
\begin{align}
i \partial_t f_{\omega' lm}&=\omega' f_{\omega' lm},\\
i \partial_t p_{\omega lm}&=\omega f_{\omega lm}.
\end{align}
Since the index of the state $i$ is uniquely determined by $\omega'$, $l$ and $m$,
we represent $f_i$ as $f_{\omega' l m}$.
The advanced time $v$ is an affine parameter at ${\cal J}^-$.
The retarded time $u$ is an affine parameter at ${\cal J}^+$.
They are defined as in (\ref{2sch07}).
The solutions $f_{\omega' lm}$ and $p_{\omega l m}$ are obtained by approximating the Klein-Gordon equation at $r \to \infty$.
Thus $F_{\omega'}(r)$ and $P_{\omega}(r)$ are integration constants containing a tiny effect depending on $r$.

By taking a continuous limit in (\ref{31pi01}), the relation (\ref{31pi01}) can be represented as
\begin{align}
p_{\omega} = \int_0 ^\infty (\alpha_{\omega \omega'} f_{\omega '}+
\beta_{\omega \omega'} f^*_{\omega '}) d \omega ',
\label{31po02}
\end{align}
where we dropped indices $l$ and $m$ 
since the wave functions with different indices $l$ and $m$ are not connected to each other in a spherically symmetric system.
In the continuous limit, the relations (\ref{31bi01}) and (\ref{31num02}) become
\begin{align}
\bm b_\omega&=\int_0 ^\infty
(\alpha_{\omega \omega'}\bm a_{\omega'} - \beta^*_{\omega \omega'}\bm a_{\omega'}^\dagger )d\omega',\\
N_\omega &= \int_0 ^\infty|\beta_{\omega \omega'}|^2d\omega'.
\label{31num03}
\end{align}
We can evaluate $\alpha_{ij}$ and $\beta_{ij}$ by performing the Fourier transform in (\ref{31po02}).
By substituting (\ref{31fo01}) into (\ref{31po02}) and multiplying the both sides by 
$\displaystyle \int_{-\infty}^\infty dv \exp(-i\omega'' v)$,
we obtain
\begin{align}
\int_{-\infty}^\infty dv  e^{-i\omega''v}p_\omega
=2\pi\int_0^{\infty}d\omega'
\left[ 
\alpha_{\omega \omega'}\frac{F_{\omega'}}{r\sqrt{2\pi \omega'}}\delta(\omega'-\omega'')
+\beta_{\omega \omega'}\frac{F_{\omega'}}{r\sqrt{2\pi \omega'}}\delta(-\omega'-\omega'')
\right].
\end{align}
The second term on the right-hand side vanishes since $(\omega'+\omega'')\neq 0$.
We thus obtain 
\begin{align}
\alpha_{\omega \omega'}=\frac{r\sqrt{\omega'}}{\sqrt{2\pi}F_{\omega'}}
\int_{-\infty}^\infty dv  e^{-i\omega'v}p_\omega.
\label{31alp01}
\end{align}
As for $\beta_{ij}$, we similarly obtain
\begin{align}
\beta_{\omega \omega'}=\frac{r\sqrt{\omega'}}{\sqrt{2\pi}F_{\omega'}}
\int_{-\infty}^\infty dv  e^{i\omega'v}p_\omega.
\label{31bet01}
\end{align}

Both (\ref{31alp01}) and (\ref{31bet01}) contain $u$ and $v$.
We can derive the relation of between $u$ and $v$ from the following connection condition:
We consider the wave function $p_\omega$ which reached ${\cal J}^+$.
When we view it backwards, we can divide the wave function into two groups by how they propagate.
The first group, which will be scattered by the Schwarzschild field outside the collapsing body, will propagate to ${\cal J}^-$.
Then the wave function $p_\omega^{(1)}$ keeps the same frequency $\omega$ and propagate at ${\cal J}^-$.
The second group will enter the collapsing body where it will be partly scattered and partly reflected through the center,
eventually emerging to ${\cal J}^-$.
It is this part $p_\omega^{(2)}$ which produces the interesting effect.
Since the retarded time $u$ is infinite at the horizon, it is considered that the effective frequency of $p_\omega^{(2)}$ is enormous near the horizon.
When the frequency is enormous, we can use the geometrical optics approximation.
It means that the scattering of the wave function by the gravitational field can be ignored.
We use the Penrose diagram in order to analyze the phase of $p_\omega^{(2)}$ (Fig. 3.4).

\newpage

%%%%%%%%%%%%%     Fig.3.4       %%%%%%%%%%
\begin{center}
\vspace{0.6cm}
\input{pen08}\\
\vspace{0.6cm}
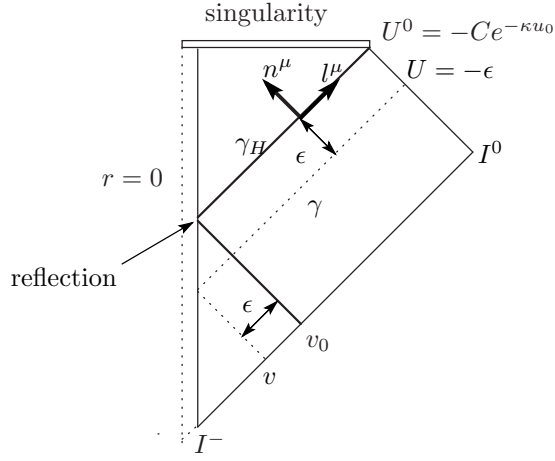
Fig. 3.4 \quad Penrose diagram for the phase analysis of $p_i$
\vspace{0.6cm}
\end{center}
%%%%%%%%%%%%%%%%%%%%%%%%%%%%%%%%%%%%%%%%%%

Although the Penrose diagram which contains the collapsing body is represented as in Fig. 2.7,
we omit the collapsing body in Fig. 3.4 since we now consider the case that the frequency is enormous near the horizon.
A coordinate $x$ is a point on the horizon outside the collapsing body,
$l^\mu$ is a null vector which is tangent to the horizon at $x$ and
$n^\mu$ is a null vector which is normal to the horizon at $x$ and is directed radially inwards.
They are normalized so that
\begin{align}
l^\mu n_\mu=-1.
\end{align}
The line ${\cal J}^+$ intersects the line ${\cal H}^+$ at a point.
We represent the point by $u_0$ which is the affine parameter on ${\cal J}^+$.
The $\gamma_{{\rm H}}$ is a null geodesic which goes back from ${\cal J}^+$ to backward.
The null geodesic $\gamma_{{\rm H}}$ goes along the future event horizon ${\cal H}^+$,
is reflected by following geometrical optics at $r=0$ and reaches ${\cal J}^-$.
We then represent the point on ${\cal J}^-$ by $v_0$ which is the affine parameter on ${\cal J}^-$.
According to the Penrose diagram, an affine parameter $v$ on ${\cal J}^-$ is larger as it goes from ${\cal I}^-$ to ${\cal I}^0$.
Thus, $v_0$ is the latest time that it leaves ${\cal J}^-$, passes through the center of the collapsing body and can escape to ${\cal J}^+$.
Since the affine parameter $v$ which is larger than $v_0$ goes into the black hole through the horizon, it cannot escape to ${\cal J}^+$.
For a very small constant $\epsilon$, a vector $-\epsilon n^\mu$ has a constant affine parameter $u$ on ${\cal J}^-$.
From (\ref{31po01}), the phase of the wave function $p_{\omega}^{(2)}$ is constant.
We can also use geometrical optics approximation near the horizon because of a very small constant $\epsilon$.
If null vectors $l^\mu$ and $n^\mu$ are translated along the null geodesic $\gamma_{{\rm H}}$ (the heavy line in Fig. 3.4),
the phase of $p_{\omega}^{(2)}$ on the null geodesic $\gamma$ generated by $-\epsilon n^\mu$ 
remains constant in the case that it enters inside the collapsing body.
The null geodesic $\gamma$ is also reflected by geometrical optics at $r=0$ and reaches ${\cal J}^-$.

The parameter $U$ is an affine parameter on the past event horizon ${\cal H}^-$.
The parameter $U$ is such that at the point of intersection of the two horizon, $U=0$ and $\displaystyle \frac{dx^\mu}{dU} =n^\mu$.
The affine parameter $U$ is related to the retarded time $u$ on the past horizon by
\begin{align}
U=-Ce^{-\kappa u},
\label{31kru01}
\end{align}
where $C$ is a constant and $\kappa$ is the surface gravity of the black hole defined by
\begin{align}
\nabla_\nu K^\mu K^\nu=- \kappa K^\mu,
\end{align}
with $K^\mu$ the time translation Killing vector.
By using this definition, we find that the surface gravity of a Schwarzschild black hole is given by
\begin{align}
\kappa = \frac{1}{4M}.
\label{31sg01}
\end{align}
The affine parameter $U$ is zero on the future horizon ${\cal H}^+$ and 
it satisfies $U=-\epsilon$ on the null geodesic $\gamma$ near the horizon.
From (\ref{31kru01}), we obtain
\begin{align}
u=-\frac{1}{\kappa}(\ln \epsilon - \ln C).
\label{31uv01}
\end{align}
The phase of the wave function $p_{\omega}^{(2)}$ is connected to a point on ${\cal J}^-$ along $\gamma$.
We represent the point by an affine parameter $v$.
As found from Fig. 3.4, we obtain $\epsilon=v_0-v$ on ${\cal J}^-$.
Since the vector $n^\mu$ on ${\cal J}^-$ is parallel to the Killing vector $K^\mu$, the vector $n^\mu$ is given by
\begin{align}
n^\mu =D K^\mu,
\end{align}
where $D$ is a constant.
We thus find that $p_{\omega}^{(2)}$ is zero for $v>v_0$ because the particle is captured by the black hole and cannot escape to ${\cal J}^+$, while the phase of $p_{\omega}^{(2)}$ is given by (\ref{31uv01}) for $v>v_0$.
The wave function then becomes
\begin{align}
p_{\omega}^{(2)}\sim 
\left\{ \begin{array}{lll}0,&\qquad(v>v_0),\\
\displaystyle \frac{P^-_{\omega}}{r\sqrt{2\pi \omega}}
\exp\left[ -i \frac{\omega}{\kappa}\ln \left( \frac{v_0-v}{CD}\right)\right],&\qquad(v \leq v_0),
\end{array}
\right.
\label{31po03}
\end{align}
where we used the fact that $v_0-v$ is small and positive, and the definition $P^-_{\omega} \equiv P_{\omega}(2M)$.
If we assume that $\omega'$ is very large, these would be determined by the asymptotic form
\begin{align}
p_{\omega}^{(2)}\sim \frac{P^-_{\omega}}{r\sqrt{2\pi \omega}}
\exp\left[ -i \frac{\omega}{\kappa}\ln \left( \frac{v_0-v}{CD}\right)\right].
\end{align}
We can actually perform integrations of both (\ref{31alp01}) and (\ref{31bet01}).
As a result, we obtain
\begin{align}
\alpha_{\omega \omega'} ^{(2)} &\approx \frac{1}{2 \pi} P_{\omega} ^-(CD)^{\frac{i \omega}{\kappa}} 
e^{-i \omega ' v_0} \left( \sqrt{\frac{\omega '}{\omega}} \right) 
\Gamma \left( 1 -\frac{i \omega }{\kappa} \right) (-i \omega ')^{-1+\frac{i\omega}{\kappa}}, 
\label{31alp02}\\
\beta_{\omega \omega'} ^{(2)} &\approx -i\alpha_{\omega (-\omega')} ^{(2)}.
\label{31bet02}
\end{align}
For details of these calculations, see Appendix E.
By expressing $\beta_{\omega \omega'}^{(2)}$ in terms of $\alpha_{\omega \omega'}^{(2)}$ from both (\ref{31alp02}) and (\ref{31bet02}),
we obtain
\begin{align}
\beta_{\omega \omega'}^{(2)}=e^{2i\omega ' v_0}e^{[i\frac{\omega-1}{\kappa}]\ln(-1)}\alpha_{\omega \omega'} ^{(2)},
\label{31bet03}
\end{align}
where we take $\ln (-1) = i\pi$ because we used the anticlockwise continuation around the singularity $\omega'=0$.
By substituting this relation into (\ref{31bet03}) and taking the absolute value,
we obtain
\begin{align}
|\beta_{\omega \omega'} ^{(2)}|
= e^{-\frac{\pi \omega}{\kappa}} |\alpha_{\omega \omega'} ^{(2)}|,
\label{31ab01}
\end{align}
where we note that this relation is valid for the large values of $\omega'$.

The expectation value of the total number of created particles at ${\cal J}^+$ in the frequency range $\omega$ to $\omega + d\omega$ is
$\displaystyle d\omega\int^{\infty}_0d\omega'|\beta_{\omega \omega'}|^2$.
Since $|\beta_{\omega \omega'}|$ behaves as $(\omega')^{-\frac{1}{2}}$ by (\ref{31bet02}),
this integral logarithmically diverges.
It is considered that this infinite total number of created particles is caused since we evaluate a finite steady rate of emission for an infinite time.
To evaluate the finite rate of emission, Hawking defined wave packets $p_{jn}$ by
\begin{align}
p_{jn}^{(2)}\equiv \varepsilon ^{-\frac{1}{2}} \int_{j\varepsilon} ^{(j+1)\varepsilon} 
e^{-\frac{2\pi i n \omega}{\varepsilon}} 
p_{\omega}^{(2)} d\omega,
\label{31pjn01}
\end{align}
where $j$ and $n$ are integers, $j \geq 0$, $\varepsilon \geq 0$.
For small $\varepsilon$ these wave packets would have frequency $j\varepsilon$ and would be peaked around retarded time $\displaystyle u=\frac{2\pi n}{\varepsilon}$.
We can expand $\{ p_{jn} \}$ in terms of $\{ f_\omega \}$
\begin{equation}
p_{jn}^{(2)}= \int_0 ^\infty (\alpha_{jn\omega'}^{(2)} f_{\omega'}+ \beta_{jn\omega'}^{(2)} f^*_{\omega '})d\omega'.
\label{31pjn02}
\end{equation}
By comparing (\ref{31pjn02}) with the relation (\ref{31pjn01}) which is obtained by using (\ref{31po01}), 
we find that the proportionality coefficient $\alpha_{jn\omega'}$ is defined by
\begin{align}
\alpha_{jn \omega'}^{(2)}=\frac{1}{\sqrt{\varepsilon}} \int_{j\epsilon} ^{(j+1)\varepsilon}
e^{-2\pi i n \omega}{\epsilon}\alpha_{\omega \omega'}^{(2)} d\omega.
\label{31ajn01}
\end{align}
By substituting (\ref{31alp02}) into (\ref{31ajn01}) for $j\gg \varepsilon$ and $n \gg \varepsilon$, we obtain
\begin{align}
|\alpha_{jn \omega'}^{(2)}|
&= \left| \frac{P_\omega ^-}{2\pi \sqrt{\omega }} 
\Gamma \left( 1- \frac{i \omega}{\kappa}\right) \frac{1}{\sqrt{\varepsilon \omega'}}\right|
\int_{j\varepsilon}^{(j+1)\varepsilon} \exp \left[ i \omega'' \left( -\frac{2\pi n}{\varepsilon} 
+ \frac{\log \omega '}{\kappa} \right) d\omega ''\right] \notag\\
&=\left| \frac{P_{\omega}^-}{\pi \sqrt{\omega}} \Gamma \left( 1- \frac{i\omega}{\kappa}\right)
\frac{ \sin{\frac{1}{2}\varepsilon z}}{z\sqrt{\varepsilon \omega'}}\right|,
\end{align}
where $\omega=j\epsilon$ and $\displaystyle z=\frac{1}{\kappa}=\ln\omega'-\frac{2\pi n}{\varepsilon}$.
In these transformations, the relation (\ref{31ab01}) remains unchanged
\begin{align}
|\beta_{jn \omega'}^{(2)}|=e^{-\frac{\pi \omega}{\kappa}}|\alpha_{jn \omega'}^{(2)}|.
\label{31ab02}
\end{align}
Since the proportionality coefficient $|\beta_{jn\omega'}|$ thus behaves as $\displaystyle \sqrt{\frac{\varepsilon}{\omega'}}$,
we can control the logarithmic divergence of the integral by an effect of $\varepsilon$.
Therefore, the expectation value of the number of particles created and emitted to infinity ${\cal J}^-$ in the wave-packet mode $p_{jn}$, is given by
\begin{align}
N_{jn}=\int_0^\infty |\beta_{jn \omega'}^{(2)}|^2 d\omega'.
\label{31njn01}
\end{align}

We have considered the wave-packet $p_{jn}$ propagating backwards from ${\cal J}^+$.
Until now, we have ignored the change in the amplitude of the wave function.
However, a fraction of the particles would actually be scattered at the horizon.
As a result, a fraction of the wave packet with $\rho(f_{jn}, f^*_{jn})=1$ as in (\ref{31nor02})
will be scattered by the static Schwarzschild field and the others will enter the collapsing body.
Then the wave packets which reach ${\cal J}^+$ would satisfy $\rho(p_{jn}, p^*_{jn})=\Gamma_{jn}<1$ where $\Gamma_{jn}$ is called 
the gray body factor.
The orthonormal condition (\ref{31nor06}) would become
\begin{align}
\Gamma_{jn} = \int_0 ^\infty \left( | \alpha _{jn\omega'} ^{(2)}|^2 -|\beta_{jn\omega'} ^{(2)}|^2\right)d\omega '.
\label{31gam01}
\end{align}
By substituting (\ref{31ab02}) into (\ref{31gam01}), we obtain
\begin{align}
\int^\infty_{0}|\beta_{jn\omega'} ^{(2)}|^2d\omega '=\frac{\Gamma_{jn}}{\exp(\frac{2\pi \omega}{\kappa})-1}.
\label{31bg01}
\end{align}
From (\ref{31njn01}), the relation (\ref{31bg01}) is written as
\begin{align}
N_{jn}=\frac{\Gamma_{jn}}{\exp(\frac{2\pi \omega}{\kappa})-1}.
\end{align}
This stands for the total number of particles created in the mode $p_{jn}^{(2)}$.
If we ignore the gray body factor, the total number of particles $N$ is given by
\begin{align}
N=\frac{1}{\exp(\frac{2\pi \omega}{\kappa})-1}.
\label{31bla01}
\end{align}

In thermodynamics, the total number of particles for the black body radiation obeying Bose-Einstein statistics is given by
\begin{align}
N=\frac{1}{\exp\left( \frac{\omega}{{\cal T}}\right)-1},
\end{align}
where $\omega$ is a frequency of the particle and ${\cal T}$ is temperature of the system.
We thus find that a black hole which has Hawking temperature ${\cal T}_{{\rm BH}}$ defined by
\begin{align}
{\cal T}_{{\rm BH}}=\frac{\kappa}{2\pi}
\label{31bht01}
\end{align}
behaves as a black body and the black hole continuously emits radiation. 
Here $\kappa$ is the surface gravity of the black hole 
and we find that the temperature of the black hole is proportional to its surface gravity as already conjectured by the corresponding relationship between black hole physics and thermodynamics.
We also find the black hole entropy $d{\cal S}_{{\rm BH}}$ from a thermodynamic consideration
\begin{align}
d{\cal S}_{{\rm BH}}= \left( \frac{dM}{{\cal T}_{{\rm BH}}} \right).
\label{31ent01}
\end{align}
By integrating Eq. (\ref{31ent01}), we obtain
\begin{align}
{\cal S}_{{\rm BH}}=\frac{A}{4},
\label{31bhe01}
\end{align}
where $A$ is the black hole area and we find that the black hole entropy is proportional to its area.
Since it was shown that a black hole can radiate matter,  we can regard that a black hole has temperature and entropy.

From the above result, Hawking suggested that a black hole can evaporate.
The temperature of a Schwarzschild black hole is given by
\begin{align}
{\cal T}_{{\rm BH}}=\frac{1}{8\pi M},
\label{31bht01}
\end{align}
where we used $\displaystyle \kappa=\frac{1}{4M}$ in (\ref{31bht01}).
Namely the temperature of the black hole is inversely proportional to its mass.
This means that the temperature is higher as the mass is smaller and the temperature is lower as the mass is larger (Tab. 3.1).
It is known that the temperature for a black hole of the solar mass is much lower than the temperature of the cosmic microwave background radiation.
Thus black holes of this size would be absorbing radiation faster than they emitted it and would be increasing its mass.
However, there might be tiny black holes in the early universe \cite{haw04,ber01}.
If the temperature of a tiny black hole is higher than the temperature of the cosmic microwave background radiation,
such tiny black holes would be radiation-dominated.
As this tiny black hole radiates matter, the mass becomes smaller, the temperature becomes higher and then
it increasingly radiates matter.
It would thus be expected that the black hole will evaporate at some point.

%%%%%%%%%%%%%%%%%       Tab. 3.1        %%%%%%%%%%%%%%%%%%%%%
\begin{center}
\vspace{0.6cm}
Tab. 3.1 \quad The behavior of black hole
\vspace{0.3cm}\\
  \begin{tabular}{|ccccc|}
    \hline
    \parbox[c][\myheight][c]{0cm}{}
    Large  & $\Longleftarrow$ & Mass & $\Longrightarrow$ & Small   \\
    \hline
    \parbox[c][\myheight][c]{0cm}{}
    Low & $\Longleftarrow$ & Temperature  & $\Longrightarrow$ & High\\
    \hline
    \parbox[c][\myheight][c]{0cm}{}
    Absorption-dominated & $\Longleftarrow$ & Behavior of black hole   & $\Longrightarrow$ & Radiation-dominant\\
    \hline
  \end{tabular}
\vspace{0.6cm}\\
   \end{center}
%%%%%%%%%%%%%%%%%       Tab. 3.1        %%%%%%%%%%%%%%%%%%%%%

Hawking radiation can also be shown for not only a Schwarzschild black hole but also for other black holes.
In the case of a Kerr-Newman black hole, the relation (\ref{31bla01}) is extended to
\begin{align}
N=\frac{1}{\exp\left[ \frac{2\pi}{\kappa} \left( \omega - m \Omega_{{\rm H}} - e \Phi_{{\rm H}} \right)\right]-1},
\end{align}
where $m$ is a magnetic quantum number of the emitted matter field, $e$ is the charge of the matter field, 
$\Omega_{{\rm H}}$ is the angular velocity of the black hole, $\Phi_{{\rm H}}$ is the electrical potential of the black hole 
and $\kappa$ is given by not $\displaystyle \frac{1}{4M}$ but by $\displaystyle \frac{4\pi (r_+ -M)}{A}$ as in (\ref{2sg01}).

Since the black holes actually have temperature and entropy, the first law of the black hole physics is written as
\begin{align}
dM={\cal T}_{{\rm BH}} d {\cal S}_{{\rm BH}}+\Omega_{{\rm H}}dL+\Phi_{{\rm H}}dQ,
\label{31fir01}
\end{align}
and the second law is given by
\begin{align}
\Delta {\cal S}_{{\rm BH}} + \Delta {\cal S}_{{\rm C}}= \Delta ({\cal S}_{{\rm BH}}+{\cal S}_{{\rm C}})\geq 0,
\label{31sec01}
\end{align}
where ${\cal S}_{{\rm C}}$ is the entropy of the matter outside the black hole.
It was shown that black holes can radiate by using quantum effects.
As was shown in Section 2.3, although a part of energy can be extracted from a rotating black hole by the Penrose process,
this cannot break the classical Hawking's black hole area theorem (\ref{25al01}).
On the other hand, Hawking radiation decreases its black hole area,
and the classical Hawking's black hole area theorem is violated \cite{haw05}.
Thus one needs to generalize the second law as in (\ref{31sec01}).
This consideration was already performed by Bekenstein \cite{bek01,bek03} before Hawking's original paper \cite{haw01}.
The generalized second law always holds in any physical process.

%%%%%%%%%%%%%%%%%%%%%%%%%%%%%%%%%%%%%%%%%%%%%%%%%%%%%%%%%%%%%%%%%%%%%%%%%%%%%%%%%%%%%%%%%%%%%%%%%
%%                                                                                             %%
%%                                        Ssection 3.2                                         %%
%%                                                                                             %%
%%%%%%%%%%%%%%%%%%%%%%%%%%%%%%%%%%%%%%%%%%%%%%%%%%%%%%%%%%%%%%%%%%%%%%%%%%%%%%%%%%%%%%%%%%%%%%%%%
\section{Previous Works on Hawking Radiation}

\quad~After Hawking's original derivation, various derivations of Hawking radiation have been suggested.
In this section, we would like to review some representative derivations of Hawking radiation
very briefly.
For sake of simplicity, we consider the case of a Schwarzschild black hole unless stated otherwise.

Firstly, we review the derivation by using the path integral \cite{gib01}, which was suggested by Gibbons and Hawking.
The basic 4-dimensional action for the gravitational field is commonly given by the Einstein-Hilbert action,
\begin{align}
S=\frac{1}{16\pi} \int_{{\cal M}} d^4x \sqrt{-g}R  +\frac{1}{8\pi} \int_{\partial {\cal M}} d^{3} x \sqrt{\pm h} K,
\label{32eha01}
\end{align}
where $R$ is the scalar curvature, $K$ is the trace of extrinsic curvature $K_{\mu\nu}$, 
$\partial {\cal M}$ is a suitable boundary of a manifold ${\cal M}$ and 
$h$ is the determinant of $h_{\mu \nu}$ which is the induced metric of the boundary $\partial {\cal M}$.
Of course, we can derive the Einstein equation (\ref{2ein01}) by considering the variation of $g_{\mu\nu}$ in (\ref{32eha01}).

Since we consider the case of a Schwarzschild background, the metric is given by the Schwarzschild metric,
\begin{align}
ds^2=-\left( 1-\frac{2M}{r}\right)dt^2+ \frac{1}{1-\frac{2M}{r}} dr^2 + r^2 d\Omega^2.
\label{32met01}
\end{align}
It is well known that the metric has singularities both at the origin and the horizon.
To remove a fictitious singularity at the horizon, the Kruskal-Szekeres coordinates are often used,
\begin{align}
ds^2=\frac{32M^3}{r} \exp \left( -\frac{r}{2M}\right)(-dz^2+dy^2) + r^2d\Omega^2,
\end{align}
where
\begin{align}
-z^2+y^2&=\left( \frac{r}{2M} -1 \right)\exp\left( \frac{r}{2M} \right),\\
\frac{y+z}{y-z}&=\exp\left( \frac{t}{2M}\right).
\label{32tper01}
\end{align}
Here we define the imaginary time by
\begin{align}
\tau \equiv it.
\end{align}
From (\ref{32tper01}), we then find that $\tau$ is periodic with the period of $8\pi M$.
By substituting the Euclidean metric associated with (\ref{32met01}) into the Euclidean action associated with (\ref{32eha01}),
we can evaluate the action integral.

According to quantum field theory, in the path integral approach to the quantization of a real scalar field $\phi$,
we can represent the transition amplitude to go from $\phi_1$ at a time $t_1$ to $\phi_2$ at a time $t_2$ as
\begin{align}
\langle \phi_2, t_2|\phi_1,t_1 \rangle =\int {\cal D} \phi e^{iS(\phi)},
\end{align}
where the path integral is over all field configurations. 
On the other hand, in the operator formulation, the transition amplitude is given by
\begin{align}
\langle \phi_2, t_2|\phi_1,t_1 \rangle=\langle \phi_2|e^{-iH(t_2-t_1)} |\phi_1  \rangle,
\end{align}
where $H$ is the Hamiltonian.
We define a Euclidean time with a certain period as $t_2-t_1=-i\beta$ and we then set $\phi_1=\phi_2$.
By taking the sums over all $\phi_1$, we obtain
\begin{align}
\sum_{\phi_1} \langle \phi_1|e^{-\beta H} |\phi_1  \rangle={\rm Tr} \left( e^{-\beta H}\right).
\end{align}
According to quantum statistical mechanics, the right-hand side of this relation just corresponds to
the partition function $Z$ for the canonical ensemble consisting of the field $\phi$ at temperature $\displaystyle T=\frac{1}{\beta}$, i.e.,
\begin{align}
Z={\rm Tr}\left( e^{-\beta H}\right).
\end{align}
We thus find that the partition function of the system is represented as the path integral with a periodic Euclidean time.
According to statistical mechanics, it is known that the entropy is represented in terms of the partition function as
\begin{align}
{\cal S}=-\left( \beta \frac{\partial}{\partial \beta} -1\right)\ln Z.
\end{align}

In the black hole background, we found that the system has the period by taking the Euclidean time in (\ref{32tper01}).
Furthermore, the Euclidean path integral presents the partition function and we can then derive the entropy
by using the approximative treatment both of the path integral and Smarr's formula \cite{sma01}.
This agrees with Hawking's original result.
In this sense, Gibbons and Hawking's derivation is very simple and provides universal picture of the black hole entropy.
However, on the other hand, it reveals mysterious results that the black hole entropy is generated simply by a Legendre transformation
from Hamiltonian picture to Lagrangian picture in the path integral.

In connection with this derivation, 
the approach which uses both the Legendre transformation and the consideration based on the change of the topology,
has been analyzed by Ba\~nados,
Teitelboim and Zanelli \cite{btz01}, and also by Hawking and Horowitz \cite{haw06}.
In particular, the derivation of Ba\~nados,
Teitelboim and Zanelli using the Gauss-Bonnet theorem, is interesting.
The Euclidean Einstein-Hilbert action is written as
\begin{align}
S_{{\rm E}} =\frac{1}{16\pi} \int_{{\cal M}} d^4 x \sqrt{g} g^{\mu\nu} R^\alpha_{\mu\alpha \nu} + \frac{1}{8\pi}\int_{\partial {\cal M}} d^3 x\sqrt{h} K,
\end{align}
while the Gauss-Bonnet theorem for a 2-dimensional manifold with a suitable boundary is written as
\begin{align}
\frac{1}{2} \int_{{\cal M}} dx^2 \sqrt{g} g^{\mu\nu} R^\alpha_{\mu\alpha \nu} + \int_{\partial {\cal M}} d x\sqrt{h} K=2\pi \chi({\cal M}),
\end{align}
where $\chi({\cal M})$ is the Euler number of ${\cal M}$ which depends solely on its topology.
For example, $\chi({\cal M})=1$ for a disk and $\chi({\cal M})=0$ for an annulus.
By using the Gauss-Bonnet theorem, the important role of topologies becomes clear.
The relationship between the action and the entropy in the above arguments is also related by the Legendre transformation.
Although this result agrees with Hawking's original one, it has not been presented in the past as an explicit manner as the  basic reason why the entropy is generated by the Legendre transformation and the change of topologies.

Secondly, the derivation of Hawking radiation from the calculation of the energy-momentum tensor in a black hole background was suggested by Christensen and Fulling \cite{chr03}.
They determined the form of the energy momentum tensor by using symmetry arguments and the conservation law of the energy-momentum tensor by taking into consideration of the trace anomaly which is given by
\begin{align}
T^\alpha_{~\alpha}=\frac{1}{24\pi}R,
\end{align}
where $R$ is the scalar curvature in a 2-dimensional theory (for example, see \cite{ber02}).
This anomaly appears as a quantum contribution to the trace $T^\alpha_{~\alpha}$ of the energy-momentum tensor.
By requiring that the energy-momentum tensor is finite as seen by a free falling observer at the horizon in a 2-dimensional Schwarzschild  background and imposing the anomalous trace equation everywhere, we can obtain the characteristic flux 
\begin{align}
F_{{\rm H}}=\frac{M}{2} \int^\infty_{2M} \frac{dr}{r^2} T^\alpha_{~\alpha}(r),
\end{align}
which corresponds to the Hawking's result.
However, this result is valid for a 2-dimensional theory only.
In a 4-dimensional theory, there remains an indeterminable function and the all energy-momentum tensor 
cannot be determined by symmetries alone.
Therefore, this derivation has the weakness that it is not valid for a 4-dimensional theory.

Thirdly, we would like to refer to the derivation of the black hole entropy based on the idea of entanglement entropy.  
The entanglement entropy is understood as a measure of the information loss due to a division of the system.
We consider that the total system can be divided into two subsystems.
The Hilbert space of the total system ${\cal H}$ can be written as
\begin{align}
{\cal H}={\cal H}_1 \otimes {\cal H}_2,
\end{align}
where ${\cal H}_1$ and ${\cal H}_2$ stand for the two subsystems and $\otimes$ is the tensor product.
Roughly speaking, if a state cannot be written as a product of the states in each subspace, namely, the state is not separable,
it is called the entangled state. 
(Recently, the criterion of entanglement was quantitatively clarified by Fujikawa \cite{fuj06,fuj07}. )
The entanglement entropy is quantitatively defined by the quantum von Neumann entropy
\begin{align}
{\cal S}_{12} =- {\rm Tr}_1 (\rho_{{\rm red}} \ln \rho_{{\rm red}} ),
\end{align}
where $\rho_{\rm red}$ is the reduced density matrix to the space ${\cal H}_1$ and the trace is taken over the states of ${\cal H}_1$.
This is understood as a generalization of usual entropy in thermodynamics.
We also note that one of the important properties of the entanglement entropy is that 
it is symmetric under an interchange of the role of ${\cal H}_1$ and ${\cal H}_2$,
\begin{align}
S_{12}=S_{21}.
\end{align}
Even if the sizes of two subspaces are different as drawn in Fig 3.5,
it means that the two entropies agree with each other.
We find that the entanglement entropy possesses properties differing from the usual entropy which is an extensive quantity in statistical mechanics.
For details, see \cite{sre01,muk01}. 

%%%%%%%%%%%%%     Fig.3.5       %%%%%%%%%%
\begin{center}
\vspace{0.6cm}
\input{ent01}\\
\vspace{0.6cm}
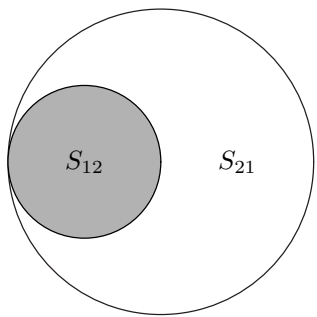
Fig. 3.5 \quad The symmetry of the entanglement entropy 
\vspace{0.6cm}
\end{center}
%%%%%%%%%%%%%%%%%%%%%%%%%%%%%%%%%%%%%%%%%%

The derivation of the black hole entropy based on the entanglement entropy was analyzed by Terashima \cite{ter01}.
When the concept of the entanglement entropy is applied to the case of a black hole,
we can understand the black hole entropy as the information loss due to a spatial separation by the appearance of the horizon.
He regarded the outside and a thin region (of the order of the Planck length) inside of the horizon as each subsystem
and evaluated the entanglement entropy.
In this derivation, the origin of the black hole entropy is very clear,
while the coefficient of the black hole entropy cannot be exactly determined.

Fourthly, although it is not the case of a Schwarzschild black hole,
we would like to state the derivation of the black hole entropy based on the consideration of string theory \cite{str01,cal01}.
This derivation was suggested by Strominger and Vafa \cite{str01}.
In the 5-dimensional extremal black hole made by ``D-branes",
they directly evaluated the entropy by counting the number of the BPS states which partly preserve supersymmetry.
In this derivation, the origin of the black hole entropy is very clear,
and furthermore this result surprisingly agrees with the previous result including the numerical coefficient of the black hole entropy.
However, this derivation is valid for quite atypical black holes only.
The method has not been extended yet to the case of a well-known black hole with a finite temperature such as a Schwarzschild black hole.

Finally, from the above discussions, we find that there are strengths and weaknesses in each derivation.
In the method directly deriving the black hole entropy from the point of view of statistical mechanics or information theory \cite{bek01,ter01},
we cannot determine the correct coefficient of the black hole entropy for black holes with a finite temperature.
The derivation by string theory \cite{str01}, where the origin of the entropy is clear and furthermore the result is correctly reproduced up to the coefficient of the black hole entropy, is valid for black holes with zero temperature (extremal) which means that the black hole does not exhibit any radiation.
In the methods which derive the temperature of the black hole \cite{haw01,gib01,chr03,par01,rob01},
although approximation procedures are used in each derivation,
the correct result including the coefficient of the temperature is derived. 
We may say that the origin of the black hole entropy is not sufficiently understood yet.
As recent attempts toward the better understanding of black hole radiation, we will discuss the derivation of Hawking radiation from the tunneling mechanism \cite{par01} in chapter 5 and from anomalies \cite{rob01} in chapter 4, respectively.

%%%%%%%%%%%%%%%%%%%%%%%%%%%%%%%%%%%%%%%%%%%%%%%%%%%%%%%%%%%%%%%%%%%%%%%%%%%%%%%%%%%%%%%%%%%%%%%%%

%%%%%%%%%%%%%%%%%%%%%%%%%%%%%%%%%%%%%%%%%%%%%%%%%%%%%%%%%%%%%%%%%%%%%%%%%%%%%%%%%%%%%%%%%%%%%%%%%
%%                                                                                             %%
%%                                        Chapter 4                                            %%
%%                                                                                             %%
%%%%%%%%%%%%%%%%%%%%%%%%%%%%%%%%%%%%%%%%%%%%%%%%%%%%%%%%%%%%%%%%%%%%%%%%%%%%%%%%%%%%%%%%%%%%%%%%%
\newpage

\chapter{Hawking Radiation and Anomalies}

\quad~ Robinson and Wilczek suggested a new method of deriving Hawking radiation 
by the consideration of anomalies \cite{rob01}.
The basic idea of their approach is that the flux of Hawking radiation is determined by 
anomaly cancellation conditions
in the Schwarzschild black hole background.
Iso, Umetsu and Wilczek, and also Murata and Soda, extended the method to
a charged Reissner-Nordstr\"om black hole \cite{iso01} and a rotating Kerr black hole \cite{iso02,mur01}, 
and they showed that the flux of Hawking radiation can also be determined 
by anomaly cancellation conditions 
and regularity conditions of currents at the horizon.
Their formulation thus gives the correct Hawking flux for all the cases at infinity 
and thus provides a new attractive method of understanding Hawking radiation.
We present some arguments which clarify this derivation \cite{ume01}.
We show that the Ward identities and boundary conditions for covariant currents,
without referring to the Wess-Zumino terms and the effective action, are sufficient to 
derive Hawking radiation.
Our method, which does not use step functions, 
thus simplifies some of the technical aspects of the original formulation described above.

The contents of this chapter are as follows.
In Section 4.1, we briefly review quantum anomalies in quantum field theory.
In Section 4.2, we would like to discuss the connection between Hawking radiation and quantum anomalies.
In Section 4.3, we clarify some arguments in previous works with respect to the derivation of Hawking radiation from anomalies and 
present a simplified derivation.

%%%%%%%%%%%%%%%%%%%%%%%%%%%%%%%%%%%%%%%%%%%%%%%%%%%%%%%%%%%%%%%%%%%%%%%%%%%%%%%%%%%%%%%%%%%%%%%%%
%%                                                                                             %%
%%                                        Section 4.1                                          %%
%%                                                                                             %%
%%%%%%%%%%%%%%%%%%%%%%%%%%%%%%%%%%%%%%%%%%%%%%%%%%%%%%%%%%%%%%%%%%%%%%%%%%%%%%%%%%%%%%%%%%%%%%%%%

\section{Quantum Anomaly}

\quad~ Quantum anomaly is one of important phenomena in quantum field theory.
This phenomenon is closely related to the concept of symmetry in field theory.
The symmetry plays a very important role in modern physics.
It is well known that if a system has symmetries, there are corresponding conserved quantities which are dictated by the N\"other's theorem \cite{noe01}.
For example, if a system has the time translation symmetry, the corresponding conserved quantity is the energy of the system and the energy conservation law holds.
Also if a system has the rotational symmetry, the angular momentum is conserved and the angular momentum conservation law holds.

From this viewpoint, a quantum anomaly represents the fact that N\"other's theorem can be broken by quantization.
Even if there is a certain symmetry and the corresponding conservation law exists in a classical theory,
it is possible that the symmetry is broken in a quantized theory.
This symmetry breaking is commonly called ``quantum anomaly" or simply ``anomaly".

Historically the quantum anomaly was first discovered in the evaluation of the two-photon decay of the neutral $\pi$ meson by Fukuda and Miyamoto \cite{fuk01}.
Afterward, it was clarified that the anomaly is an inevitable phenomenon in the local field theory by Bell and Jackiw \cite{bel01}, and Adler \cite{adl01}.
In the path integral formulation, it was shown that an anomaly is formulated by a Jacobian in the change of path integral variables by Fujikawa \cite{fuj01}. 

The chiral symmetry is known as a famous symmetry which causes the anomaly.
This symmetry is a relatively-new symmetry which was discovered by the introduction of the Dirac equation.
In this section, we would like to discuss the chiral anomaly as an example of anomalies in 4-dimensional Minkowski space-time.

N\"other's theorem and anomalies can be simply described in the path integral formulation.
The pass integral for a fermion field $\psi(x)$ is defined by
\begin{align}
\int {\cal D} \bar{\psi}{\cal D} \psi \exp \left\{ iS \right\},
\label{41path01}
\end{align}
where $S$ is the action of the system. The path integral measure is defined by 
\begin{align}
\int {\cal D}\bar{\psi} {\cal D}\psi \equiv \prod_x\frac{\delta}{\delta \bar{\psi}(x)} \prod_y\frac{\delta}{\delta \psi(y)},
\end{align}
and the fermion field $\psi(x)$ is also called the Dirac field which satisfies the anticommutation relation
(Grassmann number) in the classical level
\begin{align}
\left\{ \psi(x), \psi (y)\right\}=\left\{ \psi(x), \bar{\psi} (y)\right\}=\left\{ \bar{\psi}(x), \bar{\psi} (y)\right\}=0,
\end{align}
and the Dirac conjugate $\bar{\psi}(x)$ is defined by
\begin{align}
\bar{\psi}(x)\equiv \psi^\dagger(x) \gamma^0.
\end{align}
In quantum electrodynamics, the action for a massless Dirac field is given by
\begin{align}
S(\psi,\bar{\psi},V_\mu)=\int d^4 x  \bar{\psi} (x) \left[ i\gamma^\mu \left( \partial_\mu -ie V_\mu (x) \right) \right] \psi(x),
\label{41act01}
\end{align}
where  $V_\mu(x)$ is the gauge field, $e$ is the charge of the fermion field
and $\gamma^\mu$ is called the gamma matrix or the Dirac matrix defined by
\begin{align}
&\{ \gamma^\mu, \gamma^\nu\}=2\eta^{\mu\nu},\\
&(\gamma^0)^\dagger=\gamma^0,\\
&(\gamma^k)^\dagger=-\gamma^k,\qquad (k=1,2,3),
\end{align}
where $\eta^{\mu\nu}$ is the Minkowski metric. 
By substituting (\ref{41act01}) into (\ref{41path01}), we obtain the path integral of quantum electrodynamics
\begin{align}
\int {\cal D} \bar{\psi} {\cal D} \psi \exp \left\{ i \int d^4 x  \bar{\psi} (x) \left[ i\gamma^\mu \left( \partial_\mu -ie V_\mu (x) \right) \right] \psi(x) \right\}.
\label{41path02}
\end{align}

First we consider the local $U(1)$ gauge transformation defined by
\begin{align}
\psi'(x) &= e^{i\alpha(x)} \psi(x),\\
\bar{\psi}'(x)&=\bar{\psi}(x)e^{-i\alpha(x)},\\
V'_\mu(x)&=V_{\mu}+\frac{1}{e}\partial_\mu \alpha(x).
\label{41gau01}
\end{align}
We find that the action (\ref{41act01}) is invariant under this transformation
\begin{align}
S(\psi',\bar{\psi}',V_\mu')=S(\psi,\bar{\psi},V_\mu).
\label{41act02}
\end{align}
By the fact that the value of a definite integral does not depend on the naming of integration variables, we have the identity
\begin{align}
\int {\cal D} \bar{\psi}'{\cal D} \psi' \exp \left\{ iS(\psi',\bar{\psi}', V_\mu') \right\}
=\int {\cal D} \bar{\psi}{\cal D} \psi \exp \left\{ iS(\psi,\bar{\psi}, V_\mu') \right\}.
\label{41path03}
\end{align}
By substituting (\ref{41act02}) into (\ref{41path03}), we obtain
\begin{align}
\int {\cal D} \bar{\psi}{\cal D} \psi \exp \left\{ iS(\psi,\bar{\psi}, V_\mu) \right\}
=\int {\cal D} \bar{\psi}{\cal D} \psi \exp \left\{ iS(\psi,\bar{\psi}, V_\mu') \right\},
\label{41path04}
\end{align}
where we assumed that the integral measure is invariant under the above gauge transformation
\begin{align}
{\cal D} \bar{\psi}'{\cal D} \psi'={\cal D} \bar{\psi}{\cal D} \psi.
\end{align}
The relation (\ref{41path04}) is called the Ward identity.
By substituting (\ref{41gau01}) into (\ref{41act01}) and integrating by parts, we obtain
\begin{align}
S(\psi,\bar{\psi}, V_\mu')=\int d^4 x  \bar{\psi} (x) \left[ i\gamma^\mu \left( \partial_\mu -ie V_\mu (x) \right) \right] \psi(x)
+ \int d^4 x\alpha (x) \partial_\mu (\bar{\psi}(x) \gamma^\mu \psi(x)).
\end{align}
The Ward identity (\ref{41path04}) thus becomes
\begin{align}
i\int d^4x \alpha (x) \partial_\mu \langle (\bar{\psi}(x)\gamma^\mu \psi(x)) \rangle=0,
\label{41path05}
\end{align}
where we chose $\alpha(x)$ as an infinitely small parameter and we defined the expectation value of an operator $\hat{O}(x)$ by
\begin{align}
\langle O(x) \rangle \equiv \int {\cal D} \bar{\psi}{\cal D} \psi O(x) \exp\left[ iS\right].
\end{align}
From the identity (\ref{41path05}), we can obtain the current conservation law
\begin{align}
\partial_\mu J^\mu (x)=0,
\label{41gc01}
\end{align}
where $J^\mu(x)$ is the N\"other current defined by
\begin{align}
 J^\mu(x) \equiv \langle (\bar{\psi}(x)\gamma^\mu \psi(x) )\rangle,
\label{41gc02}
 \end{align}
which stands for the quantized quantity in the operator formalism.
From the above discussion, N\"other's theorem corresponds to the fact that the integral measure is invariant under the transformation of integration variables in the path integral formulation.

Next we consider the chiral transformation
\begin{align}
\psi'(x)=e^{i\gamma_5 \alpha(x)}\psi(x),\\
\bar{\psi}'(x)=\bar{\psi}(x)e^{i\gamma_5 \alpha(x)},
\end{align}
where $\gamma_5$ is defined by
\begin{align}
&\gamma_5 \equiv i\gamma^0\gamma^1\gamma^2\gamma^3,\\
&\{\gamma_5, \gamma^\mu \}=0.
\end{align}
It is shown that
the action becomes
\begin{align}
S(\psi',\bar{\psi}',V_\mu)=S(\psi,\bar{\psi},V_\mu)+\int d^4 x \alpha(x) \partial_\mu (\bar{\psi}(x)\gamma^\mu\gamma_5 \psi(x))
\end{align}
under the chiral transformation and the integral measure is transformed as \cite{fuj03}
\begin{align}
{\cal D} \bar{\psi}'{\cal D} \psi'={\cal D} \bar{\psi}{\cal D} \psi \exp \left[ -i \int d^4 x \alpha(x) \frac{e^2}{16\pi^2} \epsilon^{\alpha \beta\mu\nu}F_{\alpha\beta} F_{\mu\nu}\right],
\end{align}
where $\epsilon^{\alpha\beta\mu\nu}$ is the Levi-Civita symbol and $F_{\mu\nu}$ is the field strength tensor defined by
\begin{align}
F_{\mu\nu}\equiv \partial_\mu V_\nu - \partial_\nu V_\mu.
\end{align}
For details of this calculation, see, for example, \S 5.1 in \cite{fuj03}.
By using the Ward identity, it is shown that the chiral anomaly is given by
\begin{align}
\partial_\mu  J^\mu_5 (x)=\frac{e^2}{16\pi^2} \epsilon^{\alpha \beta\mu\nu}F_{\alpha\beta} F_{\mu\nu},
\label{41cc01}
\end{align}
where $J^\mu_5(x)$ stands for the chiral current defined by
\begin{align}
J^\mu_5 (x)\equiv \langle \bar{\psi}(x)\gamma^\mu \gamma_5 \psi(x)\rangle.
\label{41cc02}
\end{align}

According to classical theory, it can be shown that both the gauge current $J^\mu(x)$ and the chiral current $J^\mu_5(x)$ are conserved.
By the quantization procedure, the gauge current is conserved as in (\ref{41gc01}) and the gauge symmetry thus holds,
while the chiral current is not conserved as in (\ref{41cc01}) and the chiral symmetry is broken.
As found from the form of (\ref{41cc01}), the current conservation law is broken and we can therefore regard the ``anomaly" as ``generating a source of the current" by quantization.
This picture is useful to understand the following discussions.
 
%%%%%%%%%%%%%%%%%%%%%%%%%%%%%%%%%%%%%%%%%%%%%%%%%%%%%%%%%%%%%%%%%%%%%%%%%%%%%%%%%%%%%%%%%%%%%%%%%
%%                                                                                             %%
%%                                        Section 4.2                                          %%
%%                                                                                             %%
%%%%%%%%%%%%%%%%%%%%%%%%%%%%%%%%%%%%%%%%%%%%%%%%%%%%%%%%%%%%%%%%%%%%%%%%%%%%%%%%%%%%%%%%%%%%%%%%%

\section{Derivation of Hawking Radiation from Anomalies}

\quad~ Robinson and Wilczek demonstrated a new method of deriving Hawking radiation \cite{rob01}.
They derived Hawking radiation by the consideration of quantum anomalies.
Their derivation has an important advantage 
in localizing the source of Hawking radiation 
near the horizon where anomalies are visible.
Since both of the anomalies and Hawking radiation are typical quantum effects,
it is natural that Hawking radiation is related to anomalies.
Iso, Umetsu and Wilczek improved the approach of \cite{rob01} 
and extended the method to a charged Reissner-Nordstr\"om black hole \cite{iso01}. 
This approach was also extended to a rotating Kerr black hole 
and a charged and rotating Kerr-Newman black hole
by Murata and Soda \cite{mur01} and
by Iso, Umetsu and Wilczek \cite{iso02}.

The essential idea of Iso, Umetsu and Wilczek \cite{iso02} is the following.
They consider a quantum field in a black hole background.
As shown in Section 2.4, by using the technique of the dimensional reduction, the field can be effectively described by an infinite collection of $(1+1)$-dimensional fields on $(t, r)$ space near the horizon. 
Then the mass or potential terms of quantum fields can be suppressed near the horizon. 
Therefore we can treat the 4-dimensional theories as a collection of 2-dimensional quantum fields. 
In this 2-dimensions,
outgoing modes near the horizon behave as 
right moving modes while ingoing modes as left moving modes.
Since the horizon is a null hypersurface, 
all ingoing modes at the horizon can not classically affect physics outside the horizon (see Fig. 4.1, where we utilized the Penrose diagram in the Schwarzschild background for simplicity).
Then, if we integrate the ingoing modes to obtain the effective action in the exterior region, 
it becomes anomalous with respect to gauge or general coordinate symmetries 
since the effective theory is now chiral at the horizon. 
The underlying theory is of course invariant 
under these symmetries and these anomalies must be cancelled by quantum effects of the 
classically irrelevant ingoing modes.
They showed that the condition for anomaly cancellation 
at the horizon determines the Hawking flux of the charge and energy-momentum.
The flux is universally determined only by the value of anomalies at the horizon.

%%%%%%%%%%%%%     Fig.4.1       %%%%%%%%%%
\begin{center}
\vspace{0.6cm}
\input{pen09}\\
\vspace{0.6cm}
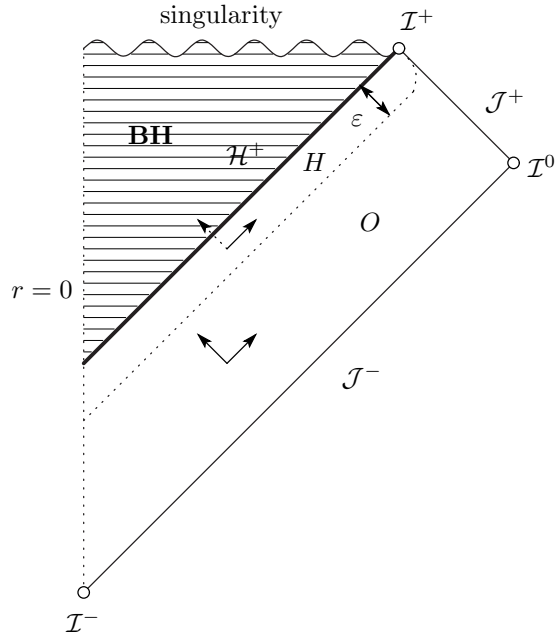
Fig. 4.1 \quad The Penrose diagram relevant to the analysis of Robinson and Wilczek
\vspace{0.6cm}
\end{center}
%%%%%%%%%%%%%%%%%%%%%%%%%%%%%%%%%%%%%%%%%%

The approach of Iso, Umetsu and Wilczek \cite{iso02} is very transparent and interesting.
However, there remain several points to be clarified.
First, Iso, Umetsu and Wilczek start by using both the consistent and covariant currents.
However, they only impose boundary conditions on covariant currents.
As discussed in \cite{iso01}, 
it is not clear why we should use covariant currents instead of consistent ones
to specify the boundary conditions at the horizon.
Banerjee and Kulkarni considered an approach using only covariant currents
without consistent currents \cite{ban01}.
However, their approach heavily relies on the Wess-Zumino terms defined
by consistent currents \cite{wes01}.
The Wess-Zumino terms are also used in the approach of Iso, Umetsu and Wilczek.
Therefore, Banerjee and Kulkarni's approach is not completely described by covariant currents only.

Second,
in the approach of Iso, Umetsu and Wilczek the region outside the horizon must be divided into two regions
because the effective theories are different near and far from the horizon.
They thus used step functions to divide these two regions.
We think that the region near the horizon 
and the region far from the horizon are continuously related.
Nevertheless, if one uses step functions,
terms with delta functions, which originate from the derivatives of step functions
when one considers the variation of the effective action, appear.
They disregarded the extra terms by claiming that these terms correspond to
the contributions of the ingoing modes.
This is the second issue that we wish to address here.
Banerjee and Kulkarni also considered an approach
without step functions \cite{ban02}. 
They obtained the Hawking flux by using the effective actions and two boundary conditions 
for covariant currents.
However, they assumed that the effective actions are 2-dimensional 
in both the region near the horizon and the region far from the horizon \cite{pol01,leu01}.
As already discussed in the approach of Iso, Umetsu and Wilczek,
the original 4-dimensional theory is the 2-dimensional effective theory 
in the region near the horizon.
However, the effective theory should be 4-dimensional 
in the region far from the horizon.

In contrast with the above approaches,
we derive the Hawking flux
using only the Ward identities and two boundary conditions for the covariant currents.
We formally perform the path integral, 
and the N\"other currents are constructed by the variational principle.
Therefore, we can naturally treat the covariant currents \cite{fuj01,fuj02,fuj04}. 
We do not use the Wess-Zumino term, the effective action or step functions.
Therefore, we do not need to define consistent currents.
Although we use the two boundary conditions used
in Banerjee and Kulkarni's method,
we use the 4-dimensional
effective theory far from the horizon 
and the 2-dimensional theory near the horizon.
In this sense,
our method corresponds to the method of Iso, Umetsu and Wilczek.
It is easier to understand the derivation of 
the Ward identities directly from the variation of matter fields 
than their derivation from the effective action
since we consider Hawking radiation as resulting from the effects of matter fields.

Our approach is essentially based on the approach of Iso, Umetsu and Wilczek.
However, we simplify the derivation of Hawking radiation by clarifying the above issues.
We only use the Ward identities and two boundary conditions for covariant currents, 
and we do not use the Wess-Zumino terms, the effective action or step functions, as stated above.
In the next section, we will show our simple derivation.

%%%%%%%%%%%%%%%%%%%%%%%%%%%%%%%%%%%%%%%%%%%%%%%%%%%%%%%%%%%%%%%%%%%%%%%%%%%%%%%%%%%%%%%%%%%%%%%%%
%%                                                                                             %%
%%                                        Section 4.3                                          %%
%%                                                                                             %%
%%%%%%%%%%%%%%%%%%%%%%%%%%%%%%%%%%%%%%%%%%%%%%%%%%%%%%%%%%%%%%%%%%%%%%%%%%%%%%%%%%%%%%%%%%%%%%%%%

\section{Ward Identity in the Derivation of Hawking Radiation from Anomalies}

\quad~ In this section, we would like to clarify some arguments in previous works and present a simple derivation of Hawking radiation from anomalies.
By using the Ward identities and two boundary conditions only, we show how to derive the Hawking flux.

The contents of this section are as follows.
In Subsection 4.3.1, we show a simple derivation of Hawking radiation from anomalies for the case of a Kerr black hole.
In Subsection 4.3.2, we discuss differences among our work and previous works.
In Subsection 4.3.3, we also show that the Hawking flux can be derived for the case of a Reissner-Nordstr\"om black hole by using our approach.
%%%%%%%%%%%%%%%%%%%%%%%%%%%%%%%%%%%%%%%%%%%%%%%%%%%%%%%%%%%%%%%%%%%%%%%%%%%%%%%%%%%%%%%%%%%%%%%%%
%%                                                                                             %%
%%                                        Subsection 4.3.1                                     %%
%%                                                                                             %%
%%%%%%%%%%%%%%%%%%%%%%%%%%%%%%%%%%%%%%%%%%%%%%%%%%%%%%%%%%%%%%%%%%%%%%%%%%%%%%%%%%%%%%%%%%%%%%%%%

\subsection{The case of a Kerr black hole}

\quad~ To compare our method with the approach of Iso, Umetsu and Wilczek \cite{iso02}, we consider the Kerr black hole background.
By taking $Q=0$ in the Kerr-Newman metric which is given by (\ref{2kn01}), we obtain the Kerr metric
\begin{align}
ds^2=&-\frac{\Delta_{{}} -a^2 \sin^2\theta}{\Sigma}dt^2-\frac{2a\sin^2\theta}{\Sigma}(r^2+a^2-\Delta)dtd\varphi \notag\\
&-\frac{a^2\Delta \sin^2\theta-(r^2+a^2)^2}{\Sigma}\sin^2\theta d\varphi^2+\frac{\Sigma}{\Delta}dr^2+\Sigma d\theta^2,
\label{42kn01}
\end{align}
with
\begin{align}
\Sigma&\equiv r^2+a^2\cos^2 \theta,
\label{42kn03}\\
\Delta&\equiv r^2-2Mr+a^2=(r-r_+)(r-r_-),
\label{42kn04}
\end{align}
where $r_{+(-)}$ is the radius of the outer (inner) horizon
\begin{align}
r_{\pm} =M\pm \sqrt{M^2-a^2}.
\end{align}
The action for a scalar field is given by 
\begin{align}
S_{(O)}=\frac{1}{2} \int d^4 x \sqrt{-g} g^{\mu\nu} \partial_\mu \phi \partial_\nu \phi + S_{{\rm int}}
\label{42act01}
\end{align}
as in (\ref{KNact01}). We note that no gauge field exists in (\ref{42act01}).
By using the technique of the dimensional reduction as shown in Section 2.4,
we obtain the effective $(1+1)$-dimensional action near the horizon
\begin{align}
S_{(H)}=-\sum_{l,m}&\int dt dr \Phi \phi_{lm}^* \Bigg[ g^{tt} \left( \partial_t -im U_t\right)^2+\partial_r g^{rr} \partial_r \Bigg] \phi_{lm},
\label{42act02}
\end{align}
with
\begin{align}
&\Phi=r^2+a^2,\\
&g_{tt}=-f(r),\quad
g_{rr}=\frac{1}{f(r)},\quad
g_{rt}=0,
\label{42ker01}
\\
&f(r)\equiv\frac{\Delta}{r^2+a^2},\\
&U_t =-\frac{a}{r^2+a^2}, \ U_r =0,
\label{42gau01}
\end{align}
where $\Phi$ is the dilaton field, $U_\mu$ is the $U(1)$ gauge field and $m$ is the $U(1)$ charge.

From (\ref{42act02}), we find that the effective theory is 
the $(1+1)$-dimensional theory near the horizon.
However, we cannot simply regard the effective theory far from the horizon 
as $(1+1)$-dimensional theory.
We need to divide the region outside the horizon into two regions
because the effective theories are different near the horizon
and far from the horizon.
We define region $O$ as the region far from the horizon and region $H$ as the region near the horizon.
Note that the action in region $O$ is
$S_{(O)}[\phi,g^{\mu\nu}_{(4)}]$
and 
the action in region $H$ is $S_{(H)}[\phi,g^{\mu\nu}_{(2)},U_\mu,\Phi]$.

We can divide the field associated with a particle into ingoing modes falling toward the horizon (left-handed) 
and outgoing modes moving away from the horizon (right-handed) 
using a Penrose diagram \cite{rob01,iso01,iso02} (Fig. 4.2).
Since the horizon is a null hypersurface, none of the ingoing modes
at the horizon are expected to affect the classical physics outside the horizon.
Thus, we ignore the ingoing modes.
Therefore, anomalies appear with respect to the gauge or general coordinate symmetries 
since the effective theory is chiral near the horizon.
Here, we do not consider the backscattering of ingoing modes, i.e., the gray body radiation.

%%%%%%%%%%%%%     Fig.4.2       %%%%%%%%%%
\begin{center}
\vspace{0.6cm}
\input{pen10}\qquad \qquad \qquad \qquad \qquad \qquad \qquad \qquad \qquad \qquad\\
\vspace{0.6cm}
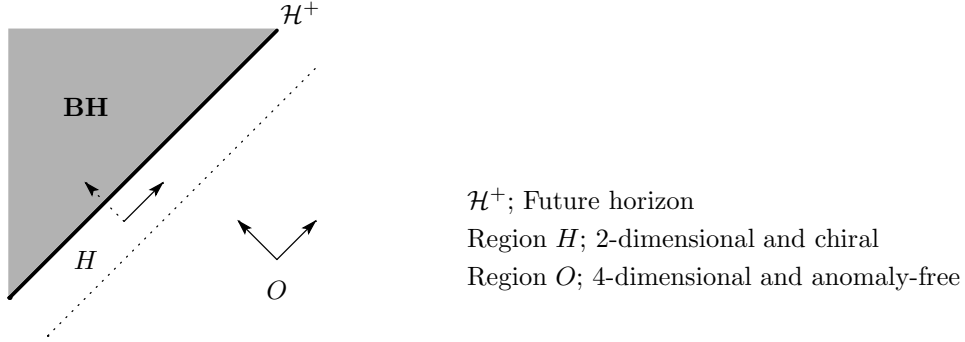
Fig. 4.2 \quad Part of the Penrose diagram relevant to our analysis.
The dashed arrow in region $H$ represents the ignored ingoing mode falling toward the horizon.
\vspace{0.6cm}
\end{center}
%%%%%%%%%%%%%%%%%%%%%%%%%%%%%%%%%%%%%%%%%%

We now present the derivation of Hawking radiation for the Kerr black hole.
First, we consider the effective theory in region $O$.
The effective theory is 4-dimensional in region $O$, which we cannot reduce to a 2-dimensional theory. 
In contrast to the case of a charged black hole, 
a 4-dimensional gauge field such as the Coulomb potential $V_t$ 
does not exist in a rotating Kerr black hole.
Therefore, we do not define the $U(1)$ gauge current in region $O$.
On the other hand, the effective theory in region $H$ is a 2-dimensional chiral theory and
we can regard a part of the metric as a gauge field such as (\ref{42gau01}),
since the action of (\ref{42act02}) is $S_{(H)}[\phi,g^{\mu\nu}_{(2)},U_\mu,\Phi]$.

Second, we consider the Ward identity for the gauge transformation 
in region $H$ near the horizon.
Here, we pretend to formally perform the path integral for
$S_{(H)}[\phi,g^{\mu\nu}_{(2)},U_\mu,\Phi]$, 
where the N\"other current is constructed by the variational principle,
although we do not perform an actual path integral.
Therefore, we can naturally treat \textit{covariant} currents \cite{fuj01}.
As a result, we obtain the Ward identity with a gauge anomaly
\begin{align}
\nabla_\mu J^{\mu}_{(H)}-{\mathscr C}=0,
\label{42jco01}
\end{align}
where we define covariant currents $J^\mu_{(H)}(r)$ 
and ${\mathscr C}$ is a covariant gauge anomaly.
This Ward identity is for right-handed fields.
The covariant form of the 2-dimensional Abelian anomaly ${\mathscr C}$ is given by
\begin{align}
{\mathscr C}=\pm \frac{m^2}{4\pi\sqrt{-g_{(2)}}}\epsilon^{\mu\nu}F_{\mu\nu},\quad(\mu,\nu=t,r)
\end{align}
where $+(-)$ corresponds to right(left)-handed matter fields, 
$\epsilon^{\mu\nu}$ is an antisymmetric tensor with $\epsilon^{tr}=1$ and
$F_{\mu\nu}$ is the field-strength tensor.
Using the 2-dimensional metric (\ref{42ker01}), the identity (\ref{42jco01}) is written as
\begin{align}
\partial_r J^r_{(H)}(r)=\frac{m^2}{2\pi}\partial_r U_t(r).
\label{42jco02}
\end{align}
By integrating Eq. (\ref{42jco02}) over $r$ from $r_+$ to $r$, we obtain
\begin{align}
J^r_{(H)}(r)=\frac{m^2}{2\pi}\left[ U_t(r)- U_t(r_+)\right],
\label{42jco03}
\end{align}
where we use the condition
\begin{align}
J^r_{(H)}(r_+)=0.
\label{42bc01}
\end{align}
The condition (\ref{42bc01}) corresponds to the statement
that free falling observers see a finite amount of the charged current at the horizon,
i.e., (\ref{42bc01}) is derived from the regularity of covariant currents.
This condition was used in the approach of Iso, Umetsu and Wilczek \cite{iso02}.
We regard (\ref{42jco03}) as a covariant $U(1)$ gauge current appearing
in region $H$ near the horizon.

Third, we consider the Ward identity for the general coordinate transformation 
in region $O$ far from the horizon.
By improving the approach of \cite{iso02},
we define
the formal 2-dimensional energy-momentum tensor in region $O$ 
from the exact 4-dimensional energy-momentum tensor in region $O$
and we connect the 2-dimensional energy-momentum tensor thus-defined in region $O$ 
with the 2-dimensional energy-momentum tensor in region $H$.
Since the action is $S_{(O)}[\phi,g^{\mu\nu}_{(4)}]$ in region $O$, 
the Ward identity for the general coordinate transformation is written as
\begin{align}
\nabla_\nu T^{\mu\nu}_{(4)}=0,
\label{42emt01}
\end{align}
where $T^{\mu\nu}_{(4)}$ is the 4-dimensional energy-momentum tensor.
Since the Kerr background is stationary and axisymmetric,
the expectation value of the energy-momentum tensor 
in the background depends only on $r$ and $\theta$,
i.e., $\langle T^{\mu\nu}\rangle=\langle T^{\mu\nu}(r,\theta) \rangle$.
The $\mu=t$ component of the conservation law (\ref{42emt01}) is written as
\begin{align}
\partial_r(\sqrt{-g} T^{~r}_{t(4)})+\partial_\theta (\sqrt{-g}T^{~\theta}_{t(4)})=0,
\label{42emt02}
\end{align}
where $\sqrt{-g}=(r^2+a^2 \cos^2\theta)\sin \theta$.
By integrating Eq. (\ref{42emt02}) over the angular coordinates $\theta$ and $\varphi$,
we obtain
\begin{align}
\partial_r T^{~r}_{t(2)}=0,
\end{align}
where we define the effective 2-dimensional tensor $T^{~r}_{t(2)}$ by
\begin{align}
T^{~r}_{t(2)}\equiv \int d\Omega_{(2)}(r^2+a^2\cos ^2\theta)T^{~r}_{t(4)}.
\label{42emt03}
\end{align}
We define $T^{~r}_{t(2)}\equiv T^{~r}_{t(O)}$ to emphasize 
region $O$ far from the horizon.
The energy-momentum tensor $T^{~r}_{t(O)}$ is conserved in region $O$;
\begin{align}
\partial_r T^{~r}_{t(O)}=0.
\label{42emt04}
\end{align}
By integrating Eq. (\ref{42emt04}), we obtain
\begin{align}
T^{~r}_{t(O)}=a_o,
\label{42emt05}
\end{align}
where $a_o$ is an integration constant.

Finally, we consider the Ward identity for the general coordinate transformation 
in region $H$ near the horizon.
The Ward identity for the general coordinate transformation in the presence of a gravitational anomaly
is 
\begin{align}
\nabla_\nu T^{~\nu}_{\mu(H)}(r)-F_{\mu\nu} J^\nu_{(H)}(r)
-\frac{\partial_\mu \Phi}{\sqrt{-g_{(2)}}}\frac{\delta S_{(H)}}{\delta \Phi}-{\mathscr A}_{\mu}(r)=0,
\label{42emt06}
\end{align}
where both of the gauge current and the energy-momentum tensor are defined
to be of the \textit{covariant} form
and ${\mathscr A}_\mu$ is the covariant form of the 2-dimensional gravitational anomaly.
This Ward identity corresponds to that of \cite{ban01} when there is no dilaton field.
The covariant form of the 2-dimensional gravitational anomaly ${\mathscr A}_\mu$ 
is given by \cite{alv01,fuj05,ber01}
\begin{align}
{\mathscr A}_\mu=\frac{1}{96\pi\sqrt{-g_{(2)}}}\epsilon_{\mu\nu}\partial^\nu R =\partial_r N^{~r}_{\mu},
\label{42ano01}
\end{align}
where we define $N^{~r}_{\mu}$ by
\begin{align}
N^{~r}_{t}\equiv \frac{ff''-(f')^2/2}{96\pi},\qquad N^{~r}_{r}\equiv 0,
\label{42ano02}
\end{align}
and \{ $'$ \} represents differentiation with respect to $r$.
The $\mu=t$ component of (\ref{42emt06}) is written as
\begin{align}
\partial_r T^{~r}_{t(H)}(r)=F_{rt}J^r_{(H)}(r)+\partial_r N^{~r}_{t}(r).
\label{42emt07}
\end{align}
Using (\ref{42jco03}) and integrating (\ref{42emt07}) over $r$ from $r_+$ to $r$, we obtain
\begin{align}
T^{~r}_{t(H)}(r)
=-\frac{m^2}{2\pi}U_t (r_+) U_t (r) + \frac{m^2}{4\pi} U^2_t(r)+N^{~r}_t(r)
%\notag\\&
+\frac{m^2}{4\pi} U^2_t(r_+)-N^{~r}_t(r_+),
\label{42emt08}
\end{align}
where we impose the condition that the energy-momentum tensor vanishes at the horizon, 
which is the same condition as (\ref{42bc01}):
\begin{align}
T^{~r}_{t(H)}(r_+)=0.
\label{42bc02}
\end{align}

We compare (\ref{42emt05}) with (\ref{42emt08}).
By following Banerjee and Kulkarni's approach \cite{ban02},
we impose the condition that the asymptotic form of (\ref{42emt08}) 
in the limit $r\rightarrow \infty$ is equal to (\ref{42emt05}):
\begin{align}
T^{~r}_{t(O)}=T^{~r}_{t(H)}(\infty).
\label{42bc03}
\end{align}
Condition (\ref{42bc03}) corresponds to the statement that
no energy flux is generated away from the horizon region.
Therefore, the asymptotic form of (\ref{42emt08}) has to agree with that of (\ref{42emt05}).
From (\ref{42bc03}), we obtain
\begin{align}
a_o=\frac{m^2\Omega_{{\rm H}}^2}{4\pi}+\frac{\pi}{12\beta^2},
\end{align}
where $\Omega_{{\rm H}}$ is the angular velocity of the black hole,
\begin{align}
\Omega_{{\rm H}} \equiv \frac{a}{r^2_+ + a^2},
\end{align}
and we used both of the surface gravity of the black hole,
\begin{align}
\kappa=\frac{2\pi}{\beta}=\frac{1}{2}f'(r_+),
\end{align}
and (\ref{42ano02}).
As a result, we obtain the flux of the energy-momentum tensor 
in the region far from the horizon from (\ref{42bc03}) as
\begin{align}
T^{~r}_{t(O)}=\frac{m^2\Omega^2_{{\rm H}}}{4\pi}+\frac{\pi}{12\beta^2}.
\label{42emt09}
\end{align}
This flux agrees with the Hawking flux.
Our result corresponds to that of \cite{iso02} in the limit $r\rightarrow\infty$.
In contrast with the case in \cite{iso02}, our result does not depend on gauge fields
in the region far from the horizon where the radial coordinate $r$ is large but finite.
As can be seen from the action (\ref{42act01}),
the gauge field does not exist in the Kerr black hole physics 
in a realistic 4-dimensional sense, and only the mass and angular momentum appear.
We thus consider that
our picture presented here is more natural than that of \cite{iso02}.

%%%%%%%%%%%%%%%%%%%%%%%%%%%%%%%%%%%%%%%%%%%%%%%%%%%%%%%%%%%%%%%%%%%%%%%%%%%%%%%%%%%%%%%%%%%%%%%%%
%%                                                                                             %%
%%                                        Subsection 4.3.2                                     %%
%%                                                                                             %%
%%%%%%%%%%%%%%%%%%%%%%%%%%%%%%%%%%%%%%%%%%%%%%%%%%%%%%%%%%%%%%%%%%%%%%%%%%%%%%%%%%%%%%%%%%%%%%%%%

\subsection{Comparison with previous works}

\quad~ In this subsection, we would like to state explicitly the differences among our approach and previous works. 
When one compares our method with that of Iso, Umetsu and Wilczek \cite{iso02},
one recognizes the following differences.
To begin with, they define the gauge current by the $\varphi$ component of 
the 4-dimensional energy-momentum tensor $T^{~r}_{\varphi(4)}$ 
in the region far from the horizon.
In contrast, we do not define the gauge current in the region far from the horizon, 
since no gauge current exists in a Kerr black hole.
This difference appears in the energy conservation condition; we use (\ref{42emt04}), whereas Iso, Umetsu and Wilczek used the equation
\begin{align}
\partial_r T^{~r}_{t(2)}-F_{rt}J^r_{(2)}=0.
\label{432iso01}
\end{align}
If we define gauge currents suitably,
we might be able to consider the Kerr black hole 
in the same way as the Reissner-Nordstr\"om black hole, as Iso, Umetsu and Wilczek attempt to do.
However, some subtle aspects are involved in such attempts to define gauge currents.

To be explicit, the authors in \cite{iso02} regard part of the metrics as the gauge field by defining 
$A^{\mu}\equiv -g^{\mu\varphi}_{(4)}$, as in Kaluza-Klein theory.
This definition is consistent with the initial definition of the gauge field (\ref{42gau01})
near the horizon, i.e.,
\begin{align}
A_t=\frac{g_{t\varphi(4)}}{g_{\varphi\varphi(4)}}
=\frac{a(r^2+a^2 -\Delta)}{a^2 \Delta\sin^2 \theta -(r^2+a^2)^2}~
\stackrel{\Delta\to 0}{\longrightarrow}~
U_t=-\frac{a}{r^2+a^2}.
\label{432iso02}
\end{align}
To maintain consistency, 
they simultaneously assume that 
the definition of (\ref{42emt03}) is modified such that
it leads to (\ref{432iso01}) by using 
the $\mu=t$ component of (\ref{42emt01}), i.e., 
\begin{align}
T_{t(2)}^{~r}=\int d \Omega_2 \left(r^2+a^2\cos^2 \theta\right) 
\left( T_{t(4)}^{~r} - U_t T^{~r}_{\varphi(4)} \right).
\end{align}
In this way they maintain consistency.
However, we consider that definition (\ref{42emt03}) is more natural 
than this modified definition,
since in definition (\ref{42emt03}) the formal 2-dimensional energy-momentum tensor 
is defined by
integrating the exact 4-dimensional energy-momentum tensor over the angular coordinates 
without introducing an artificial gauge current in the region far from the horizon. 
In our approach, which is natural for the Kerr black hole, no gauge field appears
in the region far from the horizon where the radial coordinate $r$ is large but finite,
in contrast to the formulation in \cite{iso02}.
We thus believe that our formulation is more natural than the formulation in \cite{iso02}, 
although both formulations give rise to the same physical conclusion.

Furthermore, in comparison between the derivation of Iso, Umetsu and Wilczek and ours, there are important differences.
Since they defined the gauge field away from the horizon, they can treat the gauge current there.
In the region away from the horizon, the current is conserved
\begin{align}
\partial_r J^r_{(O)}=0.
\label{432cur01}
\end{align}
On the contrary, in the near horizon region, the current obeys an anomalous equation
\begin{align}
\partial_r J^r_{(O)}=\frac{m^2}{4\pi}\partial_r U_t.
\label{432cur02}
\end{align}
The right-hand side is the gauge anomaly in a consistent form \cite{bar02}.
The current is  accordingly a consistent current which can be obtained from the variation of the effective action with respect with the gauge field.
One can solve these equations in each region as
\begin{align}
J^r_{(O)}&=c_o,
\label{432cur03}\\
J^r_{(H)}&=c_H+\frac{m^2}{4\pi}\left( U_t (r) -U_t(r_+)\right),
\label{432cur04}
\end{align}
where $c_o$ and $c_H$ are integration constants.

Here, the authors in \cite{iso02} consider the effective action $W$ without the ingoing modes in the near horizon.
The variation of the effective action under the gauge transformation is then given by
\begin{align}
-\delta W &=\int d^2x \sqrt{-g_{(2)}} \lambda \nabla_\mu J^\mu,
\label{432act01}
\end{align}
where $\lambda$ is a gauge parameter and we note that all the currents are {\it consistent} forms.
Since the effective theories are different near and far from the horizon,
they wrote the current as a sum in two regions
\begin{align}
J^r= J^r_{(O)}\Theta_{+}(r) + J^r_{(H)} H(r),
\label{432cur05}
\end{align}
where $\Theta_{+}(r)$ and $H(r)$ are step functions defined by
\begin{align}
\Theta_+(r) &\equiv \theta(r-r_+-\epsilon),\\
H(r)&\equiv 1-\Theta_+(r).
\end{align}
By substituting (\ref{432cur05}) into (\ref{432act01}) and integrating by parts, we have
\begin{align}
-\delta W=\int d^2x \lambda \Bigg[ 
\delta (r-r_+ -\epsilon) \left( J^r_{(O)} - J^r_{H} +\frac{m^2}{4\pi}U_t \right)+\partial_r \left( \frac{m^2}{4\pi}U_t H(r)\right)
\Bigg].
\end{align}
Both the coefficient of the delta function in the first term and the second term should vanish because the total effective action must be gauge invariant.
They have required that the second term should be cancelled by quantum effects of the classically irrelevant ingoing modes related to the  Wess-Zumino term.
By imposing the condition that the coefficient of the {\it covariant} current at the horizon should vanish,
they determined the current flux.
Similarly the energy-momentum tensor can also be determined.

We would like to note that they needed the quantum effects of the once ignored ingoing modes
because they used step functions for the continuously connected two regions.
Also they used the boundary condition for the covariant current in order to fix the value of the current,
and they hence used two kinds of currents (consistent and covariant), which complicate the analysis.

In other approaches, Banerjee and Kulkarni used the Ward identities for the covariant current \cite{ban01}.
However they had to define the consistent current in order to use the Wess-Zumino terms.
The Wess-Zumino terms are also used in the approach of Iso, Umetsu and Wilczek \cite{iso02}.
Therefore, their approach is not completely described by covariant currents only.
They also considered an approach without step functions \cite{ban02}.
They obtained the Hawking flux by using effective actions and two boundary conditions for the covariant current.
However, they assumed that the effective actions are 2-dimensional in both the region near the horizon the region far from the horizon.
As discussed in the paper of Iso, Umetsu and Wilczek, the effective theory should be 4-dimensional in the region far from the horizon.
If one should assume this 4-dimensional theory as an effective 2-dimensional theory (in the sense of conformal field theory), one encounters a difficulty since one cannot consider matter fields
with mass and interactions away from the horizon in conformal field theory according to our current understanding of conformal field theory.

In contrast with the above approaches,
we do not use the consistent current at any stage of our analysis
since we use neither the Wess-Zumino term nor the effective action.
Thus we only use the covariant current.
We now argue why we use the regularity conditions 
for \textit{covariant} currents instead of consistent currents.
All the physical quantities should be gauge-invariant.
Thus, physical currents should be \textit{covariant}.
This is consistent with, for example, the well-known anomalous baryon number current 
in the Weinberg-Salam theory \cite{tht01}.
Since we do not use any step function either, we need not consider the quantum effect of the ingoing modes.
Furthermore, in principle, we can incorporate matter fields with mass and interactions away from the horizon.
Therefore, we believe that our approach clarifies some essential aspects of the derivation
of Hawking flux from anomalies.

We have shown that the Ward identities and boundary conditions for covariant currents, 
without referring to the Wess-Zumino terms and the effective action, are sufficient to 
derive Hawking radiation.
The first boundary condition states 
that both the $U(1)$ gauge current and the energy-momentum tensor 
vanish at the horizon, as in (\ref{42bc01}) and (\ref{42bc02}).
This condition corresponds to the regularity condition
that a free falling observer sees a finite amount of the charged current at the horizon.
The second boundary condition is that
the asymptotic form of the energy-momentum tensor which was originally defined in the region near the horizon
is equal to the energy-momentum tensor 
in the region far from the horizon in the limit $r\rightarrow \infty$, 
as in (\ref{42bc03}).
This condition means that no energy flux is generated away from the near horizon region.

In passing, we mention that the Hawking flux is determined from (\ref{42emt08}) simply by considering the direct limit
\begin{align}
T^{~r}_{t(H)}(r\rightarrow \infty)
=
%\notag\\&
\frac{m^2}{4\pi} A^2_t(r_+)-N^{~r}_t(r_+),
\end{align}
which agrees with (\ref{42emt09}).
The physical meaning of this consideration is that Hawking radiation is induced by
quantum anomalies, which are defined in an arbitrarily small region near the horizon
since they are short-distance phenomena, 
and at any region far from the horizon the theory is anomaly-free
and thus, no further flux is generated.
Namely, we utilize an intuitive picture on the basis of the Gauss theorem, 
which is applied to a closed region surrounded 
by a surface $S$ very close to the horizon and a surface $S'$ far from the horizon
in the asymptotic region (Fig. 4.3).
If no flux is generated in this closed region,
the flux on the surface very close to the horizon
and the flux on the surface far from the horizon in the asymptotic region coincide.

%%%%%%%%%%%%%     Fig.4.3       %%%%%%%%%%
\begin{center}
\vspace{0.6cm}
\input{ward02}\\
\vspace{0.6cm}
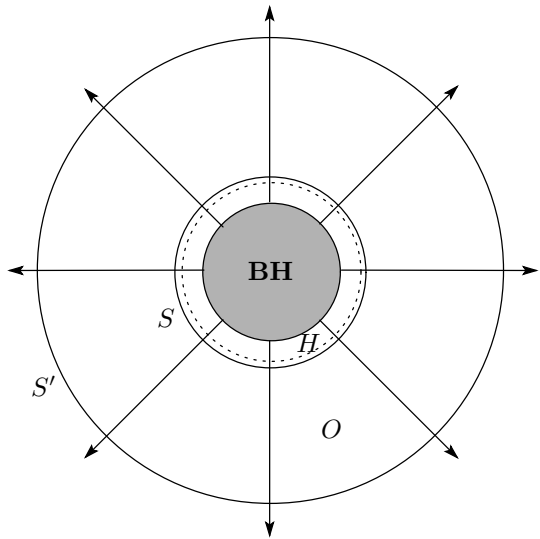
Fig. 4.3 \quad Intuitive picture on the basis of the Gauss theorem.
The total fluxes on $S$ and $S'$ are equal by the Gauss theorem.
\vspace{0.6cm}
\end{center}
%%%%%%%%%%%%%%%%%%%%%%%%%%%%%%%%%%%%%%%%%%

%For recent related works, please refer to Refs.~\citen{iso3, ind, chi}.

Finally, we mention that there recently appeared many papers about the derivation of Hawking radiation from anomalies.
Further developments associated with this derivation are given in \cite{iso03,shi01,bon01,kul01,fuj00,mor01,bon02}.
This method is capable of wide application.
For example, it has been extended to various black holes \cite{iso04,por01,nam01,por02,wei01}, and higher spin generalization of the anomaly method have been discussed \cite{iso05,iso06,iso07,iso08}. 

%%%%%%%%%%%%%%%%%%%%%%%%%%%%%%%%%%%%%%%%%%%%%%%%%%%%%%%%%%%%%%%%%%%%%%%%%%%%%%%%%%%%%%%%%%%%%%%%%
%%                                                                                             %%
%%                                        Subsection 4.3.3                                     %%
%%                                                                                             %%
%%%%%%%%%%%%%%%%%%%%%%%%%%%%%%%%%%%%%%%%%%%%%%%%%%%%%%%%%%%%%%%%%%%%%%%%%%%%%%%%%%%%%%%%%%%%%%%%%

\subsection{The case of a Reissner-Nordstr\"om black hole}

\quad~ In this subsection,
we show that Hawking flux in a charged black hole can be obtained by using our approach.
Since we consider a charged Reissner-Nordstr\"om black hole,
the external space is given by the Reissner-Nordstr\"om metric
\begin{align}
ds^2=f(r)dt^2-\frac{1}{f(r)}dr^2-r^2 d\theta^2-r^2\sin^2\theta d\varphi^2,
\end{align}
and $f(r)$ is given by
\begin{align}
f(r)=1-\frac{2M}{r}+\frac{Q^2}{r^2}=\frac{(r-r_+)(r-r_-)}{r^2},
\end{align}
where $r_\pm=M\pm \sqrt{M^2-Q^2}$
and $r_+$ is the distance from the center of the black hole to the outer horizon.
We consider quantum fields in the vicinity of the Reissner-Nordstr\"om black hole.
In 4 dimensions, the action for a complex scalar field is given by
\begin{align}
S=\int d^4x\sqrt{-g}g^{\mu\nu}(\partial_\mu+ieV_\mu)\phi^*(\partial_\nu-ieV_\nu)\phi
+S_{\rm int},
\label{srei1}
\end{align}
where the first term is the kinetic term 
and the second term $S_{\rm int}$ represents the mass, potential and interaction terms.
In contrast with the Kerr black hole background,
we note that the $U(1)$ gauge field $\displaystyle V_t=-\frac{Q}{r}$ appears
in the Reissner-Nordstr\"om black hole background.
By performing the partial wave decomposition of $\phi$ in terms of the spherical harmonics
($\displaystyle \phi=\sum_{l,m}\phi_{lm}Y_{lm}$) and using the property $f(r_+)=0$ at the horizon,
the action $S_{(H)}$ near the horizon is written as
\begin{align}
S_{(H)}=-\sum_{l,m}\int dtdr \Phi \phi^*_{lm}\left[g^{tt}(\partial_t-ieV_t)^2
+\partial_r g^{rr} \partial_r \right]\phi_{lm},
\end{align}
where we ignore $S_{\rm int}$ 
because the kinetic term dominates near the horizon in high-energy theory. 
From this action, we find that $\phi_{lm}$ can be considered as $(1+1)$-dimensional complex
scalar fields in the backgrounds of the dilaton $\Phi$, metric $g_{\mu\nu}$ 
and $U(1)$ gauge field $V_\mu$, where
\begin{align}
&\Phi=r^2,\\
&g_{tt}=f(r),\quad
g_{rr}=-\frac{1}{f(r)},\quad
g_{rt}=0,
\label{grr2}\\
&V_t=-\frac{Q}{r}, \quad V_r=0.
\end{align}
The $U(1)$ charge of the 2-dimensional field $\phi_{lm}$ is $e$.
Note that the action in the region far from the horizon 
is $S_{(O)}[\phi,g^{\mu\nu}_{(4)},V_\mu]$ and 
the action in the region near the horizon is $S_{(H)}[\phi,g^{\mu\nu}_{(2)},V_\mu,\Phi]$.

We now present the derivation of Hawking radiation for the Reissner-Nordstr\"om black hole.
First, we consider the Ward identity for the gauge transformation
in region $O$ far away from the horizon.
Here, we formally perform the path integral for $S_{(O)}[\phi,g^{\mu\nu}_{(4)},V_\mu]$,
where the N\"other current is constructed by the variational principle.
Therefore, we can naturally treat \textit{covariant} currents \cite{fuj01}.
As a result, we obtain the Ward identity
\begin{align}
\nabla_\mu J^\mu_{(4)}=0,
\label{Jcon1}
\end{align}
where $J^\mu_{(4)}$ is the 4-dimensional gauge current.
Since the Reissner-Nordstr\"om background is stationary and spherically symmetric,
the expectation value of the gauge current in the background depends only on $r$,
i.e., $\langle J^{\mu} \rangle=\langle J^{\mu}(r) \rangle$.
Using the 4-dimensional metric, the conservation law (\ref{Jcon1}) is written as
\begin{align}
\partial_r(\sqrt{-g}J^r_{(4)})+(\partial_\theta\sqrt{-g})J^\theta_{(4)}=0,
\label{Jcon2}
\end{align}
where $\sqrt{-g}=r^2 \sin \theta$.
By integrating Eq. (\ref{Jcon2}) over the angular coordinates $\theta$ and $\varphi$, 
we obtain
\begin{align}
\partial_r J^r_{(2)}=0,
\end{align}
where we define the effective 2-dimensional current $J^r_{(2)}$ by
\begin{align}
J^r_{(2)}\equiv\int d\Omega_{(2)} r^2 J^r_{(4)}.
\label{2.21}
\end{align}
We define $J^r_{(2)}\equiv J^r_{(O)}$ to emphasize region $O$ 
far from the horizon.
The gauge current $J^r_{(O)}$ is conserved in region $O$,
\begin{align}
\partial_r J^r_{(O)}=0.
\label{2.20}
\end{align}
By integrating Eq. (\ref{2.20}), we obtain
\begin{align}
J^r_{(O)}=c_o,
\label{2.23}
\end{align}
where $c_o$ is an integration constant.

Second, we consider the Ward identity for the gauge transformation 
in the region $H$ near the horizon.
When there is a gauge anomaly, the Ward identity for the gauge transformation is given by
\begin{align}
\nabla_\mu J^\mu_{(H)}-{\mathscr B}=0,
\label{2.24}
\end{align}
where we define the covariant current as $J^\mu_{(H)}$
and ${\mathscr B}$ is a covariant gauge anomaly.
The covariant form of the 2-dimensional gauge anomaly ${\mathscr B}$ is given by
\begin{align}
{\mathscr B}=\pm \frac{e^2}{4\pi\sqrt{-g_{(2)}}}\epsilon^{\mu\nu}F_{\mu\nu},\qquad (\mu,\nu=t,r)
\end{align}
where $+(-)$ corresponds to the anomaly for right(left)-handed fields.
Here $\epsilon^{\mu\nu}$ is an antisymmetric tensor with $\epsilon^{tr}=1$ and
$F_{\mu\nu}$ is the field-strength tensor defined by
\begin{align}
F_{\mu\nu} \equiv \partial_\mu V_\nu -\partial_\nu V_\mu.
\end{align}
Using the 2-dimensional metric (\ref{grr2}), (\ref{2.24}) is written as
\begin{align}
\partial_r J^r_{(H)}(r)=\frac{e^2}{2\pi}\partial_r V_t(r).
\label{a16}
\end{align}
By integrating (\ref{a16}) over $r$ from $r_+$ to $r$, we obtain
\begin{align}
J^r_{(H)}(r)=\frac{e^2}{2\pi}[ V_t(r)- V_t(r_+)],
\label{2.29}
\end{align}
where we impose the condition
\begin{align}
J^r_{(H)}(r_+)=0.
\end{align}
This condition corresponds to (\ref{42bc01}) in the present paper.
We also impose the condition that the asymptotic form of (\ref{2.29}) 
is equal to that of (\ref{2.23}),
\begin{align}
J^r_{(O)}(\infty)=J^r_{(H)}(\infty).
\label{a19}
\end{align}
From (\ref{a19}), we obtain the gauge current in region $O$ as
\begin{align}
J^r_{(O)}=-\frac{e^2}{2\pi}V_t(r_+).
\label{2.31}
\end{align}

Third, we consider the Ward identity for the general coordinate transformation 
in the region $O$ far from the horizon.
We define the formal 2-dimensional energy-momentum tensor in region $O$ 
from the exact 4-dimensional energy-momentum tensor in region $O$
and we connect the 2-dimensional energy-momentum tensor in region $O$ 
with the 2-dimensional energy-momentum tensor thus defined in region $H$.
Since the action is $S_{(O)}[\phi,g^{\mu\nu}_{(4)},V_\mu]$ in region $O$, 
the Ward identity for the general coordinate transformation, which is anomaly-free is written as
\begin{align}
\nabla_\nu T^{~\nu}_{\mu(4)}-F_{\nu\mu}J^\nu_{(4)}=0,
\label{a21}
\end{align}
where $T^{\mu\nu}_{(4)}$ is the 4-dimensional energy-momentum tensor.
Since the Reissner-Nordstr\"om background is stationary and spherically symmetric,
the expectation value of the energy-momentum tensor in the background depends only on $r$,
i.e., $\langle T^{\mu\nu}\rangle=\langle T^{\mu\nu}(r) \rangle$.
The $\mu=t$ component of the conservation law (\ref{a21}) is written as
\begin{align}
\partial_r \left( \sqrt{-g} T^{~r}_{t(4)}\right)
+\left( \partial_\theta \sqrt{-g} \right) T^{~\theta}_{t(4)}
-\sqrt{-g}F_{r t}J^r_{(4)}=0.
\label{a22}
\end{align}
By integrating (\ref{a22}) over $\theta$ and $\varphi$, we obtain
\begin{align}
\partial_r T^{~r}_{t(2)}=F_{r t}J^r_{(2)},
\label{2.34}
\end{align}
where we define the effective 2-dimensional tensor $T^{~r}_{t(2)}$ by
\begin{align}
T^{~r}_{t(2)}\equiv \int d\Omega_{(2)} r^2 T^{~r}_{t(4)},
\end{align}
and $J^r_{(2)}$ is defined by (\ref{2.21}).
To emphasize region $O$ far from the horizon,
we write (\ref{2.34}) as
\begin{align}
\partial_r T^{~r}_{t(O)}=F_{r t}J^r_{(O)}.
\label{a25}
\end{align}
By substituting (\ref{2.31}) into (\ref{a25}) and integrating it over $r$, we obtain
\begin{align}
T^{~r}_{t(O)}(r)=a_o-\frac{e^2}{2\pi}V_t(r_+)V_t(r).
\label{2.39}
\end{align}

Finally, we consider the Ward identity for the general coordinate transformation 
in region $H$ near the horizon.
When there exists a gravitational anomaly,
the Ward identity for the general coordinate transformation is given by
\begin{align}
\nabla_\nu T^{~\nu}_{\mu(H)}-F_{\nu\mu}J^\nu_{(H)}-
\frac{\partial_\mu \Phi}{\sqrt{-g}}\frac{\delta S}{\delta \Phi}-{\mathscr A}_\mu=0,
\label{2.40}
\end{align}
where both the gauge current and the energy-momentum tensor 
are defined to be of the \textit{covariant} form
and ${\mathscr A}_\mu$ is the covariant form of the 2-dimensional gravitational anomaly.
This Ward identity corresponds to that of \cite{ban01} when there is no dilaton field.
The covariant form of the 2-dimensional gravitational anomaly ${\mathscr A}_\mu$ 
agrees with (\ref{42ano01}).
Using the 2-dimensional metric (\ref{grr2}),
the $\mu=t$ component of (\ref{2.40}) is written as
\begin{align}
\partial_r T^{~r}_{t(H)}(r)=\partial_r\left[-\frac{e^2}{2\pi}V_t(r_+)V_t(r)
+\frac{e^2}{4\pi}V^2_t(r)+N^{~r}_{t}\right].
\label{a28}
\end{align}
By integrating (\ref{a28}) over $r$ from $r_+$ to $r$, we obtain
\begin{align}
T^{~r}_{t(H)}(r)=-\frac{e^2}{2\pi}V_t(r_+) V_t(r)+\frac{e^2}{4\pi}V_t^2(r)+N^{~r}_{t}(r)
+\frac{e^2}{4\pi}V^2_t(r_+)-N^{~r}_t(r_+),
\label{a29}
\end{align}
where we impose the condition that the energy-momentum tensor vanishes at the horizon, 
which is the same as (\ref{42bc02}):
\begin{align}
T^{~r}_{t(H)}(r_+)=0.
\end{align}
As for (\ref{42bc03}),
we impose the condition that the asymptotic form of (\ref{a29}) 
in the limit $r\rightarrow \infty$ is equal to that of (\ref{2.39}),
\begin{align}
T^{~r}_{t(O)}(\infty)=T^{~r}_{t(H)}(\infty).
\label{a31}
\end{align}
From (\ref{a31}), we obtain
\begin{align}
a_o=\frac{e^2}{4\pi}V^2_t(r_+)-N^{~r}_t(r_+).
\end{align}
We thus obtain the flux of the energy-momentum tensor 
in the region far from the horizon as 
\begin{align}
T^{~r}_{t(O)}(r)=\frac{e^2 Q^2}{4\pi r^2_+}+\frac{\pi}{12\beta^2}+\frac{e^2Q}{2\pi r_+}V_t(r).
\end{align}
This result agrees with that of \cite{iso01}.
In contrast with the case of a rotating Kerr black hole,
the energy flux depends on the gauge field 
in the region far from the horizon, 
where the radial coordinate $r$ is large but still finite,
since the gauge field exists in a charged Reissner-Nordstr\"om black hole background.
However, in the evaluation of Hawking radiation by setting $r\rightarrow \infty$,
the effect of the gauge field $V_t(r)$ disappears.

%%%%%%%%%%%%%%%%%%%%%%%%%%%%%%%%%%%%%%%%%%%%%%%%%%%%%%%%%%%%%%%%%%%%%%%%%%%%%%%%%%%%%%%%%%%%%%%%%
%%                                                                                             %%
%%                                        Chapter 5                                            %%
%%                                                                                             %%
%%%%%%%%%%%%%%%%%%%%%%%%%%%%%%%%%%%%%%%%%%%%%%%%%%%%%%%%%%%%%%%%%%%%%%%%%%%%%%%%%%%%%%%%%%%%%%%%%

\newpage

\chapter{Hawking Radiation and Tunneling Mechanism}

\quad~ Parikh and Wilczek proposed a method of deriving Hawking radiation based on quantum tunneling \cite{par01}.
This derivation using the tunneling mechanism is intuitive and it is also capable of wide application.
The essential idea of the tunneling mechanism is that a particle-antiparticle pair is formed close to the horizon.
The ingoing mode is trapped inside the horizon while the outgoing mode can quantum mechanically tunnel through the horizon
and it is observed at infinity as the Hawking flux (Fig. 5.1).
As a background of this derivation,
we might consider that for the outgoing particles inside a black hole, the horizon plays a role as an infinite barrier.
This infinite barrier may be written as a potential of the delta-function type.
The particles cannot classically pass through the potential.
According to quantum theory, it is well known that 
a part of particles can pass through the potential by the quantum tunneling effect.
By applying the above discussion to the case of a black hole,
we can regard the particles appeared outside the horizon as the radiation from the black hole.
Since both of the tunneling effect and Hawking radiation are typical quantum effects, 
it is note that the quantum tunneling effect is related to Hawking radiation.

%%%%%%%%%%%%%     Fig.5.1       %%%%%%%%%%
\begin{center}
\vspace{0.6cm}
\input{tun01}\\
\vspace{0.6cm}
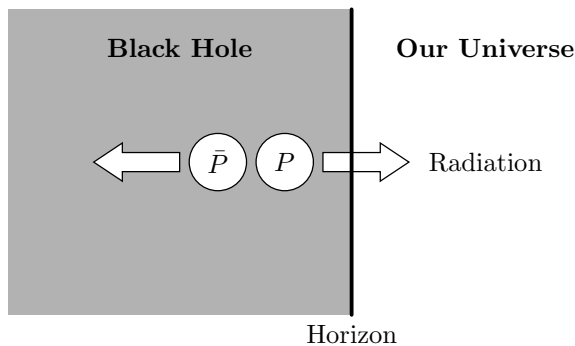
Fig. 5.1 \quad Intuitive picture of the tunneling mechanism
\vspace{0.6cm}
\end{center}
%%%%%%%%%%%%%%%%%%%%%%%%%%%%%%%%%%%%%%%%%%

Parikh and Wilczek calculated the WKB amplitudes for classically forbidden paths.
The first order calculation is given by
\begin{align}
\Gamma \sim e^{-2{\rm Im} S}\sim e^{-\frac{2\pi \omega}{\kappa}},
\label{5amp01}
\end{align}
where $\Gamma$ is the tunneling probability, $S$ is the action of the system, $\omega$ is a frequency and $\kappa$ is the surface gravity of the black hole respectively.
In comparison with the Boltzmann factor in a thermal equilibrium state at a temperature ${\cal T}$,
\begin{align}
\Gamma_{{\rm B}}=e^{-\frac{\omega}{{\cal T}}}, 
\label{5amp02}
\end{align}
it is confined that the temperature of (\ref{5amp01}) agrees with the Hawking temperature,
\begin{align}
{\cal T}_{{\rm BH}} =\frac{\kappa}{2\pi}.
\label{5tem01}
\end{align}
However, the analysis is confined to the derivation of the Hawking temperature
only by comparing the tunneling probability of an outgoing particle with the Boltzmann factor.
There exists no discussion of the spectrum.
Therefore, there remains the possibility that the black hole is not the black but merely the thermal body.
This problem was pointed out by Banerjee and Majhi \cite{maj01}.
They directly showed how to reproduce the black body spectrum with the Hawking temperature from the expectation value of number operator
by using the properties of the tunneling mechanism.
Thus the derivation from the tunneling mechanism became more satisfactory.

Their result is valid for black holes with spherically symmetric geometry such as Schwarzschild or Reissner-Nordstr\"om black holes in the 4-dimensional theory.
However, 4-dimensional black holes have not only a mass and a charge but also angular momentum
according to the black hole uniqueness theorem (the no hair theorem) \cite{car01,ruf01}.
In 4 dimensions, the Kerr-Newman black hole, which has both the charge and angular momentum, is the most general black hole
and its geometry becomes spherically asymmetric because of its own rotation.
There exist several previous works for a rotating black hole in the tunneling method (see for example \cite{zha01,zha02,jia01,che01,mod01}), 
but they are mathematically very involved.

We would like to extend the simplified derivation of Hawking radiation by Banerjee and Majhi on the basis of the tunneling mechanism to the case of the Kerr-Newman black hole.
In Section 2.4, we have shown that the 4-dimensional Kerr-Newman metric effectively becomes a 2-dimensional spherically symmetric metric by using the technique of the dimensional reduction near the horizon.
This technique was often used in the derivation of Hawking radiation from anomalies \cite{iso02,ume01}.

We note that this technique is valid only for the region very close to the horizon.
The use of the same technique in the tunneling mechanism is justified
since the tunneling effect is also the quantum effect arising within the Planck length near the horizon region.
By this procedure, the metric for the Kerr-Newman black hole becomes an effectively 2-dimensional spherically symmetric metric,
and we can use the approach of Banerjee and Majhi which is valid for black holes with spherically symmetric geometry. 
We can thus derive the black body spectrum and Hawking flux for the Kerr-Newman black hole in the tunneling mechanism.

The contents of this chapter are as follows.
In Section 5.1, we review the derivation of black hole radiation by the tunneling mechanism due to Parikh and Wilczek.
In Section 5.2, we discuss the method of Parikh and Wilczek from a point of view of the canonical theory.
In Section 5.3, we review a variant of the derivation from the tunneling mechanism by Banerjee and Majhi.
In Section 5.4, we extend the method of Banerjee and Majhi to the case of a Kerr-Newman black hole.

%%%%%%%%%%%%%%%%%%%%%%%%%%%%%%%%%%%%%%%%%%%%%%%%%%%%%%%%%%%%%%%%%%%%%%%%%%%%%%%%%%%%%%%%%%%%%%%%%
%%                                                                                             %%
%%                                        Section 5.1                                          %%
%%                                                                                             %%
%%%%%%%%%%%%%%%%%%%%%%%%%%%%%%%%%%%%%%%%%%%%%%%%%%%%%%%%%%%%%%%%%%%%%%%%%%%%%%%%%%%%%%%%%%%%%%%%%

\section{Hawking Radiation as Tunneling}

\quad~ Parikh and Wilczek proposed a method of deriving Hawking radiation based on quantum tunneling \cite{par01}.
In this section, we would like to review the derivation by Parikh and Wilczek.
For sake of simplicity, we consider the case of a 4-dimensional Schwarzschild black hole.
It is well known that the Schwarzschild metric is given by
\begin{align}
ds^2= -\left( 1-\frac{2M}{r}\right)dt^2+\frac{1}{1-\frac{2M}{r}}dr^2+r^2d\Omega^2,
\label{51sch01}
\end{align}
where $d\Omega^2$ is a 2-dimensional unit sphere.
This metric is singular at $r=0$ and $r=2M$.
A singularity at $r=0$ is the curvature singularity which cannot be removed, while the other singularity at $r=2M$ is a fictitious singularity arising merely from an improper choice of coordinates.

To remove the fictitious singularity, Parikh and Wilczek introduced the Painlev\'e coordinates \cite{pai01}
\begin{align}
t_p=t+2\sqrt{2Mr}+2M\ln \left( \frac{\sqrt{r}-\sqrt{2M}}{\sqrt{r}+\sqrt{2M}}\right),
\label{51pai01}
\end{align}
with
\begin{align}
dt_p=dt+\sqrt{\frac{2M}{r}} \frac{1}{1-\frac{2M}{r}}dr.
\label{51pai02}
\end{align}

By substituting (\ref{51pai02}) into (\ref{51sch01}), the Painlev\'e metric is given by
\begin{align}
ds^2=-\left( 1- \frac{2M}{r} \right)dt_p^2 + 2\sqrt{\frac{2M}{r}} dt_p dr +dr^2+r^2d\Omega^2,
\label{51pai03}
\end{align}
and we can confirm that there is no singularity at $r=2M$.
The radial null geodesics are given by
\begin{align}
\dot{r}_{p}\equiv \frac{dr}{dt_{p}}=\pm 1-\sqrt{\frac{2M}{r}},
\label{51pai04}
\end{align}
where the positive (negative) sign corresponds to the outgoing (ingoing) geodesic, under the implicit assumption that $t_p$ increases towards the future.

These equations are modified when the particle's self-gravitation is taken into account.
Self-gravitating shells in Hamiltonian gravity were studied by Kraus and Wilczek \cite{kra01}.
Now we consider that a particle with a positive energy $\omega$ inside a black hole quantum mechanically tunnels through the horizon and it appears outside the black hole (Fig. 5.2).
By the energy conservation law, the black hole energy decreases when the particle escapes from the black hole, namely,
\begin{align}
M=(M-\omega) + \omega,
\end{align}
where $M$ is the total ADM mass \cite{adm01,adm02,adm03} of the initial black hole, the first term $(M-\omega)$ of the right-hand side is the mass of the final black hole and the second term $\omega$ is the energy of particle.
After the particle escapes from the black hole, the metric of the black hole is given by
\begin{align}
ds^2=-\left( 1- \frac{2(M-\omega)}{r} \right)dt_p^2 + 2\sqrt{\frac{2(M-\omega)}{r}} dt_p dr +dr^2+r^2d\Omega^2,
\label{51pai05}
\end{align}
from (\ref{51pai03}).
Strictly speaking, it seems that we need to consider the time dependence of the black hole mass in this process. 
However, since it is involute, Parikh and Wilczek used this static metric as the background metric.

%%%%%%%%%%%%%     Fig.5.2       %%%%%%%%%%
\begin{center}
\vspace{0.6cm}
\input{tun02}\\
\vspace{0.6cm}
Fig. 5.2 \quad Particle emission by tunneling
\vspace{0.6cm}
\end{center}
%%%%%%%%%%%%%%%%%%%%%%%%%%%%%%%%%%%%%%%%%%

Since the typical wavelength of the radiation is of the order of the size of the black hole,
one might doubt whether a point particle description is appropriate.
However, when the outgoing wave is traced back towards the horizon, its wavelength, as measured by local fiducial observers,
is ever-increasingly blue-shifted.
Thus they considered that the radial wave number approaches infinity and the point particle or WKB approximation is justified near the horizon.

The imaginary part of the action for an s-wave outgoing positive energy particle which crosses the horizon outwards from $r_{{\rm in}}$ to $r_{{\rm out}}$ can be expressed as
\begin{align}
{\rm Im}~S ={\rm Im}\int^{r_{{\rm out}}}_{r_{{\rm in}}} p_p dr = {\rm Im}\int^{r_{{\rm out}}}_{r_{{\rm in}}}\int^{p_p}_0 d p'_p dr,
\label{51act01}
\end{align}
where $p_p$ is the canonical momentum for the radial coordinate $r$.
By using the Hamilton equation
\begin{align}
\dot{r}_p = \frac{dH_p}{dp_p},
\label{51ham01}
\end{align}
where $H_p$ is the Hamiltonian, the relation (\ref{51act01}) becomes
\begin{align}
{\rm Im}~S={\rm Im}\int^{M-\omega}_{M} \int^{r_{{\rm out}}}_{r_{{\rm in}}} \frac{1}{\dot{r}_p} dH_p dr.
\label{51act02}
\end{align}
By substituting the outgoing mode of the radial null geodesics associated with (\ref{51pai05}) into (\ref{51act02}) and using the Hamiltonian $H_p=M'-\omega$,
we obtain
\begin{align}
{\rm Im}~S={\rm Im}\int^{M-\omega}_{M} \int^{r_{{\rm out}}}_{r_{{\rm in}}} \frac{1}{1-\sqrt{\frac{2(M-\omega)}{r}}}dM'dr,
\label{51act03}
\end{align}
where we regarded $\omega$ as a constant and $M'$ as a variable.
By using Feynman's $i\epsilon$ prescription for positive energy solutions $\omega\to \omega-i\epsilon$, we obtain $M\to M-i\epsilon$ and
\begin{align}
{\rm Im}~S&={\rm Im}\int^{M-\omega}_{M} \int^{r_{{\rm out}}}_{r_{{\rm in}}} \frac{1}{1-\sqrt{\frac{2M'}{r}} +i\epsilon}dM'dr\\
&={\rm Im}\int^{r_{{\rm out}}}_{r_{{\rm in}}}
\left[ P \frac{1}{1-\sqrt{\frac{2M'}{r}}} +\int^{M-\omega}_{M}-i\pi \delta \left( 1-\sqrt{\frac{2M'}{r}} \right)dM'\right]dr,
\label{51act04}
\end{align}
where $P$ stands for the principal value.
We can ignore the first term because it is a real part.
Now since the mass ranges from $M$ to $M-\omega$, 
the radial coordinate ranges from $r_{{\rm in}}=2M$ to $r_{{\rm out}}=2(M-\omega)$.
Thus we can obtain the imaginary part of the action 
\begin{align}
{\rm Im}~S&=\int^{r_{{\rm out}}}_{r_{{\rm in}}}\int^{M-\omega}_{M}-\pi\delta \left( 1-\sqrt{\frac{2M'}{r}} \right)dM'dr\\
&=\int^{2(M-\omega)}_{2M} (-\pi r) dr\\
&=4\pi \omega M- 2\pi \omega^2.
\label{51act05}
\end{align}
The radially inward motion has a classically forbidden trajectory because the apparent horizon is itself contracting.
Thus, the limits on the integral indicate that, over the course of the classically forbidden trajectory, the outgoing particle starts from 
$r=2M-\epsilon$, just inside the initial position of the horizon, and traverses the contracting horizon to materialize at $r=2(M-\omega)+\epsilon$, just outside the final position of the horizon (Fig. 5.3).

%%%%%%%%%%%%%     Fig.5.3       %%%%%%%%%%
\begin{center}
\vspace{0.6cm}
\input{tun03}\\
\vspace{0.6cm}
Fig. 5.3 \quad The contracting horizon
\vspace{0.6cm}
\end{center}
%%%%%%%%%%%%%%%%%%%%%%%%%%%%%%%%%%%%%%%%%%

By using (\ref{51act05}), we find the semi-classical WKB probability as
\begin{align}
\Gamma\sim e^{-2{\rm Im}~S}=e^{-8\pi \omega M +4\pi \omega^2}.
\label{51amp01}
\end{align}
Note that the probability is given by an absolute square of the amplitude.
When we can ignore the quadratic term of $\omega$ in the case of $M\gg \omega$, the probability (\ref{51amp01}) can be written as
\begin{align}
\Gamma \sim e^{-8\pi \omega M}=e^{-\frac{2\pi\omega}{\kappa}},
\label{51amp02}
\end{align}
where we used the surface gravity of the Schwarzschild black hole $\displaystyle\kappa=\frac{1}{4M}$.
Here we recall thermodynamics.
It is well known that the Boltzmann factor in a thermal equilibrium state at a temperature ${\cal T}$ is given by
\begin{align}
\Gamma_{\rm B}=e^{-\frac{\omega}{{\cal T}}}.
\label{51amp03}
\end{align}
By comparing between (\ref{51amp02}) and (\ref{51amp03}),
we find that the temperature of the black hole is obtained by
\begin{align}
{\cal T}_{{\rm BH}}=\frac{\kappa}{2\pi}.
\end{align}
This result agrees with the result of previous works as in (\ref{31bht01}).

%%%%%%%%%%%%%%%%%%%%%%%%%%%%%%%%%%%%%%%%%%%%%%%%%%%%%%%%%%%%%%%%%%%%%%%%%%%%%%%%%%%%%%%%%%%%%%%%%
%%                                                                                             %%
%%                                        Section 5.2                                          %%
%%                                                                                             %%
%%%%%%%%%%%%%%%%%%%%%%%%%%%%%%%%%%%%%%%%%%%%%%%%%%%%%%%%%%%%%%%%%%%%%%%%%%%%%%%%%%%%%%%%%%%%%%%%%

\section{Tunneling Mechanism in the Canonical Theory}

\quad~ The tunneling mechanism by Parikh and Wilczek is explicit and straightforward in the canonical theory.
In this subsection, we would like to review the method of Parikh and Wilczek by using the canonical theory.

To begin with, we consider the action of the system.
The action is defined by
\begin{align}
S=\int L dt=-\mu_m \int ds = \int -\mu_m \frac{ds}{dt} dt, 
\label{52act01}
\end{align}
where $L$ is Lagrangian defined by
\begin{align}
L=-\mu_m \frac{ds}{dt},
\label{52lag01}
\end{align}
and $\mu_m$ is a mass of a particle.
It is convenient to take the time component of the Painlev\'e metric positive, namely,
\begin{align}
ds^2=\left( 1- \frac{2M}{r} \right)dt_p^2 - 2\sqrt{\frac{2M}{r}} dt_p dr -dr^2,
\label{52pai03}
\end{align}
where the Painlev\'e time $t_p$ is defined by
\begin{align}
t_p=t+2\sqrt{2Mr}+2M\ln \left( \frac{\sqrt{r}-\sqrt{2M}}{\sqrt{r}+\sqrt{2M}}\right),
\label{52pai01}
\end{align}
and $t$ is the Schwarzschild time. 
From the metric (\ref{52pai03}), we obtain
\begin{align}
\frac{ds}{dt_p}=\pm \sqrt{\left( 1- \frac{2M}{r} \right) - 2\sqrt{\frac{2M}{r}} \frac{dr}{dt_p} 
-\left(\frac{dr}{dt_p}\right)^2},
\label{52pai04}
\end{align}
where we adopt $+$.
We thus obtain the Lagrangian
\begin{align}
L_p(r,\dot{r}_p)=-\mu_m \sqrt{\left( 1- \frac{2M}{r} \right) - 2\sqrt{\frac{2M}{r}} \dot{r}_p 
-\dot{r}_p^2},
\end{align}
where $\displaystyle\dot{r}_p\equiv \frac{dr}{dt_p}$.
The canonical momentum for the radial coordinate $r$ is defined by
\begin{align}
p_p\equiv \frac{\partial L_p}{\partial \dot{r}_p}=
\frac{ \mu_m\left( \dot{r}_p+\sqrt{\frac{2M}{r}} \right)}{\sqrt{1-\left( \dot{r}_p +\sqrt{\frac{2M}{r}}\right)^2}},
\end{align}
and by solving for $\dot{r}_p$, we obtain
\begin{align}
\dot{r}_p=\pm \sqrt{\frac{p_p^2}{p_p^2+\mu_m^2}}-\sqrt{\frac{2M}{r}},
\label{52rp01}
\end{align}
where $+(-)$ represents the outgoing (ingoing) mode.
We adopt $+$, because we consider the outgoing mode.

In canonical theory, the Hamiltonian is defined by
\begin{align}
H=p \dot{r} -L(r,\dot{r}).
\label{52ham01}
\end{align}
By substituting (\ref{52rp01}) into (\ref{52ham01}), we obtain
\begin{align}
H_p(r,p_p)=p_p \left( \sqrt{\frac{p_p^2}{p_p^2+\mu_m^2}} - \sqrt{\frac{2M}{r}} \right) + \mu_m\sqrt{\frac{\mu_m^2}{p_p^2+\mu_m^2}}.
\label{52ham02}
\end{align}
In the derivation of Parikh and Wilczek, they used the null geodesic equation.
This can be reproduced by taking $\mu_m=0$ in (\ref{52rp01}).
By substituting $\mu_m=0$ into the Hamiltonian (\ref{52ham02}) and solving for $p_p$, we obtain
\begin{align}
p_p = \frac{H_p}{1- \sqrt{ \frac{2M}{r} }}.
\label{52mom01}
\end{align}

From both (\ref{52ham01}) and (\ref{52mom01}), the action (\ref{52act01}) is written as
\begin{align}
S=\int L_p dt_p = \int \left[ p_p \dot{r}_p - H_p \right] dt_p=\int p_p dr - \int H_p dt_p
=\int \frac{H_p}{1- \sqrt{ \frac{2M}{r} }} dr -\int H_p dt_p.
\label{52act02}
\end{align}
Now since the metric is stationary, it has a time-like Killing vector and 
there exists an energy as the corresponding conserved quantity.
The energy is defined as $\omega$ and it is the eigenvalue of the Hamiltonian $H_p=\omega$.
The action (\ref{52act02}) is written as
\begin{align}
S=\int^{r_{{\rm out}}} _{r_{{\rm in}}} \frac{\omega}{1- \sqrt{ \frac{2M}{r} }} dr -\int_{ t_{p(\rm in)} } ^{ t_{p(\rm out)} } \omega dt_p.
\label{52act03}
\end{align}
Now, we consider that an outgoing positive energy particle arising by the pair creation at $r_{{\rm in}}$ close to the horizon inside the black hole,
appears at $r_{{\rm out}}$ close to the horizon outside the black hole through the horizon $r_{\rm H}=2M$.
The Painlev\'e-time coordinates corresponding to these coordinates are respectively defined as $t_{p(\rm in)}$ and $t_{p(\rm out)}$.

Since Hawking radiation is a quantum effect, we have only to evaluate the classically hidden action i.e., the imaginary part of the action
\begin{align}
{\rm Im} ~S = {\rm Im} \int^{r_{{\rm out}}} _{r_{{\rm in}}} \frac{\omega}{1- \sqrt{ \frac{2M}{r} }} dr.
\label{52act04}
\end{align}
In the second term of (\ref{52act03}), the Painlev\'e-time coordinate is finite on the horizon
and it has no discontinuous point between $t_{p(\rm in)}$ and $t_{p(\rm out)}$.
This can be understood from a naive discussion that the Painlev\'e coordinate can be kept to finite values
even by substituting the future event horizon $(t,r)=(+\infty,2M)$ into (\ref{52pai01}).
We can ignore the second term in (\ref{52act03}) because it gives the real part only.
Since the first term in (\ref{52act03}) is singular at the horizon $r=2M$,
there is a possibility that the imaginary part appears, and we need to evaluate it.

By using the Feynman's $i\epsilon$ prescription for a real particle, we can obtain 
\begin{align}
{\rm Im} ~S &= {\rm Im} \int^{r_{{\rm out}}} _{r_{{\rm in}}} \frac{\omega}{1- \sqrt{ \frac{2M}{r} }-i\epsilon} dr \\
&={\rm Im} \left[ P\frac{\omega}{1- \sqrt{ \frac{2M}{r} }} + i \pi \omega\int^{r_{{\rm out}}} _{r_{{\rm in}}} \delta \left(1- \sqrt{ \frac{2M}{r} }\right)dr \right]\\
&=4\pi \omega M,
\label{52act04}
\end{align}
where $P$ stands for the principal value (the real part).
This result agrees with (\ref{51act05}) to the first order.

%%%%%%%%%%%%%%%%%%%%%%%%%%%%%%%%%%%%%%%%%%%%%%%%%%%%%%%%%%%%%%%%%%%%%%%%%%%%%%%%%%%%%%%%%%%%%%%%%
%%                                                                                             %%
%%                                        Section 5.3                                          %%
%%                                                                                             %%
%%%%%%%%%%%%%%%%%%%%%%%%%%%%%%%%%%%%%%%%%%%%%%%%%%%%%%%%%%%%%%%%%%%%%%%%%%%%%%%%%%%%%%%%%%%%%%%%%

\section{Hawking Black Body Spectrum from Tunneling Mechanism}

\quad~ A method of deriving Hawking radiation based on quantum tunneling was originally proposed by Parikh and Wilczek \cite{par01}.
After Parikh and Wilczek's derivation, a lot of papers on the tunneling mechanism have been published. 
However the analysis has been confined to obtain the Hawking temperature only by comparing the tunneling probability of an outgoing particle with the Boltzmann factor.
The discussion of the spectrum was not transparent.
Therefore, there remains the possibility that the black hole is not the black but merely the thermal body.
In this sense the tunneling method, presented so far, is not satisfactory yet.
This problem was emphasized by Banerjee and Majhi \cite{maj01}.
They showed how to reproduce the black body spectrum with the Hawking temperature directly from the expectation value of number operator
by using the properties of the tunneling mechanism.
Thus the derivation by the tunneling mechanism became more satisfactory.

In this subsection, we would like to review Banerjee and Majhi's method \cite{maj01}.
For sake of simplicity, we consider the case of a Schwarzschild black hole background.
The metric is then given by
\begin{align}
ds^2=-f(r)dt^2+\frac{1}{f(r)}dr^2+r^2d\Omega^2,
\label{53met01}
\end{align}
where $f(r)$ is defined by
\begin{align}
f(r)\equiv 1-\frac{2M}{r}.
\label{53met02}
\end{align}
We would like to note that they used not the Painlev\'e metric (\ref{51pai03}) but the Schwarzschild metric (\ref{53met01}).

Here we consider the Klein-Gordon equation for a massless scalar field
\begin{align}
g^{\mu\nu}\nabla_\mu\nabla_\nu \phi=0,
\label{53kge01}
\end{align}
where $\nabla_\mu$ is the covariant derivative defined by (\ref{31cov02}) and (\ref{31cov03}).
By using the $(r-t)$ sector of the metric (\ref{53met01}) in (\ref{53kge01}),
\begin{align}
ds^2=-f(r)dt^2+\frac{1}{f(r)}dr^2,
\label{53met03}
\end{align}
we obtain
\begin{align}
\left[ \frac{1}{f(r)}\partial_t^2-f(r)\partial_r^2 -f'(r)\partial_r\right]\phi=0.
\label{53kge00}
\end{align}
Taking the standard WKB ansatz
\begin{align}
\phi(r,t)=e^{\frac{i}{\hbar}S(r,t)},
\end{align}
and substituting the expansion for $S(r,t)$ in terms of the Planck constant $\hbar$
\begin{align}
S(r,t) =S_0 (r,t) +\sum_{i=1}^\infty \hbar^i S_i(r,t),
\end{align}
in (\ref{53kge00}) where we write the Planck constant explicitly, we obtain, in the semiclassical limit (i.e., keeping only $S_0$),
\begin{align}
\partial_t S_0 (r,t)=\pm f(r) \partial_r S_0 (r,t).
\label{53kge02}
\end{align}

Now we consider the classical Hamilton-Jacobi equation
\begin{align}
\frac{\partial S_0}{\partial t} +H=0,
\label{53hje01}
\end{align}
where $S_0$ is the classical action and $H$ is the Hamiltonian.
Since the metric (\ref{53met03}) is stationary, it has a timelike Killing vector.
Thus the Hamiltonian is given by
\begin{align}
H=\omega,
\end{align}
where $\omega$ is a constant and the conserved quantity corresponding to the timelike Killing vector.
This is identified as the effective energy experienced by the particle at asymptotic infinity.
By the Hamilton-Jacobi equation (\ref{53hje01}), we obtain
\begin{align}
S_0=-\omega t +\tilde{S}_0(r),
\label{53hje02}
\end{align}
where $\tilde{S}_0(r)$ is a time-independent arbitrary function.
By substituting (\ref{53hje02}) into (\ref{53kge02}), we obtain
\begin{align}
\omega=\pm f(r) \partial_r \tilde{S}_0(r).
\label{53hje03}
\end{align}
By using the relation of the tortoise coordinate defined by (\ref{2sch04}),
\begin{align}
\partial_{r_*}=f(r)\partial_r,
\end{align}
the Hamilton-Jacobi equation (\ref{53hje03}) becomes
\begin{align}
\partial_{r_*}\tilde{S}_0(r)=\pm \omega.
\label{53hje04}
\end{align}
By integrating (\ref{53hje04}) over $r_*$, we obtained
\begin{align}
\tilde{S}_0 (r) =\pm \omega r_* + C,
\end{align}
where $C$ is an integration constant and we ignore it since it is included in a normalization constant of the wave function.
By substituting this into (\ref{53hje02}), the classical action becomes
\begin{align}
S_0(r,t)=-\omega(t \mp r_*).
\label{53act01}
\end{align}
Thus we can obtain the semiclassical solution for the scalar field
\begin{align}
\phi(r,t)=\exp\left[-\frac{i}{\hbar}\omega(t\mp r_*)\right].
\label{53phi01}
\end{align}

Here we introduce both the retarded time $u$ and the advanced time $v$ defined by (\ref{2sch07}),
\begin{align}
u\equiv t-r_*,\qquad v\equiv t+r_*,
\label{53uv}
\end{align}
where we can regard $u$ $(v)$ as the outgoing (ingoing) modes of particles.
We can then separate the scalar field (\ref{53phi01}) into the ingoing (left handed) modes
and outgoing (right handed) modes.
Since the tunneling effect is the quantum effect arising within the Planck length in the near horizon region,
we have to consider both the inside and outside regions which are very close to the horizon.
In the regions $r_+-\varepsilon < r <r_+$, and $r_+ \leq r <r_+-\varepsilon$, respectively, we express the field $\phi$ as
\begin{align}
&\left. \begin{array}{ccc}
&\displaystyle \phi^{R} _{{\rm in}}=\exp \left( -\frac{i}{\hbar} \omega u_{{\rm in}}\right)\\
&\displaystyle \phi^{L}_{{\rm in}}=\exp \left( -\frac{i}{\hbar} \omega v_{{\rm in}}\right) 
\end{array}\right\} \qquad (r_+-\varepsilon < r <r_+),
\label{53phiin}\\
&\left. \begin{array}{ccc}
&\displaystyle \phi^{R}_{{\rm out}}=\exp \left( -\frac{i}{\hbar} \omega' u_{{\rm out}}\right)\\
&\displaystyle \phi^{L} _{{\rm out}}=\exp \left( -\frac{i}{\hbar} \omega' v_{{\rm out}}\right)
\end{array}\right\}
\quad (r_+ < r <r_++\varepsilon),
\label{53phiout}
\end{align}
where $\varepsilon$ is an arbitrarily small constant, ``R (L)" stands for the right (left) modes and ``in (out)" stands for the inside (outside) of the black hole, respectively (Fig. 5.4).
Here we note that these fields are defined both in the inside and outside regions which are very close to the horizon.

%%%%%%%%%%%%%     Fig.5.4       %%%%%%%%%%
\begin{center}
\vspace{0.6cm}
\input{tun04}\\
\vspace{0.6cm}
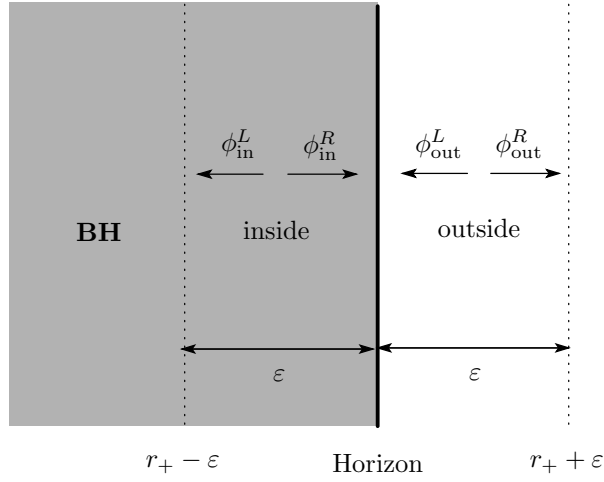
Fig. 5.4 \quad Intuitive picture of the scalar field near the horizon
\vspace{0.6cm}
\end{center}
%%%%%%%%%%%%%%%%%%%%%%%%%%%%%%%%%%%%%%%%%%

Now we consider that the outgoing particles inside the black hole quantum mechanically tunnel through the horizon.
However, it is well known that both the Schwarzschild coordinates $(t,r)$ and the Eddington-Finkelstein coordinates $(u,v)$
are singular at the horizon.
To describe horizon-crossing phenomena, Parikh and Wilczek used the Painlev\'e coordinates which have no singularity at the horizon
as already stated in Subsection 5.1.
On the other hand, Banerjee and Majhi used the Kruskal-Szekeres coordinates defined by
\begin{align}
\left. \begin{array}{ccc}
T&=\displaystyle \left( \frac{r}{2M} -1 \right)^{\frac{1}{2}} e^{\frac{r}{4M}} \sinh \left( \frac{t}{4M}\right)\\
R&=\displaystyle\left( \frac{r}{2M} -1 \right)^{\frac{1}{2}} e^{\frac{r}{4M}} \cosh \left( \frac{t}{4M}\right)
\end{array}
\right\} \quad {\rm when} \quad r >2M,
\label{53kru03}\\
\left. \begin{array}{ccc}
T&=\displaystyle\left( 1-\frac{r}{2M}  \right)^{\frac{1}{2}} e^{\frac{r}{4M}} \cosh \left( \frac{t}{4M}\right)\\
R&=\displaystyle\left( 1-\frac{r}{2M}  \right)^{\frac{1}{2}} e^{\frac{r}{4M}} \sinh \left( \frac{t}{4M}\right)
\end{array}
\right\} \quad {\rm when} \quad r <2M.
\label{53kru04}
\end{align}
It is well known that these coordinates also have no singularity at the horizon.
The Kruskal time $T$ and radial $R$ coordinates inside and outside the horizon are represented as
\begin{align}
&\left. \begin{array}{ccc}
&\displaystyle T_{\rm{out}}=\exp \left[ \kappa (r_*)_{\rm{out}}\right] \sinh \left( \kappa t_{\rm{out}} \right)\\
&\displaystyle R_{\rm{out}}=\exp \left[ \kappa (r_*)_{\rm{out}}\right] \cosh \left( \kappa t_{\rm{out}} \right)
\end{array}\right\},
\label{53kru02}\\
&\left. \begin{array}{ccc}
&\displaystyle T_{\rm{in}}=\exp \left[ \kappa (r_*)_{\rm{in}}\right] \cosh \left( \kappa t_{\rm{in}} \right)\\
&\displaystyle R_{\rm{in}}=\exp \left[ \kappa (r_*)_{\rm{in}}\right] \sinh \left( \kappa t_{\rm{in}} \right)
\end{array}\right\},
\label{53kru01}
\end{align}
where we used the tortoise coordinates as in (\ref{2sch06}) and the surface gravity of the Schwarzschild black hole (\ref{31sg01}) 
in both (\ref{53kru03}) and (\ref{53kru04}).

In general, the Schwarzschild metric describes the behavior outside the black hole.
Consequently, the readers may wonder if these metrics defined by the Kruskal-Szekeres coordinates
can really describe the behavior inside the black hole.
In our case, however, we study the tunneling effect across the horizon
and thus we study the behavior in the very small regions near the horizon.
In such an analysis, due to the reasons of continuity, it may not be unnatural to assume that
the Kruskal-Szekeres coordinates (\ref{53kru02}) and (\ref{53kru01}) can be used in both of outside and inside regions of near the horizon.
We note that these coordinates was used in the description of the Penrose diagram as already stated in Subsection 2.2.

A set of coordinates (\ref{53kru01}) are connected with the other coordinates (\ref{53kru02}) by the relations,
\begin{align}
t_{\rm{in}}\to t_{\rm{out}}-i\frac{\pi}{2\kappa},\quad
(r_*)_{\rm{in}}\to (r_*)_{\rm{out}}+i\frac{\pi}{2\kappa},
\label{53tr*}
\end{align}
so that, with this mapping, $T_{\rm{in}} \to T_{\rm{out}}$ and $R_{\rm{in}} \to R_{\rm{out}}$ smoothly.
Following the definition (\ref{53uv}), we obtain the relations connecting the null coordinates defined inside and outside the horizon,
\begin{align}
u_{\rm{in}}&\equiv t_{\rm{in}} -(r_*)_{\rm{in}} \to u_{\rm{out}}-i\frac{\pi}{\kappa},
\label{53u}\\
v_{\rm{in}}&\equiv t_{\rm{in}} +(r_*)_{\rm{in}}\to v_{\rm{out}}.
\label{53v}
\end{align}
This mapping is not defined if these coordinates are restricted to be real numbers.  
However, we can define it by extending these coordinates to complex numbers as in (\ref{53tr*}) and (\ref{53u}). 
We regard the appearance of the complex coordinates as a manifestation of quantum tunneling. 
In analogy to the quantum tunneling in ordinary quantum mechanics, 
our picture is that the inside solutions (\ref{53phiin}) and the outside solutions (\ref{53phiout}) of the infinitely high but very thin barrier, 
which is located on top of the horizon, are connected via the complex coordinates (i.e., tunneling);
the precise connection of two coordinates is determined by asking the smooth connection of the Kruskal-Szekeres coordinates  at the horizon.
Under these transformations the inside and outside modes are connected by,
\begin{align}
\phi^{R}_{\rm{in}}&\equiv \exp \left( -\frac{i}{\hbar} \omega u_{\rm{in}}\right) 
\to \exp\left( - \frac{\pi\omega}{\hbar \kappa} \right) \phi^{R}_{\rm{out}},
\label{53phiR}\\
\phi^{L}_{\rm{in}}&\equiv\exp \left( -\frac{i}{\hbar} \omega v_{\rm{in}}\right) \to \phi^{L}_{\rm{out}},
\label{53phiL}
\end{align}
where $\displaystyle \exp\left( - \frac{\pi\omega}{\hbar \kappa} \right)$ of (\ref{53phiR}) stands for the effect of the tunneling mechanism.
We would like to note that these scalar fields are still identified with Schr\"odinger amplitude for a single particle state.
Therefore, the squared absolute value of the wave function represents the probability.
The probability of the tunneling effect is given by
\begin{align}
\Gamma =\exp \left( -\frac{2\pi \omega}{\hbar\kappa}\right).
\end{align}
We find that this result in the natural system of units $\hbar=1$ agrees with Parikh and Wilczek's result as in (\ref{51amp02}).

To find the black body spectrum and the Hawking flux, Banerjee and Majhi considered $n$ number of non-interacting virtual pairs that 
are created inside the black hole.
According to quantum field theory, a wave function of one-particle system $\phi$ is related to the second-quantized field operator $\hat{\psi}$ by
\begin{align}
\phi=\langle 0| \hat{\psi} |\omega \rangle.
\end{align}
By following the approach of Banerjee and Majhi \cite{maj01}, each of these pairs is represented by the modes defined in the first set of (\ref{53phiin}) and (\ref{53phiout}).
Then they defined the physical state of the system, observed from outside, as
\begin{align}
| \Psi \rangle =N\sum_n |n_{\rm{in}}^L \rangle \otimes |n_{\rm{in}}^R \rangle,
\label{53sta}
\end{align}
where $|n_{\rm{in}}^{L(R)} \rangle$ is the number state of left (right) going modes inside the black hole and
$N$ is a normalization constant. From the transformations of both (\ref{53phiR}) and (\ref{53phiL}), we obtain
\begin{align}
|\Psi \rangle =N \sum_n e^{-\frac{\pi n \omega}{\hbar \kappa}} |n_{\rm{out}}^L \rangle \otimes |n_{\rm{out}}^R \rangle.
\label{53psi1}
\end{align}
Here $N$ can be determined by using the normalization condition $\langle \Psi | \Psi \rangle=1$.
It is natural to determine the normalization constant $N$ for the state outside the black hole, 
because the observer exists outside the black hole.
Thus we obtain
\begin{align}
N=  \frac{1}{\left( \displaystyle\sum _{n} e^{-\frac{2\pi n \omega}{\hbar \kappa}} \right) ^{\frac{1}{2}}} .
\end{align}
For bosons ($n=0, 1, 2, \cdots $), $N_{(\rm{boson})}$ is calculated as
\begin{align}
N_{(\rm{boson})}=\left( 1 - e^{-\frac{2\pi \omega}{\hbar \kappa}} \right)^{\frac{1}{2}}.
\label{53Nbos}
\end{align}
For fermions ($n=0, 1$), $N_{(\rm{fermion})}$ is similarly calculated as
\begin{align}
N_{(\rm{fermion})}=\left( 1 + e^{-\frac{2\pi \omega}{\hbar \kappa}} \right)^{-\frac{1}{2}}.
\label{53Nfer}
\end{align}
By substituting (\ref{53Nbos}) or (\ref{53Nfer}) into (\ref{53psi1}), we obtain the normalized physical states of a system of bosons or fermions
\begin{align}
|\Psi \rangle_{(\rm{boson})}&=\left( 1- e^{-\frac{2\pi \omega}{\hbar \kappa}} \right)^{\frac{1}{2}}\sum_{n} 
e^{-\frac{\pi n \omega}{\hbar \kappa}} |n_{\rm{out}}^{L}\rangle \otimes |n_{\rm{out}}^{R}\rangle,\\
|\Psi \rangle_{(\rm{fermion})}&=\left( 1+ e^{-\frac{2\pi \omega}{\hbar \kappa}} \right)^{-\frac{1}{2}}\sum_{n} 
e^{-\frac{\pi n \omega}{\hbar \kappa}} |n_{\rm{out}}^{L}\rangle \otimes |n_{\rm{out}}^{R}\rangle.
\end{align}
From here on this analysis will be only for bosons since the analysis for fermions is identical.
For bosons, the density matrix operator of the system is given by
\begin{align}
\hat{\rho}_{(\rm{boson})}\equiv& |\Psi \rangle_{(\rm{boson})}\langle \Psi |_{(\rm{boson})}\notag\\
=&\left( 1- e^{-\frac{2\pi \omega}{\hbar \kappa}} \right)\sum_{n,m}
e^{-\frac{\pi \omega}{\hbar \kappa}(n+m)}
|n_{\rm{out}}^{L}\rangle \otimes |n_{\rm{out}}^{R}\rangle\langle m_{\rm{out}}^{R} | \otimes \langle m_{\rm{out}}^{L}|.
\end{align}
By tracing out the left going modes, we obtain the reduced density matrix for the right going modes, 
\begin{align}
\hat{\rho}_{(\rm{boson})}^{(R)}
&={\rm Tr} \left( \hat{\rho}_{(\rm{boson})}^{(R)} \right)\\
&=\sum_{n}\langle n_{\rm{out}}^{L} | \hat{\rho}_{(\rm{boson})} |n_{\rm{out}}^{L}\rangle\\
&=\left( 1- e^{-\frac{2\pi \omega}{\hbar \kappa}} \right)\sum_{n}e^{-\frac{2\pi n \omega}{\hbar \kappa}}  |n_{\rm{out}}^{R}\rangle\langle n_{\rm{out}}^{R} |.
\label{red}
\end{align}
Therefore the average number of particles detected at asymptotic infinity is given by
\begin{align}
\langle n \rangle_{{\rm boson}} &= {\rm Tr} \left( \hat{n} \hat{\rho}_{(\rm{boson})}^{(R)} \right)\\
&=\left( 1- e^{-\frac{2\pi \omega}{\hbar \kappa}} \right) \sum_{n} n e^{-\frac{2\pi n \omega}{\hbar \kappa}}\\
&=\left( 1- e^{-\frac{2\pi \omega}{\hbar \kappa}} \right)\left( -\frac{\hbar \kappa}{2\pi}\right)\frac{\partial}{\partial \omega}\left( \sum_{n=0}^{\infty} e^{-\frac{2\pi n \omega}{\hbar \kappa}}\right)\\
&=\left( 1- e^{-\frac{2\pi \omega}{\hbar \kappa}} \right)\left( -\frac{\hbar \kappa}{2\pi}\right)\frac{\partial}{\partial \omega}\left( \frac{1}{1-e^{-\frac{2\pi \omega}{\hbar \kappa}} }
\right)\\
&=\frac{1}{e^{\frac{2\pi \omega}{\hbar \kappa}} -1},
\label{num3}
\end{align}
where the trace is taken over all $|n_{{\rm out}}^{(R)}\rangle$ eigenstates.
This result in the natural system of units $(\hbar=1)$ agrees with the result of Hawking's original derivation as in (\ref{31bla01}).
Similar analysis for fermions leads to the Fermi distribution given by
\begin{align}
\langle n \rangle_{{\rm fermion}}=\frac{1}{e^{\frac{2\pi \omega}{\hbar \kappa}}+1}.
\end{align}
Correspondingly, the Hawking flux $F_{{\rm H}}$ can be obtained by integrating the above distribution functions over all $\omega$'s.
By following the discussion in Subsection 5.4.3, we would like to evaluate the Hawking flux for fermions here.
It is given by
\begin{align}
F_{{\rm H}}=\frac{1}{\pi}\int _0^\infty \frac{\omega}{e^{\frac{2\pi \omega}{\hbar \kappa}} +1}d\omega=\frac{\hbar^2\kappa^2}{48\pi}=\frac{\pi}{12\beta^2},
\end{align}
where we used 
\begin{align}
\beta\equiv \frac{1}{{\cal T}_{{BH}}}\equiv\frac{2\pi}{\hbar\kappa}.
\end{align}
This result agrees with the case of $a=0$ in (\ref{42emt09}), i.e., the second term of (\ref{42emt09}).

Thus Banerjee and Majhi directly showed how to reproduce the black body spectrum and the Hawking flux with the Hawking temperature from the expectation value of the number operator by using the properties of the tunneling mechanism.
Therefore, it is shown that black holes are not merely the thermal body but the black body in the derivation from the tunneling mechanism.
In this sense, it may be said that the derivation on the basis of the tunneling mechanism became more satisfactory.

%%%%%%%%%%%%%%%%%%%%%%%%%%%%%%%%%%%%%%%%%%%%%%%%%%%%%%%%%%%%%%%%%%%%%%%%%%%%%%%%%%%%%%%%%%%%%%%%%
%%                                                                                             %%
%%                                        Section 5.4                                          %%
%%                                                                                             %%
%%%%%%%%%%%%%%%%%%%%%%%%%%%%%%%%%%%%%%%%%%%%%%%%%%%%%%%%%%%%%%%%%%%%%%%%%%%%%%%%%%%%%%%%%%%%%%%%%

\section{Hawking Radiation from Kerr-Newman Black Hole and Tunneling Mechanism}

\quad~ A variant of the approach to the derivation of Hawking radiation from the tunneling mechanism 
was suggested by Banerjee and Majhi \cite{maj01} as described in the preceding section.
However, as stated in their paper, their result is valid only for black holes with spherically symmetric geometry such as Schwarzschild or Reissner-Nordstr\"om black holes in the 4-dimensional theory.
According to the black hole uniqueness theorem (see Section 2.1), 4-dimensional black holes have not only a mass and a charge but also angular momentum.
In 4 dimensions, the Kerr-Newman black hole, which has both the charge and angular momentum, is the most general black hole
and its geometry becomes spherically asymmetric because of its own rotation.
There exist several previous works on a rotating black hole in the framework of the tunneling method (see for example \cite{zha01,zha02,jia01,che01,mod01}), 
but they are mathematically very involved.

In this section, we would like to extend the method of Banerjee and Majhi based the tunneling mechanism to the case of the Kerr-Newman black hole \cite{ume02}.
As shown in Section 2.4, we recall that the 4-dimensional Kerr-Newman metric effectively becomes a 2-dimensional spherically symmetric metric by using the technique of the dimensional reduction near the horizon.
To the best of my knowledge, there is no derivation of the spectrum by using the technique of the dimensional reduction in the tunneling mechanism.
Therefore, we believe that this derivation clarifies some aspects of the tunneling mechanism.
The essential idea is as follows: We consider the action for a scalar field. 
We can then ignore the mass, potential and interaction terms in the action because the kinetic term dominates in the high-energy theory near the horizon.
By expanding the scalar field in terms of the spherical harmonics and using properties at horizon,
we find that the integrand in the action dose not depend on angular variables.
Thus we find that the 4-dimensional action with the Kerr-Newman metric effectively becomes a 2-dimensional action with the spherically symmetric metric.

We note that this technique is valid only for the region near the horizon.
The use of the above technique in the tunneling mechanism is justified
since the tunneling effect is also the quantum effect arising within the Planck length near the horizon region.
By this procedure, the metric for the Kerr-Newman black hole becomes an effectively 2-dimensional spherically symmetric metric,
and we can use the approach of Banerjee and Majhi which is valid for black holes with spherically symmetric geometry. 
We can thus derive the black body spectrum and Hawking flux for the Kerr-Newman black hole in the tunneling mechanism.
We would like to suggest that the technique of the dimensional reduction is also valid for Parikh and Wilczek's original method in the tunneling mechanism.

The contents of this section are as follows.
In Subsection 5.4.1, we show how to define the Kruskal-like coordinate for the effective 2-dimensional metric.
In Subsection 5.4.2, we discuss the tunneling mechanism for the case of a Kerr-Newman black hole.
In Subsection 5.4.3, we show the black body spectrum and the Hawking flux for the case of the Kerr-Newman black hole.

%%%%%%%%%%%%%%%%%%%%%%%%%%%%%%%%%%%%%%%%%%%%%%%%%%%%%%%%%%%%%%%%%%%%%%%%%%%%%%%%%%%%%%%%%%%%%%%%%
%%                                                                                             %%
%%                                        Subsection 5.4.1                                     %%
%%                                                                                             %%
%%%%%%%%%%%%%%%%%%%%%%%%%%%%%%%%%%%%%%%%%%%%%%%%%%%%%%%%%%%%%%%%%%%%%%%%%%%%%%%%%%%%%%%%%%%%%%%%%

\subsection{Kruskal-like coordinates for the effective 2-dimensional metric}

\quad~ In this subsection, we briefly explain that the 4-dimensional Kerr-Newman metric becomes a 2-dimensional spherically symmetric metric 
by using technique of the dimensional reduction near the horizon.
Then the Kruskal-like coordinates for the reduced metric are required
instead of the Kruskal-Szekeres coordinates for the 2-dimensional Schwarzschild metric as in (\ref{53kru03}) and (\ref{53kru04}).
We would like to show how to obtain the Kruskal-like coordinates from the reduced metric.

For a rotating and charged black hole, the metric is given by the Kerr-Newman metric (\ref{2kn01})
\begin{align}
ds^2=&-\frac{\Delta -a^2 \sin^2\theta}{\Sigma}dt^2-\frac{2a\sin^2\theta}{\Sigma}(r^2+a^2-\Delta)dtd\varphi \notag\\
&-\frac{a^2\Delta \sin^2\theta-(r^2+a^2)^2}{\Sigma}\sin^2\theta d\varphi^2+\frac{\Sigma}{\Delta}dr^2+\Sigma d\theta^2,
\label{541kn01}
\end{align}
where notations were respectively defined in Section 2.1.
It follows from this expression that the Kerr-Newman metric is spherically asymmetric geometry.

In the Kerr-Newman black hole background, the 4-dimensional action for a complex scalar field is given by (\ref{KNact01})
\begin{align}
S=\int d^4x \sqrt{-g}g^{\mu\nu}( \partial_\mu +ieV_\mu ) \phi^* ( \partial_\nu -ieV_\nu )\phi +S_{{\rm int}},
\label{541act01}
\end{align}
where the first term is the kinetic term, the second term $S_{{\rm int}}$ represents the mass, potential and interaction terms and $V_\mu$ is a gauge field associated with the Coulomb potential of the black hole.

By using the technique of the dimensional reduction near the horizon, it can be shown that the 4-dimensional action (\ref{541act01}) becomes 
\begin{align}
S_{({\rm H})}=-\sum_{l,m}&\int dt dr \Phi \phi_{lm}^* \Bigg[ g^{tt}\left( \partial_t -iA_t \right)^2+\partial_r g^{rr} \partial_r \Bigg] \phi_{lm},
\end{align}
as shown in (\ref{kerract10}).
Then the effective metric near the horizon is given by (\ref{kerr5})
\begin{align}
ds^2=-f(r)dt^2+\frac{1}{f(r)}dr^2,
\label{541met01}
\end{align}
where $f(r)$ is defined by (\ref{fdef})
\begin{align}
f(r)\equiv \frac{\Delta}{r^2+a^2}.
\end{align}
We note that this function $f(r)$ certainly contains the effect of the rotating black hole expressed by $a$.
From the form of (\ref{541met01}), we find that a 4-dimensional Kerr-Newman metric (\ref{541kn01}) effectively becomes the 2-dimensional 
spherically symmetric metric near the horizon.
This expression also shows that it is reasonable to consider only the $(r-t)$ sector of the 4-dimensional metric 
and massless particles without interactions, which were used in previous works \cite{par01, maj01}.

Banerjee and Majhi used Kruskal-Szekeres coordinates for a spherically symmetric metric such as Schwarzschild or Reissner-Nordstr\"om metrics as in (\ref{53kru02}) and (\ref{53kru01}).
In general, the concrete forms of Kruskal-Szekeres coordinates for Schwarzschild or Reissner-Nordstr\"om metrics are well known (as for the case of a Reissner-Nordstr\"om metric, for example, see \S 3.1 in \cite{tow01}).
In order to extend Banerjee and Majhi's method to the case of a Kerr-Newman black hole,
we need to derive the Kruskal-like coordinates for the effective reduced metric (\ref{541met01}) following the derivation of the Kruskal-Szekeres coordinates in the standard textbook.
In this derivation, we apply a series of coordinate transformations.
Note that all the variables which appear in our transformations are defined to be real numbers.

Now the metric is given by (\ref{541met01}).
By using (\ref{21hor02}), $f(r)$ is also written as
\begin{align}
f(r)=\frac{(r-r_+)(r-r_-)}{r^2+a^2},
\label{541fdef01}
\end{align}
and $r_{+(-)}$ is the outer (inner) horizon given by (\ref{2hor01}).

As a first step of coordinate transformations, we use the tortoise coordinate defined by (\ref{fdef})
\begin{align}
dr_*\equiv \frac{1}{f(r)}dr.
\label{r*def01}
\end{align}
The metric (\ref{541met01}) is then written by
\begin{align}
ds^2=-f(r)(dt-dr_*)(dt+dr_*).
\label{ds02}
\end{align}
By integrating (\ref{r*def01}) over $r$ from 0 to r, we obtain
\begin{align}
r_*=r+\frac{1}{2\kappa_+}\ln \frac{|r-r_+|}{r_+}+\frac{1}{2\kappa_-}\ln\frac{|r-r_-|}{r_-}+C,
\label{r*def03}
\end{align}
where $\kappa_{+(-)}$ is the surface gravity on the outer (inner) horizon and $C$ is generally a pure imaginary integration constant which appears in the analytic continuation.
However, as already stated, we would like to treat the case where all the variables (or parameters) are defined in the range of real numbers.
Thus we need to consider the three cases $r>r_+$, $r_+<r<r_-$ and $r_-<r<0$ with respect to the range of $r$.
Actually, we have only to consider the two cases $r>r_+$ and $r_+<r<r_-$ because of the consideration near the outer horizon.
When $r>r_+$, the relation (\ref{r*def03}) becomes
\begin{align}
r_*=r+\frac{1}{2\kappa_+}\ln \frac{r-r_+}{r_+}+\frac{1}{2\kappa_-}\ln\frac{r-r_-}{r_-}.
\label{r*def04}
\end{align}
As the second step we use the retarded time $u$ and the advanced time $v$ defined by
\begin{align}
\left.
\begin{array}{ccc}
u&\displaystyle \equiv t-r_*=t-r-\frac{1}{2\kappa_+}\ln \frac{r-r_+}{r_+}-\frac{1}{2\kappa_-}\ln\frac{r-r_-}{r_-},\\
v&\displaystyle \equiv t+r_*=t+r+\frac{1}{2\kappa_+}\ln \frac{r-r_+}{r_+}+\frac{1}{2\kappa_-}\ln\frac{r-r_-}{r_-}.
\end{array}
\right\} \quad {\rm when} \quad r>r_+.
\end{align}
The metric (\ref{ds02}) is then written as
\begin{align}
ds^2=-f(r)dudv.
\label{ds03}
\end{align}
As the third step we use the following coordinate transformations $U$ and $V$ defined by
\begin{align}
\left.
\begin{array}{ccc}
\displaystyle U\equiv -e^{-\kappa_+ u}=-\left( \frac{r-r_+}{r_+} \right)^{\frac{1}{2}} \left( \frac{r-r_-}{r_-}\right)^{\frac{\kappa_+}{2\kappa_-}}e^{\kappa_+ r} e^{-\kappa_+t},\\
\displaystyle V\equiv e^{\kappa_+ v}=\left( \frac{r-r_+}{r_+} \right)^{\frac{1}{2}} \left( \frac{r-r_-}{r_-}\right)^{\frac{\kappa_+}{2\kappa_-}}
e^{\kappa_+ r} e^{\kappa_+t}.
\qquad \quad
\end{array}\right\} \quad {\rm when} \quad r>r_+.
\end{align}
The metric (\ref{ds03}) is then written as
\begin{align}
ds^2=-\frac{r_+r_-}{\kappa_+^2} \frac{e^{-2\kappa_+ r}}{r^2+a^2} \left( \frac{r_-}{r-r_-}\right)^{\frac{\kappa_+}{\kappa_-} -1} dUdV.
\label{ds04}
\end{align}
As the final step we use the following coordinate transformations $T$, $R$ defined by
\begin{align}
\left.
\begin{array}{ccc}
\displaystyle T \equiv \frac{1}{2}(V+U)=\left( \frac{r-r_+}{r_+} \right)^{\frac{1}{2}} \left( \frac{r-r_-}{r_-}\right)^{\frac{\kappa_+}{2\kappa_-}}
e^{\kappa_+ r} \sinh (\kappa_+ t),\\
\displaystyle R \equiv \frac{1}{2}(V-U)=\left( \frac{r-r_+}{r_+} \right)^{\frac{1}{2}} \left( \frac{r-r_-}{r_-}\right)^{\frac{\kappa_+}{2\kappa_-}}e^{\kappa_+ r} \cosh (\kappa_+ t).
\end{array}\right\} \quad {\rm when} \quad r>r_+.
\label{apkru01}
\end{align}
The metric (\ref{ds04}) is then written as
\begin{align}
ds^2=\frac{r_+r_-}{\kappa_+^2} \frac{e^{-2\kappa_+ r}}{r^2+a^2} \left( \frac{r_-}{r-r_-}\right)^{\frac{\kappa_+}{\kappa_-} -1} (-dT^2+dR^2).
\label{ds05}
\end{align}
Similarly, we consider the case of $r_+>r>r_-$.
When $r_+>r>r_-$, the relation (\ref{r*def03}) becomes
\begin{align}
r_*=r+\frac{1}{2\kappa_+}\ln \frac{r_+-r}{r_+}+\frac{1}{2\kappa_-}\ln\frac{r-r_-}{r_-}.
\label{r*def05}
\end{align}
As for the remaining coordinate transformations, we use the following ones
\begin{align}
&\left.
\begin{array}{ccc}
u&\equiv \displaystyle t-r_*=t-r-\frac{1}{2\kappa_+}\ln \frac{r_+-r}{r_+}-\frac{1}{2\kappa_-}\ln\frac{r-r_-}{r_-},\\
v&\equiv \displaystyle t+r_*=t+r+\frac{1}{2\kappa_+}\ln \frac{r_+-r}{r_+}+\frac{1}{2\kappa_-}\ln\frac{r-r_-}{r_-},
\end{array}
\right\} \quad {\rm when} \quad r_+>r>r_-,\\
&\left.
\begin{array}{ccc}
&U\equiv \displaystyle e^{-\kappa_+ u}=\left( \frac{r_+-r}{r_+} \right)^{\frac{1}{2}} \left( \frac{r-r_-}{r_-}\right)^{\frac{\kappa_+}{2\kappa_-}}e^{\kappa_+ r} e^{-\kappa_+t},\\
&V\equiv \displaystyle e^{\kappa_+ v}=\left( \frac{r_+-r}{r_+} \right)^{\frac{1}{2}} \left( \frac{r-r_-}{r_-}\right)^{\frac{\kappa_+}{2\kappa_-}}e^{\kappa_+ r} e^{\kappa_+t},\quad ~
\end{array}
\right\} \quad {\rm when} \quad r_+>r>r_-,\\
&\left.
\begin{array}{ccc}
\displaystyle  T \equiv \frac{1}{2}(V+U)=\left( \frac{r_+-r}{r_+} \right)^{\frac{1}{2}} \left( \frac{r-r_-}{r_-}\right)^{\frac{\kappa_+}{2\kappa_-}}e^{\kappa_+ r} \cosh (\kappa_+ t),\\
\displaystyle R \equiv \frac{1}{2}(V-U)=\left( \frac{r_+-r}{r_+} \right)^{\frac{1}{2}} \left( \frac{r-r_-}{r_-}\right)^{\frac{\kappa_+}{2\kappa_-}}e^{\kappa_+ r} \sinh (\kappa_+ t).
\end{array}
\right\} \quad {\rm when} \quad r_+>r>r_-.
\label{apkru02}
\end{align}
Of course, by performing these coordinate transformations, the corresponding metrics (\ref{ds03}), (\ref{ds04}) and (\ref{ds05}) 
do not change.
The two sets of coordinates introduced here, both of (\ref{apkru01}) and (\ref{apkru02}), are in fact the Kruskal-like coordinates.
In the Schwarzshild case ($a=Q=0$), (\ref{apkru01}) and (\ref{apkru02}) respectively agree with (\ref{53kru03}) and (\ref{53kru04}).
Finally, by rewriting the expressions written in terms of $r$ to the ones in terms of $r_*$ in the formulas (\ref{apkru01}) and (\ref{apkru02}),
we obtain
\begin{align}
&\left. \begin{array}{ccc}
&\displaystyle T=\exp \left[ \kappa_+ r_* \right] \sinh \left( \kappa_+ t \right)\\
&\displaystyle R=\exp \left[ \kappa_+ r_*\right] \cosh \left( \kappa_+ t \right)
\end{array}\right\}\quad {\rm when} \quad r>r_+,
\label{541kru02}\\
&\left. \begin{array}{ccc}
&\displaystyle T=\exp \left[ \kappa_+ r_*\right] \cosh \left( \kappa_+ t \right)\\
&\displaystyle R=\exp \left[ \kappa_+ r_*\right] \sinh \left( \kappa_+ t \right)
\end{array}\right\} \quad {\rm when} \quad r_+>r>r_-.
\label{541kru01}
\end{align}
These results agree with (\ref{53kru02}) and (\ref{53kru01}).
Thus we can regard these coordinate variables as the Kruskal-like coordinate variables for the effective reduced metric.

%%%%%%%%%%%%%%%%%%%%%%%%%%%%%%%%%%%%%%%%%%%%%%%%%%%%%%%%%%%%%%%%%%%%%%%%%%%%%%%%%%%%%%%%%%%%%%%%%
%%                                                                                             %%
%%                                        Subsection 5.4.2                                     %%
%%                                                                                             %%
%%%%%%%%%%%%%%%%%%%%%%%%%%%%%%%%%%%%%%%%%%%%%%%%%%%%%%%%%%%%%%%%%%%%%%%%%%%%%%%%%%%%%%%%%%%%%%%%%

\subsection{Tunneling mechanism}

\quad~ In this subsection, we discuss the connection between states inside and outside the black hole to analyze the tunneling effect in the induced metric.
We consider the Klein-Gordon equation near the horizon. 
In Section 2.4, 
we showed that we can regard the 4-dimensional Kerr metric as the 2-dimensional spherically symmetric metric in the region near the horizon.
As already stated, 
since the kinetic term dominates in the high-energy theory near the horizon, we can ignore the mass, potential and interaction terms.
We obtain the Klein-Gordon equation with the gauge field from the action (\ref{kerract9})
\begin{align}
\left[ \frac{1}{f(r)}\left( \partial_t -iA_t\right)^2-f(r)\partial_r^2 - f'(r) \partial_r \right] \phi=0,
\label{KGeq01}
\end{align}
where $A_t$ is defined in (\ref{gauge02}) and $f(r)$ is defined in (\ref{541fdef01}).
Of course, this equation can be obtained from the general Klein-Gordon equation for a free particle with the gauge field in 2-dimensional space-time
\begin{align}
g^{\mu\nu} (\nabla_\mu-iA_\mu) (\nabla_\nu-iA_\nu) \phi=0,
\end{align}
where $\nabla_\mu$ is the covariant derivative as in (\ref{31cov01}).
In a manner similar to the procedure explained in Section 5.3, we adopt the standard WKB ansatz
\begin{align}
\phi(r,t)=e^{\frac{i}{\hbar}S(r,t)},
\end{align}
and substituting the expansion of $S(r,t)$
\begin{align}
S(r,t)=S_0(r,t)+\sum^\infty_{i=1}\hbar^i S_i(r,t),
\end{align}
in (\ref{KGeq01}), we obtain, in the semiclassical limit (i.e., keeping only $S_0$),
\begin{align}
\partial_t S_0 (r,t)=\pm f(r) \partial_r S_0 (r,t).
\label{KGeq02}
\end{align}
We find that terms including the gauge field vanished in the semiclassical limit.
This equation completely agrees with the equation in \cite{maj01} although the content of $f(r)$ is different from that used in \cite{maj01}.
From the Hamilton-Jacobi equation as in (\ref{53hje01}) and (\ref{KGeq02}), we obtain 
\begin{align}
S_0(r,t)=- \left( \omega -e\Phi_{{\rm H}} - m\Omega_{{\rm H}} \right)(t\pm r_*)\equiv -\omega'(t\pm r_*),
\label{ome'}
\end{align}
where $r_*$ is the tortoise coordinate defined by (\ref{fdef}), $\omega$ is the characteristic frequency;
$\Phi_{{\rm H}}$ is the electric potential defined by (\ref{2ep01}),
and $\Omega_{{\rm H}}$ is the angular frequency on the horizon respectively defined by (\ref{2av01}),
\begin{align}
\Phi_{{\rm H}}=\frac{Qr_+}{r_+^2+a^2},\quad \Omega_{{\rm H}}=\frac{a}{r_+^2+a^2}.
\end{align}
We also used the fact that the Hamiltonian is asymptotically given by
\begin{align}
H=\omega -e\Phi_{{\rm H}} - m\Omega_{{\rm H}}.
\end{align}
Thus we obtain the semiclassical solution for the scalar field
\begin{align}
\phi(r,t)=\exp\left[ -\frac{i}{\hbar} \omega' (t \pm r_*) \right].
\label{phi01}
\end{align}

If one considers the metric at all regions for the charged and rotating black hole,
the exterior metric of the horizon is given by the 4-dimensional Kerr-Newman metric (\ref{541kn01}).
However, we showed that the metric near the horizon can be regarded as the 2-dimensional spherically symmetric metric as in (\ref{541met01}).
We thus use the metric (\ref{541met01}) in the region near the horizon.
But we do not know the interior metric of the black hole.
However, if the interior metric is smoothly connected to the exterior metric through the horizon,
we may be able to identify the interior metric near the horizon to be the same as the exterior metric near the horizon.
We thus suppose that the interior metric near the horizon is given by (\ref{541met01}).
Furthermore, since the tunneling effect is the quantum effect arising within the Planck length in the near horizon region,
we have to consider both the inside and outside regions which are very close to the horizon.

Here we use both the retarded time $u$ and the advanced time $v$ defined by
\begin{align}
u\equiv t-r_*,\qquad v\equiv t+r_*.
\label{uv}
\end{align}
We can then separate the scalar field (\ref{phi01}) into the ingoing (left handed) modes and outgoing (right handed) modes.
In the regions $r_+-\varepsilon < r <r_+$, and $r_+ \leq r <r_+-\varepsilon$, respectively, we express the field $\phi$ as
\begin{align}
&\left. \begin{array}{ccc}
&\displaystyle \phi^{R} _{\rm{in}}=\exp \left( -\frac{i}{\hbar} \omega' u_{\rm{in}}\right)\\
&\displaystyle \phi^{L}_{\rm{in}}=\exp \left( -\frac{i}{\hbar} \omega' v_{\rm{in}}\right) 
\end{array}\right\} \qquad (r_+-\varepsilon < r <r_+),
\label{phiin}\\
&\left. \begin{array}{ccc}
&\displaystyle \phi^{R}_{\rm{out}}=\exp \left( -\frac{i}{\hbar} \omega' u_{\rm{out}}\right)\\
&\displaystyle \phi^{L} _{\rm{out}}=\exp \left( -\frac{i}{\hbar} \omega' v_{\rm{out}}\right)
\end{array}\right\}
\quad (r_+ < r <r_++\varepsilon),
\label{phiout}
\end{align}
where $\varepsilon$ is an arbitrarily small constant, ``R (L)" stands for the right (left) modes and ``in (out)" stands for the inside (outside) of the black hole, respectively (see Fig. 5.4).
Here we note that these fields are defined both the inside and outside regions which are very close to the horizon.
As for the definition of the fields in the region close to the horizon, there are related discussions in the literatures \cite{ful01,ful02}.

As shown in Subsection 5.4.1, we use the Kruskal-like coordinate variables
\begin{align}
&\left. \begin{array}{ccc}
&T_{\rm{out}}=\exp \left[ \kappa_+ (r_*)_{\rm{out}}\right] \sinh \left( \kappa_+ t_{\rm{out}} \right)\\
&R_{\rm{out}}=\exp \left[ \kappa_+ (r_*)_{\rm{out}}\right] \cosh \left( \kappa_+ t_{\rm{out}} \right)
\end{array}\right\}.
\label{Kru01}\\
&\left.
\begin{array}{ccc}
&T_{\rm{in}}=\exp \left[ \kappa_+ (r_*)_{\rm{in}}\right] \cosh \left( \kappa_+ t_{\rm{in}} \right)\\
&R_{\rm{in}}=\exp \left[ \kappa_+ (r_*)_{\rm{in}}\right] \sinh \left( \kappa_+ t_{\rm{in}} \right)
\end{array}\right\},
\label{Kru02}
\end{align}
Both of the relations (\ref{Kru01}) and (\ref{Kru02}) agree with the relations in \cite{maj01}.

In general, the Schwarzschild and Kerr-Newman metrics describe the behavior outside the black hole.
Consequently, the readers may wonder if these metrics defined by the Kruskal coordinates
can really describe the behavior inside the black hole.
In our case, however, we study the tunneling effect across the horizon
and thus we study the behavior in the very small regions near the horizon.
In such analysis, due to the reasons of continuity, it may not be unnatural to assume that
the Kruskal coordinates (\ref{Kru01}) and (\ref{Kru02}) can be used in both of outside and inside regions of near the horizon.

A set of coordinates (\ref{Kru01}) are connected with the other coordinates (\ref{Kru02}) by the relations,
\begin{align}
t_{\rm{in}}\to t_{\rm{out}}-i\frac{\pi}{2\kappa_+},\quad
(r_*)_{\rm{in}}\to (r_*)_{\rm{out}}+i\frac{\pi}{2\kappa_+},
\label{tr*}
\end{align}
so that, with this mapping, $T_{\rm{in}} \to T_{\rm{out}}$ and $R_{\rm{in}} \to R_{\rm{out}}$ smoothly.
Following the definition (\ref{uv}), we obtain the relations connecting the null coordinates defined inside and outside the horizon,
\begin{align}
u_{\rm{in}}&\equiv t_{\rm{in}} -(r_*)_{\rm{in}} \to u_{\rm{out}}-i\frac{\pi}{\kappa_+},
\label{u}\\
v_{\rm{in}}&\equiv t_{\rm{in}} +(r_*)_{\rm{in}}\to v_{\rm{out}}.
\label{v}
\end{align}
This mapping is not defined if these coordinates are restricted to be real numbers.  
However, we can define it by extending these coordinates to complex numbers as in (\ref{tr*}) and (\ref{u}). 
We regard the appearance of the complex coordinates as a manifestation of quantum tunneling. 
In analogy to the quantum tunneling in ordinary quantum mechanics, 
our picture is that the inside solutions (\ref{phiin}) and the outside solutions  (\ref{phiout}) of the infinitely high but very thin barrier, 
which is located on top of the horizon, are connected via the complex coordinates (i.e., tunneling);
the precise connection of two coordinates is determined by asking the smooth connection of the Kruskal-like coordinates  at the horizon.
Under these transformations the inside and outside modes are connected by,
\begin{align}
\phi^{R}_{\rm{in}}&\equiv \exp \left( -\frac{i}{\hbar} \omega' u_{\rm{in}}\right) 
\to \exp\left( - \frac{\pi\omega'}{\hbar \kappa_+} \right) \phi^{R}_{\rm{out}},
\label{phiR}\\
\phi^{L}_{\rm{in}}&\equiv\exp \left( -\frac{i}{\hbar} \omega' v_{\rm{in}}\right) \to \phi^{L}_{\rm{out}}.
\label{phiL}
\end{align}
As already discussed by Banerjee and Majhi in \cite{maj01}, the essential idea of the tunneling mechanism is that a particle-antiparticle pair is formed close to the horizon.
This pair creation may arise inside the black hole ( in the region close to the horizon ), 
since the space-time is locally flat.
The ingoing mode is trapped inside the horizon while the outgoing mode can quantum mechanically tunnel through the horizon.
The outgoing mode is then observed at infinity as the Hawking flux. 
We find that the effect of the ingoing mode inside the horizon do not appear outside the horizon as in (\ref{phiL})
since $v_{\rm{in}}$ changes to $v_{\rm{out}}$ without an extra term under the transformation connecting the null coordinates defined inside and outside the horizon as in (\ref{v}).
On the other hand, we find that the effect of the outgoing mode inside the horizon appear with a non-negligible probability  
by tunneling through the horizon quantum mechanically as in (\ref{phiR}).
This consideration agrees with the concept of tunneling mechanism.
Furthermore, we showed that we can treat the Kerr-Newman metric as a 2-dimensional spherically symmetric metric 
with a 2-dimensional effective gauge field just as in the case of
the simplest Schwarzschild metric in the tunneling mechanism.

%%%%%%%%%%%%%%%%%%%%%%%%%%%%%%%%%%%%%%%%%%%%%%%%%%%%%%%%%%%%%%%%%%%%%%%%%%%%%%%%%%%%%%%%%%%%%%%%%
%%                                                                                             %%
%%                                        Subsection 5.4.2                                     %%
%%                                                                                             %%
%%%%%%%%%%%%%%%%%%%%%%%%%%%%%%%%%%%%%%%%%%%%%%%%%%%%%%%%%%%%%%%%%%%%%%%%%%%%%%%%%%%%%%%%%%%%%%%%%

\subsection{Black body spectrum and Hawking flux}

\quad~ In this subsection, we show how to derive the Hawking black body spectrum for a Kerr-Newman black hole by following the approach of Banerjee and Majhi as shown in Section 5.3.
First, we consider $n$ number of non-interacting virtual pairs that are created inside the black hole.
Then the physical state of the system is conventionally written as
\begin{align}
| \Psi \rangle =N\sum_n |n_{\rm{in}}^L \rangle \otimes |n_{\rm{in}}^R \rangle,
\label{sta}
\end{align}
where $|n_{\rm{in}}^{L(R)} \rangle$ is the number state of left (right) going modes inside the black hole and
$N$ is a normalization constant. From the transformations of both (\ref{phiR}) and (\ref{phiL}), we obtain
\begin{align}
|\Psi \rangle =N \sum_n e^{-\frac{\pi n \omega'}{\hbar \kappa_+}} |n_{\rm{out}}^L \rangle \otimes |n_{\rm{out}}^R \rangle.
\label{5psi1}
\end{align}
Here $N$ can be determined by using the normalization condition $\langle \Psi | \Psi \rangle=1$.
It is natural to determine the normalization constant $N$ for the state outside the black hole, 
because the observer exists outside the black hole.
Thus we obtain
\begin{align}
N=  \frac{1}{\left( \displaystyle\sum _{n} e^{-\frac{2\pi n \omega'}{\hbar \kappa_+}} \right) ^{\frac{1}{2}}} .
\end{align}
For bosons ($n=0, 1, 2, \cdots $), $N_{(\rm{boson})}$ is calculated as
\begin{align}
N_{(\rm{boson})}=\left( 1 - e^{-\frac{2\pi \omega'}{\hbar \kappa_+}} \right)^{\frac{1}{2}}.
\label{5Nbos}
\end{align}
For fermions ($n=0, 1, 2$), $N_{(\rm{fermion})}$ is also calculated as
\begin{align}
N_{(\rm{fermion})}=\left( 1 + e^{-\frac{2\pi \omega'}{\hbar \kappa_+}} \right)^{-\frac{1}{2}}.
\label{5Nfer}
\end{align}
By substituting (\ref{5Nbos}) or (\ref{5Nfer}) into (\ref{5psi1}), we obtain the normalized physical states of a system of bosons or fermions
\begin{align}
|\Psi \rangle_{(\rm{boson})}&=\left( 1- e^{-\frac{2\pi \omega'}{\hbar \kappa_+}} \right)^{\frac{1}{2}}\sum_{n} 
e^{-\frac{\pi n \omega'}{\hbar \kappa_+}} |n_{\rm{out}}^{L}\rangle \otimes |n_{\rm{out}}^{R}\rangle,\\
|\Psi \rangle_{(\rm{fermion})}&=\left( 1+ e^{-\frac{2\pi \omega'}{\hbar \kappa_+}} \right)^{-\frac{1}{2}}\sum_{n} 
e^{-\frac{\pi n \omega'}{\hbar \kappa_+}} |n_{\rm{out}}^{L}\rangle \otimes |n_{\rm{out}}^{R}\rangle.
\end{align}
Similarly to Section 5.3, we will present the analysis for bosons only.
The density matrix operator of the system is given by
\begin{align}
\hat{\rho}_{(\rm{boson})}\equiv& |\Psi \rangle_{(\rm{boson})}\langle \Psi |_{(\rm{boson})}\notag\\
=&\left( 1- e^{-\frac{2\pi \omega'}{\hbar \kappa_+}} \right)\sum_{n,m}
e^{-\frac{\pi \omega'}{\hbar \kappa_+}(n+m)}\notag\\
&\times
|n_{\rm{out}}^{L}\rangle \otimes |n_{\rm{out}}^{R}\rangle\langle m_{\rm{out}}^{R} | \otimes \langle m_{\rm{out}}^{L}|.
\end{align}
By tracing out the left going modes, we obtain the reduced density matrix for the right going modes, 
\begin{align}
\hat{\rho}_{(\rm{boson})}^{(R)}
&={\rm Tr} \left( \hat{\rho}_{(\rm{boson})}^{(R)} \right)\\
&=\sum_{n}\langle n_{\rm{out}}^{L} | \hat{\rho}_{(\rm{boson})} |n_{\rm{out}}^{L}\rangle\\
&=\left( 1- e^{-\frac{2\pi \omega'}{\hbar \kappa_+}} \right)\sum_{n}e^{-\frac{2\pi n \omega'}{\hbar \kappa_+}}  |n_{\rm{out}}^{R}\rangle\langle n_{\rm{out}}^{R} |.
\label{red}
\end{align}
Then, the expectation value of the number operator $\hat{n}$ is given by
\begin{align}
\langle n \rangle_{{\rm boson}} &= {\rm Tr} \left( \hat{n} \hat{\rho}_{(\rm{boson})}^{(R)} \right)\\
&=\frac{1}{e^{\frac{2\pi \omega'}{\hbar \kappa_+}} -1}\\
&=\frac{1}{e^{\beta \left( \omega -e \Phi - m \Omega \right)} -1}
\label{num3}
\end{align}
where in the last line we used the definition (\ref{ome'}) and we identify the Hawking temperature ${\cal T}_{{\rm BH}}$ by
\begin{align}
\beta \equiv \frac{1}{{\cal T}_{{\rm BH}}} \equiv \frac{2\pi}{\hbar \kappa_+}.
\end{align}
This result corresponds to the black body spectrum with the Hawking temperature
and agrees with previous works in the Kerr-Newman black hole background \cite{haw01}.
Similar analysis for fermions leads to the Fermi distribution
\begin{align}
\langle n \rangle_{{\rm fermion}}=\frac{1}{e^{\beta \left( \omega -e \Phi - m \Omega \right)} +1}.
\end{align}
Moreover, the Hawking flux $F_H$ is derived by integrating the sum of the distribution function for a particle with a quantum number ($e$, $m$) and its antiparticle with ($-e$, $-m$) over all $\omega$'s.
However, as shown in Section 2.3, boson fields display the superradiance provided that they have 
frequency in a certain range whereas fermion fields do not.
We therefore evaluate the Hawking flux for fermions.
It is given by
\begin{align}
F_{{\rm H}}&= \int^\infty _0 \frac{d\omega}{2\pi} \omega \left[ \frac{1}{e^{\beta(\omega-e\Phi-m\Omega)}+1}
+\frac{1}{e^{\beta(\omega+e\Phi +m\Omega)}+1} \right]\\
&=\frac{\pi}{12\beta^2}+\frac{1}{4\pi} (e\Phi+m\Omega)^2.
\end{align}
%where we ignored the superradiance. In the fermionic case also, one can derive the Hawking flux following the above derivation. 
This result agrees with the previous result \cite{iso02} (see Appendix A in \cite{iso02}).

One might be surprised by the sudden appearance of fermions.
However, we can explicitly present an answer to the question.
As shown in Section 2.4, we know that the effective theory near the horizon becomes a 2-dimensional theory.
According to 2-dimensional quantum field theory, it is known that there exists the boson-fermion duality \cite{col01}.
Namely, the 2-dimensional boson theory can be treated as the 2-dimensional fermion theory by the fermionization.
We can also discuss the tunneling mechanism for fermions in a manner similar to bosons.

Before closing, we discuss the black hole entropy.
Since a particle emitted by the black hole has the Hawking temperature, it is natural to consider that the black hole itself has the same temperature.
Thus we can obtain the black hole entropy $dS_{{\rm BH}}$ from a thermodynamic consideration
\begin{align}
dS_{{\rm BH}}=\left( \frac{dM}{{\cal T}_{BH}} \right).
\label{ent01}
\end{align}
By integrating Eq. (\ref{ent01}), the entropy agrees with the Bekenstein-Hawking entropy
\begin{align}
S_{{\rm BH}}=\frac{A_{{\rm BH}}}{4\hbar},
\label{ent02}
\end{align}
where $A_{{\rm BH}}$ is the surface area of the black hole
\begin{align}
A_{{\rm BH}} =4\pi (r_+^2+a^2).
\end{align}
This result agrees with the Bekenstein-Hawking entropy (\ref{ent02}).
On the other hand, we defined quantum states as in (\ref{sta}) and obtained the reduced density matrix as in (\ref{red}).
We can evaluate the entropy for these states by using the von Neumann entropy formula.
However, we must mention that it is not the entropy of the black hole itself but rather the entropy of the boson field.
Our method does not allow us to derive the entropy for a black hole with a finite temperature by counting the number of quantum states associated with the black hole.
We thus simply derived the black hole entropy from thermodynamic considerations as in (\ref{ent01}) following previous works.
The derivation of the black hole entropy by counting the number of black hole quantum states remains as one of future problems.

Finally, we mention that further recent developments associated with this derivation are given in \cite{maj02,mod02,deb01}.

%%%%%%%%%%%%%%%%%%%%%%%%%%%%%%%%%%%%%%%%%%%%%%%%%%%%%%%%%%%%%%%%%%%%%%%%%%%%%%%%%%%%%%%%%%%%%%%%%
%%                                                                                             %%
%%                                        Chapter 6                                            %%
%%                                                                                             %%
%%%%%%%%%%%%%%%%%%%%%%%%%%%%%%%%%%%%%%%%%%%%%%%%%%%%%%%%%%%%%%%%%%%%%%%%%%%%%%%%%%%%%%%%%%%%%%%%%

\newpage

\chapter{Discussion and Conclusion}

\quad~ In this thesis, we investigated the black hole radiation which is commonly called Hawking radiation.
Hawking radiation is one of very interesting phenomena where both of general relativity and quantum theory play a role at the same time
since Hawking radiation is derived by taking into account the quantum effects in the framework of general relativity.
Hawking radiation is widely accepted by now because the same result is derived by several different methods.
At the same time, there remain several aspects which have yet to be clarified.
We attempted to clarify some arguments in previous works and present more satisfactory derivations of Hawking radiation.

In Chapter 2, we reviewed basic facts and various properties of black holes as the necessary preparation to discuss Hawking radiation.
We showed that both of the black hole solutions and their types are given as a result of general relativity,
and that Penrose diagrams are useful to understand the global structure of black hole space-time;
a part of energy can be extracted from a rotating black hole by the Penrose process
and the technique of the dimensional reduction plays an important role to understand the behavior of matter fields near the event horizon.
We also discussed analogies between black hole physics and thermodynamics, and 
we explained a method to derive the black hole entropy which was suggested by Bekenstein.
In particular, to understand the properties of black holes, it is very useful to consider black hole physics in terms of 
well-known thermodynamics.
However, the corresponding relationships are no more than analogies in classical theory.
If we would like to show the complete correspondence between black hole physics and thermodynamics, namely,
to show that black holes actually have entropy and temperature,
we need to explain black hole radiation.
Although Bekenstein suggested that black holes can have entropy,
the complete corresponding relationships was not established because Bekenstein was not able to explain the mechanism of black hole radiation.

In Chapter 3, we discussed several previous works on Hawking radiation.
Hawking radiation which is derived by using quantum effects in black hole physics.
By following Hawking's original derivation, we calculated the expectation value of the particle number by using the Bogoliubov transformations.
As is well known, the result agrees with the black body spectrum with a certain temperature.
By defining the temperature as the black hole temperature,
it is confirmed that a black hole behaves as a black body and we can thus explain the black hole radiation.
The existence of black hole radiation implies that
the Hawking area theorem is violated.
However, the second law of black hole physics holds in a suitably generalized form.
From these considerations, we can understand the complete corresponding relationships between black hole physics and thermodynamics, 
and we find that the radiation-dominated tiny black hole will eventually evaporate at some point.
We also briefly reviewed some representative derivations of Hawking radiation and stated that there remain some aspects to be clarified.

In Chapter 4, we discussed the derivation of Hawking radiation based on anomalies.
The essential idea is as follows:
We can divide the field associated with a particle into ingoing modes falling toward the horizon and outgoing modes moving away from the horizon near the horizon.
Since none of ingoing modes at horizon are expected to affect the classical physics outside the horizon, we ignore the ingoing modes.
Anomalies then appear since the effective theory is chiral near the horizon.
Anomalies mean the symmetry breaking by the quantization, i.e., the corresponding conservation law is violated, and it is known that a source of energy is generated.
We can consider that this source of energy corresponds to the characteristic energy flux of Hawking radiation.
Since both of the anomalies and Hawking radiation are typical quantum effects,
it is natural that Hawking radiation is related to anomalies.
A method of deriving Hawking radiation based on the consideration of anomalies was first suggested by Robinson and Wilczek and generalized by Iso, Umetsu and Wilczek.
However, there remained some aspects to be clarified.
We clarified some arguments in previous works on this approach and presented a simple derivation of Hawking radiation from anomalies.
We showed how to derive the Hawking flux by using the Ward identities for covariant currents and two physically-meaningful boundary conditions.
This derivation has a merit that we can potentially incorporate matter fields with mass and interactions away from the horizon.

In Chapter 5, we discussed the derivation of Hawking radiation on the basis of quantum tunneling.
The basic idea of the tunneling mechanism is as follows: 
We imagine that a particle-antiparticle pair is formed close to the horizon.
The ingoing mode is trapped inside the horizon
while the outgoing mode can quantum mechanically tunnel through the horizon and it is observed
at infinity as the Hawking flux.
A method of deriving Hawking radiation based on the tunneling mechanism was suggested by Parikh and Wilczek.
In this derivation, the discussion of the spectrum was not transparent.
Recently, by using the tunneling mechanism, a method of deriving the black body spectrum directly was suggested by Banerjee and Majhi.
But, their derivation is valid only for black holes with spherically symmetric geometry such as Schwarzschild
or Reissner-Nordstr\"om black holes.
We extended the simple derivation of Hawking radiation by Banerjee and Majhi
on the basis of the tunneling mechanism to the case of the Kerr-Newman black hole.
By using the technique of the dimensional reduction near the horizon,
it is shown that the 4-dimensional Kerr-Newman metric effectively becomes a 2-dimensional
spherically symmetric metric near the horizon.
The use of this technique in the tunneling mechanism is justified since the tunneling effect
is also the quantum effect arising near the horizon region.
To discuss the behavior of matter fields near the event horizon, we defined the Kruskal-like coordinates for the effective reduced metric.
We showed that our final result of the black hole radiation from a rotating black hole agrees with the previous result which is based on more elaborate analyses of the tunneling mechanism.

In conclusion, we presented several arguments which clarify two of the recent derivations of Hawking radiation based on anomalies and tunneling.
To be specific, we presented a simple derivation of Hawking radiation on the basis of anomalies,
and we extended Banerjee and Majhi's tunneling mechanism to a Kerr-Newman black hole by using the technique of the dimensional reduction near the horizon.
A unified interpretation of various derivations of the black hole radiation,
which include two derivations analyzed here,
remains as an interesting problem.

\newpage

\section*{Acknowledgements}

\quad~ I would like to thank my supervisor, Professor Kazuo Fujikawa, for teaching me the profound aspects of physics and for helpful support and encouragement throughout my graduate study.

I also thank Professor Shigefumi Naka and Professor Shinichi Deguchi for a careful reading of the manuscript and continuous encouragement.
I am very grateful to Professor Katsuyoshi Tsai, Professor Takashi Mishima and Doctor Takeshi Nihei for useful comments and encouragement.
I am also grateful to all the other members of elementary particles theory group at Nihon University for many discussions and encouragement.

I am thankful to Professor Satoshi Iso and Doctor Hiroshi Umetsu for valuable discussions and helpful comments.
I am happy to thank Professor Rabin Banerjee, Doctor Shailesh Kulkarni, Doctor Bibhas Ranjan Majhi, Doctor Sujoy Kumar Modak, Doctor Debraj Roy and other members of S. N. Bose National Centre for Basic Sciences for their hospitality during my stay there and for many useful discussions.
I would like to appreciate my family and my friends for their encouragement and helpful support.

Finally, I gratefully acknowledge the financial support by the Strategic Project for Academic Research of Nihon University.

%%%%%%%%%%%%%%%%%%%%%%%%%%%%%%%%%%%%%%%%%%%%%%%%%%%%%%%%%%%%%%%%%%%%%%%%%%%%%%%%%%%%%%%%%%%%%%%%%
%%                                                                                             %%
%%                                       Appendix                                              %%
%%                                                                                             %%
%%%%%%%%%%%%%%%%%%%%%%%%%%%%%%%%%%%%%%%%%%%%%%%%%%%%%%%%%%%%%%%%%%%%%%%%%%%%%%%%%%%%%%%%%%%%%%%%%
\newpage
\renewcommand{\thesection}{\Alph{section}}
\setcounter{section}{0}
\begin{LARGE}
\textbf{Appendix}\\
\end{LARGE}
%\appendix
\addcontentsline{toc}{section}{Appendix}
%%%%%%%%%%%%%%%%%%%%%%%%%%%%%%%%%%%%%%%%%%%%%%%%%%%%%%%%%%%%%%%%%%%%%%%%%%%%%%%%%%%%%%%%%%%%%%%%%

%%%%%%%%%%%%%%%%%%%%%%%%%%%%%%%%%%%%%%%%%%%%%%%%%%%%%%%%%%%%%%%%%%%%%%%%%%%%%%%%%%%%%%%%%%%%%%%%%
%%                                                                                             %%
%%                                       Appendix A                                            %%
%%                                                                                             %%
%%%%%%%%%%%%%%%%%%%%%%%%%%%%%%%%%%%%%%%%%%%%%%%%%%%%%%%%%%%%%%%%%%%%%%%%%%%%%%%%%%%%%%%%%%%%%%%%%
\section{Killing Vectors and Null Hypersurfaces}

\quad~ In this appendix, we review basic properties of both Killing vectors and null hypersurfaces.
To begin with, we discuss Killing vectors.
On a 3-dimensional spacelike hypersurface $\Sigma$, the total energy-momentum vector is given by
\begin{align}
P^\mu = \int_\Sigma T^{\mu\nu} d\Sigma_\nu,
\label{a01}
\end{align}
where $T^{\mu\nu}$ is the energy-momentum tensor.
This definition loses a physical meaning in the curved space.
In general, the global energy or momentum conservation laws cannot be maintained.
However, when there exist particular vectors, the corresponding conservation laws can be maintained.

To find these laws, we consider a quantity given by
\begin{align}
P_\xi (\Sigma)=\int_\Sigma \xi_\mu T^{\mu\nu} d\Sigma_\nu,
\label{a02}
\end{align}
where $\xi^\mu$ is an arbitrary vector.
We note that this is a scalar quantity.
Now, we consider the volume $V$ enclosed by two surfaces $\Sigma$ and $\Sigma'$ (see Fig. 3.2 in Section 3.1).
According to the Gauss theorem, we obtain
\begin{align}
P_\xi (\Sigma') - P_\xi (\Sigma)&=\int_V \nabla_\nu (\xi_\mu T^{\mu\nu})dV\\
&=\int_V \left[ (\nabla_\nu \xi_\mu) T^{\mu\nu}+ \xi_\mu (\nabla_\nu T^{\mu\nu}) \right] dV\\
&=\frac{1}{2} \int_V (\nabla_\nu \xi_\mu +\nabla_\mu \xi_\nu ) T^{\mu\nu} dV,
\end{align}
where we used both the local conservation law of the energy-momentum tensor
\begin{align}
\nabla_\nu  T^{\mu\nu}=0,
\end{align}
and the fact that $T^{\mu\nu}$ is a symmetric tensor, i.e.,
\begin{align}
(\nabla_\nu \xi_\mu) T^{\mu\nu}= \frac{1}{2} \left( \nabla_\nu \xi_\mu +\nabla_\mu \xi_\nu \right) T^{\mu\nu}.
\end{align}
If we choose the vector $\xi_\mu$ as that the vector satisfies 
\begin{align}
\nabla_\nu \xi_\mu +\nabla_\mu \xi_\nu=0,
\label{a08}
\end{align}
the quantity $P_\xi (\Sigma)$ is conserved, and we thus find that there is a corresponding symmetry in the system.
The particular vector and the corresponding equation (\ref{a08}) are respectively called the ``Killing vector" and the ``Killing equation".
In other words, when there are symmetries in the system, there exist the corresponding conserved quantities and the corresponding Killing vectors. The conserved quantities are identified at asymptotic infinity.

Next we would like to discuss hypersurfaces.
To begin with, we define ${\mathscr S}(x)$ as a smooth function of the space-time coordinates $x^\mu$
and consider a family of hyperspaces
\begin{align}
{\mathscr S}={\rm constant}.
\end{align}
The vector normal to the hypersurface is given by
\begin{align}
l=F(x) \left( g^{\mu\nu} \partial _\nu {\mathscr S} \right) \frac{\partial}{\partial x^\mu},
\label{a10}
\end{align}
where $F(r)$ is an arbitrary non-zero function.
If the relation
\begin{align}
l^2=0,
\end{align}
is satisfied for a particular hypersurface ${\mathscr N}$, then ${\mathscr N}$ is called a ``null hypersurface".

As an example, we consider the case of Schwarzschild background.
The metric is given by
\begin{align}
ds^2=-\left( 1-\frac{2M}{r}\right) dt^2 + \frac{1}{1-\frac{2M}{r}}dr^2 + r^2 d\theta ^2 + r^2 \sin \theta d\varphi^2.
\label{a12}
\end{align}
By using the advanced time
\begin{align}
v=t+r_*,
\end{align}
the Schwarzschild metric (\ref{a12}) in the ingoing Eddington-Finkelstein coordinates $(v,r,\theta,\varphi)$
is rewritten as
\begin{align}
ds^2=-\left( 1-\frac{2M}{r}\right) dv^2+2drdv+r^2d\Omega^2.
\label{a14}
\end{align}
Then a surface is defined by
\begin{align}
{\mathscr S}=r-2M.
\label{a15}
\end{align}
By substituting (\ref{a15}) into (\ref{a10}), we obtain
\begin{align}
l&=F(r) \left[ \left( 1-\frac{2M}{r}\right)\frac{\partial}{\partial r}+ \frac{\partial}{\partial v}\right],
\label{a16}
\end{align}
and
\begin{align}
l^2&=g^{\mu\nu}\partial_\mu {\mathscr S} \partial_\nu {\mathscr S} F^2(r)\\
&=\left( 1-\frac{2M}{r}\right)F^2(r).
\end{align}
Thus we find that $r=2M$ is a null hypersurface and the relation (\ref{a16}) then becomes
\begin{align}
l|_{r=2M} = F(r) \frac{\partial}{\partial v}.
\end{align}
Thus we find that interesting properties of null hypersurfaces as follows:
We define ${\mathscr N}$ as a null hypersurface with a normal vector $l$.
A vector $t$ which is tangent to ${\mathscr N}$ satisfies $l\cdot t=0$.
However, since ${\mathscr N}$ is null, the relation $l\cdot l =0$ is satisfied.
Therefore, the vector $l$ is itself a tangent vector, i.e., 
we have 
\begin{align}
l^\mu =\frac{dx^\mu}{d\lambda},
\end{align}
where $x^\mu(\lambda)$ is geodesic.

According to the definition of a Killing horizon,
it is known that a Killing horizon is a null hypersurface ${\mathscr N}$ with a Killing vector $\xi$ which is normal to ${\mathscr N}$.

%%%%%%%%%%%%%%%%%%%%%%%%%%%%%%%%%%%%%%%%%%%%%%%%%%%%%%%%%%%%%%%%%%%%%%%%%%%%%%%%%%%%%%%%%%%%%%%%%
%%                                                                                             %%
%%                                       Appendix B                                            %%
%%                                                                                             %%
%%%%%%%%%%%%%%%%%%%%%%%%%%%%%%%%%%%%%%%%%%%%%%%%%%%%%%%%%%%%%%%%%%%%%%%%%%%%%%%%%%%%%%%%%%%%%%%%%
\section{The First Integral by Carter}

\quad~ In this appendix, we show that the first integral is represented by
\begin{align}
E^2[r^4+a^2 (r^2 + 2Mr -Q^2 )] -2E(2Mr-Q^2)ap_{\varphi}-(r^2 -2Mr+Q^2) p_{\varphi} ^2
- (\mu ^2 r^2 + q) \Delta = ( p_r \Delta )^2
\end{align}
for the Kerr-Newman metric given by
\begin{align}
ds^2=&-\frac{\Delta -a^2 \sin^2\theta}{\Sigma}dt^2-\frac{2a\sin^2\theta}{\Sigma}(r^2+a^2-\Delta)dtd\varphi \notag\\
&-\frac{a^2\Delta \sin^2\theta-(r^2+a^2)^2}{\Sigma}\sin^2\theta d\varphi^2+\frac{\Sigma}{\Delta}dr^2+\Sigma d\theta^2,
\label{bmet01}
\end{align}
where notations are the same as in Section 2.7.
This derivation was first shown by Carter \cite{car02}.
In this derivation, he adopted the metric given by
\begin{align}
ds^2=&\Sigma d\theta^2-2a\sin^2\theta dr d\tilde{\varphi}+2drdu
+\frac{1}{\Sigma}[(r^2+a^2)^2-\Delta a^2 \sin ^2\theta]\sin^2 \theta d\tilde{\varphi}^2\notag\\
&-\frac{2a}{\Sigma}(2Mr-Q^2)\sin^2\theta d\tilde{\varphi} du 
-\left[ 1-\frac{2Mr-Q^2}{\Sigma}\right]du^2,
\label{bmet02}
\end{align}
where $u$ is the retarded time.
This metric agrees with the Kerr-Newman metric (\ref{bmet01}) under the transformations given by
\begin{align}
\left\{ \begin{array}{lll}
du &=& \displaystyle dt+ \frac{r^2 + a^2}{\Delta} dr \\
d \tilde{\varphi} &=& \displaystyle d \varphi + \frac{a}{\Delta} dr.
\end{array}\right.
\label{bmet03}
\end{align}
Carter considered the behavior of a particle with a mass $\mu$ and an electrical charge $e$ in the background with the metric (\ref{bmet02}).
In this case, the general form of the Hamilton-Jacobi equation is represented as
\begin{align}
\frac{\partial S}{\partial \lambda}=\frac{1}{2}g^{ij}
\left[ 
\frac{\partial S}{\partial x^i}-e A_i
\right]
\left[ 
\frac{\partial S}{\partial x^j}-e A_j
\right],
\end{align}
where $\lambda$ is an affine parameter associated with the proper time $\tau$ and defined by
\begin{align}
\tau  \equiv \mu \lambda,
\end{align}
$A_\mu$ stands for the gauge field (the electrical potential) and $S$ is the Jacobi action.
If there is a separable solution, then in terms of the already known constants of the motion it must take the form
\begin{align}
S=-\frac{1}{2}\mu^2 \lambda -Eu +L \tilde{\varphi}+S_\theta+S_r,
\label{bact01}
\end{align}
where $E$ and $L$ are given by
\begin{align}
p_u = -E ,\quad
p_{\tilde\varphi} = L,
\end{align}
where $p_{\mu}$ stands for the momentum component in the direction of each coordinate variable.
Here $S_\theta$ and $S_r$ are respectively functions of $\theta$ and $r$ only.
In this case, it can be shown that the first integral is given by
\begin{align}
p_\theta ^2 +\left( aE \sin \theta -\frac{L}{\sin \theta} \right)^2+a^2\mu^2\cos^2\theta ={\mathcal K},
\label{bene01}\\
\Delta p_r^2-2\left[ (r^2+a^2)E -a L + e Qr \right]p_r+\mu^2 r^2=-{\mathcal K},
\label{bene02}
\end{align}
where ${\mathcal K}$ is a constant.

Here we would like to know how the relations (\ref{bene01}) and (\ref{bene02}) are changed when we use the Kerr-Newman metric (\ref{bmet01}) instead of the metric (\ref{bmet02}).
Since we consider an electrically neutral particle, we have $e=0$.
By differentiating (\ref{bact01}), we obtain
\begin{align}
dS=- \frac{1}{2} \mu ^2 d\lambda -E du + L d \tilde{\varphi} 
+\left( \frac{\partial S_\theta}{\partial \theta} \right) d \theta + \left( \frac{\partial S_r}{\partial r} \right) dr.
\label{bact02}
\end{align}
Now we recall the transformations (\ref{bmet03}) which is used in order to derive the Kerr-Newman metric from (\ref{bmet02}).
By substituting (\ref{bmet03}) into (\ref{bact02}), we obtain
\begin{align}
dS&=- \frac{1}{2} \mu ^2 d\lambda -E \left( dt+ \frac{r^2 + a^2}{\Delta} dr \right) 
+ L \left( d \varphi + \frac{a}{\Delta} dr \right) 
+\left( \frac{\partial S_\theta}{\partial \theta} \right) d \theta 
+ \left( \frac{\partial S_r}{\partial r} \right) dr \nonumber\\
&=- \frac{1}{2} \mu ^2 d\lambda -E dt+ L d \varphi 
+\left( \frac{\partial S_\theta}{\partial \theta} \right) d \theta 
+ \left( -\frac{r^2 + a^2}{\Delta}E + \frac{a}{\Delta}\Phi+ \frac{\partial S_r}{\partial r} \right) dr.
\label{bact03}
\end{align}
Then the momenta conjugate to $\theta$ and $r$ are given by
\begin{align}
p_\theta = \frac{\partial S_\theta}{\partial \theta},
\qquad
p_r =\frac{\partial S_r}{\partial r}.
\end{align}
By using these relations, the relation (\ref{bact03}) becomes
\begin{align}
dS=- \frac{1}{2} \mu ^2 d\lambda -E dt+ L d \varphi 
+p_\theta d \theta 
+ \left( - \frac{r^2 + a^2}{\Delta}E + \frac{a}{\Delta}L+ p_r \right) dr.
\label{bact04}
\end{align}
By comparing between (\ref{bact02}) and (\ref{bact04}),
we find that a new radial momentum $p'_{r}$ is given by
\begin{align}
p'_{r} = p_r - \frac{r^2 + a^2}{\Delta}E + \frac{a}{\Delta}L,
\label{bpr01}
\end{align}
under the transformations (\ref{bmet03}).
The relation (\ref{bpr01}) is also written as 
\begin{align}
p_{r} = p'_r + \frac{1}{\Delta} \left[ (r^2 + a^2)E - a L \right].
\label{bpr02}
\end{align}
If the constant ${\cal K}$ in the relation (\ref{bene01}) is rewritten as
\begin{align}
{\mathcal K}= q + (L -a E)^2,
\label{bk01}
\end{align}
we obtain
\begin{align}
&p_\theta ^2 + \left( aE \sin \theta - \frac{p_\varphi}{\sin \theta} \right) ^2 
+ a^2 \mu ^2 \cos ^2 \theta = q + (p_\varphi -aE)^2\\
\Leftrightarrow \qquad&q
= \cos^2 \theta \left[ a^2(\mu^2 -E^2) + \frac{p_\varphi ^2}{\sin^2 \theta} \right] + p_\theta ^2.
\end{align}
Thus Carter's kinetic constant is used in (\ref{262q01}) in Section 2.6.
By substituting (\ref{bk01}) into (\ref{bene02}), we obtain
\begin{align}
\Delta p_r ^2 -2 [ (r^2 +a^2)E -a L ]p_r + \mu^2 r^2 = -q - p_\varphi ^2 + 2aEp_\varphi -a^2E^2.
\label{bene03}
\end{align}
By using the transformation (\ref{bpr02}) in (\ref{bene03}), we can obtain
\begin{align}
&\Delta \left( p'_r + \frac{1}{\Delta} \left[ (r^2 + a^2)E - a p_{\varphi} \right] \right)^2
-2[(r^2+a^2)E-a p_\varphi ]\left( p'_r + \frac{1}{\Delta} \left[ (r^2 + a^2)E - a p_{\varphi} \right] \right)\nonumber\\
&
+\mu^2 r^2+ q + p_\varphi ^2 - 2aEp_\varphi +a^2E^2=0 \\
\Leftrightarrow\quad 
&E^2[ r ^4 + a^2 ( r^2 + 2Mr - Q^2) ] 
-2E(2Mr-Q^2)a p_\varphi -(r^2 - 2Mr + Q ^2 ) p_\varphi ^2-( \mu ^2 r^2+ q ) \Delta 
=(\Delta {p'_r})^2
\end{align}

%%%%%%%%%%%%%%%%%%%%%%%%%%%%%%%%%%%%%%%%%%%%%%%%%%%%%%%%%%%%%%%%%%%%%%%%%%%%%%%%%%%%%%%%%%%%%%%%%
%%                                                                                             %%
%%                                       Appendix C                                            %%
%%                                                                                             %%
%%%%%%%%%%%%%%%%%%%%%%%%%%%%%%%%%%%%%%%%%%%%%%%%%%%%%%%%%%%%%%%%%%%%%%%%%%%%%%%%%%%%%%%%%%%%%%%%%

\section{Bogoliubov Transformations}

\quad~ In this appendix, we show that the Bogoliubov transformations (\ref{31bi01}) and (\ref{31bi02}),
\begin{align}
\bm{b}_i &= \sum_j (\alpha^* _{ij} \bm{a}_j - \beta^* _{ij} \bm{a}_j^\dagger ),
\label{cbi01}\\
\bm{b}_i^\dagger&=\sum_j (\alpha_{ij} \bm a_j^\dagger - \beta_{ij} \bm{a}_j),
\label{cbi02}
\end{align}
are the inverse transforms of (\ref{31ai01}) and (\ref{31ai02}) given by
\begin{align}
\bm a_i&=\sum_j (\bm b_j \alpha_{ij} +\bm b_j^\dagger \beta^*_{ij}),
\label{cai01}\\
\bm a_i^\dagger&=\sum_j (\bm b_j \beta_{ij}+\bm b_j^\dagger \alpha^*_{ij}).
\label{cai02}
\end{align}
By substituting both (\ref{cai01}) and (\ref{cai02}) into the right-hand side of (\ref{cbi01}), we obtain
\begin{align}
&\sum_j (\alpha^* _{ij} \bm a_j - \beta^* _{ij} \bm a_j^\dagger )\\
=&\sum_{j,k}\left[ \alpha^*_{ij}\left(\alpha_{jk}\bm b_k+\beta^*_{ik}\bm b_k^\dagger\right)
-\beta^*_{ij}\left(\beta_{jk}\bm b_k+\alpha^*_{ik}\bm b_k^\dagger\right)\right]\\
=&\sum_k\left[ 
\sum_j\left( 
\alpha^*_{ij}\alpha_{jk}-\beta^*_{ij}\beta_{jk}\right)\bm b_k 
+\sum_j\left( \alpha^*_{ij}\beta^*_{jk}
-\beta^*_{ij}\alpha^*_{jk}\right) \bm b_k^\dagger
\right].
\label{c07}
\end{align}
Now we recall the orthonormal condition (\ref{31nor02}) for $\{ f_i \}$ and $\{ f^*_i \}$, i.e.,
\begin{align}
\rho(f_i, f^*_j)=\frac{1}{2}i \int_\Sigma 
( f_i \nabla_{\mu}f^*_{j} - f^*_j \nabla_\mu f_{i})d \Sigma^\mu=\delta_{ij},
\label{c08}
\end{align}
which implies the following relations
\begin{align}
\rho(f^*_i,{f}_j)&=-\delta_{ij},
\label{c09}\\
\rho({f}_i,{f}_j)&=\rho(f^*_i,f^*_j)=0.
\label{c10}
\end{align}
The orthonormal condition (\ref{31nor06}) for $\{ p_i \}$ and $\{ p^*_i \}$ is also satisfied 
\begin{align}
\rho(p_i,p^*_j)=\delta_{ij}.
\label{c11}
\end{align}
By substituting the relations between $\{ p_i \}$ and $\{ f_i \}$,
\begin{align}
p_i &= \sum_k (\alpha _{ik} f_k + \beta _{ik} f^*_k ),
\label{cpi01}\\
p^*_j&=\sum_l(\alpha^* _{jl} f^*_l + \beta^* _{jl} {f}_l),
\label{cpi02}
\end{align}
into (\ref{c11}), we obtain
\begin{align}
\rho(p_i,p^*_j)
&=\rho\left( 
\sum_k(\alpha_{ik}f_k+\beta_{ik}f^*_k),
~\sum_l(\alpha^*_{jl} f^*_l+ \beta^*_{jl}f_l))
\right)\\
&=\sum_{k,l}
\left[ 
\alpha_{ik}\alpha^*_{jl}\rho(f_k, f^*_l)
+\alpha_{ik} \beta^*_{jl}\rho({f}_k,{f}_l)
+\beta_{ik} \alpha^*_{jl}\rho(f^*_k, f^*_l)
+\beta_{ik} \beta^*_{jl}\rho(f^*_k,{f}_l)
\right],
\label{c15}
\end{align}
By using (\ref{c08}), (\ref{c09}) and (\ref{c10}) in (\ref{c15}), we obtain
\begin{align}
\rho(p_i,p^*_j)&=\sum_{k,l}
\left[
\alpha_{ik}\alpha^*_{jl}\delta_{kl}+\beta_{ik}\beta^*_{jl}(-\delta_{kl})
\right]\\
&=\sum_k(\alpha_{ik}\alpha^*_{jk}-\beta_{ik}\beta^*_{jk}).
\end{align}
From (\ref{c11}), we thus obtain
\begin{align}
\sum_k\left( \alpha_{ik}\alpha^*_{jk}-\beta_{ik}\beta^*_{jk} \right)=\delta_{ij}.
\label{c18}
\end{align}
Similarly, when we consider the case of $\rho(p^*_i, p^*_j)=0$, we obtain 
\begin{align}
\sum_k\left( \beta^*_{ik} \alpha^*_{jk}-\alpha^*_{ik}\beta^*_{jk} \right)=0.
\label{c19}
\end{align}
By substituting (\ref{c18}) and (\ref{c19}) into (\ref{c07}), we obtain
\begin{align}
\sum_k\left[ 
\sum_j\left( 
\alpha^*_{ij}\alpha_{jk}-\beta^*_{ij}\beta_{jk}\right)\bm b_k 
+\sum_j\left( \alpha^*_{ij} \beta^*_{jk}
- \beta^*_{ij} \alpha^*_{jk}\right) \bm b_k^\dagger
\right]
&=\sum_k\delta_{ik} \bm b_k\\
&=\bm b_i
\end{align}
Thus we can reproduce $\bm b_i$ from the right-hand side of (\ref{cbi01}).
We can similarly calculate as for $\bm b_i^\dagger$.
Therefore we can confirm that the relations (\ref{cbi01}) and (\ref{cbi02}) are the inverse transforms of (\ref{cai01}) and (\ref{cai02}).

%%%%%%%%%%%%%%%%%%%%%%%%%%%%%%%%%%%%%%%%%%%%%%%%%%%%%%%%%%%%%%%%%%%%%%%%%%%%%%%%%%%%%%%%%%%%%%%%%
%%                                                                                             %%
%%                                       Appendix D                                            %%
%%                                                                                             %%
%%%%%%%%%%%%%%%%%%%%%%%%%%%%%%%%%%%%%%%%%%%%%%%%%%%%%%%%%%%%%%%%%%%%%%%%%%%%%%%%%%%%%%%%%%%%%%%%%

\section{Solutions of Klein-Gordon Equation in Schwarzschild Space-time}

\quad~ In this appendix, we show that the solutions of the Klein-Gordon equation, in (\ref{31kg01})
\begin{align}
g^{\mu\nu}\nabla_\mu \nabla_\nu \phi=0,
\label{d01}
\end{align}
are given by (\ref{31fo01}) and (\ref{31po01})
\begin{align}
f_{\omega' lm}&=\frac{F_{\omega'}(r)}{r\sqrt{2\pi \omega'}}e^{i\omega' v}Y_{lm}(\theta,\varphi),
\label{d02}\\
p_{\omega l m}&=\frac{P_{\omega}(r)}{r\sqrt{2\pi \omega}}e^{i\omega u}Y_{lm}(\theta,\varphi).
\label{d03}
\end{align}
By using a property of the covariant derivative for a scalar field $\phi$,
\begin{align}
\nabla_\nu \phi =\partial_\nu \phi,
\label{d04}
\end{align}
the equation (\ref{d01}) becomes
\begin{align}
\nabla_\mu \left( g^{\mu\nu} \partial_\nu \phi\right)=0,
\label{d05}
\end{align}
where $\partial_\nu$ and $\nabla_\mu$ respectively stand for an ordinary derivative and a covariant derivative.
The covariant derivative for a vector field $A^\mu$ is also given by
\begin{align}
\nabla_\mu A^\mu=\partial_\mu A^\mu + \Gamma^{\mu}_{\nu\mu}A^\nu,
\label{d06}
\end{align}
where $\Gamma^{\mu}_{\nu\rho}$ is the Christoffel symbol and $\Gamma^{\mu}_{\nu\mu}$ is written as
\begin{align}
\Gamma^\mu_{\nu\mu}&=\frac{1}{2}g^{\mu\rho}(\partial_\mu g_{\rho\nu}+\partial_\nu g_{\rho\mu}-\partial_\rho g_{\nu\mu})\\
&=\frac{1}{2}(g^{\mu\rho}\partial_\mu g_{\rho\nu}+g^{\mu\rho}\partial_\nu g_{\rho\mu}-g^{\mu\rho}\partial_\rho g_{\nu\mu}).
\label{d08}
\end{align}
Since the third term is canceled by the first term by exchanging $\mu$ and $\rho$ in the third term of (\ref{d08}), the relation (\ref{d08})
becomes
\begin{align}
\Gamma^\mu_{\nu\mu}=\frac{1}{2} g^{\mu\rho}\partial_\nu g_{\rho\mu}.
\label{d09}
\end{align}

Now we would like to show that $\partial_\nu g_{\rho\mu}$ is given by
\begin{align}
\partial_\nu g_{\rho\mu}=\left( g g^{\rho\mu}\right)^{-1} \partial_\nu g,
\label{d10}
\end{align}
where $g$ is defined by
\begin{align}
g\equiv \det \left( g_{\mu\rho} \right).
\label{d11}
\end{align}
The definition (\ref{d11}) can also be written as
\begin{align}
g&=\exp\left( \ln \det g_{\mu\rho}\right)\\
&=\exp\left( {\rm Tr} \ln g_{\mu\rho} \right).
\end{align}
Then, the small variation of $g$ is given by
\begin{align}
\delta g=\exp \left\{ {\rm Tr} \left[ \ln(g_{\mu \rho}+ \delta g_{\mu\rho}) \right] \right\}-\exp ({\rm Tr}\ln g_{\mu\rho}).
\label{d14}
\end{align}
By using the Taylor expansion for a matrix $X$
\begin{align}
\ln (X+\delta X) \approx \ln X + X^{-1} \delta X,
\end{align}
the relation (\ref{d14}) becomes
\begin{align}
\delta g &\approx \exp \left[ {\rm Tr} \left( \ln g_{\mu\rho} 
+ g^{\mu\nu} \delta g_{\nu \rho}\right) \right] -\exp\left[ {\rm Tr} (\ln g_{\mu\rho}) \right]\\
&=\exp \left[ {\rm Tr} (\ln g_{\mu\rho} ) \right] 
\exp\left[ {\rm Tr} (g^{\mu\nu} \delta g_{\nu\rho}) \right] -\exp \left[ {\rm Tr} (\ln g_{\mu\rho} )\right].
\end{align}
By performing the Taylor expansion for $\exp\left[ {\rm Tr} (g^{\mu\nu} \delta g_{\nu\rho}) \right]$ and retaining the first order of terms, we obtain
\begin{align}
\delta g&\approx\exp \left[ {\rm Tr} (\ln g_{\mu\rho} ) \right]\left( 1+ {\rm Tr} g^{\mu\nu}\delta g_{\nu \rho} \right)
-\exp \left[ {\rm Tr} (\ln g_{\mu\rho} ) \right]\\
&=\exp\left[ {\rm Tr}(\ln g_{\mu\rho}) \right] {\rm Tr}(g^{\mu\nu}\delta g_{\nu\rho})\\
&=g \cdot {\rm Tr}(g^{\mu\nu} \delta g_{\nu\rho})\\
&=g \cdot g^{\mu\nu} \delta g_{\nu\mu}.
\label{d21}
\end{align}
By replacing $\nu$ to $\rho$ in (\ref{d21}), we obtain
\begin{align}
\delta g= g \cdot g^{\mu\rho} \delta g_{\rho \mu}.
\end{align}
From this, we find
\begin{align}
&\partial_\nu g = g \cdot g^{\mu\rho}\partial_\nu g_{\rho \mu}\\
\Leftrightarrow \qquad & \partial_\nu g_{\rho \mu}=(g \cdot g^{\mu \rho})^{-1}\partial_\nu g.
\label{d24}
\end{align}
Thus we establish the relation (\ref{d10}).

By substituting (\ref{d10}) into (\ref{d09}), we obtain
\begin{align}
\Gamma^\mu_{\nu\mu}&=\frac{1}{2} g^{\mu\rho} \frac{1}{g\cdot g^{\mu\rho}}\partial_\nu g\\
&=\frac{1}{2g} \partial_\nu g.
\label{d26}
\end{align}
Since we have
\begin{align}
\frac{1}{\sqrt{-g}}\partial_\nu(\sqrt{-g})=\frac{1}{2g} \partial_\nu g,
\end{align}
the relation (\ref{d26}) becomes
\begin{align}
\Gamma^\mu_{\nu\mu}=\frac{1}{\sqrt{-g}}\partial_\nu(\sqrt{-g}).
\label{d28}
\end{align}
By substituting (\ref{d28}) into (\ref{d06}), we obtain
\begin{align}
\nabla_\mu A^\mu=\partial_\mu A^{\mu}+\frac{1}{\sqrt{-g}}\partial_\nu(\sqrt{-g})A^\nu
\label{d29}
\end{align}
By replacing $\nu$ by $\mu$, the relation (\ref{d29}) becomes
\begin{align}
\nabla_\mu A^{\mu}=\frac{1}{\sqrt{-g}}\partial_\mu(\sqrt{-g}A^\mu).
\end{align}
Thus the Klein-Gordon equation (\ref{d05}) is written as
\begin{align}
\frac{1}{\sqrt{-g}}\partial_\mu(\sqrt{-g} \cdot g^{\mu\nu} \partial_\nu \phi)=0.
\label{d31}
\end{align}
Since we consider physics in the Schwarzschild background, the metric is given by (\ref{2sch01})
\begin{align}
\left( g_{\mu\nu} \right)
=\left( \begin{array}{ccccc}
-\left( 1-\frac{2M}{r} \right)&0&0&0\\
0&\left( 1-\frac{2M}{r} \right)^{-1}&0&0\\
0&0&r^2&0\\
0&0&0&r^2\sin^2 \theta
\end{array}
\right)
\label{d32}
\end{align}
From this metric, $g$ is given by
\begin{align}
g=-r^4\sin^2 \theta
\label{d33}
\end{align}
By substituting (\ref{d32}) and (\ref{d33}) into (\ref{d31}), the Klein-Gordon equation becomes
\begin{align}
\left[-\frac{r}{r-2M}\frac{\partial^2}{\partial t^2}+\frac{1}{r^2}\frac{\partial}{\partial r}
\left\{ r^2 \left( \frac{r-2M}{r}\right)\frac{\partial }{\partial r}\right\}
-\frac{1}{r^2}\left( -\frac{\partial}{\partial \theta} \sin \theta \frac{\partial}{\partial \theta}
-\frac{1}{\sin^2\theta} \frac{\partial^2}{\partial \varphi^2}
\right)\right]\phi=0.
\label{d34}
\end{align}
Here we define a quadratic angular momentum by
\begin{align}
\hat{L}^2\equiv -\frac{\partial}{\partial \theta} \sin \theta \frac{\partial}{\partial \theta}
-\frac{1}{\sin^2\theta} \frac{\partial^2}{\partial \varphi^2}.
\label{d35}
\end{align}
Then (\ref{d34}) is written as
\begin{align}
\left[-\frac{r}{r-2M}\frac{\partial^2}{\partial t^2}+\frac{1}{r^2}\frac{\partial}{\partial r}
\left\{ r^2 \left( \frac{r-2M}{r}\right)\frac{\partial }{\partial r}\right\}
-\frac{1}{r^2}\hat{L}^2\right]\phi=0.
\label{d36}
\end{align}
Now we rewrite $\phi$ as 
\begin{align}
\phi=\left( Ae^{-i\omega t}+A^* e^{i\omega t} \right)R(r)\Theta(\theta, \varphi).
\label{d37}
\end{align}
By substituting (\ref{d37}) into (\ref{d36}), we obtain
\begin{align}
\left[-\frac{r}{r-2M}(i\omega)^2+\frac{1}{r^2}\frac{\partial}{\partial r}
\left\{ r^2 \left( \frac{r-2M}{r}\right)\frac{\partial }{\partial r}\right\}
-\frac{1}{r^2}\hat{L}^2\right]\left( Ae^{-i\omega t}+A^* e^{i\omega t} \right)R(r)\Theta(\theta, \varphi)=0.
\end{align}
By dividing the both sides by $\left( Ae^{-i\omega t}+A^* e^{i\omega t} \right)$ and transferring the third term to the right-hand side,
we obtain
\begin{align}
\left[\frac{r}{r-2M}\omega^2+\frac{1}{r^2}\frac{\partial}{\partial r}
\left\{ r^2 \left( \frac{r-2M}{r}\right)\frac{\partial }{\partial r}\right\}
\right]R(r)\Theta(\theta, \varphi)=\frac{R(r)}{r^2}\hat{L}^2 \Theta(\theta, \varphi).
\label{d39}
\end{align}
By dividing the both sides by $\displaystyle \frac{R(r)}{r^2}\Theta(\theta, \varphi)$, the equation (\ref{d39}) becomes
\begin{align}
\frac{r^2}{R(r)}\left[\frac{r}{r-2M}\omega^2+\frac{1}{r^2}\frac{\partial}{\partial r}
\left\{ r^2 \left( \frac{r-2M}{r}\right)\frac{\partial }{\partial r}\right\}
\right]R(r)=\frac{1}{\Theta(\theta,\varphi)}\hat{L}^2 \Theta(\theta, \varphi).
\label{d40}
\end{align}
Since (\ref{d40}) is represented in a separable form, the equation is set to be a constant.
By writing the constant as $l(l+1)$, we obtain
\begin{align}
&\hat{L}^2 \Theta(\theta, \varphi)=l(l+1)\Theta(\theta, \varphi),
\label{d41}\\
&\left[\frac{r}{r-2M}\omega^2+\frac{1}{r^2}\frac{\partial}{\partial r}
\left\{ r^2 \left( \frac{r-2M}{r}\right)\frac{\partial }{\partial r}\right\}
-\frac{l(l+1)}{r^2}
\right]R(r)=0.
\label{d42}
\end{align}
From (\ref{d41}), we can expand $\Theta(\theta, \varphi)$ by
\begin{align}
\Theta(\theta, \varphi)=\sum_m B_{lm}Y_{lm}(\theta, \varphi),
\label{d43}
\end{align}
where $B_{lm}$ is an integration constant and $Y_{lm}(\theta, \varphi)$ stands for the spherical harmonics.
In the equation (\ref{d42}), we use the definition
\begin{align}
R'(r_*)\equiv rR(r),
\label{d44}
\end{align}
where $r_*$ is the tortoise coordinate defined by
\begin{align}
r_*\equiv r+2M\ln \left|\frac{r}{2M}-1\right|.
\label{d45}
\end{align}
Then for this transformation we
\begin{align}
\frac{\partial}{\partial r}=\left(\frac{r}{r-2M}\right)\frac{\partial}{\partial r_*}.
\label{d46}
\end{align}
Under these transformations, the equation (\ref{d42}) becomes
\begin{align}
\frac{1}{r}\left[ \frac{r}{r-2M}\omega^2 +\left(\frac{r}{r-2M}\right) \frac{\partial^2}{\partial r_*^2}
-\frac{2M}{r^3}-\frac{1}{r^2}l(l+1)
\right]R'(r_*)=0.
\label{d47}
\end{align}
By dividing the both sides of (\ref{d47}) by $\displaystyle \frac{1}{r}\left( \frac{r}{r-2M} \right)$, we obtain
\begin{align}
\frac{\partial^2}{\partial \tilde{r}^2}R'(r_*) 
+\left[ \omega^2 -\frac{1}{r^2}\left\{ \frac{2M}{r}+ l(l+1) \right\} \left( 1-\frac{2M}{r}\right)\right]
R'(r_*)=0.
\label{d48}
\end{align}
As stated in Section 3.1, the solutions $f_{\omega' l m}$ and $p_{\omega l m}$ are partial waves at $r \to \infty$.
Thus, by taking $r \to \infty$ in (\ref{d48}), we obtain
\begin{align}
\frac{\partial}{\partial r_*^2}R'(r_*) +\omega^2 R'(r_*)=0.
\end{align}
The solution for this equation is given by
\begin{align}
R'(r_*)=C_{\omega l} e^{-i\omega r_*}+C^*_{\omega l}e^{i\omega r_*},
\label{d50}
\end{align}
where $C_{\omega l}$ is an integration constant.
By substituting (\ref{d44}) into (\ref{d50}), we obtain
\begin{align}
R(r)=\frac{1}{r}\left( C_{\omega l} e^{-i\omega r_*}+ C^*_{\omega l} e^{i\omega r_*}\right).
\label{d51}
\end{align}
By substituting both (\ref{d43}) and (\ref{d51}) into (\ref{d37}), we thus obtain
\begin{align}
\phi=\sum_{\omega, l, m}\frac{1}{r}
\left( AC_{\omega l}e^{-i\omega (t+r_*)} 
+A C^*_{\omega l}e^{-i\omega(t-r_*)}
+ A^* C_{\omega l} e^{i\omega (t-r_*)}
+ A^* C^* _{\omega l} e^{i\omega(t+ r_*)}
\right)B_{l m}Y_{lm}(\theta, \varphi).
\label{d52}
\end{align}
We use affine parameters defined by
\begin{align}
v&\equiv t+r_*,\\
u&\equiv t-r_*,
\end{align}
where $v$ and $u$ are respectively called the advanced time and the retarded time.
By putting together the integration constants in (\ref{d52}) except for the normalization constant $\displaystyle \frac{1}{\sqrt{2\pi \omega}}$, which
is often used in the Klein-Gordon equation,
we can thus write the partial waves as
\begin{align}
f_{\omega' l m}&=\frac{F_{\omega'}(r)}{r\sqrt{2\pi \omega'}}e^{i\omega' v}Y_{lm}(\theta, \varphi),\\
p_{\omega l m}&=\frac{P_{\omega}(r)}{r\sqrt{2\pi \omega'}}e^{i\omega u}Y_{lm}(\theta, \varphi),
\end{align}
where we regarded $F_{\omega'}(r)$ and $P_{\omega}(r)$ as not integration constants but rather ``integration variables" which depend on $r$,
because we want to take into account the small effect of $r$ arising from the approximation by setting $r \to \infty$.

%%%%%%%%%%%%%%%%%%%%%%%%%%%%%%%%%%%%%%%%%%%%%%%%%%%%%%%%%%%%%%%%%%%%%%%%%%%%%%%%%%%%%%%%%%%%%%%%%
%%                                                                                             %%
%%                                       Appendix E                                            %%
%%                                                                                             %%
%%%%%%%%%%%%%%%%%%%%%%%%%%%%%%%%%%%%%%%%%%%%%%%%%%%%%%%%%%%%%%%%%%%%%%%%%%%%%%%%%%%%%%%%%%%%%%%%%

\section{Calculation of Bogoliubov Coefficients}

\quad~ The Bogoliubov coefficients $\alpha_{\omega \omega'}$ and $\beta_{\omega \omega'}$ are given by (\ref{31alp01}) and (\ref{31bet01})
\begin{align}
\alpha_{\omega \omega'}&=\frac{r\sqrt{\omega'}}{\sqrt{2\pi}F_{\omega'}}
\int_{-\infty}^\infty dv  e^{-i\omega'v}p_\omega,
\label{e01}\\
\beta_{\omega \omega'}&=\frac{r\sqrt{\omega'}}{\sqrt{2\pi}F_{\omega'}}
\int_{-\infty}^\infty dv  e^{i\omega'v}p_\omega.
\label{e02}
\end{align}
When the partial wave $p_\omega$ is represented as (\ref{31po03}), 
\begin{align}
p_{\omega}^{(2)}\sim 
\left\{ \begin{array}{lll}0,&\quad(v>v_0),\\
\displaystyle \frac{P^-_{\omega}}{r\sqrt{2\pi \omega}}
\exp\left[ -i \frac{\omega}{\kappa}\ln \left( \frac{v_0-v}{CD}\right)\right],&\quad(v \leq v_0),
\end{array}
\right.
\label{e03}
\end{align}
we show that the corresponding $\alpha_{\omega \omega'}^{(2)}$ and $\beta_{\omega \omega'}^{(2)}$ are given by
(\ref{31alp02}) and (\ref{31bet02}), i.e.,
\begin{align}
\alpha_{\omega \omega'} ^{(2)} &\approx \frac{1}{2 \pi} P_{\omega} ^-(CD)^{\frac{i \omega}{\kappa}} 
e^{-i \omega ' v_0} \left( \sqrt{\frac{\omega '}{\omega}} \right) 
\Gamma \left( 1 -\frac{i \omega }{\kappa} \right) (-i \omega ')^{-1+\frac{i\omega}{\kappa}}, 
\label{e04}\\
\beta_{\omega \omega'} ^{(2)} &\approx -i\alpha_{\omega (-\omega')} ^{(2)}.
\label{e05}
\end{align}
To begin with, by substituting (\ref{e03}) into (\ref{e01})
\begin{align}
\alpha_{\omega \omega'}^{(2)}=\frac{r\sqrt{\omega'}}{\sqrt{2\pi}F_{\omega'}(r)}
\int^{v_0}_{-\infty}dv \frac{P_\omega(r)}{r\sqrt{2\pi \omega}}
\exp \left[ -i\frac{\omega}{\kappa} \ln\left( \frac{v_0-v}{CD}\right)\right]e^{-i\omega'v},
\label{e06}
\end{align}
where $F_{\omega'}(r)$ and $P_\omega(r)$ are integration variables which take into account the small effect of $r$,
and we collectively rewrote them as $P_\omega(r)$.
Since we now consider the region near the horizon $r=2M$
we use $P_\omega^- \equiv P_\omega(2M)$.
Thus the relation (\ref{e06}) becomes
\begin{align}
\alpha_{\omega \omega'}^{(2)}=\frac{1}{2\pi}P_\omega^-(CD)^{i\frac{\omega}{\kappa}}\sqrt{\frac{\omega'}{\omega}}
\int^{v_0}_{-\infty}dv(v_0-v)^{-i\frac{\omega}{\kappa}}e^{-i\omega' v}.
\end{align}
By integrating over the variable defined as $v_0-v=x$, we obtain
\begin{align}
\alpha_{\omega \omega'}^{(2)}=\frac{1}{2\pi}P_\omega^-(CD)^{i\frac{\omega}{\kappa}}\sqrt{\frac{\omega'}{\omega}}
e^{-i\omega' v_0}\int^{\infty}_{0}dx x^{-i\frac{\omega}{\kappa}} e^{-(-i\omega') x}.
\label{e08}
\end{align}
By using a formula of the gamma function,
\begin{align}
\Gamma(\varepsilon)t^{-\varepsilon}=\int^\infty_0 ds s^{\varepsilon-1}e^{-ts},
\end{align}
we finally obtain (\ref{e04}),
\begin{align}
\alpha_{\omega \omega'}^{(2)}=\frac{1}{2\pi}P_\omega^-(CD)^{i\frac{\omega}{\kappa}}\sqrt{\frac{\omega'}{\omega}}
e^{-i\omega' v_0}\Gamma(1-i\frac{\omega}{\kappa})(-i\omega')^{1-i\frac{\omega}{\kappa}}.
\end{align}
Similarly, we can show the relation (\ref{e05}) for $\beta_{\omega\omega'}^{(2)}$.

%%%%%%%%%%%%%%%%%%%%%%%%%%%%%%%%%%%%%%%%%%%%%%%%%%%%%%%%%%%%%%%%%%%%%%%%%%%%%%%%%%%%%%%%%%%%%%%%%

%%%%%%%%%%%%%%%%%%%%%%%%%%%%%%%%%%%%%%%%%%%%%%%%%%%%%%%%%%%%%%%%%%%%%%%%%%%%%%%%%%%%%%%%%%%%%%%%%
%%                                                                                             %%
%%                                       References                                            %%
%%                                                                                             %%
%%%%%%%%%%%%%%%%%%%%%%%%%%%%%%%%%%%%%%%%%%%%%%%%%%%%%%%%%%%%%%%%%%%%%%%%%%%%%%%%%%%%%%%%%%%%%%%%%

\end{document}

%% file: pen01.tex
%WinTpicVersion3.08
\unitlength 0.1in
\begin{picture}( 32.8000, 20.9000)(  3.6000,-22.4500)
% CIRCLE 2 0 3 0
% 4 2800 600 2800 640 2800 640 2800 640
% 
\special{pn 8}%
\special{ar 2800 600 40 40  0.0000000 6.2831853}%
% CIRCLE 2 0 3 0
% 4 3600 1400 3600 1440 3600 1440 3600 1440
% 
\special{pn 8}%
\special{ar 3600 1400 40 40  0.0000000 6.2831853}%
% CIRCLE 2 0 3 0
% 4 2800 2200 2800 2240 2800 2240 2800 2240
% 
\special{pn 8}%
\special{ar 2800 2200 40 40  0.0000000 6.2831853}%
% CIRCLE 2 0 3 0
% 4 1200 2200 1200 2240 1200 2240 1200 2240
% 
\special{pn 8}%
\special{ar 1200 2200 40 40  0.0000000 6.2831853}%
% CIRCLE 2 0 3 0
% 4 400 1400 400 1440 400 1440 400 1440
% 
\special{pn 8}%
\special{ar 400 1400 40 40  0.0000000 6.2831853}%
% CIRCLE 2 0 3 0
% 4 1200 600 1200 640 1200 640 1200 640
% 
\special{pn 8}%
\special{ar 1200 600 40 40  0.0000000 6.2831853}%
% LINE 2 0 3 0
% 2 1230 2170 2770 2170
% 
\special{pn 8}%
\special{pa 1230 2170}%
\special{pa 2770 2170}%
\special{fp}%
% LINE 2 0 3 0
% 2 2820 620 3570 1370
% 
\special{pn 8}%
\special{pa 2820 620}%
\special{pa 3570 1370}%
\special{fp}%
% LINE 2 0 3 0
% 2 430 1430 1180 2180
% 
\special{pn 8}%
\special{pa 430 1430}%
\special{pa 1180 2180}%
\special{fp}%
% LINE 2 0 3 0
% 2 3570 1430 2820 2180
% 
\special{pn 8}%
\special{pa 3570 1430}%
\special{pa 2820 2180}%
\special{fp}%
% LINE 2 0 3 0
% 2 1170 630 420 1380
% 
\special{pn 8}%
\special{pa 1170 630}%
\special{pa 420 1380}%
\special{fp}%
% LINE 0 0 3 0
% 2 1230 630 2770 2170
% 
\special{pn 20}%
\special{pa 1230 630}%
\special{pa 2770 2170}%
\special{fp}%
% LINE 2 0 3 0
% 2 1230 2230 2770 2230
% 
\special{pn 8}%
\special{pa 1230 2230}%
\special{pa 2770 2230}%
\special{fp}%
% LINE 2 0 3 0
% 2 1240 630 2780 630
% 
\special{pn 8}%
\special{pa 1240 630}%
\special{pa 2780 630}%
\special{fp}%
% LINE 2 0 3 0
% 2 1220 570 2760 570
% 
\special{pn 8}%
\special{pa 1220 570}%
\special{pa 2760 570}%
\special{fp}%
% LINE 0 0 3 0
% 2 2770 630 1230 2170
% 
\special{pn 20}%
\special{pa 2770 630}%
\special{pa 1230 2170}%
\special{fp}%
% STR 2 0 3 0
% 3 3800 1310 3800 1410 5 0
% ${\mathcal I}^0$
\put(38.0000,-14.1000){\makebox(0,0){${\mathcal I}^0$}}%
% STR 2 0 3 0
% 3 3000 410 3000 510 5 0
% ${\mathcal I}^+$
\put(30.0000,-5.1000){\makebox(0,0){${\mathcal I}^+$}}%
% STR 2 0 3 0
% 3 3000 2230 3000 2330 5 0
% ${\mathcal I}^-$
\put(30.0000,-23.3000){\makebox(0,0){${\mathcal I}^-$}}%
% STR 2 0 3 0
% 3 2000 380 2000 480 5 0
% ${\mathcal R}$
\put(20.0000,-4.8000){\makebox(0,0){${\mathcal R}$}}%
% STR 2 0 3 0
% 3 3400 820 3400 920 5 0
% ${\mathcal J^+}$
\put(34.0000,-9.2000){\makebox(0,0){${\mathcal J^+}$}}%
% STR 2 0 3 0
% 3 3400 1720 3400 1820 5 0
% ${\mathcal J^-}$
\put(34.0000,-18.2000){\makebox(0,0){${\mathcal J^-}$}}%
% STR 2 0 3 0
% 3 2500 1070 2500 1170 5 0
% ${\mathcal H}^+$
\put(25.0000,-11.7000){\makebox(0,0){${\mathcal H}^+$}}%
% STR 2 0 3 0
% 3 1590 1080 1590 1180 5 0
% ${\mathcal H}^-$
\put(15.9000,-11.8000){\makebox(0,0){${\mathcal H}^-$}}%
% SPLINE 2 0 3 0
% 3 2800 2200 3000 1410 2800 600
% 
\special{pn 8}%
\special{pa 2800 2200}%
\special{pa 2812 2170}%
\special{pa 2824 2138}%
\special{pa 2836 2108}%
\special{pa 2846 2076}%
\special{pa 2858 2046}%
\special{pa 2870 2014}%
\special{pa 2880 1984}%
\special{pa 2890 1952}%
\special{pa 2902 1922}%
\special{pa 2912 1890}%
\special{pa 2922 1860}%
\special{pa 2930 1828}%
\special{pa 2940 1798}%
\special{pa 2948 1766}%
\special{pa 2956 1736}%
\special{pa 2964 1704}%
\special{pa 2970 1674}%
\special{pa 2976 1642}%
\special{pa 2982 1612}%
\special{pa 2986 1580}%
\special{pa 2992 1550}%
\special{pa 2994 1518}%
\special{pa 2998 1488}%
\special{pa 3000 1456}%
\special{pa 3000 1424}%
\special{pa 3000 1394}%
\special{pa 3000 1362}%
\special{pa 2998 1332}%
\special{pa 2996 1300}%
\special{pa 2992 1270}%
\special{pa 2988 1238}%
\special{pa 2984 1208}%
\special{pa 2978 1176}%
\special{pa 2972 1146}%
\special{pa 2966 1114}%
\special{pa 2960 1084}%
\special{pa 2952 1052}%
\special{pa 2944 1022}%
\special{pa 2934 990}%
\special{pa 2926 960}%
\special{pa 2916 928}%
\special{pa 2906 896}%
\special{pa 2896 866}%
\special{pa 2886 834}%
\special{pa 2874 804}%
\special{pa 2864 772}%
\special{pa 2852 742}%
\special{pa 2842 710}%
\special{pa 2830 680}%
\special{pa 2818 648}%
\special{pa 2806 618}%
\special{pa 2800 600}%
\special{sp}%
% SPLINE 2 0 3 0
% 3 2800 600 2000 800 1200 600
% 
\special{pn 8}%
\special{pa 2800 600}%
\special{pa 2770 612}%
\special{pa 2738 624}%
\special{pa 2708 636}%
\special{pa 2676 646}%
\special{pa 2646 658}%
\special{pa 2614 670}%
\special{pa 2584 680}%
\special{pa 2552 690}%
\special{pa 2522 702}%
\special{pa 2490 712}%
\special{pa 2460 720}%
\special{pa 2428 730}%
\special{pa 2396 740}%
\special{pa 2366 748}%
\special{pa 2334 756}%
\special{pa 2304 762}%
\special{pa 2272 770}%
\special{pa 2242 776}%
\special{pa 2210 782}%
\special{pa 2180 786}%
\special{pa 2148 790}%
\special{pa 2118 794}%
\special{pa 2086 798}%
\special{pa 2056 800}%
\special{pa 2024 800}%
\special{pa 1994 800}%
\special{pa 1962 800}%
\special{pa 1932 798}%
\special{pa 1900 796}%
\special{pa 1870 792}%
\special{pa 1838 788}%
\special{pa 1808 784}%
\special{pa 1776 780}%
\special{pa 1744 774}%
\special{pa 1714 766}%
\special{pa 1682 760}%
\special{pa 1652 752}%
\special{pa 1620 744}%
\special{pa 1590 734}%
\special{pa 1558 726}%
\special{pa 1528 716}%
\special{pa 1496 706}%
\special{pa 1466 696}%
\special{pa 1434 686}%
\special{pa 1404 674}%
\special{pa 1372 664}%
\special{pa 1342 652}%
\special{pa 1310 642}%
\special{pa 1280 630}%
\special{pa 1248 618}%
\special{pa 1218 606}%
\special{pa 1200 600}%
\special{sp}%
% LINE 2 2 3 0
% 4 2440 705 2420 325 2420 325 2930 985
% 
\special{pn 8}%
\special{pa 2440 706}%
\special{pa 2420 326}%
\special{dt 0.045}%
\special{pa 2420 326}%
\special{pa 2930 986}%
\special{dt 0.045}%
% STR 2 0 3 0
% 3 2450 140 2450 240 5 0
% $r={\rm const.}$
\put(24.5000,-2.4000){\makebox(0,0){$r={\rm const.}$}}%
\end{picture}%

%% file: pen02.tex
%WinTpicVersion3.08
\unitlength 0.1in
\begin{picture}( 24.0000, 16.3000)(  8.0000,-20.0000)
% STR 2 0 3 0
% 3 2650 700 2650 800 5 0
% ${\mathcal I}^+$
\put(26.5000,-8.0000){\makebox(0,0){${\mathcal I}^+$}}%
% VECTOR 2 0 3 0
% 2 1800 2000 1800 600
% 
\special{pn 8}%
\special{pa 1800 2000}%
\special{pa 1800 600}%
\special{fp}%
\special{sh 1}%
\special{pa 1800 600}%
\special{pa 1780 668}%
\special{pa 1800 654}%
\special{pa 1820 668}%
\special{pa 1800 600}%
\special{fp}%
% LINE 2 2 3 0
% 4 1800 1000 2600 1000 2600 1000 2600 1800
% 
\special{pn 8}%
\special{pa 1800 1000}%
\special{pa 2600 1000}%
\special{dt 0.045}%
\special{pa 2600 1000}%
\special{pa 2600 1800}%
\special{dt 0.045}%
% STR 2 0 3 0
% 3 1720 440 1720 540 2 0
% $\tilde{T}$
\put(17.2000,-5.4000){\makebox(0,0)[lb]{$\tilde{T}$}}%
% VECTOR 2 0 3 0
% 2 800 1800 3200 1800
% 
\special{pn 8}%
\special{pa 800 1800}%
\special{pa 3200 1800}%
\special{fp}%
\special{sh 1}%
\special{pa 3200 1800}%
\special{pa 3134 1780}%
\special{pa 3148 1800}%
\special{pa 3134 1820}%
\special{pa 3200 1800}%
\special{fp}%
% STR 2 0 3 0
% 3 1600 1900 1600 2000 2 0
% $0$
\put(16.0000,-20.0000){\makebox(0,0)[lb]{$0$}}%
% STR 2 0 3 0
% 3 3400 1700 3400 1800 5 0
% $\tilde{R}$
\put(34.0000,-18.0000){\makebox(0,0){$\tilde{R}$}}%
% CIRCLE 2 0 3 0
% 4 2600 1000 2600 1030 2600 1030 2600 1030
% 
\special{pn 8}%
\special{ar 2600 1000 30 30  0.0000000 6.2831853}%
% STR 2 0 3 0
% 3 1650 900 1650 1000 5 0
% $\frac{\pi}{4}$
\put(16.5000,-10.0000){\makebox(0,0){$\frac{\pi}{4}$}}%
% STR 2 0 3 0
% 3 2600 1830 2600 1930 5 0
% $\frac{\pi}{4}$
\put(26.0000,-19.3000){\makebox(0,0){$\frac{\pi}{4}$}}%
% STR 2 0 3 0
% 3 2730 900 2730 1000 2 0
% $(\frac{\pi}{4},\frac{\pi}{4})$
\put(27.3000,-10.0000){\makebox(0,0)[lb]{$(\frac{\pi}{4},\frac{\pi}{4})$}}%
\end{picture}%

%% file: pen03.tex
%WinTpicVersion3.08
\unitlength 0.1in
\begin{picture}( 28.7500, 21.7000)(  9.2500,-20.0000)
% LINE 2 0 3 0
% 2 2600 1000 3400 1800
% 
\special{pn 8}%
\special{pa 2600 1000}%
\special{pa 3400 1800}%
\special{fp}%
% STR 2 0 3 0
% 3 3230 1250 3230 1350 5 0
% ${\mathcal J^+}$
\put(32.3000,-13.5000){\makebox(0,0){${\mathcal J^+}$}}%
% LINE 2 2 3 0
% 2 1800 200 2600 1000
% 
\special{pn 8}%
\special{pa 1800 200}%
\special{pa 2600 1000}%
\special{dt 0.045}%
% VECTOR 2 0 3 0
% 2 1800 2000 1800 0
% 
\special{pn 8}%
\special{pa 1800 2000}%
\special{pa 1800 0}%
\special{fp}%
\special{sh 1}%
\special{pa 1800 0}%
\special{pa 1780 68}%
\special{pa 1800 54}%
\special{pa 1820 68}%
\special{pa 1800 0}%
\special{fp}%
% LINE 2 2 3 0
% 2 3400 1800 3600 2000
% 
\special{pn 8}%
\special{pa 3400 1800}%
\special{pa 3600 2000}%
\special{dt 0.045}%
% LINE 2 2 3 0
% 4 1800 1000 2600 1000 2600 1000 2600 1800
% 
\special{pn 8}%
\special{pa 1800 1000}%
\special{pa 2600 1000}%
\special{dt 0.045}%
\special{pa 2600 1000}%
\special{pa 2600 1800}%
\special{dt 0.045}%
% STR 2 0 3 0
% 3 1730 -100 1730 0 2 0
% $\tilde{T}$
\put(17.3000,0.0000){\makebox(0,0)[lb]{$\tilde{T}$}}%
% VECTOR 2 0 3 0
% 2 1400 1800 3800 1800
% 
\special{pn 8}%
\special{pa 1400 1800}%
\special{pa 3800 1800}%
\special{fp}%
\special{sh 1}%
\special{pa 3800 1800}%
\special{pa 3734 1780}%
\special{pa 3748 1800}%
\special{pa 3734 1820}%
\special{pa 3800 1800}%
\special{fp}%
% STR 2 0 3 0
% 3 1600 1900 1600 2000 2 0
% $0$
\put(16.0000,-20.0000){\makebox(0,0)[lb]{$0$}}%
% STR 2 0 3 0
% 3 4000 1700 4000 1800 5 0
% $\tilde{R}$
\put(40.0000,-18.0000){\makebox(0,0){$\tilde{R}$}}%
% STR 2 0 3 0
% 3 2640 860 2640 960 2 0
% $(\frac{\pi}{4},\frac{\pi}{4})$
\put(26.4000,-9.6000){\makebox(0,0)[lb]{$(\frac{\pi}{4},\frac{\pi}{4})$}}%
% STR 2 0 3 0
% 3 2600 1910 2600 2010 5 0
% $\frac{\pi}{4}$
\put(26.0000,-20.1000){\makebox(0,0){$\frac{\pi}{4}$}}%
% STR 2 0 3 0
% 3 1600 900 1600 1000 5 0
% $\frac{\pi}{4}$
\put(16.0000,-10.0000){\makebox(0,0){$\frac{\pi}{4}$}}%
% STR 2 0 3 0
% 3 1610 100 1610 200 5 0
% $\frac{\pi}{2}$
\put(16.1000,-2.0000){\makebox(0,0){$\frac{\pi}{2}$}}%
% STR 2 0 3 0
% 3 3400 1910 3400 2010 5 0
% $\frac{\pi}{2}$
\put(34.0000,-20.1000){\makebox(0,0){$\frac{\pi}{2}$}}%
\end{picture}%

%% file: pen04.tex
%WinTpicVersion3.08
\unitlength 0.1in
\begin{picture}( 37.2000, 25.7500)(  4.8000,-26.0000)
% CIRCLE 2 0 3 0
% 4 2800 600 2800 640 2800 640 2800 640
% 
\special{pn 8}%
\special{ar 2800 600 40 40  0.0000000 6.2831853}%
% CIRCLE 2 0 3 0
% 4 3600 1400 3600 1440 3600 1440 3600 1440
% 
\special{pn 8}%
\special{ar 3600 1400 40 40  0.0000000 6.2831853}%
% CIRCLE 2 0 3 0
% 4 2800 2200 2800 2240 2800 2240 2800 2240
% 
\special{pn 8}%
\special{ar 2800 2200 40 40  0.0000000 6.2831853}%
% CIRCLE 2 0 3 0
% 4 1200 600 1200 640 1200 640 1200 640
% 
\special{pn 8}%
\special{ar 1200 600 40 40  0.0000000 6.2831853}%
% LINE 2 0 3 0
% 2 2820 620 3570 1370
% 
\special{pn 8}%
\special{pa 2820 620}%
\special{pa 3570 1370}%
\special{fp}%
% LINE 2 0 3 0
% 2 3570 1430 2820 2180
% 
\special{pn 8}%
\special{pa 3570 1430}%
\special{pa 2820 2180}%
\special{fp}%
% LINE 0 0 3 0
% 2 1230 630 2770 2170
% 
\special{pn 20}%
\special{pa 1230 630}%
\special{pa 2770 2170}%
\special{fp}%
% LINE 2 0 3 0
% 2 1240 630 2780 630
% 
\special{pn 8}%
\special{pa 1240 630}%
\special{pa 2780 630}%
\special{fp}%
% LINE 2 0 3 0
% 2 1220 570 2760 570
% 
\special{pn 8}%
\special{pa 1220 570}%
\special{pa 2760 570}%
\special{fp}%
% LINE 0 0 3 0
% 2 2770 630 1230 2170
% 
\special{pn 20}%
\special{pa 2770 630}%
\special{pa 1230 2170}%
\special{fp}%
% STR 2 0 3 0
% 3 3710 1180 3710 1280 5 0
% ${\mathcal I}^0$
\put(37.1000,-12.8000){\makebox(0,0){${\mathcal I}^0$}}%
% STR 2 0 3 0
% 3 3000 410 3000 510 5 0
% ${\mathcal I}^+$
\put(30.0000,-5.1000){\makebox(0,0){${\mathcal I}^+$}}%
% STR 2 0 3 0
% 3 3000 2230 3000 2330 5 0
% ${\mathcal I}^-$
\put(30.0000,-23.3000){\makebox(0,0){${\mathcal I}^-$}}%
% STR 2 0 3 0
% 3 1780 340 1780 440 5 0
% ${\mathcal R}$
\put(17.8000,-4.4000){\makebox(0,0){${\mathcal R}$}}%
% STR 2 0 3 0
% 3 3400 820 3400 920 5 0
% ${\mathcal J^+}$
\put(34.0000,-9.2000){\makebox(0,0){${\mathcal J^+}$}}%
% STR 2 0 3 0
% 3 3400 1720 3400 1820 5 0
% ${\mathcal J^-}$
\put(34.0000,-18.2000){\makebox(0,0){${\mathcal J^-}$}}%
% STR 2 0 3 0
% 3 2500 1070 2500 1170 5 0
% ${\mathcal H}^+$
\put(25.0000,-11.7000){\makebox(0,0){${\mathcal H}^+$}}%
% STR 2 0 3 0
% 3 1590 1080 1590 1180 5 0
% ${\mathcal H}^-$
\put(15.9000,-11.8000){\makebox(0,0){${\mathcal H}^-$}}%
% VECTOR 2 0 3 0
% 2 600 1400 4200 1400
% 
\special{pn 8}%
\special{pa 600 1400}%
\special{pa 4200 1400}%
\special{fp}%
\special{sh 1}%
\special{pa 4200 1400}%
\special{pa 4134 1380}%
\special{pa 4148 1400}%
\special{pa 4134 1420}%
\special{pa 4200 1400}%
\special{fp}%
% VECTOR 2 0 3 0
% 2 2000 2600 2000 200
% 
\special{pn 8}%
\special{pa 2000 2600}%
\special{pa 2000 200}%
\special{fp}%
\special{sh 1}%
\special{pa 2000 200}%
\special{pa 1980 268}%
\special{pa 2000 254}%
\special{pa 2020 268}%
\special{pa 2000 200}%
\special{fp}%
% STR 2 0 3 0
% 3 2000 10 2000 110 5 0
% $\tilde{T}$
\put(20.0000,-1.1000){\makebox(0,0){$\tilde{T}$}}%
% STR 2 0 3 0
% 3 4320 1300 4320 1400 5 0
% $\tilde{R}$
\put(43.2000,-14.0000){\makebox(0,0){$\tilde{R}$}}%
% STR 2 0 3 0
% 3 2780 350 2780 450 5 0
% ${\mathcal R}^+$
\put(27.8000,-4.5000){\makebox(0,0){${\mathcal R}^+$}}%
% STR 2 0 3 0
% 3 1200 350 1200 450 5 0
% ${\mathcal R}^-$
\put(12.0000,-4.5000){\makebox(0,0){${\mathcal R}^-$}}%
\end{picture}%

%% file: pen05.tex
%WinTpicVersion3.08
\unitlength 0.1in
\begin{picture}( 38.8000, 18.5000)( -2.4000,-22.4500)
% CIRCLE 2 0 3 0
% 4 2800 600 2800 640 2800 640 2800 640
% 
\special{pn 8}%
\special{ar 2800 600 40 40  0.0000000 6.2831853}%
% CIRCLE 2 0 3 0
% 4 3600 1400 3600 1440 3600 1440 3600 1440
% 
\special{pn 8}%
\special{ar 3600 1400 40 40  0.0000000 6.2831853}%
% CIRCLE 2 0 3 0
% 4 2800 2200 2800 2240 2800 2240 2800 2240
% 
\special{pn 8}%
\special{ar 2800 2200 40 40  0.0000000 6.2831853}%
% CIRCLE 2 0 3 0
% 4 1200 2200 1200 2240 1200 2240 1200 2240
% 
\special{pn 8}%
\special{ar 1200 2200 40 40  0.0000000 6.2831853}%
% CIRCLE 2 0 3 0
% 4 400 1400 400 1440 400 1440 400 1440
% 
\special{pn 8}%
\special{ar 400 1400 40 40  0.0000000 6.2831853}%
% CIRCLE 2 0 3 0
% 4 1200 600 1200 640 1200 640 1200 640
% 
\special{pn 8}%
\special{ar 1200 600 40 40  0.0000000 6.2831853}%
% LINE 2 0 3 0
% 2 1230 2170 2770 2170
% 
\special{pn 8}%
\special{pa 1230 2170}%
\special{pa 2770 2170}%
\special{fp}%
% LINE 2 0 3 0
% 2 2820 620 3570 1370
% 
\special{pn 8}%
\special{pa 2820 620}%
\special{pa 3570 1370}%
\special{fp}%
% LINE 2 0 3 0
% 2 430 1430 1180 2180
% 
\special{pn 8}%
\special{pa 430 1430}%
\special{pa 1180 2180}%
\special{fp}%
% LINE 2 0 3 0
% 2 3570 1430 2820 2180
% 
\special{pn 8}%
\special{pa 3570 1430}%
\special{pa 2820 2180}%
\special{fp}%
% LINE 2 0 3 0
% 2 1170 630 420 1380
% 
\special{pn 8}%
\special{pa 1170 630}%
\special{pa 420 1380}%
\special{fp}%
% LINE 0 0 3 0
% 2 1230 630 2770 2170
% 
\special{pn 20}%
\special{pa 1230 630}%
\special{pa 2770 2170}%
\special{fp}%
% LINE 2 0 3 0
% 2 1230 2230 2770 2230
% 
\special{pn 8}%
\special{pa 1230 2230}%
\special{pa 2770 2230}%
\special{fp}%
% LINE 2 0 3 0
% 2 1240 630 2780 630
% 
\special{pn 8}%
\special{pa 1240 630}%
\special{pa 2780 630}%
\special{fp}%
% LINE 2 0 3 0
% 2 1220 570 2760 570
% 
\special{pn 8}%
\special{pa 1220 570}%
\special{pa 2760 570}%
\special{fp}%
% LINE 0 0 3 0
% 2 2770 630 1230 2170
% 
\special{pn 20}%
\special{pa 2770 630}%
\special{pa 1230 2170}%
\special{fp}%
% STR 2 0 3 0
% 3 3800 1310 3800 1410 5 0
% ${\mathcal I}^0$
\put(38.0000,-14.1000){\makebox(0,0){${\mathcal I}^0$}}%
% STR 2 0 3 0
% 3 3000 410 3000 510 5 0
% ${\mathcal I}^+$
\put(30.0000,-5.1000){\makebox(0,0){${\mathcal I}^+$}}%
% STR 2 0 3 0
% 3 3000 2230 3000 2330 5 0
% ${\mathcal I}^-$
\put(30.0000,-23.3000){\makebox(0,0){${\mathcal I}^-$}}%
% STR 2 0 3 0
% 3 2000 380 2000 480 5 0
% $r=0$
\put(20.0000,-4.8000){\makebox(0,0){$r=0$}}%
% STR 2 0 3 0
% 3 3400 820 3400 920 5 0
% ${\mathcal J^+}$
\put(34.0000,-9.2000){\makebox(0,0){${\mathcal J^+}$}}%
% STR 2 0 3 0
% 3 3400 1720 3400 1820 5 0
% ${\mathcal J^-}$
\put(34.0000,-18.2000){\makebox(0,0){${\mathcal J^-}$}}%
% STR 2 0 3 0
% 3 2500 1070 2500 1170 5 0
% ${\mathcal H}^+$
\put(25.0000,-11.7000){\makebox(0,0){${\mathcal H}^+$}}%
% STR 2 0 3 0
% 3 1590 1080 1590 1180 5 0
% ${\mathcal H}^-$
\put(15.9000,-11.8000){\makebox(0,0){${\mathcal H}^-$}}%
% STR 2 0 3 0
% 3 2000 900 2000 1000 5 0
% I\hspace{-.1em}I 
\put(20.0000,-10.0000){\makebox(0,0){I\hspace{-.1em}I }}%
% STR 2 0 3 0
% 3 2800 1310 2800 1410 5 0
% I
\put(28.0000,-14.1000){\makebox(0,0){I}}%
% STR 2 0 3 0
% 3 1200 1300 1200 1400 5 0
% I\hspace{-.1em}I\hspace{-.1em}I 
\put(12.0000,-14.0000){\makebox(0,0){I\hspace{-.1em}I\hspace{-.1em}I }}%
% STR 2 0 3 0
% 3 1990 1820 1990 1920 5 0
% I\hspace{-.1em}V
\put(19.9000,-19.2000){\makebox(0,0){I\hspace{-.1em}V}}%
% VECTOR 2 0 3 0
% 2 2800 2000 3000 1800
% 
\special{pn 8}%
\special{pa 2800 2000}%
\special{pa 3000 1800}%
\special{fp}%
\special{sh 1}%
\special{pa 3000 1800}%
\special{pa 2940 1834}%
\special{pa 2962 1838}%
\special{pa 2968 1862}%
\special{pa 3000 1800}%
\special{fp}%
% VECTOR 2 0 3 0
% 8 2800 2000 2600 1800 2600 1800 2600 1800 2600 1800 2600 1800 2600 1800 2600 1800
% 
\special{pn 8}%
\special{pa 2800 2000}%
\special{pa 2600 1800}%
\special{fp}%
\special{sh 1}%
\special{pa 2600 1800}%
\special{pa 2634 1862}%
\special{pa 2638 1838}%
\special{pa 2662 1834}%
\special{pa 2600 1800}%
\special{fp}%
\special{pa 2600 1800}%
\special{pa 2600 1800}%
\special{fp}%
\special{pa 2600 1800}%
\special{pa 2600 1800}%
\special{fp}%
\special{pa 2600 1800}%
\special{pa 2600 1800}%
\special{fp}%
% VECTOR 2 0 3 0
% 2 2270 1000 2470 800
% 
\special{pn 8}%
\special{pa 2270 1000}%
\special{pa 2470 800}%
\special{fp}%
\special{sh 1}%
\special{pa 2470 800}%
\special{pa 2410 834}%
\special{pa 2432 838}%
\special{pa 2438 862}%
\special{pa 2470 800}%
\special{fp}%
% VECTOR 2 0 3 0
% 8 2270 1000 2070 800 2070 800 2070 800 2070 800 2070 800 2070 800 2070 800
% 
\special{pn 8}%
\special{pa 2270 1000}%
\special{pa 2070 800}%
\special{fp}%
\special{sh 1}%
\special{pa 2070 800}%
\special{pa 2104 862}%
\special{pa 2108 838}%
\special{pa 2132 834}%
\special{pa 2070 800}%
\special{fp}%
\special{pa 2070 800}%
\special{pa 2070 800}%
\special{fp}%
\special{pa 2070 800}%
\special{pa 2070 800}%
\special{fp}%
\special{pa 2070 800}%
\special{pa 2070 800}%
\special{fp}%
\end{picture}%

%% file: pen06.tex
%WinTpicVersion3.08
\unitlength 0.1in
\begin{picture}( 32.8000, 18.5000)(  3.6000,-22.4500)
% CIRCLE 2 0 3 0
% 4 2800 600 2800 640 2800 640 2800 640
% 
\special{pn 8}%
\special{ar 2800 600 40 40  0.0000000 6.2831853}%
% CIRCLE 2 0 3 0
% 4 3600 1400 3600 1440 3600 1440 3600 1440
% 
\special{pn 8}%
\special{ar 3600 1400 40 40  0.0000000 6.2831853}%
% CIRCLE 2 0 3 0
% 4 2800 2200 2800 2240 2800 2240 2800 2240
% 
\special{pn 8}%
\special{ar 2800 2200 40 40  0.0000000 6.2831853}%
% CIRCLE 2 0 3 0
% 4 1200 2200 1200 2240 1200 2240 1200 2240
% 
\special{pn 8}%
\special{ar 1200 2200 40 40  0.0000000 6.2831853}%
% CIRCLE 2 0 3 0
% 4 400 1400 400 1440 400 1440 400 1440
% 
\special{pn 8}%
\special{ar 400 1400 40 40  0.0000000 6.2831853}%
% CIRCLE 2 0 3 0
% 4 1200 600 1200 640 1200 640 1200 640
% 
\special{pn 8}%
\special{ar 1200 600 40 40  0.0000000 6.2831853}%
% LINE 2 0 3 0
% 2 1230 2170 2770 2170
% 
\special{pn 8}%
\special{pa 1230 2170}%
\special{pa 2770 2170}%
\special{fp}%
% LINE 2 0 3 0
% 2 2820 620 3570 1370
% 
\special{pn 8}%
\special{pa 2820 620}%
\special{pa 3570 1370}%
\special{fp}%
% LINE 2 0 3 0
% 2 430 1430 1180 2180
% 
\special{pn 8}%
\special{pa 430 1430}%
\special{pa 1180 2180}%
\special{fp}%
% LINE 2 0 3 0
% 2 3570 1430 2820 2180
% 
\special{pn 8}%
\special{pa 3570 1430}%
\special{pa 2820 2180}%
\special{fp}%
% LINE 2 0 3 0
% 2 1170 630 420 1380
% 
\special{pn 8}%
\special{pa 1170 630}%
\special{pa 420 1380}%
\special{fp}%
% LINE 0 0 3 0
% 2 1230 630 2770 2170
% 
\special{pn 20}%
\special{pa 1230 630}%
\special{pa 2770 2170}%
\special{fp}%
% LINE 2 0 3 0
% 2 1230 2230 2770 2230
% 
\special{pn 8}%
\special{pa 1230 2230}%
\special{pa 2770 2230}%
\special{fp}%
% LINE 2 0 3 0
% 2 1240 630 2780 630
% 
\special{pn 8}%
\special{pa 1240 630}%
\special{pa 2780 630}%
\special{fp}%
% LINE 2 0 3 0
% 2 1220 570 2760 570
% 
\special{pn 8}%
\special{pa 1220 570}%
\special{pa 2760 570}%
\special{fp}%
% LINE 0 0 3 0
% 2 2770 630 1230 2170
% 
\special{pn 20}%
\special{pa 2770 630}%
\special{pa 1230 2170}%
\special{fp}%
% STR 2 0 3 0
% 3 3800 1310 3800 1410 5 0
% ${\mathcal I}^0$
\put(38.0000,-14.1000){\makebox(0,0){${\mathcal I}^0$}}%
% STR 2 0 3 0
% 3 3000 410 3000 510 5 0
% ${\mathcal I}^+$
\put(30.0000,-5.1000){\makebox(0,0){${\mathcal I}^+$}}%
% STR 2 0 3 0
% 3 3000 2230 3000 2330 5 0
% ${\mathcal I}^-$
\put(30.0000,-23.3000){\makebox(0,0){${\mathcal I}^-$}}%
% STR 2 0 3 0
% 3 2000 380 2000 480 5 0
% $r=0$
\put(20.0000,-4.8000){\makebox(0,0){$r=0$}}%
% STR 2 0 3 0
% 3 3400 820 3400 920 5 0
% ${\mathcal J^+}$
\put(34.0000,-9.2000){\makebox(0,0){${\mathcal J^+}$}}%
% STR 2 0 3 0
% 3 3400 1720 3400 1820 5 0
% ${\mathcal J^-}$
\put(34.0000,-18.2000){\makebox(0,0){${\mathcal J^-}$}}%
% LINE 2 0 3 0
% 2 2210 650 2770 2160
% 
\special{pn 8}%
\special{pa 2210 650}%
\special{pa 2770 2160}%
\special{fp}%
% LINE 2 0 3 0
% 52 1700 1120 910 1910 1730 1150 940 1940 1760 1180 970 1970 1790 1210 1000 2000 1820 1240 1030 2030 1850 1270 1060 2060 1880 1300 1090 2090 1910 1330 1120 2120 1940 1360 1150 2150 1970 1390 1200 2160 1670 1090 880 1880 1640 1060 850 1850 1610 1030 820 1820 1580 1000 790 1790 1550 970 760 1760 1520 940 730 1730 1490 910 700 1700 1460 880 670 1670 1430 850 640 1640 1400 820 610 1610 1370 790 580 1580 1340 760 550 1550 1310 730 520 1520 1280 700 490 1490 1250 670 460 1460 1220 640 430 1430
% 
\special{pn 8}%
\special{pa 1700 1120}%
\special{pa 910 1910}%
\special{fp}%
\special{pa 1730 1150}%
\special{pa 940 1940}%
\special{fp}%
\special{pa 1760 1180}%
\special{pa 970 1970}%
\special{fp}%
\special{pa 1790 1210}%
\special{pa 1000 2000}%
\special{fp}%
\special{pa 1820 1240}%
\special{pa 1030 2030}%
\special{fp}%
\special{pa 1850 1270}%
\special{pa 1060 2060}%
\special{fp}%
\special{pa 1880 1300}%
\special{pa 1090 2090}%
\special{fp}%
\special{pa 1910 1330}%
\special{pa 1120 2120}%
\special{fp}%
\special{pa 1940 1360}%
\special{pa 1150 2150}%
\special{fp}%
\special{pa 1970 1390}%
\special{pa 1200 2160}%
\special{fp}%
\special{pa 1670 1090}%
\special{pa 880 1880}%
\special{fp}%
\special{pa 1640 1060}%
\special{pa 850 1850}%
\special{fp}%
\special{pa 1610 1030}%
\special{pa 820 1820}%
\special{fp}%
\special{pa 1580 1000}%
\special{pa 790 1790}%
\special{fp}%
\special{pa 1550 970}%
\special{pa 760 1760}%
\special{fp}%
\special{pa 1520 940}%
\special{pa 730 1730}%
\special{fp}%
\special{pa 1490 910}%
\special{pa 700 1700}%
\special{fp}%
\special{pa 1460 880}%
\special{pa 670 1670}%
\special{fp}%
\special{pa 1430 850}%
\special{pa 640 1640}%
\special{fp}%
\special{pa 1400 820}%
\special{pa 610 1610}%
\special{fp}%
\special{pa 1370 790}%
\special{pa 580 1580}%
\special{fp}%
\special{pa 1340 760}%
\special{pa 550 1550}%
\special{fp}%
\special{pa 1310 730}%
\special{pa 520 1520}%
\special{fp}%
\special{pa 1280 700}%
\special{pa 490 1490}%
\special{fp}%
\special{pa 1250 670}%
\special{pa 460 1460}%
\special{fp}%
\special{pa 1220 640}%
\special{pa 430 1430}%
\special{fp}%
% LINE 2 0 3 0
% 48 2230 710 1780 1160 2250 750 1810 1190 2260 800 1840 1220 2280 840 1870 1250 2300 880 1900 1280 2310 930 1930 1310 2330 970 1960 1340 2340 1020 1990 1370 2210 670 1750 1130 2190 630 1720 1100 2130 630 1690 1070 2070 630 1660 1040 2010 630 1630 1010 1950 630 1600 980 1890 630 1570 950 1830 630 1540 920 1770 630 1510 890 1710 630 1480 860 1650 630 1450 830 1590 630 1420 800 1530 630 1390 770 1470 630 1360 740 1410 630 1330 710 1350 630 1300 680
% 
\special{pn 8}%
\special{pa 2230 710}%
\special{pa 1780 1160}%
\special{fp}%
\special{pa 2250 750}%
\special{pa 1810 1190}%
\special{fp}%
\special{pa 2260 800}%
\special{pa 1840 1220}%
\special{fp}%
\special{pa 2280 840}%
\special{pa 1870 1250}%
\special{fp}%
\special{pa 2300 880}%
\special{pa 1900 1280}%
\special{fp}%
\special{pa 2310 930}%
\special{pa 1930 1310}%
\special{fp}%
\special{pa 2330 970}%
\special{pa 1960 1340}%
\special{fp}%
\special{pa 2340 1020}%
\special{pa 1990 1370}%
\special{fp}%
\special{pa 2210 670}%
\special{pa 1750 1130}%
\special{fp}%
\special{pa 2190 630}%
\special{pa 1720 1100}%
\special{fp}%
\special{pa 2130 630}%
\special{pa 1690 1070}%
\special{fp}%
\special{pa 2070 630}%
\special{pa 1660 1040}%
\special{fp}%
\special{pa 2010 630}%
\special{pa 1630 1010}%
\special{fp}%
\special{pa 1950 630}%
\special{pa 1600 980}%
\special{fp}%
\special{pa 1890 630}%
\special{pa 1570 950}%
\special{fp}%
\special{pa 1830 630}%
\special{pa 1540 920}%
\special{fp}%
\special{pa 1770 630}%
\special{pa 1510 890}%
\special{fp}%
\special{pa 1710 630}%
\special{pa 1480 860}%
\special{fp}%
\special{pa 1650 630}%
\special{pa 1450 830}%
\special{fp}%
\special{pa 1590 630}%
\special{pa 1420 800}%
\special{fp}%
\special{pa 1530 630}%
\special{pa 1390 770}%
\special{fp}%
\special{pa 1470 630}%
\special{pa 1360 740}%
\special{fp}%
\special{pa 1410 630}%
\special{pa 1330 710}%
\special{fp}%
\special{pa 1350 630}%
\special{pa 1300 680}%
\special{fp}%
% LINE 2 0 3 0
% 46 2430 1230 2140 1520 2440 1280 2170 1550 2460 1320 2200 1580 2470 1370 2230 1610 2490 1410 2260 1640 2510 1450 2290 1670 2520 1500 2320 1700 2540 1540 2350 1730 2550 1590 2380 1760 2570 1630 2410 1790 2590 1670 2440 1820 2600 1720 2470 1850 2620 1760 2500 1880 2640 1800 2530 1910 2650 1850 2560 1940 2670 1890 2590 1970 2680 1940 2620 2000 2700 1980 2650 2030 2720 2020 2680 2060 2730 2070 2710 2090 2410 1190 2110 1490 2390 1150 2080 1460 2380 1100 2050 1430
% 
\special{pn 8}%
\special{pa 2430 1230}%
\special{pa 2140 1520}%
\special{fp}%
\special{pa 2440 1280}%
\special{pa 2170 1550}%
\special{fp}%
\special{pa 2460 1320}%
\special{pa 2200 1580}%
\special{fp}%
\special{pa 2470 1370}%
\special{pa 2230 1610}%
\special{fp}%
\special{pa 2490 1410}%
\special{pa 2260 1640}%
\special{fp}%
\special{pa 2510 1450}%
\special{pa 2290 1670}%
\special{fp}%
\special{pa 2520 1500}%
\special{pa 2320 1700}%
\special{fp}%
\special{pa 2540 1540}%
\special{pa 2350 1730}%
\special{fp}%
\special{pa 2550 1590}%
\special{pa 2380 1760}%
\special{fp}%
\special{pa 2570 1630}%
\special{pa 2410 1790}%
\special{fp}%
\special{pa 2590 1670}%
\special{pa 2440 1820}%
\special{fp}%
\special{pa 2600 1720}%
\special{pa 2470 1850}%
\special{fp}%
\special{pa 2620 1760}%
\special{pa 2500 1880}%
\special{fp}%
\special{pa 2640 1800}%
\special{pa 2530 1910}%
\special{fp}%
\special{pa 2650 1850}%
\special{pa 2560 1940}%
\special{fp}%
\special{pa 2670 1890}%
\special{pa 2590 1970}%
\special{fp}%
\special{pa 2680 1940}%
\special{pa 2620 2000}%
\special{fp}%
\special{pa 2700 1980}%
\special{pa 2650 2030}%
\special{fp}%
\special{pa 2720 2020}%
\special{pa 2680 2060}%
\special{fp}%
\special{pa 2730 2070}%
\special{pa 2710 2090}%
\special{fp}%
\special{pa 2410 1190}%
\special{pa 2110 1490}%
\special{fp}%
\special{pa 2390 1150}%
\special{pa 2080 1460}%
\special{fp}%
\special{pa 2380 1100}%
\special{pa 2050 1430}%
\special{fp}%
% LINE 2 0 3 0
% 48 2330 1750 1910 2170 2300 1720 1850 2170 2270 1690 1790 2170 2240 1660 1730 2170 2210 1630 1670 2170 2180 1600 1610 2170 2150 1570 1550 2170 2120 1540 1490 2170 2090 1510 1430 2170 2060 1480 1370 2170 2030 1450 1310 2170 2360 1780 1970 2170 2390 1810 2030 2170 2420 1840 2090 2170 2450 1870 2150 2170 2480 1900 2210 2170 2510 1930 2270 2170 2540 1960 2330 2170 2570 1990 2390 2170 2600 2020 2450 2170 2630 2050 2510 2170 2660 2080 2570 2170 2690 2110 2630 2170 2720 2140 2690 2170
% 
\special{pn 8}%
\special{pa 2330 1750}%
\special{pa 1910 2170}%
\special{fp}%
\special{pa 2300 1720}%
\special{pa 1850 2170}%
\special{fp}%
\special{pa 2270 1690}%
\special{pa 1790 2170}%
\special{fp}%
\special{pa 2240 1660}%
\special{pa 1730 2170}%
\special{fp}%
\special{pa 2210 1630}%
\special{pa 1670 2170}%
\special{fp}%
\special{pa 2180 1600}%
\special{pa 1610 2170}%
\special{fp}%
\special{pa 2150 1570}%
\special{pa 1550 2170}%
\special{fp}%
\special{pa 2120 1540}%
\special{pa 1490 2170}%
\special{fp}%
\special{pa 2090 1510}%
\special{pa 1430 2170}%
\special{fp}%
\special{pa 2060 1480}%
\special{pa 1370 2170}%
\special{fp}%
\special{pa 2030 1450}%
\special{pa 1310 2170}%
\special{fp}%
\special{pa 2360 1780}%
\special{pa 1970 2170}%
\special{fp}%
\special{pa 2390 1810}%
\special{pa 2030 2170}%
\special{fp}%
\special{pa 2420 1840}%
\special{pa 2090 2170}%
\special{fp}%
\special{pa 2450 1870}%
\special{pa 2150 2170}%
\special{fp}%
\special{pa 2480 1900}%
\special{pa 2210 2170}%
\special{fp}%
\special{pa 2510 1930}%
\special{pa 2270 2170}%
\special{fp}%
\special{pa 2540 1960}%
\special{pa 2330 2170}%
\special{fp}%
\special{pa 2570 1990}%
\special{pa 2390 2170}%
\special{fp}%
\special{pa 2600 2020}%
\special{pa 2450 2170}%
\special{fp}%
\special{pa 2630 2050}%
\special{pa 2510 2170}%
\special{fp}%
\special{pa 2660 2080}%
\special{pa 2570 2170}%
\special{fp}%
\special{pa 2690 2110}%
\special{pa 2630 2170}%
\special{fp}%
\special{pa 2720 2140}%
\special{pa 2690 2170}%
\special{fp}%
\end{picture}%

%% file: pen07.tex
%WinTpicVersion3.08
\unitlength 0.1in
\begin{picture}( 27.3000, 25.1500)(  4.7000,-30.9000)
% BOX 2 0 3 0
% 2 1510 790 2600 750
% 
\special{pn 8}%
\special{pa 1510 790}%
\special{pa 2600 790}%
\special{pa 2600 750}%
\special{pa 1510 750}%
\special{pa 1510 790}%
\special{fp}%
% LINE 2 0 3 0
% 2 2600 800 3200 1400
% 
\special{pn 8}%
\special{pa 2600 800}%
\special{pa 3200 1400}%
\special{fp}%
% LINE 2 0 3 0
% 2 3200 1400 1600 3000
% 
\special{pn 8}%
\special{pa 3200 1400}%
\special{pa 1600 3000}%
\special{fp}%
% LINE 2 0 3 0
% 2 1600 800 1600 3000
% 
\special{pn 8}%
\special{pa 1600 800}%
\special{pa 1600 3000}%
\special{fp}%
% LINE 2 2 3 0
% 6 1510 790 1510 3090 1510 3070 1590 3000 1370 3040 1370 3040
% 
\special{pn 8}%
\special{pa 1510 790}%
\special{pa 1510 3090}%
\special{dt 0.045}%
\special{pa 1510 3070}%
\special{pa 1590 3000}%
\special{dt 0.045}%
\special{pa 1370 3040}%
\special{pa 1370 3040}%
\special{dt 0.045}%
% SPLINE 2 0 3 0
% 5 1590 3000 2190 1210 1870 790 1840 790 1840 790
% 
\special{pn 8}%
\special{pa 1590 3000}%
\special{pa 1610 2974}%
\special{pa 1630 2946}%
\special{pa 1650 2918}%
\special{pa 1670 2890}%
\special{pa 1690 2864}%
\special{pa 1710 2836}%
\special{pa 1728 2808}%
\special{pa 1748 2780}%
\special{pa 1768 2754}%
\special{pa 1786 2726}%
\special{pa 1806 2698}%
\special{pa 1824 2670}%
\special{pa 1842 2642}%
\special{pa 1862 2614}%
\special{pa 1880 2586}%
\special{pa 1898 2558}%
\special{pa 1914 2530}%
\special{pa 1932 2502}%
\special{pa 1950 2474}%
\special{pa 1966 2446}%
\special{pa 1982 2418}%
\special{pa 1998 2388}%
\special{pa 2014 2360}%
\special{pa 2030 2332}%
\special{pa 2044 2302}%
\special{pa 2058 2274}%
\special{pa 2072 2244}%
\special{pa 2086 2214}%
\special{pa 2100 2186}%
\special{pa 2112 2156}%
\special{pa 2124 2126}%
\special{pa 2136 2096}%
\special{pa 2146 2066}%
\special{pa 2158 2036}%
\special{pa 2166 2006}%
\special{pa 2176 1974}%
\special{pa 2186 1944}%
\special{pa 2194 1914}%
\special{pa 2200 1882}%
\special{pa 2208 1850}%
\special{pa 2214 1820}%
\special{pa 2220 1788}%
\special{pa 2224 1756}%
\special{pa 2228 1724}%
\special{pa 2232 1690}%
\special{pa 2234 1658}%
\special{pa 2236 1626}%
\special{pa 2236 1592}%
\special{pa 2236 1558}%
\special{pa 2236 1524}%
\special{pa 2234 1492}%
\special{pa 2232 1456}%
\special{pa 2228 1422}%
\special{pa 2224 1388}%
\special{pa 2220 1352}%
\special{pa 2214 1318}%
\special{pa 2208 1282}%
\special{pa 2200 1246}%
\special{pa 2190 1210}%
\special{pa 2180 1174}%
\special{pa 2170 1138}%
\special{pa 2158 1102}%
\special{pa 2144 1066}%
\special{pa 2130 1032}%
\special{pa 2116 998}%
\special{pa 2098 966}%
\special{pa 2080 936}%
\special{pa 2062 908}%
\special{pa 2042 882}%
\special{pa 2018 858}%
\special{pa 1996 838}%
\special{pa 1970 820}%
\special{pa 1944 808}%
\special{pa 1916 798}%
\special{pa 1886 792}%
\special{pa 1854 790}%
\special{pa 1840 790}%
\special{sp}%
% LINE 2 0 3 0
% 60 2170 1130 1600 1700 2180 1180 1600 1760 2190 1230 1600 1820 2200 1280 1600 1880 2210 1330 1600 1940 2220 1380 1600 2000 2230 1430 1600 2060 2230 1490 1600 2120 2240 1540 1600 2180 2240 1600 1600 2240 2230 1670 1600 2300 2220 1740 1600 2360 2220 1800 1600 2420 2190 1890 1600 2480 2180 1960 1600 2540 2150 2050 1600 2600 2110 2150 1600 2660 2060 2260 1600 2720 1990 2390 1600 2780 1920 2520 1600 2840 1810 2690 1600 2900 1660 2900 1600 2960 2150 1090 1600 1640 2140 1040 1600 1580 2120 1000 1600 1520 2090 970 1600 1460 2070 930 1600 1400 2040 900 1600 1340 2020 860 1600 1280 1990 830 1600 1220
% 
\special{pn 8}%
\special{pa 2170 1130}%
\special{pa 1600 1700}%
\special{fp}%
\special{pa 2180 1180}%
\special{pa 1600 1760}%
\special{fp}%
\special{pa 2190 1230}%
\special{pa 1600 1820}%
\special{fp}%
\special{pa 2200 1280}%
\special{pa 1600 1880}%
\special{fp}%
\special{pa 2210 1330}%
\special{pa 1600 1940}%
\special{fp}%
\special{pa 2220 1380}%
\special{pa 1600 2000}%
\special{fp}%
\special{pa 2230 1430}%
\special{pa 1600 2060}%
\special{fp}%
\special{pa 2230 1490}%
\special{pa 1600 2120}%
\special{fp}%
\special{pa 2240 1540}%
\special{pa 1600 2180}%
\special{fp}%
\special{pa 2240 1600}%
\special{pa 1600 2240}%
\special{fp}%
\special{pa 2230 1670}%
\special{pa 1600 2300}%
\special{fp}%
\special{pa 2220 1740}%
\special{pa 1600 2360}%
\special{fp}%
\special{pa 2220 1800}%
\special{pa 1600 2420}%
\special{fp}%
\special{pa 2190 1890}%
\special{pa 1600 2480}%
\special{fp}%
\special{pa 2180 1960}%
\special{pa 1600 2540}%
\special{fp}%
\special{pa 2150 2050}%
\special{pa 1600 2600}%
\special{fp}%
\special{pa 2110 2150}%
\special{pa 1600 2660}%
\special{fp}%
\special{pa 2060 2260}%
\special{pa 1600 2720}%
\special{fp}%
\special{pa 1990 2390}%
\special{pa 1600 2780}%
\special{fp}%
\special{pa 1920 2520}%
\special{pa 1600 2840}%
\special{fp}%
\special{pa 1810 2690}%
\special{pa 1600 2900}%
\special{fp}%
\special{pa 1660 2900}%
\special{pa 1600 2960}%
\special{fp}%
\special{pa 2150 1090}%
\special{pa 1600 1640}%
\special{fp}%
\special{pa 2140 1040}%
\special{pa 1600 1580}%
\special{fp}%
\special{pa 2120 1000}%
\special{pa 1600 1520}%
\special{fp}%
\special{pa 2090 970}%
\special{pa 1600 1460}%
\special{fp}%
\special{pa 2070 930}%
\special{pa 1600 1400}%
\special{fp}%
\special{pa 2040 900}%
\special{pa 1600 1340}%
\special{fp}%
\special{pa 2020 860}%
\special{pa 1600 1280}%
\special{fp}%
\special{pa 1990 830}%
\special{pa 1600 1220}%
\special{fp}%
% LINE 2 0 3 1
% 12 1950 810 1600 1160 1910 790 1600 1100 1850 790 1600 1040 1790 790 1600 980 1730 790 1600 920 1670 790 1600 860
% 
\special{pn 8}%
\special{pa 1950 810}%
\special{pa 1600 1160}%
\special{fp}%
\special{pa 1910 790}%
\special{pa 1600 1100}%
\special{fp}%
\special{pa 1850 790}%
\special{pa 1600 1040}%
\special{fp}%
\special{pa 1790 790}%
\special{pa 1600 980}%
\special{fp}%
\special{pa 1730 790}%
\special{pa 1600 920}%
\special{fp}%
\special{pa 1670 790}%
\special{pa 1600 860}%
\special{fp}%
% LINE 0 0 3 0
% 2 2600 790 1600 1780
% 
\special{pn 20}%
\special{pa 2600 790}%
\special{pa 1600 1780}%
\special{fp}%
% STR 2 0 3 0
% 3 2000 560 2000 660 5 0
% singularity
\put(20.0000,-6.6000){\makebox(0,0){singularity}}%
% STR 2 0 3 0
% 3 3000 900 3000 1000 5 0
% ${\mathcal J}^+$
\put(30.0000,-10.0000){\makebox(0,0){${\mathcal J}^+$}}%
% STR 2 0 3 0
% 3 2600 2090 2600 2190 5 0
% ${\mathcal J}^-$
\put(26.0000,-21.9000){\makebox(0,0){${\mathcal J}^-$}}%
% STR 2 0 3 0
% 3 1760 2980 1760 3080 5 0
% ${\mathcal I}^-$
\put(17.6000,-30.8000){\makebox(0,0){${\mathcal I}^-$}}%
% STR 2 0 3 0
% 3 1130 1550 1130 1650 3 0
% $r=0$
\put(11.3000,-16.5000){\makebox(0,0)[rb]{$r=0$}}%
% STR 2 0 3 0
% 3 2510 1020 2510 1120 5 0
% ${\mathcal H}^+$
\put(25.1000,-11.2000){\makebox(0,0){${\mathcal H}^+$}}%
% STR 2 0 3 0
% 3 1280 1740 1280 1840 3 0
% origin of
\put(12.8000,-18.4000){\makebox(0,0)[rb]{origin of}}%
% STR 2 0 3 0
% 3 1430 1880 1430 1980 3 0
% coordinate
\put(14.3000,-19.8000){\makebox(0,0)[rb]{coordinate}}%
\end{picture}%

%% file: ergo01.tex
%WinTpicVersion3.08
\unitlength 0.1in
\begin{picture}( 34.0000, 21.8500)(  0.0000,-24.0000)
% ELLIPSE 2 0 1 0
% 4 1800 1400 2800 2000 2800 2000 2800 2000
% 
\special{pn 8}%
\special{sh 0.300}%
\special{ar 1800 1400 1000 600  0.0000000 6.2831853}%
% CIRCLE 2 0 0 0
% 4 1800 1400 1800 2000 1800 2000 1800 2000
% 
\special{pn 8}%
\special{sh 0.600}%
\special{ar 1800 1400 600 600  0.0000000 6.2831853}%
% ELLIPSE 2 0 3 0
% 4 1800 600 2000 690 1700 530 1920 530
% 
\special{pn 8}%
\special{ar 1800 600 200 90  5.3680846 6.2831853}%
\special{ar 1800 600 200 90  0.0000000 4.1423485}%
% VECTOR 2 0 3 0
% 6 2600 690 2600 690 2600 690 2600 690 1920 520 1870 500
% 
\special{pn 8}%
\special{pa 2600 690}%
\special{pa 2600 690}%
\special{fp}%
\special{pa 2600 690}%
\special{pa 2600 690}%
\special{fp}%
\special{pa 1920 520}%
\special{pa 1870 500}%
\special{fp}%
\special{sh 1}%
\special{pa 1870 500}%
\special{pa 1924 544}%
\special{pa 1920 520}%
\special{pa 1940 506}%
\special{pa 1870 500}%
\special{fp}%
% STR 2 0 3 0
% 3 585 1985 585 2085 5 0
% event horizon
\put(5.8500,-20.8500){\makebox(0,0){event horizon}}%
% STR 2 0 3 0
% 3 575 2175 575 2275 5 0
% $r=r_+$
\put(5.7500,-22.7500){\makebox(0,0){$r=r_+$}}%
% STR 2 0 3 0
% 3 1800 200 1800 300 5 0
% $\varphi$
\put(18.0000,-3.0000){\makebox(0,0){$\varphi$}}%
% VECTOR 2 0 3 0
% 2 3000 800 2600 1200
% 
\special{pn 8}%
\special{pa 3000 800}%
\special{pa 2600 1200}%
\special{fp}%
\special{sh 1}%
\special{pa 2600 1200}%
\special{pa 2662 1168}%
\special{pa 2638 1162}%
\special{pa 2634 1140}%
\special{pa 2600 1200}%
\special{fp}%
% STR 2 0 3 0
% 3 3000 600 3000 700 5 0
% ergoregion
\put(30.0000,-7.0000){\makebox(0,0){ergoregion}}%
% STR 2 0 3 0
% 3 3285 2185 3285 2285 5 0
% $r=M+\sqrt{M^2-a^2\cos^2 \theta}$
\put(32.8500,-22.8500){\makebox(0,0){$r=M+\sqrt{M^2-a^2\cos^2 \theta}$}}%
% STR 2 0 3 0
% 3 3285 1985 3285 2085 5 0
% ergosphere
\put(32.8500,-20.8500){\makebox(0,0){ergosphere}}%
% VECTOR 2 0 3 0
% 2 600 2000 1200 1400
% 
\special{pn 8}%
\special{pa 600 2000}%
\special{pa 1200 1400}%
\special{fp}%
\special{sh 1}%
\special{pa 1200 1400}%
\special{pa 1140 1434}%
\special{pa 1162 1438}%
\special{pa 1168 1462}%
\special{pa 1200 1400}%
\special{fp}%
% VECTOR 2 0 3 0
% 2 3400 2000 2800 1400
% 
\special{pn 8}%
\special{pa 3400 2000}%
\special{pa 2800 1400}%
\special{fp}%
\special{sh 1}%
\special{pa 2800 1400}%
\special{pa 2834 1462}%
\special{pa 2838 1438}%
\special{pa 2862 1434}%
\special{pa 2800 1400}%
\special{fp}%
% STR 2 0 3 0
% 3 1800 1300 1800 1400 5 0
% {\bf BH}
\put(18.0000,-14.0000){\makebox(0,0){{\bf BH}}}%
% VECTOR 2 0 3 0
% 2 1800 2400 1800 400
% 
\special{pn 8}%
\special{pa 1800 2400}%
\special{pa 1800 400}%
\special{fp}%
\special{sh 1}%
\special{pa 1800 400}%
\special{pa 1780 468}%
\special{pa 1800 454}%
\special{pa 1820 468}%
\special{pa 1800 400}%
\special{fp}%
\end{picture}%

%% file: ergo02.tex
%WinTpicVersion3.08
\unitlength 0.1in
\begin{picture}( 27.6000, 22.8000)(  8.0000,-23.0000)
% ELLIPSE 2 0 1 0
% 4 1800 1400 2800 2000 2800 2000 2800 2000
% 
\special{pn 8}%
\special{sh 0.300}%
\special{ar 1800 1400 1000 600  0.0000000 6.2831853}%
% CIRCLE 2 0 0 0
% 4 1800 1400 1800 2000 1800 2000 1800 2000
% 
\special{pn 8}%
\special{sh 0.600}%
\special{ar 1800 1400 600 600  0.0000000 6.2831853}%
% STR 2 0 3 0
% 3 1800 1300 1800 1400 5 0
% {\bf BH}
\put(18.0000,-14.0000){\makebox(0,0){{\bf BH}}}%
% VECTOR 2 0 3 0
% 2 3400 210 2600 1200
% 
\special{pn 8}%
\special{pa 3400 210}%
\special{pa 2600 1200}%
\special{fp}%
\special{sh 1}%
\special{pa 2600 1200}%
\special{pa 2658 1162}%
\special{pa 2634 1160}%
\special{pa 2626 1136}%
\special{pa 2600 1200}%
\special{fp}%
% VECTOR 2 0 3 0
% 2 2600 1210 3200 2210
% 
\special{pn 8}%
\special{pa 2600 1210}%
\special{pa 3200 2210}%
\special{fp}%
\special{sh 1}%
\special{pa 3200 2210}%
\special{pa 3184 2144}%
\special{pa 3174 2164}%
\special{pa 3150 2164}%
\special{pa 3200 2210}%
\special{fp}%
% VECTOR 2 0 3 0
% 2 2600 1200 2200 1300
% 
\special{pn 8}%
\special{pa 2600 1200}%
\special{pa 2200 1300}%
\special{fp}%
\special{sh 1}%
\special{pa 2200 1300}%
\special{pa 2270 1304}%
\special{pa 2252 1288}%
\special{pa 2260 1264}%
\special{pa 2200 1300}%
\special{fp}%
% STR 2 0 3 0
% 3 2200 1350 2200 1450 5 0
% $E_1$
\put(22.0000,-14.5000){\makebox(0,0){$E_1$}}%
% CIRCLE 2 0 3 0
% 4 3230 2260 3230 2300 3230 2310 3230 2310
% 
\special{pn 8}%
\special{ar 3230 2260 40 40  0.0000000 6.2831853}%
% CIRCLE 2 0 2 0
% 4 2150 1300 2150 1340 2150 1350 2150 1350
% 
\special{pn 8}%
\special{sh 0}%
\special{ar 2150 1300 40 40  0.0000000 6.2831853}%
% CIRCLE 2 0 3 0
% 4 3480 100 3480 180 3480 180 3480 180
% 
\special{pn 8}%
\special{ar 3480 100 80 80  0.0000000 6.2831853}%
% STR 2 0 3 0
% 3 3550 200 3550 300 5 0
% $E_0$
\put(35.5000,-3.0000){\makebox(0,0){$E_0$}}%
% STR 2 0 3 0
% 3 3400 2150 3400 2250 5 0
% $E_2$
\put(34.0000,-22.5000){\makebox(0,0){$E_2$}}%
\end{picture}%

%% file: gauss01.tex
%WinTpicVersion3.08
\unitlength 0.1in
\begin{picture}( 36.1500, 22.0000)(  4.8500,-24.0000)
% LINE 2 0 3 0
% 2 3400 1200 3400 1600
% 
\special{pn 8}%
\special{pa 3400 1200}%
\special{pa 3400 1600}%
\special{fp}%
% LINE 2 0 3 0
% 2 3400 1200 4000 1000
% 
\special{pn 8}%
\special{pa 3400 1200}%
\special{pa 4000 1000}%
\special{fp}%
% LINE 2 0 3 0
% 2 4000 1000 4000 1400
% 
\special{pn 8}%
\special{pa 4000 1000}%
\special{pa 4000 1400}%
\special{fp}%
% LINE 2 0 3 0
% 2 3400 1600 4000 1400
% 
\special{pn 8}%
\special{pa 3400 1600}%
\special{pa 4000 1400}%
\special{fp}%
% LINE 2 0 3 0
% 2 1000 1200 1600 1000
% 
\special{pn 8}%
\special{pa 1000 1200}%
\special{pa 1600 1000}%
\special{fp}%
% SPLINE 2 0 3 0
% 8 1000 1200 1400 1180 1800 1210 2200 1220 2600 1170 3000 1240 3400 1200 3400 1200
% 
\special{pn 8}%
\special{pa 1000 1200}%
\special{pa 1032 1198}%
\special{pa 1064 1196}%
\special{pa 1096 1192}%
\special{pa 1128 1190}%
\special{pa 1160 1188}%
\special{pa 1192 1186}%
\special{pa 1224 1184}%
\special{pa 1256 1182}%
\special{pa 1288 1182}%
\special{pa 1320 1180}%
\special{pa 1352 1180}%
\special{pa 1384 1180}%
\special{pa 1416 1180}%
\special{pa 1448 1182}%
\special{pa 1480 1184}%
\special{pa 1512 1184}%
\special{pa 1544 1186}%
\special{pa 1576 1190}%
\special{pa 1608 1192}%
\special{pa 1640 1194}%
\special{pa 1672 1198}%
\special{pa 1704 1200}%
\special{pa 1734 1204}%
\special{pa 1766 1208}%
\special{pa 1798 1210}%
\special{pa 1830 1214}%
\special{pa 1862 1216}%
\special{pa 1894 1220}%
\special{pa 1926 1222}%
\special{pa 1958 1224}%
\special{pa 1990 1226}%
\special{pa 2022 1226}%
\special{pa 2054 1228}%
\special{pa 2086 1228}%
\special{pa 2118 1226}%
\special{pa 2150 1224}%
\special{pa 2182 1222}%
\special{pa 2214 1218}%
\special{pa 2246 1214}%
\special{pa 2278 1210}%
\special{pa 2310 1204}%
\special{pa 2342 1198}%
\special{pa 2374 1192}%
\special{pa 2406 1188}%
\special{pa 2438 1182}%
\special{pa 2470 1178}%
\special{pa 2500 1174}%
\special{pa 2532 1172}%
\special{pa 2564 1170}%
\special{pa 2596 1170}%
\special{pa 2628 1172}%
\special{pa 2658 1176}%
\special{pa 2690 1180}%
\special{pa 2722 1186}%
\special{pa 2754 1192}%
\special{pa 2784 1200}%
\special{pa 2816 1206}%
\special{pa 2848 1214}%
\special{pa 2880 1220}%
\special{pa 2910 1228}%
\special{pa 2942 1232}%
\special{pa 2974 1238}%
\special{pa 3006 1240}%
\special{pa 3038 1242}%
\special{pa 3068 1244}%
\special{pa 3100 1244}%
\special{pa 3132 1242}%
\special{pa 3164 1238}%
\special{pa 3196 1236}%
\special{pa 3228 1232}%
\special{pa 3260 1226}%
\special{pa 3292 1222}%
\special{pa 3324 1216}%
\special{pa 3356 1210}%
\special{pa 3388 1204}%
\special{pa 3400 1200}%
\special{sp}%
% SPLINE 2 0 3 0
% 8 1000 1600 1400 1580 1800 1610 2200 1620 2600 1570 3000 1640 3400 1600 3400 1600
% 
\special{pn 8}%
\special{pa 1000 1600}%
\special{pa 1032 1598}%
\special{pa 1064 1596}%
\special{pa 1096 1592}%
\special{pa 1128 1590}%
\special{pa 1160 1588}%
\special{pa 1192 1586}%
\special{pa 1224 1584}%
\special{pa 1256 1582}%
\special{pa 1288 1582}%
\special{pa 1320 1580}%
\special{pa 1352 1580}%
\special{pa 1384 1580}%
\special{pa 1416 1580}%
\special{pa 1448 1582}%
\special{pa 1480 1584}%
\special{pa 1512 1584}%
\special{pa 1544 1586}%
\special{pa 1576 1590}%
\special{pa 1608 1592}%
\special{pa 1640 1594}%
\special{pa 1672 1598}%
\special{pa 1704 1600}%
\special{pa 1734 1604}%
\special{pa 1766 1608}%
\special{pa 1798 1610}%
\special{pa 1830 1614}%
\special{pa 1862 1616}%
\special{pa 1894 1620}%
\special{pa 1926 1622}%
\special{pa 1958 1624}%
\special{pa 1990 1626}%
\special{pa 2022 1626}%
\special{pa 2054 1628}%
\special{pa 2086 1628}%
\special{pa 2118 1626}%
\special{pa 2150 1624}%
\special{pa 2182 1622}%
\special{pa 2214 1618}%
\special{pa 2246 1614}%
\special{pa 2278 1610}%
\special{pa 2310 1604}%
\special{pa 2342 1598}%
\special{pa 2374 1592}%
\special{pa 2406 1588}%
\special{pa 2438 1582}%
\special{pa 2470 1578}%
\special{pa 2500 1574}%
\special{pa 2532 1572}%
\special{pa 2564 1570}%
\special{pa 2596 1570}%
\special{pa 2628 1572}%
\special{pa 2658 1576}%
\special{pa 2690 1580}%
\special{pa 2722 1586}%
\special{pa 2754 1592}%
\special{pa 2784 1600}%
\special{pa 2816 1606}%
\special{pa 2848 1614}%
\special{pa 2880 1620}%
\special{pa 2910 1628}%
\special{pa 2942 1632}%
\special{pa 2974 1638}%
\special{pa 3006 1640}%
\special{pa 3038 1642}%
\special{pa 3068 1644}%
\special{pa 3100 1644}%
\special{pa 3132 1642}%
\special{pa 3164 1638}%
\special{pa 3196 1636}%
\special{pa 3228 1632}%
\special{pa 3260 1626}%
\special{pa 3292 1622}%
\special{pa 3324 1616}%
\special{pa 3356 1610}%
\special{pa 3388 1604}%
\special{pa 3400 1600}%
\special{sp}%
% SPLINE 2 0 3 0
% 8 1600 1000 2000 980 2400 1010 2800 1020 3200 970 3600 1040 4000 1000 4000 1000
% 
\special{pn 8}%
\special{pa 1600 1000}%
\special{pa 1632 998}%
\special{pa 1664 996}%
\special{pa 1696 992}%
\special{pa 1728 990}%
\special{pa 1760 988}%
\special{pa 1792 986}%
\special{pa 1824 984}%
\special{pa 1856 982}%
\special{pa 1888 982}%
\special{pa 1920 980}%
\special{pa 1952 980}%
\special{pa 1984 980}%
\special{pa 2016 980}%
\special{pa 2048 982}%
\special{pa 2080 984}%
\special{pa 2112 984}%
\special{pa 2144 986}%
\special{pa 2176 990}%
\special{pa 2208 992}%
\special{pa 2240 994}%
\special{pa 2272 998}%
\special{pa 2304 1000}%
\special{pa 2334 1004}%
\special{pa 2366 1008}%
\special{pa 2398 1010}%
\special{pa 2430 1014}%
\special{pa 2462 1016}%
\special{pa 2494 1020}%
\special{pa 2526 1022}%
\special{pa 2558 1024}%
\special{pa 2590 1026}%
\special{pa 2622 1026}%
\special{pa 2654 1028}%
\special{pa 2686 1028}%
\special{pa 2718 1026}%
\special{pa 2750 1024}%
\special{pa 2782 1022}%
\special{pa 2814 1018}%
\special{pa 2846 1014}%
\special{pa 2878 1010}%
\special{pa 2910 1004}%
\special{pa 2942 998}%
\special{pa 2974 992}%
\special{pa 3006 988}%
\special{pa 3038 982}%
\special{pa 3070 978}%
\special{pa 3100 974}%
\special{pa 3132 972}%
\special{pa 3164 970}%
\special{pa 3196 970}%
\special{pa 3228 972}%
\special{pa 3258 976}%
\special{pa 3290 980}%
\special{pa 3322 986}%
\special{pa 3354 992}%
\special{pa 3384 1000}%
\special{pa 3416 1006}%
\special{pa 3448 1014}%
\special{pa 3480 1020}%
\special{pa 3510 1028}%
\special{pa 3542 1032}%
\special{pa 3574 1038}%
\special{pa 3606 1040}%
\special{pa 3638 1042}%
\special{pa 3668 1044}%
\special{pa 3700 1044}%
\special{pa 3732 1042}%
\special{pa 3764 1038}%
\special{pa 3796 1036}%
\special{pa 3828 1032}%
\special{pa 3860 1026}%
\special{pa 3892 1022}%
\special{pa 3924 1016}%
\special{pa 3956 1010}%
\special{pa 3988 1004}%
\special{pa 4000 1000}%
\special{sp}%
% LINE 2 2 3 0
% 2 1000 400 1000 2400
% 
\special{pn 8}%
\special{pa 1000 400}%
\special{pa 1000 2400}%
\special{dt 0.045}%
% LINE 2 0 3 0
% 2 1000 1200 1000 1600
% 
\special{pn 8}%
\special{pa 1000 1200}%
\special{pa 1000 1600}%
\special{fp}%
% LINE 2 2 3 0
% 2 1600 210 1600 2210
% 
\special{pn 8}%
\special{pa 1600 210}%
\special{pa 1600 2210}%
\special{dt 0.045}%
% LINE 2 2 3 0
% 2 3400 400 3400 2400
% 
\special{pn 8}%
\special{pa 3400 400}%
\special{pa 3400 2400}%
\special{dt 0.045}%
% LINE 2 2 3 0
% 2 4000 200 4000 2200
% 
\special{pn 8}%
\special{pa 4000 200}%
\special{pa 4000 2200}%
\special{dt 0.045}%
% VECTOR 1 0 3 0
% 2 2600 1100 2600 700
% 
\special{pn 13}%
\special{pa 2600 1100}%
\special{pa 2600 700}%
\special{fp}%
\special{sh 1}%
\special{pa 2600 700}%
\special{pa 2580 768}%
\special{pa 2600 754}%
\special{pa 2620 768}%
\special{pa 2600 700}%
\special{fp}%
% VECTOR 1 0 3 0
% 2 3700 1310 4100 1310
% 
\special{pn 13}%
\special{pa 3700 1310}%
\special{pa 4100 1310}%
\special{fp}%
\special{sh 1}%
\special{pa 4100 1310}%
\special{pa 4034 1290}%
\special{pa 4048 1310}%
\special{pa 4034 1330}%
\special{pa 4100 1310}%
\special{fp}%
% LINE 1 2 3 0
% 2 2600 1550 2600 1480
% 
\special{pn 13}%
\special{pa 2600 1550}%
\special{pa 2600 1480}%
\special{dt 0.045}%
% VECTOR 1 0 3 0
% 2 2600 1570 2600 1900
% 
\special{pn 13}%
\special{pa 2600 1570}%
\special{pa 2600 1900}%
\special{fp}%
\special{sh 1}%
\special{pa 2600 1900}%
\special{pa 2620 1834}%
\special{pa 2600 1848}%
\special{pa 2580 1834}%
\special{pa 2600 1900}%
\special{fp}%
% LINE 1 2 3 0
% 2 1300 1300 1000 1300
% 
\special{pn 13}%
\special{pa 1300 1300}%
\special{pa 1000 1300}%
\special{dt 0.045}%
% VECTOR 1 0 3 0
% 2 1000 1300 900 1300
% 
\special{pn 13}%
\special{pa 1000 1300}%
\special{pa 900 1300}%
\special{fp}%
\special{sh 1}%
\special{pa 900 1300}%
\special{pa 968 1320}%
\special{pa 954 1300}%
\special{pa 968 1280}%
\special{pa 900 1300}%
\special{fp}%
% LINE 2 2 3 0
% 2 1000 1600 1600 1400
% 
\special{pn 8}%
\special{pa 1000 1600}%
\special{pa 1600 1400}%
\special{dt 0.045}%
% SPLINE 2 2 3 0
% 8 1600 1400 2000 1380 2400 1410 2800 1420 3200 1370 3600 1440 4000 1400 4000 1400
% 
\special{pn 8}%
\special{pa 1600 1400}%
\special{pa 1632 1398}%
\special{pa 1664 1396}%
\special{pa 1696 1392}%
\special{pa 1728 1390}%
\special{pa 1760 1388}%
\special{pa 1792 1386}%
\special{pa 1824 1384}%
\special{pa 1856 1382}%
\special{pa 1888 1382}%
\special{pa 1920 1380}%
\special{pa 1952 1380}%
\special{pa 1984 1380}%
\special{pa 2016 1380}%
\special{pa 2048 1382}%
\special{pa 2080 1384}%
\special{pa 2112 1384}%
\special{pa 2144 1386}%
\special{pa 2176 1390}%
\special{pa 2208 1392}%
\special{pa 2240 1394}%
\special{pa 2272 1398}%
\special{pa 2304 1400}%
\special{pa 2334 1404}%
\special{pa 2366 1408}%
\special{pa 2398 1410}%
\special{pa 2430 1414}%
\special{pa 2462 1416}%
\special{pa 2494 1420}%
\special{pa 2526 1422}%
\special{pa 2558 1424}%
\special{pa 2590 1426}%
\special{pa 2622 1426}%
\special{pa 2654 1428}%
\special{pa 2686 1428}%
\special{pa 2718 1426}%
\special{pa 2750 1424}%
\special{pa 2782 1422}%
\special{pa 2814 1418}%
\special{pa 2846 1414}%
\special{pa 2878 1410}%
\special{pa 2910 1404}%
\special{pa 2942 1398}%
\special{pa 2974 1392}%
\special{pa 3006 1388}%
\special{pa 3038 1382}%
\special{pa 3070 1378}%
\special{pa 3100 1374}%
\special{pa 3132 1372}%
\special{pa 3164 1370}%
\special{pa 3196 1370}%
\special{pa 3228 1372}%
\special{pa 3258 1376}%
\special{pa 3290 1380}%
\special{pa 3322 1386}%
\special{pa 3354 1392}%
\special{pa 3384 1400}%
\special{pa 3416 1406}%
\special{pa 3448 1414}%
\special{pa 3480 1420}%
\special{pa 3510 1428}%
\special{pa 3542 1432}%
\special{pa 3574 1438}%
\special{pa 3606 1440}%
\special{pa 3638 1442}%
\special{pa 3668 1444}%
\special{pa 3700 1444}%
\special{pa 3732 1442}%
\special{pa 3764 1438}%
\special{pa 3796 1436}%
\special{pa 3828 1432}%
\special{pa 3860 1426}%
\special{pa 3892 1422}%
\special{pa 3924 1416}%
\special{pa 3956 1410}%
\special{pa 3988 1404}%
\special{pa 4000 1400}%
\special{sp -0.045}%
% STR 2 0 3 0
% 3 2600 500 2600 600 5 0
% $n^\mu$
\put(26.0000,-6.0000){\makebox(0,0){$n^\mu$}}%
% STR 2 0 3 0
% 3 4200 1200 4200 1300 5 0
% $n^\mu$
\put(42.0000,-13.0000){\makebox(0,0){$n^\mu$}}%
% STR 2 0 3 0
% 3 2600 1900 2600 2000 5 0
% $n^\mu$
\put(26.0000,-20.0000){\makebox(0,0){$n^\mu$}}%
% STR 2 0 3 0
% 3 800 1200 800 1300 5 0
% $n^\mu$
\put(8.0000,-13.0000){\makebox(0,0){$n^\mu$}}%
% STR 2 0 3 0
% 3 2000 750 2000 850 5 0
% $\Sigma_2(t+\delta t)$
\put(20.0000,-8.5000){\makebox(0,0){$\Sigma_2(t+\delta t)$}}%
% STR 2 0 3 0
% 3 2000 1650 2000 1750 5 0
% $\Sigma_1(t)$
\put(20.0000,-17.5000){\makebox(0,0){$\Sigma_1(t)$}}%
% STR 2 0 3 0
% 3 3060 1350 3060 1450 5 0
% $\delta t$
\put(30.6000,-14.5000){\makebox(0,0){$\delta t$}}%
% VECTOR 2 0 3 0
% 2 3060 1350 3060 1250
% 
\special{pn 8}%
\special{pa 3060 1350}%
\special{pa 3060 1250}%
\special{fp}%
\special{sh 1}%
\special{pa 3060 1250}%
\special{pa 3040 1318}%
\special{pa 3060 1304}%
\special{pa 3080 1318}%
\special{pa 3060 1250}%
\special{fp}%
% VECTOR 2 0 3 0
% 2 3060 1550 3060 1650
% 
\special{pn 8}%
\special{pa 3060 1550}%
\special{pa 3060 1650}%
\special{fp}%
\special{sh 1}%
\special{pa 3060 1650}%
\special{pa 3080 1584}%
\special{pa 3060 1598}%
\special{pa 3040 1584}%
\special{pa 3060 1650}%
\special{fp}%
% STR 2 0 3 0
% 3 2500 1300 2500 1400 5 0
% ${\cal K}$
\put(25.0000,-14.0000){\makebox(0,0){${\cal K}$}}%
% STR 2 0 3 0
% 3 3700 2300 3700 2400 5 0
% ${\mathscr S}(\infty)$
\put(37.0000,-24.0000){\makebox(0,0){${\mathscr S}(\infty)$}}%
% STR 2 0 3 0
% 3 1300 2300 1300 2400 5 0
% ${\cal H}(r_+)$
\put(13.0000,-24.0000){\makebox(0,0){${\cal H}(r_+)$}}%
\end{picture}%

%% file: gauss02.tex
%WinTpicVersion3.08
\unitlength 0.1in
\begin{picture}( 22.4500, 18.1500)( 11.8500,-20.0000)
% LINE 2 0 3 0
% 2 1400 1600 2600 400
% 
\special{pn 8}%
\special{pa 1400 1600}%
\special{pa 2600 400}%
\special{fp}%
% LINE 2 0 3 0
% 2 2600 400 3400 1200
% 
\special{pn 8}%
\special{pa 2600 400}%
\special{pa 3400 1200}%
\special{fp}%
% STR 2 0 3 0
% 3 2600 170 2600 270 5 0
% ${\cal I}^+$
\put(26.0000,-2.7000){\makebox(0,0){${\cal I}^+$}}%
% STR 2 0 3 0
% 3 3540 1100 3540 1200 5 0
% ${\cal I}^0$
\put(35.4000,-12.0000){\makebox(0,0){${\cal I}^0$}}%
% STR 2 0 3 0
% 3 2200 500 2200 600 5 0
% ${\cal H}^+$
\put(22.0000,-6.0000){\makebox(0,0){${\cal H}^+$}}%
% VECTOR 1 0 3 0
% 2 1800 1200 1600 1400
% 
\special{pn 13}%
\special{pa 1800 1200}%
\special{pa 1600 1400}%
\special{fp}%
\special{sh 1}%
\special{pa 1600 1400}%
\special{pa 1662 1368}%
\special{pa 1638 1362}%
\special{pa 1634 1340}%
\special{pa 1600 1400}%
\special{fp}%
% LINE 2 0 3 0
% 2 2600 2000 3400 1200
% 
\special{pn 8}%
\special{pa 2600 2000}%
\special{pa 3400 1200}%
\special{fp}%
% POLYLINE 2 0 3 0
% 7 1910 1110 2200 1090 2520 1140 2770 1110 3400 1200 3400 1200 3400 1200
% 
\special{pn 8}%
\special{pa 1910 1110}%
\special{pa 2200 1090}%
\special{pa 2520 1140}%
\special{pa 2770 1110}%
\special{pa 3400 1200}%
\special{pa 3400 1200}%
\special{pa 3400 1200}%
\special{fp}%
% SPLINE 2 0 3 0
% 7 1560 1450 1890 1390 2230 1430 2590 1330 2800 1360 3400 1200 3400 1200
% 
\special{pn 8}%
\special{pa 1560 1450}%
\special{pa 1592 1442}%
\special{pa 1624 1432}%
\special{pa 1654 1424}%
\special{pa 1686 1414}%
\special{pa 1718 1408}%
\special{pa 1748 1402}%
\special{pa 1780 1396}%
\special{pa 1812 1392}%
\special{pa 1844 1390}%
\special{pa 1876 1390}%
\special{pa 1908 1392}%
\special{pa 1938 1396}%
\special{pa 1970 1400}%
\special{pa 2002 1406}%
\special{pa 2034 1412}%
\special{pa 2066 1418}%
\special{pa 2098 1424}%
\special{pa 2130 1428}%
\special{pa 2162 1432}%
\special{pa 2194 1432}%
\special{pa 2224 1432}%
\special{pa 2256 1426}%
\special{pa 2286 1420}%
\special{pa 2318 1410}%
\special{pa 2348 1400}%
\special{pa 2380 1388}%
\special{pa 2410 1378}%
\special{pa 2440 1366}%
\special{pa 2472 1354}%
\special{pa 2502 1346}%
\special{pa 2532 1338}%
\special{pa 2564 1332}%
\special{pa 2596 1330}%
\special{pa 2626 1332}%
\special{pa 2658 1334}%
\special{pa 2690 1340}%
\special{pa 2722 1346}%
\special{pa 2754 1352}%
\special{pa 2786 1358}%
\special{pa 2816 1362}%
\special{pa 2848 1364}%
\special{pa 2880 1366}%
\special{pa 2912 1364}%
\special{pa 2942 1362}%
\special{pa 2974 1356}%
\special{pa 3004 1352}%
\special{pa 3036 1344}%
\special{pa 3066 1336}%
\special{pa 3098 1328}%
\special{pa 3128 1318}%
\special{pa 3160 1306}%
\special{pa 3190 1294}%
\special{pa 3220 1282}%
\special{pa 3252 1270}%
\special{pa 3282 1256}%
\special{pa 3312 1242}%
\special{pa 3342 1228}%
\special{pa 3374 1214}%
\special{pa 3400 1200}%
\special{sp}%
% VECTOR 1 0 3 0
% 2 2430 1120 2490 920
% 
\special{pn 13}%
\special{pa 2430 1120}%
\special{pa 2490 920}%
\special{fp}%
\special{sh 1}%
\special{pa 2490 920}%
\special{pa 2452 978}%
\special{pa 2476 972}%
\special{pa 2490 990}%
\special{pa 2490 920}%
\special{fp}%
% VECTOR 1 0 3 0
% 2 2320 1410 2410 1610
% 
\special{pn 13}%
\special{pa 2320 1410}%
\special{pa 2410 1610}%
\special{fp}%
\special{sh 1}%
\special{pa 2410 1610}%
\special{pa 2402 1542}%
\special{pa 2388 1562}%
\special{pa 2364 1558}%
\special{pa 2410 1610}%
\special{fp}%
% STR 2 0 3 0
% 3 2550 700 2550 800 5 0
% $n^\mu$
\put(25.5000,-8.0000){\makebox(0,0){$n^\mu$}}%
% STR 2 0 3 0
% 3 2500 1600 2500 1700 5 0
% $n^\mu$
\put(25.0000,-17.0000){\makebox(0,0){$n^\mu$}}%
% STR 2 0 3 0
% 3 1500 1200 1500 1300 5 0
% $n^\mu$
\put(15.0000,-13.0000){\makebox(0,0){$n^\mu$}}%
% STR 2 0 3 0
% 3 2200 900 2200 1000 5 0
% $\Sigma_2$
\put(22.0000,-10.0000){\makebox(0,0){$\Sigma_2$}}%
% STR 2 0 3 0
% 3 2000 1400 2000 1500 5 0
% $\Sigma_1$
\put(20.0000,-15.0000){\makebox(0,0){$\Sigma_1$}}%
% STR 2 0 3 0
% 3 2200 1150 2200 1250 5 0
% ${\cal K}$
\put(22.0000,-12.5000){\makebox(0,0){${\cal K}$}}%
% CIRCLE 2 0 2 0
% 4 3400 1200 3400 1230 3400 1230 3400 1230
% 
\special{pn 8}%
\special{sh 0}%
\special{ar 3400 1200 30 30  0.0000000 6.2831853}%
% CIRCLE 2 0 2 0
% 4 2600 400 2600 430 2600 430 2600 430
% 
\special{pn 8}%
\special{sh 0}%
\special{ar 2600 400 30 30  0.0000000 6.2831853}%
\end{picture}%

%% file: area01.tex
%WinTpicVersion3.08
\unitlength 0.1in
\begin{picture}( 49.6500, 13.1500)(  4.3500,-15.1500)
% CIRCLE 2 0 1 0
% 4 1200 800 1200 1200 1200 1200 1200 1200
% 
\special{pn 8}%
\special{sh 0.300}%
\special{ar 1200 800 400 400  0.0000000 6.2831853}%
% CIRCLE 2 0 1 0
% 4 2800 800 2800 1200 2800 1200 2800 1200
% 
\special{pn 8}%
\special{sh 0.300}%
\special{ar 2800 800 400 400  0.0000000 6.2831853}%
% CIRCLE 2 0 1 0
% 4 4800 800 4800 1400 4800 1400 4800 1400
% 
\special{pn 8}%
\special{sh 0.300}%
\special{ar 4800 800 600 600  0.0000000 6.2831853}%
% VECTOR 2 0 3 0
% 2 3500 800 3900 800
% 
\special{pn 8}%
\special{pa 3500 800}%
\special{pa 3900 800}%
\special{fp}%
\special{sh 1}%
\special{pa 3900 800}%
\special{pa 3834 780}%
\special{pa 3848 800}%
\special{pa 3834 820}%
\special{pa 3900 800}%
\special{fp}%
% STR 2 0 3 0
% 3 2000 700 2000 800 5 0
% $+$
\put(20.0000,-8.0000){\makebox(0,0){$+$}}%
% STR 2 0 3 0
% 3 1200 1500 1200 1600 5 0
% $A_1=16\pi M_1^2$
\put(12.0000,-16.0000){\makebox(0,0){$A_1=16\pi M_1^2$}}%
% STR 2 0 3 0
% 3 2800 1500 2800 1600 5 0
% $A_2=16\pi M_2^2$
\put(28.0000,-16.0000){\makebox(0,0){$A_2=16\pi M_2^2$}}%
% STR 2 0 3 0
% 3 4800 1500 4800 1600 5 0
% $A=16\pi (M_1+M_2)^2$
\put(48.0000,-16.0000){\makebox(0,0){$A=16\pi (M_1+M_2)^2$}}%
% STR 2 0 3 0
% 3 1200 700 1200 800 5 0
% $M_1$
\put(12.0000,-8.0000){\makebox(0,0){$M_1$}}%
% STR 2 0 3 0
% 3 2800 700 2800 800 5 0
% $M_2$
\put(28.0000,-8.0000){\makebox(0,0){$M_2$}}%
% STR 2 0 3 0
% 3 4800 700 4800 800 5 0
% $M$
\put(48.0000,-8.0000){\makebox(0,0){$M$}}%
\end{picture}%

%% file: beke01.tex
%WinTpicVersion3.08
\unitlength 0.1in
\begin{picture}( 36.4000, 15.5800)(  5.9800,-19.5600)
% CIRCLE 2 0 3 0
% 4 773 573 798 746 798 746 798 746
% 
\special{pn 8}%
\special{ar 774 574 176 176  0.0000000 6.2831853}%
% ELLIPSE 2 0 1 0
% 4 1373 1186 1956 1774 1373 1774 1373 1774
% 
\special{pn 8}%
\special{sh 0.300}%
\special{ar 1374 1186 584 588  0.0000000 6.2831853}%
% ELLIPSE 2 0 1 0
% 4 3378 1213 4134 1956 3416 1954 3416 1954
% 
\special{pn 8}%
\special{sh 0.300}%
\special{ar 3378 1214 756 744  0.0000000 6.2831853}%
% ELLIPSE 2 1 1 0
% 4 3370 1213 3953 1785 3370 1785 3370 1785
% 
\special{pn 8}%
\special{sh 0.300}%
\special{ia 3370 1214 584 572  0.0000000 6.2831853}%
\special{ar 3370 1214 584 572  0.0000000 0.1038961}%
\special{ar 3370 1214 584 572  0.1662338 0.2701299}%
\special{ar 3370 1214 584 572  0.3324675 0.4363636}%
\special{ar 3370 1214 584 572  0.4987013 0.6025974}%
\special{ar 3370 1214 584 572  0.6649351 0.7688312}%
\special{ar 3370 1214 584 572  0.8311688 0.9350649}%
\special{ar 3370 1214 584 572  0.9974026 1.1012987}%
\special{ar 3370 1214 584 572  1.1636364 1.2675325}%
\special{ar 3370 1214 584 572  1.3298701 1.4337662}%
\special{ar 3370 1214 584 572  1.4961039 1.6000000}%
\special{ar 3370 1214 584 572  1.6623377 1.7662338}%
\special{ar 3370 1214 584 572  1.8285714 1.9324675}%
\special{ar 3370 1214 584 572  1.9948052 2.0987013}%
\special{ar 3370 1214 584 572  2.1610390 2.2649351}%
\special{ar 3370 1214 584 572  2.3272727 2.4311688}%
\special{ar 3370 1214 584 572  2.4935065 2.5974026}%
\special{ar 3370 1214 584 572  2.6597403 2.7636364}%
\special{ar 3370 1214 584 572  2.8259740 2.9298701}%
\special{ar 3370 1214 584 572  2.9922078 3.0961039}%
\special{ar 3370 1214 584 572  3.1584416 3.2623377}%
\special{ar 3370 1214 584 572  3.3246753 3.4285714}%
\special{ar 3370 1214 584 572  3.4909091 3.5948052}%
\special{ar 3370 1214 584 572  3.6571429 3.7610390}%
\special{ar 3370 1214 584 572  3.8233766 3.9272727}%
\special{ar 3370 1214 584 572  3.9896104 4.0935065}%
\special{ar 3370 1214 584 572  4.1558442 4.2597403}%
\special{ar 3370 1214 584 572  4.3220779 4.4259740}%
\special{ar 3370 1214 584 572  4.4883117 4.5922078}%
\special{ar 3370 1214 584 572  4.6545455 4.7584416}%
\special{ar 3370 1214 584 572  4.8207792 4.9246753}%
\special{ar 3370 1214 584 572  4.9870130 5.0909091}%
\special{ar 3370 1214 584 572  5.1532468 5.2571429}%
\special{ar 3370 1214 584 572  5.3194805 5.4233766}%
\special{ar 3370 1214 584 572  5.4857143 5.5896104}%
\special{ar 3370 1214 584 572  5.6519481 5.7558442}%
\special{ar 3370 1214 584 572  5.8181818 5.9220779}%
\special{ar 3370 1214 584 572  5.9844156 6.0883117}%
\special{ar 3370 1214 584 572  6.1506494 6.2545455}%
% VECTOR 2 0 3 0
% 2 2138 1246 2475 1246
% 
\special{pn 8}%
\special{pa 2138 1246}%
\special{pa 2476 1246}%
\special{fp}%
\special{sh 1}%
\special{pa 2476 1246}%
\special{pa 2408 1226}%
\special{pa 2422 1246}%
\special{pa 2408 1266}%
\special{pa 2476 1246}%
\special{fp}%
% STR 2 0 3 0
% 3 2742 1160 2742 1246 0 0
% 
\put(27.4200,-12.4600){\makebox(0,0)[lb]{}}%
% STR 2 0 3 0
% 3 2501 487 2501 573 5 0
% $\ln 2$
\put(25.0100,-5.7300){\makebox(0,0){$\ln 2$}}%
% STR 2 0 3 0
% 3 1400 1114 1400 1200 5 0
% {\bf BH}
\put(14.0000,-12.0000){\makebox(0,0){{\bf BH}}}%
% STR 2 0 3 0
% 3 3400 1114 3400 1200 5 0
% {\bf BH}
\put(34.0000,-12.0000){\makebox(0,0){{\bf BH}}}%
% VECTOR 2 0 3 0
% 2 4238 746 4005 970
% 
\special{pn 8}%
\special{pa 4238 746}%
\special{pa 4006 970}%
\special{fp}%
\special{sh 1}%
\special{pa 4006 970}%
\special{pa 4068 938}%
\special{pa 4044 934}%
\special{pa 4040 910}%
\special{pa 4006 970}%
\special{fp}%
% STR 2 0 3 0
% 3 4221 633 4221 720 2 0
% $\triangle S_{BH}$
\put(42.2100,-7.2000){\makebox(0,0)[lb]{$\triangle S_{BH}$}}%
% CIRCLE 2 2 3 0
% 4 2501 573 2501 746 2501 746 2501 746
% 
\special{pn 8}%
\special{ar 2502 574 174 174  0.0000000 0.0693642}%
\special{ar 2502 574 174 174  0.2774566 0.3468208}%
\special{ar 2502 574 174 174  0.5549133 0.6242775}%
\special{ar 2502 574 174 174  0.8323699 0.9017341}%
\special{ar 2502 574 174 174  1.1098266 1.1791908}%
\special{ar 2502 574 174 174  1.3872832 1.4566474}%
\special{ar 2502 574 174 174  1.6647399 1.7341040}%
\special{ar 2502 574 174 174  1.9421965 2.0115607}%
\special{ar 2502 574 174 174  2.2196532 2.2890173}%
\special{ar 2502 574 174 174  2.4971098 2.5664740}%
\special{ar 2502 574 174 174  2.7745665 2.8439306}%
\special{ar 2502 574 174 174  3.0520231 3.1213873}%
\special{ar 2502 574 174 174  3.3294798 3.3988439}%
\special{ar 2502 574 174 174  3.6069364 3.6763006}%
\special{ar 2502 574 174 174  3.8843931 3.9537572}%
\special{ar 2502 574 174 174  4.1618497 4.2312139}%
\special{ar 2502 574 174 174  4.4393064 4.5086705}%
\special{ar 2502 574 174 174  4.7167630 4.7861272}%
\special{ar 2502 574 174 174  4.9942197 5.0635838}%
\special{ar 2502 574 174 174  5.2716763 5.3410405}%
\special{ar 2502 574 174 174  5.5491329 5.6184971}%
\special{ar 2502 574 174 174  5.8265896 5.8959538}%
\special{ar 2502 574 174 174  6.1040462 6.1734104}%
\end{picture}%

%% file: haw01.tex
%WinTpicVersion3.08
\unitlength 0.1in
\begin{picture}( 52.8000, 12.9600)(  0.0000,-14.7600)
% CIRCLE 2 0 1 0
% 4 640 836 1280 852 1280 852 1280 852
% 
\special{pn 8}%
\special{sh 0.300}%
\special{ar 640 836 640 640  0.0000000 6.2831853}%
% STR 2 0 3 0
% 3 640 756 640 836 5 0
% {\bf BH}
\put(6.4000,-8.3600){\makebox(0,0){{\bf BH}}}%
% CIRCLE 2 0 3 0
% 4 1600 676 1600 836 1600 836 1600 836
% 
\special{pn 8}%
\special{ar 1600 676 160 160  0.0000000 6.2831853}%
% CIRCLE 2 0 3 0
% 4 1600 996 1600 1156 1600 1156 1600 1156
% 
\special{pn 8}%
\special{ar 1600 996 160 160  0.0000000 6.2831853}%
% STR 2 0 3 0
% 3 1600 916 1600 996 5 0
% $-E$
\put(16.0000,-9.9600){\makebox(0,0){$-E$}}%
% STR 2 0 3 0
% 3 1600 596 1600 676 5 0
% $E$
\put(16.0000,-6.7600){\makebox(0,0){$E$}}%
% VECTOR 2 0 3 0
% 2 1760 676 2080 676
% 
\special{pn 8}%
\special{pa 1760 676}%
\special{pa 2080 676}%
\special{fp}%
\special{sh 1}%
\special{pa 2080 676}%
\special{pa 2014 656}%
\special{pa 2028 676}%
\special{pa 2014 696}%
\special{pa 2080 676}%
\special{fp}%
% VECTOR 2 0 3 0
% 2 1440 1004 1120 1004
% 
\special{pn 8}%
\special{pa 1440 1004}%
\special{pa 1120 1004}%
\special{fp}%
\special{sh 1}%
\special{pa 1120 1004}%
\special{pa 1188 1024}%
\special{pa 1174 1004}%
\special{pa 1188 984}%
\special{pa 1120 1004}%
\special{fp}%
% VECTOR 0 0 3 0
% 2 2240 836 2720 836
% 
\special{pn 20}%
\special{pa 2240 836}%
\special{pa 2720 836}%
\special{fp}%
\special{sh 1}%
\special{pa 2720 836}%
\special{pa 2654 816}%
\special{pa 2668 836}%
\special{pa 2654 856}%
\special{pa 2720 836}%
\special{fp}%
% CIRCLE 2 2 3 0
% 4 3528 820 4168 820 4168 820 4168 820
% 
\special{pn 8}%
\special{ar 3528 820 640 640  0.0000000 0.0187500}%
\special{ar 3528 820 640 640  0.0750000 0.0937500}%
\special{ar 3528 820 640 640  0.1500000 0.1687500}%
\special{ar 3528 820 640 640  0.2250000 0.2437500}%
\special{ar 3528 820 640 640  0.3000000 0.3187500}%
\special{ar 3528 820 640 640  0.3750000 0.3937500}%
\special{ar 3528 820 640 640  0.4500000 0.4687500}%
\special{ar 3528 820 640 640  0.5250000 0.5437500}%
\special{ar 3528 820 640 640  0.6000000 0.6187500}%
\special{ar 3528 820 640 640  0.6750000 0.6937500}%
\special{ar 3528 820 640 640  0.7500000 0.7687500}%
\special{ar 3528 820 640 640  0.8250000 0.8437500}%
\special{ar 3528 820 640 640  0.9000000 0.9187500}%
\special{ar 3528 820 640 640  0.9750000 0.9937500}%
\special{ar 3528 820 640 640  1.0500000 1.0687500}%
\special{ar 3528 820 640 640  1.1250000 1.1437500}%
\special{ar 3528 820 640 640  1.2000000 1.2187500}%
\special{ar 3528 820 640 640  1.2750000 1.2937500}%
\special{ar 3528 820 640 640  1.3500000 1.3687500}%
\special{ar 3528 820 640 640  1.4250000 1.4437500}%
\special{ar 3528 820 640 640  1.5000000 1.5187500}%
\special{ar 3528 820 640 640  1.5750000 1.5937500}%
\special{ar 3528 820 640 640  1.6500000 1.6687500}%
\special{ar 3528 820 640 640  1.7250000 1.7437500}%
\special{ar 3528 820 640 640  1.8000000 1.8187500}%
\special{ar 3528 820 640 640  1.8750000 1.8937500}%
\special{ar 3528 820 640 640  1.9500000 1.9687500}%
\special{ar 3528 820 640 640  2.0250000 2.0437500}%
\special{ar 3528 820 640 640  2.1000000 2.1187500}%
\special{ar 3528 820 640 640  2.1750000 2.1937500}%
\special{ar 3528 820 640 640  2.2500000 2.2687500}%
\special{ar 3528 820 640 640  2.3250000 2.3437500}%
\special{ar 3528 820 640 640  2.4000000 2.4187500}%
\special{ar 3528 820 640 640  2.4750000 2.4937500}%
\special{ar 3528 820 640 640  2.5500000 2.5687500}%
\special{ar 3528 820 640 640  2.6250000 2.6437500}%
\special{ar 3528 820 640 640  2.7000000 2.7187500}%
\special{ar 3528 820 640 640  2.7750000 2.7937500}%
\special{ar 3528 820 640 640  2.8500000 2.8687500}%
\special{ar 3528 820 640 640  2.9250000 2.9437500}%
\special{ar 3528 820 640 640  3.0000000 3.0187500}%
\special{ar 3528 820 640 640  3.0750000 3.0937500}%
\special{ar 3528 820 640 640  3.1500000 3.1687500}%
\special{ar 3528 820 640 640  3.2250000 3.2437500}%
\special{ar 3528 820 640 640  3.3000000 3.3187500}%
\special{ar 3528 820 640 640  3.3750000 3.3937500}%
\special{ar 3528 820 640 640  3.4500000 3.4687500}%
\special{ar 3528 820 640 640  3.5250000 3.5437500}%
\special{ar 3528 820 640 640  3.6000000 3.6187500}%
\special{ar 3528 820 640 640  3.6750000 3.6937500}%
\special{ar 3528 820 640 640  3.7500000 3.7687500}%
\special{ar 3528 820 640 640  3.8250000 3.8437500}%
\special{ar 3528 820 640 640  3.9000000 3.9187500}%
\special{ar 3528 820 640 640  3.9750000 3.9937500}%
\special{ar 3528 820 640 640  4.0500000 4.0687500}%
\special{ar 3528 820 640 640  4.1250000 4.1437500}%
\special{ar 3528 820 640 640  4.2000000 4.2187500}%
\special{ar 3528 820 640 640  4.2750000 4.2937500}%
\special{ar 3528 820 640 640  4.3500000 4.3687500}%
\special{ar 3528 820 640 640  4.4250000 4.4437500}%
\special{ar 3528 820 640 640  4.5000000 4.5187500}%
\special{ar 3528 820 640 640  4.5750000 4.5937500}%
\special{ar 3528 820 640 640  4.6500000 4.6687500}%
\special{ar 3528 820 640 640  4.7250000 4.7437500}%
\special{ar 3528 820 640 640  4.8000000 4.8187500}%
\special{ar 3528 820 640 640  4.8750000 4.8937500}%
\special{ar 3528 820 640 640  4.9500000 4.9687500}%
\special{ar 3528 820 640 640  5.0250000 5.0437500}%
\special{ar 3528 820 640 640  5.1000000 5.1187500}%
\special{ar 3528 820 640 640  5.1750000 5.1937500}%
\special{ar 3528 820 640 640  5.2500000 5.2687500}%
\special{ar 3528 820 640 640  5.3250000 5.3437500}%
\special{ar 3528 820 640 640  5.4000000 5.4187500}%
\special{ar 3528 820 640 640  5.4750000 5.4937500}%
\special{ar 3528 820 640 640  5.5500000 5.5687500}%
\special{ar 3528 820 640 640  5.6250000 5.6437500}%
\special{ar 3528 820 640 640  5.7000000 5.7187500}%
\special{ar 3528 820 640 640  5.7750000 5.7937500}%
\special{ar 3528 820 640 640  5.8500000 5.8687500}%
\special{ar 3528 820 640 640  5.9250000 5.9437500}%
\special{ar 3528 820 640 640  6.0000000 6.0187500}%
\special{ar 3528 820 640 640  6.0750000 6.0937500}%
\special{ar 3528 820 640 640  6.1500000 6.1687500}%
\special{ar 3528 820 640 640  6.2250000 6.2437500}%
% CIRCLE 2 0 1 0
% 4 3528 820 4008 500 4008 500 4008 500
% 
\special{pn 8}%
\special{sh 0.300}%
\special{ar 3528 820 578 578  0.0000000 6.2831853}%
% STR 2 0 3 0
% 3 3528 740 3528 820 0 0
% 
\put(35.2800,-8.2000){\makebox(0,0)[lb]{}}%
% STR 2 0 3 0
% 3 3528 740 3528 820 5 0
% {\bf BH}
\put(35.2800,-8.2000){\makebox(0,0){{\bf BH}}}%
% CIRCLE 2 0 3 0
% 4 4800 676 4800 836 4800 836 4800 836
% 
\special{pn 8}%
\special{ar 4800 676 160 160  0.0000000 6.2831853}%
% STR 2 0 3 0
% 3 4800 596 4800 676 5 0
% $E$
\put(48.0000,-6.7600){\makebox(0,0){$E$}}%
% VECTOR 2 0 3 0
% 2 4960 676 5280 676
% 
\special{pn 8}%
\special{pa 4960 676}%
\special{pa 5280 676}%
\special{fp}%
\special{sh 1}%
\special{pa 5280 676}%
\special{pa 5214 656}%
\special{pa 5228 676}%
\special{pa 5214 696}%
\special{pa 5280 676}%
\special{fp}%
% STR 2 0 3 0
% 3 5040 916 5040 996 5 0
% Hawking radiation
\put(50.4000,-9.9600){\makebox(0,0){Hawking radiation}}%
\end{picture}%

%% file: nor01.tex
%WinTpicVersion3.08
\unitlength 0.1in
\begin{picture}( 29.5000,  6.2000)(  7.9000,-14.9500)
% STR 2 0 3 0
% 3 2350 860 2350 960 5 0
% $\Sigma'$
\put(23.5000,-9.6000){\makebox(0,0){$\Sigma'$}}%
% STR 2 0 3 0
% 3 2360 1480 2360 1580 5 0
% $\Sigma$
\put(23.6000,-15.8000){\makebox(0,0){$\Sigma$}}%
% SPLINE 2 0 3 0
% 12 790 1280 990 1200 1280 1280 1600 1200 2000 990 2320 1090 2600 990 3000 1200 3310 1280 3610 1200 3740 1240 3740 1240
% 
\special{pn 8}%
\special{pa 790 1280}%
\special{pa 820 1264}%
\special{pa 850 1248}%
\special{pa 880 1232}%
\special{pa 908 1220}%
\special{pa 938 1210}%
\special{pa 968 1202}%
\special{pa 998 1200}%
\special{pa 1028 1202}%
\special{pa 1060 1208}%
\special{pa 1090 1216}%
\special{pa 1120 1228}%
\special{pa 1152 1238}%
\special{pa 1182 1250}%
\special{pa 1214 1262}%
\special{pa 1244 1272}%
\special{pa 1276 1280}%
\special{pa 1308 1284}%
\special{pa 1340 1286}%
\special{pa 1372 1284}%
\special{pa 1402 1280}%
\special{pa 1434 1274}%
\special{pa 1466 1266}%
\special{pa 1496 1254}%
\special{pa 1528 1240}%
\special{pa 1558 1226}%
\special{pa 1588 1208}%
\special{pa 1616 1190}%
\special{pa 1646 1170}%
\special{pa 1674 1150}%
\special{pa 1704 1128}%
\special{pa 1732 1108}%
\special{pa 1760 1086}%
\special{pa 1788 1066}%
\special{pa 1816 1048}%
\special{pa 1842 1032}%
\special{pa 1870 1018}%
\special{pa 1898 1006}%
\special{pa 1926 996}%
\special{pa 1956 992}%
\special{pa 1984 990}%
\special{pa 2014 992}%
\special{pa 2044 1000}%
\special{pa 2072 1010}%
\special{pa 2104 1022}%
\special{pa 2134 1036}%
\special{pa 2164 1050}%
\special{pa 2196 1062}%
\special{pa 2226 1074}%
\special{pa 2258 1084}%
\special{pa 2288 1090}%
\special{pa 2320 1090}%
\special{pa 2350 1086}%
\special{pa 2382 1078}%
\special{pa 2412 1068}%
\special{pa 2442 1054}%
\special{pa 2472 1038}%
\special{pa 2502 1024}%
\special{pa 2532 1010}%
\special{pa 2562 1000}%
\special{pa 2592 992}%
\special{pa 2620 988}%
\special{pa 2648 990}%
\special{pa 2676 996}%
\special{pa 2704 1004}%
\special{pa 2732 1016}%
\special{pa 2760 1030}%
\special{pa 2788 1048}%
\special{pa 2816 1066}%
\special{pa 2844 1086}%
\special{pa 2872 1108}%
\special{pa 2900 1130}%
\special{pa 2930 1150}%
\special{pa 2958 1172}%
\special{pa 2988 1192}%
\special{pa 3018 1210}%
\special{pa 3048 1228}%
\special{pa 3078 1242}%
\special{pa 3108 1256}%
\special{pa 3140 1266}%
\special{pa 3170 1276}%
\special{pa 3202 1282}%
\special{pa 3234 1284}%
\special{pa 3264 1286}%
\special{pa 3296 1282}%
\special{pa 3328 1276}%
\special{pa 3358 1268}%
\special{pa 3390 1258}%
\special{pa 3422 1246}%
\special{pa 3452 1234}%
\special{pa 3482 1222}%
\special{pa 3514 1212}%
\special{pa 3544 1206}%
\special{pa 3576 1200}%
\special{pa 3606 1200}%
\special{pa 3636 1204}%
\special{pa 3668 1212}%
\special{pa 3698 1222}%
\special{pa 3728 1236}%
\special{pa 3740 1240}%
\special{sp}%
% SPLINE 2 0 3 0
% 6 1400 1280 1610 1390 1960 1270 2330 1430 2900 1370 3140 1280
% 
\special{pn 8}%
\special{pa 1400 1280}%
\special{pa 1428 1302}%
\special{pa 1456 1322}%
\special{pa 1484 1342}%
\special{pa 1512 1360}%
\special{pa 1540 1374}%
\special{pa 1570 1384}%
\special{pa 1598 1390}%
\special{pa 1628 1390}%
\special{pa 1658 1386}%
\special{pa 1688 1378}%
\special{pa 1718 1366}%
\special{pa 1748 1354}%
\special{pa 1778 1338}%
\special{pa 1810 1324}%
\special{pa 1840 1308}%
\special{pa 1870 1294}%
\special{pa 1902 1284}%
\special{pa 1932 1274}%
\special{pa 1962 1270}%
\special{pa 1992 1270}%
\special{pa 2020 1276}%
\special{pa 2050 1284}%
\special{pa 2078 1294}%
\special{pa 2108 1308}%
\special{pa 2136 1324}%
\special{pa 2166 1340}%
\special{pa 2194 1358}%
\special{pa 2224 1376}%
\special{pa 2254 1392}%
\special{pa 2284 1408}%
\special{pa 2314 1424}%
\special{pa 2344 1436}%
\special{pa 2374 1446}%
\special{pa 2406 1452}%
\special{pa 2438 1458}%
\special{pa 2468 1460}%
\special{pa 2500 1462}%
\special{pa 2532 1462}%
\special{pa 2566 1458}%
\special{pa 2598 1456}%
\special{pa 2630 1450}%
\special{pa 2662 1444}%
\special{pa 2694 1436}%
\special{pa 2726 1428}%
\special{pa 2760 1418}%
\special{pa 2792 1408}%
\special{pa 2822 1398}%
\special{pa 2854 1386}%
\special{pa 2886 1376}%
\special{pa 2916 1364}%
\special{pa 2948 1354}%
\special{pa 2978 1342}%
\special{pa 3008 1332}%
\special{pa 3038 1320}%
\special{pa 3066 1308}%
\special{pa 3096 1298}%
\special{pa 3126 1286}%
\special{pa 3140 1280}%
\special{sp}%
% STR 2 0 3 0
% 3 2380 1170 2380 1270 5 0
% $V$
\put(23.8000,-12.7000){\makebox(0,0){$V$}}%
% LINE 3 0 3 0
% 28 2930 1150 2630 1450 2860 1100 2500 1460 2790 1050 2390 1450 2710 1010 2300 1420 2610 990 2230 1370 2410 1070 2150 1330 2270 1090 2070 1290 2180 1060 1970 1270 2100 1020 1790 1330 2010 990 1610 1390 1850 1030 1520 1360 1520 1240 1450 1310 3000 1200 2790 1410 3080 1240 2990 1330
% 
\special{pn 4}%
\special{pa 2930 1150}%
\special{pa 2630 1450}%
\special{fp}%
\special{pa 2860 1100}%
\special{pa 2500 1460}%
\special{fp}%
\special{pa 2790 1050}%
\special{pa 2390 1450}%
\special{fp}%
\special{pa 2710 1010}%
\special{pa 2300 1420}%
\special{fp}%
\special{pa 2610 990}%
\special{pa 2230 1370}%
\special{fp}%
\special{pa 2410 1070}%
\special{pa 2150 1330}%
\special{fp}%
\special{pa 2270 1090}%
\special{pa 2070 1290}%
\special{fp}%
\special{pa 2180 1060}%
\special{pa 1970 1270}%
\special{fp}%
\special{pa 2100 1020}%
\special{pa 1790 1330}%
\special{fp}%
\special{pa 2010 990}%
\special{pa 1610 1390}%
\special{fp}%
\special{pa 1850 1030}%
\special{pa 1520 1360}%
\special{fp}%
\special{pa 1520 1240}%
\special{pa 1450 1310}%
\special{fp}%
\special{pa 3000 1200}%
\special{pa 2790 1410}%
\special{fp}%
\special{pa 3080 1240}%
\special{pa 2990 1330}%
\special{fp}%
\end{picture}%

%% file: bog01.tex
%WinTpicVersion3.08
\unitlength 0.1in
\begin{picture}( 46.7500, 15.1500)(  3.2500,-19.1500)
% VECTOR 2 0 3 0
% 2 600 800 1400 800
% 
\special{pn 8}%
\special{pa 600 800}%
\special{pa 1400 800}%
\special{fp}%
\special{sh 1}%
\special{pa 1400 800}%
\special{pa 1334 780}%
\special{pa 1348 800}%
\special{pa 1334 820}%
\special{pa 1400 800}%
\special{fp}%
% VECTOR 2 0 3 0
% 2 1400 1200 600 1200
% 
\special{pn 8}%
\special{pa 1400 1200}%
\special{pa 600 1200}%
\special{fp}%
\special{sh 1}%
\special{pa 600 1200}%
\special{pa 668 1220}%
\special{pa 654 1200}%
\special{pa 668 1180}%
\special{pa 600 1200}%
\special{fp}%
% CIRCLE 2 0 3 0
% 4 2800 1000 2800 1600 2800 1600 2800 1600
% 
\special{pn 8}%
\special{ar 2800 1000 600 600  0.0000000 6.2831853}%
% LINE 2 0 3 0
% 56 3240 600 2400 1440 3270 630 2430 1470 3290 670 2470 1490 3320 700 2500 1520 3340 740 2540 1540 3360 780 2580 1560 3370 830 2630 1570 3390 870 2670 1590 3390 930 2730 1590 3400 980 2780 1600 3400 1040 2850 1590 3390 1110 2910 1590 3360 1200 3000 1560 3310 1310 3110 1510 3210 570 2370 1410 3180 540 2340 1380 3150 510 2310 1350 3110 490 2290 1310 3070 470 2270 1270 3030 450 2250 1230 2990 430 2230 1190 2940 420 2220 1140 2890 410 2210 1090 2840 400 2200 1040 2780 400 2200 980 2710 410 2210 910 2630 430 2230 830 2530 470 2270 730
% 
\special{pn 8}%
\special{pa 3240 600}%
\special{pa 2400 1440}%
\special{fp}%
\special{pa 3270 630}%
\special{pa 2430 1470}%
\special{fp}%
\special{pa 3290 670}%
\special{pa 2470 1490}%
\special{fp}%
\special{pa 3320 700}%
\special{pa 2500 1520}%
\special{fp}%
\special{pa 3340 740}%
\special{pa 2540 1540}%
\special{fp}%
\special{pa 3360 780}%
\special{pa 2580 1560}%
\special{fp}%
\special{pa 3370 830}%
\special{pa 2630 1570}%
\special{fp}%
\special{pa 3390 870}%
\special{pa 2670 1590}%
\special{fp}%
\special{pa 3390 930}%
\special{pa 2730 1590}%
\special{fp}%
\special{pa 3400 980}%
\special{pa 2780 1600}%
\special{fp}%
\special{pa 3400 1040}%
\special{pa 2850 1590}%
\special{fp}%
\special{pa 3390 1110}%
\special{pa 2910 1590}%
\special{fp}%
\special{pa 3360 1200}%
\special{pa 3000 1560}%
\special{fp}%
\special{pa 3310 1310}%
\special{pa 3110 1510}%
\special{fp}%
\special{pa 3210 570}%
\special{pa 2370 1410}%
\special{fp}%
\special{pa 3180 540}%
\special{pa 2340 1380}%
\special{fp}%
\special{pa 3150 510}%
\special{pa 2310 1350}%
\special{fp}%
\special{pa 3110 490}%
\special{pa 2290 1310}%
\special{fp}%
\special{pa 3070 470}%
\special{pa 2270 1270}%
\special{fp}%
\special{pa 3030 450}%
\special{pa 2250 1230}%
\special{fp}%
\special{pa 2990 430}%
\special{pa 2230 1190}%
\special{fp}%
\special{pa 2940 420}%
\special{pa 2220 1140}%
\special{fp}%
\special{pa 2890 410}%
\special{pa 2210 1090}%
\special{fp}%
\special{pa 2840 400}%
\special{pa 2200 1040}%
\special{fp}%
\special{pa 2780 400}%
\special{pa 2200 980}%
\special{fp}%
\special{pa 2710 410}%
\special{pa 2210 910}%
\special{fp}%
\special{pa 2630 430}%
\special{pa 2230 830}%
\special{fp}%
\special{pa 2530 470}%
\special{pa 2270 730}%
\special{fp}%
% VECTOR 2 0 3 0
% 2 4200 1000 5000 1000
% 
\special{pn 8}%
\special{pa 4200 1000}%
\special{pa 5000 1000}%
\special{fp}%
\special{sh 1}%
\special{pa 5000 1000}%
\special{pa 4934 980}%
\special{pa 4948 1000}%
\special{pa 4934 1020}%
\special{pa 5000 1000}%
\special{fp}%
% STR 2 0 3 0
% 3 1000 510 1000 610 5 0
% $f_i$
\put(10.0000,-6.1000){\makebox(0,0){$f_i$}}%
% STR 2 0 3 0
% 3 1000 1300 1000 1400 5 0
% $f^*_i$
\put(10.0000,-14.0000){\makebox(0,0){$f^*_i$}}%
% STR 2 0 3 0
% 3 4600 700 4600 800 5 0
% $p_i$
\put(46.0000,-8.0000){\makebox(0,0){$p_i$}}%
% STR 2 0 3 0
% 3 2800 1700 2800 1800 5 0
% The collapsing body
\put(28.0000,-18.0000){\makebox(0,0){The collapsing body}}%
% STR 2 0 3 0
% 3 1000 1700 1000 1800 5 0
% $t~\to~-\infty$
\put(10.0000,-18.0000){\makebox(0,0){$t~\to~-\infty$}}%
% STR 2 0 3 0
% 3 1000 1900 1000 2000 5 0
% Minkowski space
\put(10.0000,-20.0000){\makebox(0,0){Minkowski space}}%
% STR 2 0 3 0
% 3 4600 1900 4600 2000 5 0
% Minkowski space
\put(46.0000,-20.0000){\makebox(0,0){Minkowski space}}%
% STR 2 0 3 0
% 3 4600 1700 4600 1800 5 0
% $t~\to~\infty$
\put(46.0000,-18.0000){\makebox(0,0){$t~\to~\infty$}}%
\end{picture}%

%% file: pen08.tex
%WinTpicVersion3.08
\unitlength 0.1in
\begin{picture}( 27.9000, 23.2600)(  1.4000,-28.3300)
% BOX 2 0 3 0
% 2 1382 763 2363 727
% 
\special{pn 8}%
\special{pa 1382 764}%
\special{pa 2364 764}%
\special{pa 2364 728}%
\special{pa 1382 728}%
\special{pa 1382 764}%
\special{fp}%
% LINE 2 0 3 0
% 2 2363 772 2903 1312
% 
\special{pn 8}%
\special{pa 2364 772}%
\special{pa 2904 1312}%
\special{fp}%
% LINE 2 0 3 0
% 2 2903 1312 1463 2752
% 
\special{pn 8}%
\special{pa 2904 1312}%
\special{pa 1464 2752}%
\special{fp}%
% LINE 2 0 3 0
% 2 1463 772 1463 2752
% 
\special{pn 8}%
\special{pa 1464 772}%
\special{pa 1464 2752}%
\special{fp}%
% LINE 2 2 3 0
% 6 1382 763 1382 2833 1382 2815 1454 2752 1256 2788 1256 2788
% 
\special{pn 8}%
\special{pa 1382 764}%
\special{pa 1382 2834}%
\special{dt 0.045}%
\special{pa 1382 2816}%
\special{pa 1454 2752}%
\special{dt 0.045}%
\special{pa 1256 2788}%
\special{pa 1256 2788}%
\special{dt 0.045}%
% LINE 1 0 3 0
% 2 2363 763 1463 1654
% 
\special{pn 13}%
\special{pa 2364 764}%
\special{pa 1464 1654}%
\special{fp}%
% STR 2 0 3 0
% 3 1823 501 1823 592 5 0
% singularity
\put(18.2300,-5.9200){\makebox(0,0){singularity}}%
% STR 2 0 3 0
% 3 1445 2689 1445 2779 1 0
% $I^-$
\put(14.4500,-27.7900){\makebox(0,0)[lt]{$I^-$}}%
% STR 2 0 3 0
% 3 1283 1402 1283 1492 3 0
% $r=0$
\put(12.8300,-14.9200){\makebox(0,0)[rb]{$r=0$}}%
% STR 2 0 3 0
% 3 2426 646 2426 736 2 0
% $U^0=-Ce^{-\kappa u_0}$
\put(24.2600,-7.3600){\makebox(0,0)[lb]{$U^0=-Ce^{-\kappa u_0}$}}%
% LINE 1 0 3 0
% 2 1463 1672 2003 2212
% 
\special{pn 13}%
\special{pa 1464 1672}%
\special{pa 2004 2212}%
\special{fp}%
% STR 2 0 3 0
% 3 2030 2266 2030 2356 2 0
% $v_0$
\put(20.3000,-23.5600){\makebox(0,0)[lb]{$v_0$}}%
% LINE 2 2 3 0
% 2 1463 2041 1814 2392
% 
\special{pn 8}%
\special{pa 1464 2042}%
\special{pa 1814 2392}%
\special{dt 0.045}%
% LINE 2 2 3 0
% 2 2543 961 1463 2032
% 
\special{pn 8}%
\special{pa 2544 962}%
\special{pa 1464 2032}%
\special{dt 0.045}%
% VECTOR 0 0 3 0
% 2 2003 1132 1823 952
% 
\special{pn 20}%
\special{pa 2004 1132}%
\special{pa 1824 952}%
\special{fp}%
\special{sh 1}%
\special{pa 1824 952}%
\special{pa 1856 1014}%
\special{pa 1862 990}%
\special{pa 1884 986}%
\special{pa 1824 952}%
\special{fp}%
% VECTOR 0 0 3 0
% 2 2003 1132 2183 952
% 
\special{pn 20}%
\special{pa 2004 1132}%
\special{pa 2184 952}%
\special{fp}%
\special{sh 1}%
\special{pa 2184 952}%
\special{pa 2122 986}%
\special{pa 2146 990}%
\special{pa 2150 1014}%
\special{pa 2184 952}%
\special{fp}%
% STR 2 0 3 0
% 3 2786 790 2786 880 5 0
% $U=-\epsilon$
\put(27.8600,-8.8000){\makebox(0,0){$U=-\epsilon$}}%
% STR 2 0 3 0
% 3 1805 826 1805 916 2 0
% $n^\mu$
\put(18.0500,-9.1600){\makebox(0,0)[lb]{$n^\mu$}}%
% STR 2 0 3 0
% 3 2111 853 2111 943 2 0
% $l^\mu$
\put(21.1100,-9.4300){\makebox(0,0)[lb]{$l^\mu$}}%
% STR 2 0 3 0
% 3 1661 1240 1661 1330 2 0
% $\gamma_H$
\put(16.6100,-13.3000){\makebox(0,0)[lb]{$\gamma_H$}}%
% STR 2 0 3 0
% 3 2039 1564 2039 1654 2 0
% $\gamma$
\put(20.3900,-16.5400){\makebox(0,0)[lb]{$\gamma$}}%
% STR 2 0 3 0
% 3 1805 2437 1805 2527 2 0
% $v$
\put(18.0500,-25.2700){\makebox(0,0)[lb]{$v$}}%
% VECTOR 2 0 3 0
% 4 1877 2095 1697 2275 1697 2275 1877 2095
% 
\special{pn 8}%
\special{pa 1878 2096}%
\special{pa 1698 2276}%
\special{fp}%
\special{sh 1}%
\special{pa 1698 2276}%
\special{pa 1758 2242}%
\special{pa 1736 2238}%
\special{pa 1730 2214}%
\special{pa 1698 2276}%
\special{fp}%
\special{pa 1698 2276}%
\special{pa 1878 2096}%
\special{fp}%
\special{sh 1}%
\special{pa 1878 2096}%
\special{pa 1816 2128}%
\special{pa 1840 2134}%
\special{pa 1844 2156}%
\special{pa 1878 2096}%
\special{fp}%
% VECTOR 2 0 3 0
% 4 2003 1141 2183 1321 2183 1321 2003 1141
% 
\special{pn 8}%
\special{pa 2004 1142}%
\special{pa 2184 1322}%
\special{fp}%
\special{sh 1}%
\special{pa 2184 1322}%
\special{pa 2150 1260}%
\special{pa 2146 1284}%
\special{pa 2122 1288}%
\special{pa 2184 1322}%
\special{fp}%
\special{pa 2184 1322}%
\special{pa 2004 1142}%
\special{fp}%
\special{sh 1}%
\special{pa 2004 1142}%
\special{pa 2036 1202}%
\special{pa 2042 1180}%
\special{pa 2064 1174}%
\special{pa 2004 1142}%
\special{fp}%
% STR 2 0 3 0
% 3 1976 1276 1976 1366 2 0
% $\epsilon$
\put(19.7600,-13.6600){\makebox(0,0)[lb]{$\epsilon$}}%
% STR 2 0 3 0
% 3 1697 2059 1697 2149 2 0
% $\epsilon$
\put(16.9700,-21.4900){\makebox(0,0)[lb]{$\epsilon$}}%
% STR 2 0 3 0
% 3 2930 1276 2930 1366 2 0
% $I^0$
\put(29.3000,-13.6600){\makebox(0,0)[lb]{$I^0$}}%
% VECTOR 2 0 3 0
% 2 1067 1897 1436 1681
% 
\special{pn 8}%
\special{pa 1068 1898}%
\special{pa 1436 1682}%
\special{fp}%
\special{sh 1}%
\special{pa 1436 1682}%
\special{pa 1368 1698}%
\special{pa 1390 1708}%
\special{pa 1390 1732}%
\special{pa 1436 1682}%
\special{fp}%
% STR 2 0 3 0
% 3 1040 1825 1040 1915 4 0
% reflection
\put(10.4000,-19.1500){\makebox(0,0)[rt]{reflection}}%
\end{picture}%

%% file: ent01.tex
%WinTpicVersion3.08
\unitlength 0.1in
\begin{picture}( 16.0000, 16.0000)(  4.0000,-18.0000)
% CIRCLE 2 0 3 0
% 4 1200 1000 1200 1800 1200 1800 1200 1800
% 
\special{pn 8}%
\special{ar 1200 1000 800 800  0.0000000 6.2831853}%
% CIRCLE 2 0 1 0
% 4 800 1000 1200 1000 1200 1000 1200 1000
% 
\special{pn 8}%
\special{sh 0.300}%
\special{ar 800 1000 400 400  0.0000000 6.2831853}%
% STR 2 0 3 0
% 3 800 900 800 1000 5 0
% $S_{12}$
\put(8.0000,-10.0000){\makebox(0,0){$S_{12}$}}%
% STR 2 0 3 0
% 3 1600 900 1600 1000 5 0
% $S_{21}$
\put(16.0000,-10.0000){\makebox(0,0){$S_{21}$}}%
\end{picture}%

%% file: pen09.tex
%WinTpicVersion3.08
\unitlength 0.1in
\begin{picture}( 30.1500, 31.7000)( -0.7000,-36.6500)
% SPLINE 2 0 3 0
% 27 650 750 719 704 791 750 856 793 925 750 990 704 1062 750 1132 795 1200 750 1269 704 1338 750 1406 795 1475 750 1544 704 1612 750 1681 795 1746 750 1818 704 1888 750 1956 795 2025 750 2087 704 2166 750 2231 795 2300 750 2300 750 2300 750
% 
\special{pn 8}%
\special{pa 650 750}%
\special{pa 678 726}%
\special{pa 704 708}%
\special{pa 730 706}%
\special{pa 758 720}%
\special{pa 784 744}%
\special{pa 812 770}%
\special{pa 838 788}%
\special{pa 864 792}%
\special{pa 892 780}%
\special{pa 920 756}%
\special{pa 946 730}%
\special{pa 972 710}%
\special{pa 998 704}%
\special{pa 1024 718}%
\special{pa 1052 740}%
\special{pa 1078 766}%
\special{pa 1106 788}%
\special{pa 1132 796}%
\special{pa 1160 786}%
\special{pa 1186 764}%
\special{pa 1212 738}%
\special{pa 1240 716}%
\special{pa 1266 704}%
\special{pa 1292 712}%
\special{pa 1320 732}%
\special{pa 1346 758}%
\special{pa 1372 782}%
\special{pa 1400 794}%
\special{pa 1426 790}%
\special{pa 1452 772}%
\special{pa 1480 746}%
\special{pa 1506 722}%
\special{pa 1532 706}%
\special{pa 1560 708}%
\special{pa 1586 724}%
\special{pa 1612 750}%
\special{pa 1640 776}%
\special{pa 1666 792}%
\special{pa 1692 794}%
\special{pa 1718 776}%
\special{pa 1746 752}%
\special{pa 1772 726}%
\special{pa 1798 708}%
\special{pa 1826 706}%
\special{pa 1852 718}%
\special{pa 1880 742}%
\special{pa 1906 768}%
\special{pa 1932 788}%
\special{pa 1960 796}%
\special{pa 1986 784}%
\special{pa 2014 762}%
\special{pa 2040 736}%
\special{pa 2064 714}%
\special{pa 2090 704}%
\special{pa 2118 712}%
\special{pa 2146 732}%
\special{pa 2174 758}%
\special{pa 2200 782}%
\special{pa 2226 794}%
\special{pa 2252 790}%
\special{pa 2280 770}%
\special{pa 2300 750}%
\special{sp}%
% LINE 2 0 3 0
% 2 2300 750 2900 1350
% 
\special{pn 8}%
\special{pa 2300 750}%
\special{pa 2900 1350}%
\special{fp}%
% LINE 2 0 3 0
% 2 2900 1350 650 3600
% 
\special{pn 8}%
\special{pa 2900 1350}%
\special{pa 650 3600}%
\special{fp}%
% LINE 2 2 3 0
% 2 650 750 650 3600
% 
\special{pn 8}%
\special{pa 650 750}%
\special{pa 650 3600}%
\special{dt 0.045}%
% LINE 0 0 3 0
% 2 2300 750 650 2400
% 
\special{pn 20}%
\special{pa 2300 750}%
\special{pa 650 2400}%
\special{fp}%
% LINE 2 2 3 0
% 2 650 2700 2375 975
% 
\special{pn 8}%
\special{pa 650 2700}%
\special{pa 2376 976}%
\special{dt 0.045}%
% VECTOR 2 0 3 0
% 4 2098 945 2248 1095 2248 1095 2098 945
% 
\special{pn 8}%
\special{pa 2098 946}%
\special{pa 2248 1096}%
\special{fp}%
\special{sh 1}%
\special{pa 2248 1096}%
\special{pa 2216 1034}%
\special{pa 2210 1058}%
\special{pa 2188 1062}%
\special{pa 2248 1096}%
\special{fp}%
\special{pa 2248 1096}%
\special{pa 2098 946}%
\special{fp}%
\special{sh 1}%
\special{pa 2098 946}%
\special{pa 2132 1006}%
\special{pa 2136 984}%
\special{pa 2160 978}%
\special{pa 2098 946}%
\special{fp}%
% STR 2 0 3 0
% 3 2082 1050 2082 1125 5 0
% $\varepsilon$
\put(20.8200,-11.2500){\makebox(0,0){$\varepsilon$}}%
% VECTOR 2 2 3 0
% 2 1400 1800 1250 1650
% 
\special{pn 8}%
\special{pa 1400 1800}%
\special{pa 1250 1650}%
\special{dt 0.045}%
\special{sh 1}%
\special{pa 1250 1650}%
\special{pa 1284 1712}%
\special{pa 1288 1688}%
\special{pa 1312 1684}%
\special{pa 1250 1650}%
\special{fp}%
% VECTOR 2 0 3 0
% 2 1400 1800 1550 1650
% 
\special{pn 8}%
\special{pa 1400 1800}%
\special{pa 1550 1650}%
\special{fp}%
\special{sh 1}%
\special{pa 1550 1650}%
\special{pa 1490 1684}%
\special{pa 1512 1688}%
\special{pa 1518 1712}%
\special{pa 1550 1650}%
\special{fp}%
% VECTOR 2 0 3 0
% 2 1400 2400 1550 2250
% 
\special{pn 8}%
\special{pa 1400 2400}%
\special{pa 1550 2250}%
\special{fp}%
\special{sh 1}%
\special{pa 1550 2250}%
\special{pa 1490 2284}%
\special{pa 1512 2288}%
\special{pa 1518 2312}%
\special{pa 1550 2250}%
\special{fp}%
% VECTOR 2 0 3 0
% 2 1400 2400 1250 2250
% 
\special{pn 8}%
\special{pa 1400 2400}%
\special{pa 1250 2250}%
\special{fp}%
\special{sh 1}%
\special{pa 1250 2250}%
\special{pa 1284 2312}%
\special{pa 1288 2288}%
\special{pa 1312 2284}%
\special{pa 1250 2250}%
\special{fp}%
% STR 2 0 3 0
% 3 1858 1268 1858 1342 5 0
% $H$
\put(18.5800,-13.4200){\makebox(0,0){$H$}}%
% STR 2 0 3 0
% 3 1500 1226 1500 1300 5 0
% ${\mathcal H}^+$
\put(15.0000,-13.0000){\makebox(0,0){${\mathcal H}^+$}}%
% STR 2 0 3 0
% 3 2750 1012 2750 1088 2 0
% ${\mathcal J}^+$
\put(27.5000,-10.8800){\makebox(0,0)[lb]{${\mathcal J}^+$}}%
% STR 2 0 3 0
% 3 2000 2332 2000 2408 1 0
% ${\mathcal J}^-$
\put(20.0000,-24.0800){\makebox(0,0)[lt]{${\mathcal J}^-$}}%
% STR 2 0 3 0
% 3 2150 1582 2150 1658 5 0
% $O$
\put(21.5000,-16.5800){\makebox(0,0){$O$}}%
% STR 2 0 3 0
% 3 2945 1268 2945 1342 0 0
% 
\put(29.4500,-13.4200){\makebox(0,0)[lb]{}}%
% STR 2 0 3 0
% 3 3050 1290 3050 1365 5 0
% ${\mathcal I}^0$
\put(30.5000,-13.6500){\makebox(0,0){${\mathcal I}^0$}}%
% STR 2 0 3 0
% 3 650 3676 650 3750 5 0
% ${\mathcal I}^-$
\put(6.5000,-37.5000){\makebox(0,0){${\mathcal I}^-$}}%
% STR 2 0 3 0
% 3 432 1935 432 2010 5 0
% $r=0$
\put(4.3200,-20.1000){\makebox(0,0){$r=0$}}%
% STR 2 0 3 0
% 3 1370 505 1370 580 5 0
% singularity
\put(13.7000,-5.8000){\makebox(0,0){singularity}}%
% SPLINE 2 2 3 0
% 4 2375 975 2390 885 2300 750 2300 750
% 
\special{pn 8}%
\special{pa 2376 976}%
\special{pa 2384 944}%
\special{pa 2390 912}%
\special{pa 2390 882}%
\special{pa 2382 852}%
\special{pa 2366 826}%
\special{pa 2346 800}%
\special{pa 2324 774}%
\special{pa 2300 750}%
\special{sp -0.045}%
% CIRCLE 2 0 2 0
% 4 2300 750 2300 780 2300 780 2300 780
% 
\special{pn 8}%
\special{sh 0}%
\special{ar 2300 750 30 30  0.0000000 6.2831853}%
% CIRCLE 2 0 2 0
% 4 2900 1350 2900 1380 2900 1380 2900 1380
% 
\special{pn 8}%
\special{sh 0}%
\special{ar 2900 1350 30 30  0.0000000 6.2831853}%
% CIRCLE 2 0 2 0
% 4 650 3600 650 3630 650 3630 650 3630
% 
\special{pn 8}%
\special{sh 0}%
\special{ar 650 3600 30 30  0.0000000 6.2831853}%
% STR 2 0 3 0
% 3 2400 525 2400 600 5 0
% ${\mathcal I}^+$
\put(24.0000,-6.0000){\makebox(0,0){${\mathcal I}^+$}}%
% LINE 3 0 3 0
% 60 1590 1440 650 1440 1530 1500 650 1500 1470 1560 650 1560 1410 1620 650 1620 1350 1680 1310 1680 1260 1680 650 1680 1290 1740 650 1740 1230 1800 650 1800 1170 1860 650 1860 1110 1920 650 1920 1050 1980 650 1980 990 2040 650 2040 930 2100 650 2100 870 2160 650 2160 810 2220 650 2220 750 2280 650 2280 690 2340 650 2340 1650 1380 650 1380 1710 1320 650 1320 1770 1260 650 1260 1830 1200 650 1200 1890 1140 650 1140 1950 1080 650 1080 2010 1020 650 1020 2070 960 650 960 2130 900 650 900 2190 840 650 840 820 780 650 780 750 720 690 720 1090 780 900 780
% 
\special{pn 4}%
\special{pa 1590 1440}%
\special{pa 650 1440}%
\special{fp}%
\special{pa 1530 1500}%
\special{pa 650 1500}%
\special{fp}%
\special{pa 1470 1560}%
\special{pa 650 1560}%
\special{fp}%
\special{pa 1410 1620}%
\special{pa 650 1620}%
\special{fp}%
\special{pa 1350 1680}%
\special{pa 1310 1680}%
\special{fp}%
\special{pa 1260 1680}%
\special{pa 650 1680}%
\special{fp}%
\special{pa 1290 1740}%
\special{pa 650 1740}%
\special{fp}%
\special{pa 1230 1800}%
\special{pa 650 1800}%
\special{fp}%
\special{pa 1170 1860}%
\special{pa 650 1860}%
\special{fp}%
\special{pa 1110 1920}%
\special{pa 650 1920}%
\special{fp}%
\special{pa 1050 1980}%
\special{pa 650 1980}%
\special{fp}%
\special{pa 990 2040}%
\special{pa 650 2040}%
\special{fp}%
\special{pa 930 2100}%
\special{pa 650 2100}%
\special{fp}%
\special{pa 870 2160}%
\special{pa 650 2160}%
\special{fp}%
\special{pa 810 2220}%
\special{pa 650 2220}%
\special{fp}%
\special{pa 750 2280}%
\special{pa 650 2280}%
\special{fp}%
\special{pa 690 2340}%
\special{pa 650 2340}%
\special{fp}%
\special{pa 1650 1380}%
\special{pa 650 1380}%
\special{fp}%
\special{pa 1710 1320}%
\special{pa 650 1320}%
\special{fp}%
\special{pa 1770 1260}%
\special{pa 650 1260}%
\special{fp}%
\special{pa 1830 1200}%
\special{pa 650 1200}%
\special{fp}%
\special{pa 1890 1140}%
\special{pa 650 1140}%
\special{fp}%
\special{pa 1950 1080}%
\special{pa 650 1080}%
\special{fp}%
\special{pa 2010 1020}%
\special{pa 650 1020}%
\special{fp}%
\special{pa 2070 960}%
\special{pa 650 960}%
\special{fp}%
\special{pa 2130 900}%
\special{pa 650 900}%
\special{fp}%
\special{pa 2190 840}%
\special{pa 650 840}%
\special{fp}%
\special{pa 820 780}%
\special{pa 650 780}%
\special{fp}%
\special{pa 750 720}%
\special{pa 690 720}%
\special{fp}%
\special{pa 1090 780}%
\special{pa 900 780}%
\special{fp}%
% LINE 3 0 3 1
% 18 1020 720 960 720 1370 780 1170 780 1300 720 1240 720 1640 780 1440 780 1570 720 1510 720 1920 780 1720 780 1850 720 1790 720 2200 780 2000 780 2130 720 2060 720
% 
\special{pn 4}%
\special{pa 1020 720}%
\special{pa 960 720}%
\special{fp}%
\special{pa 1370 780}%
\special{pa 1170 780}%
\special{fp}%
\special{pa 1300 720}%
\special{pa 1240 720}%
\special{fp}%
\special{pa 1640 780}%
\special{pa 1440 780}%
\special{fp}%
\special{pa 1570 720}%
\special{pa 1510 720}%
\special{fp}%
\special{pa 1920 780}%
\special{pa 1720 780}%
\special{fp}%
\special{pa 1850 720}%
\special{pa 1790 720}%
\special{fp}%
\special{pa 2200 780}%
\special{pa 2000 780}%
\special{fp}%
\special{pa 2130 720}%
\special{pa 2060 720}%
\special{fp}%
% STR 2 0 3 0
% 3 1000 1100 1000 1200 5 0
% {\bf BH}
\put(10.0000,-12.0000){\makebox(0,0){{\bf BH}}}%
\end{picture}%

%% file: pen10.tex
%WinTpicVersion3.08
\unitlength 0.1in
\begin{picture}( 24.0000, 18.1000)( 10.0000,-20.1000)
% POLYGON 2 5 1 0
% 5 2400 400 1000 400 1000 1800 1000 1800 2400 400
% 
\special{pn 8}%
\special{sh 0.300}%
\special{pa 2400 400}%
\special{pa 1000 400}%
\special{pa 1000 1800}%
\special{pa 1000 1800}%
\special{pa 2400 400}%
\special{ip}%
% LINE 0 0 3 0
% 8 2405 410 1005 1810 1005 1810 1005 1810 1005 1810 1005 1810 1005 1810 1005 1810
% 
\special{pn 20}%
\special{pa 2406 410}%
\special{pa 1006 1810}%
\special{fp}%
\special{pa 1006 1810}%
\special{pa 1006 1810}%
\special{fp}%
\special{pa 1006 1810}%
\special{pa 1006 1810}%
\special{fp}%
\special{pa 1006 1810}%
\special{pa 1006 1810}%
\special{fp}%
% STR 2 0 3 0
% 3 2425 270 2425 370 2 0
% ${\cal H}^+$
\put(24.2500,-3.7000){\makebox(0,0)[lb]{${\cal H}^+$}}%
% LINE 2 2 3 0
% 8 2605 610 1205 2010 1205 2010 1205 2010 1205 2010 1205 2010 1205 2010 1205 2010
% 
\special{pn 8}%
\special{pa 2606 610}%
\special{pa 1206 2010}%
\special{dt 0.045}%
\special{pa 1206 2010}%
\special{pa 1206 2010}%
\special{dt 0.045}%
\special{pa 1206 2010}%
\special{pa 1206 2010}%
\special{dt 0.045}%
\special{pa 1206 2010}%
\special{pa 1206 2010}%
\special{dt 0.045}%
% VECTOR 2 0 3 0
% 6 2405 1610 2605 1410 2605 1410 2605 1410 2605 1410 2605 1410
% 
\special{pn 8}%
\special{pa 2406 1610}%
\special{pa 2606 1410}%
\special{fp}%
\special{sh 1}%
\special{pa 2606 1410}%
\special{pa 2544 1444}%
\special{pa 2568 1448}%
\special{pa 2572 1472}%
\special{pa 2606 1410}%
\special{fp}%
\special{pa 2606 1410}%
\special{pa 2606 1410}%
\special{fp}%
\special{pa 2606 1410}%
\special{pa 2606 1410}%
\special{fp}%
% VECTOR 2 0 3 0
% 6 2405 1610 2205 1410 2205 1410 2205 1410 2205 1410 2205 1410
% 
\special{pn 8}%
\special{pa 2406 1610}%
\special{pa 2206 1410}%
\special{fp}%
\special{sh 1}%
\special{pa 2206 1410}%
\special{pa 2238 1472}%
\special{pa 2244 1448}%
\special{pa 2266 1444}%
\special{pa 2206 1410}%
\special{fp}%
\special{pa 2206 1410}%
\special{pa 2206 1410}%
\special{fp}%
\special{pa 2206 1410}%
\special{pa 2206 1410}%
\special{fp}%
% VECTOR 2 0 3 0
% 6 1605 1410 1805 1210 1805 1210 1805 1210 1805 1210 1805 1210
% 
\special{pn 8}%
\special{pa 1606 1410}%
\special{pa 1806 1210}%
\special{fp}%
\special{sh 1}%
\special{pa 1806 1210}%
\special{pa 1744 1244}%
\special{pa 1768 1248}%
\special{pa 1772 1272}%
\special{pa 1806 1210}%
\special{fp}%
\special{pa 1806 1210}%
\special{pa 1806 1210}%
\special{fp}%
\special{pa 1806 1210}%
\special{pa 1806 1210}%
\special{fp}%
% VECTOR 2 2 3 0
% 6 1605 1410 1405 1210 1405 1210 1405 1210 1405 1210 1405 1210
% 
\special{pn 8}%
\special{pa 1606 1410}%
\special{pa 1406 1210}%
\special{dt 0.045}%
\special{sh 1}%
\special{pa 1406 1210}%
\special{pa 1438 1272}%
\special{pa 1444 1248}%
\special{pa 1466 1244}%
\special{pa 1406 1210}%
\special{fp}%
\special{pa 1406 1210}%
\special{pa 1406 1210}%
\special{dt 0.045}%
\special{pa 1406 1210}%
\special{pa 1406 1210}%
\special{dt 0.045}%
% STR 2 0 3 0
% 3 2405 1670 2405 1770 5 0
% {\bf $O$}
\put(24.0500,-17.7000){\makebox(0,0){{\bf $O$}}}%
% STR 2 0 3 0
% 3 3400 1270 3400 1370 2 0
% ${\cal H}^+$; Future horizon
\put(34.0000,-13.7000){\makebox(0,0)[lb]{${\cal H}^+$; Future horizon}}%
% STR 2 0 3 0
% 3 3400 1470 3400 1570 2 0
% Region $H$; 2-dimensional and chiral
\put(34.0000,-15.7000){\makebox(0,0)[lb]{Region $H$; 2-dimensional and chiral}}%
% STR 2 0 3 0
% 3 3400 1670 3400 1770 2 0
% Region $O$; 4-dimensional and anomaly-free
\put(34.0000,-17.7000){\makebox(0,0)[lb]{Region $O$; 4-dimensional and anomaly-free}}%
% STR 2 0 3 0
% 3 1405 1510 1405 1610 5 0
% {\bf $H$}
\put(14.0500,-16.1000){\makebox(0,0){{\bf $H$}}}%
% STR 2 0 3 0
% 3 1405 710 1405 810 5 0
% {\bf BH}
\put(14.0500,-8.1000){\makebox(0,0){{\bf BH}}}%
\end{picture}%

%% file: ward02.tex
%WinTpicVersion3.08
\unitlength 0.1in
\begin{picture}( 28.7100, 27.6700)( 11.2000,-30.2700)
% CIRCLE 2 0 1 0
% 4 2605 1648 2605 2008 2605 2008 2605 2008
% 
\special{pn 8}%
\special{sh 0.300}%
\special{ar 2606 1648 360 360  0.0000000 6.2831853}%
% CIRCLE 2 2 3 0
% 4 2605 1648 2605 2116 2605 2116 2605 2116
% 
\special{pn 8}%
\special{ar 2606 1648 468 468  0.0000000 0.0256410}%
\special{ar 2606 1648 468 468  0.1025641 0.1282051}%
\special{ar 2606 1648 468 468  0.2051282 0.2307692}%
\special{ar 2606 1648 468 468  0.3076923 0.3333333}%
\special{ar 2606 1648 468 468  0.4102564 0.4358974}%
\special{ar 2606 1648 468 468  0.5128205 0.5384615}%
\special{ar 2606 1648 468 468  0.6153846 0.6410256}%
\special{ar 2606 1648 468 468  0.7179487 0.7435897}%
\special{ar 2606 1648 468 468  0.8205128 0.8461538}%
\special{ar 2606 1648 468 468  0.9230769 0.9487179}%
\special{ar 2606 1648 468 468  1.0256410 1.0512821}%
\special{ar 2606 1648 468 468  1.1282051 1.1538462}%
\special{ar 2606 1648 468 468  1.2307692 1.2564103}%
\special{ar 2606 1648 468 468  1.3333333 1.3589744}%
\special{ar 2606 1648 468 468  1.4358974 1.4615385}%
\special{ar 2606 1648 468 468  1.5384615 1.5641026}%
\special{ar 2606 1648 468 468  1.6410256 1.6666667}%
\special{ar 2606 1648 468 468  1.7435897 1.7692308}%
\special{ar 2606 1648 468 468  1.8461538 1.8717949}%
\special{ar 2606 1648 468 468  1.9487179 1.9743590}%
\special{ar 2606 1648 468 468  2.0512821 2.0769231}%
\special{ar 2606 1648 468 468  2.1538462 2.1794872}%
\special{ar 2606 1648 468 468  2.2564103 2.2820513}%
\special{ar 2606 1648 468 468  2.3589744 2.3846154}%
\special{ar 2606 1648 468 468  2.4615385 2.4871795}%
\special{ar 2606 1648 468 468  2.5641026 2.5897436}%
\special{ar 2606 1648 468 468  2.6666667 2.6923077}%
\special{ar 2606 1648 468 468  2.7692308 2.7948718}%
\special{ar 2606 1648 468 468  2.8717949 2.8974359}%
\special{ar 2606 1648 468 468  2.9743590 3.0000000}%
\special{ar 2606 1648 468 468  3.0769231 3.1025641}%
\special{ar 2606 1648 468 468  3.1794872 3.2051282}%
\special{ar 2606 1648 468 468  3.2820513 3.3076923}%
\special{ar 2606 1648 468 468  3.3846154 3.4102564}%
\special{ar 2606 1648 468 468  3.4871795 3.5128205}%
\special{ar 2606 1648 468 468  3.5897436 3.6153846}%
\special{ar 2606 1648 468 468  3.6923077 3.7179487}%
\special{ar 2606 1648 468 468  3.7948718 3.8205128}%
\special{ar 2606 1648 468 468  3.8974359 3.9230769}%
\special{ar 2606 1648 468 468  4.0000000 4.0256410}%
\special{ar 2606 1648 468 468  4.1025641 4.1282051}%
\special{ar 2606 1648 468 468  4.2051282 4.2307692}%
\special{ar 2606 1648 468 468  4.3076923 4.3333333}%
\special{ar 2606 1648 468 468  4.4102564 4.4358974}%
\special{ar 2606 1648 468 468  4.5128205 4.5384615}%
\special{ar 2606 1648 468 468  4.6153846 4.6410256}%
\special{ar 2606 1648 468 468  4.7179487 4.7435897}%
\special{ar 2606 1648 468 468  4.8205128 4.8461538}%
\special{ar 2606 1648 468 468  4.9230769 4.9487179}%
\special{ar 2606 1648 468 468  5.0256410 5.0512821}%
\special{ar 2606 1648 468 468  5.1282051 5.1538462}%
\special{ar 2606 1648 468 468  5.2307692 5.2564103}%
\special{ar 2606 1648 468 468  5.3333333 5.3589744}%
\special{ar 2606 1648 468 468  5.4358974 5.4615385}%
\special{ar 2606 1648 468 468  5.5384615 5.5641026}%
\special{ar 2606 1648 468 468  5.6410256 5.6666667}%
\special{ar 2606 1648 468 468  5.7435897 5.7692308}%
\special{ar 2606 1648 468 468  5.8461538 5.8717949}%
\special{ar 2606 1648 468 468  5.9487179 5.9743590}%
\special{ar 2606 1648 468 468  6.0512821 6.0769231}%
\special{ar 2606 1648 468 468  6.1538462 6.1794872}%
\special{ar 2606 1648 468 468  6.2564103 6.2820513}%
% STR 2 0 3 0
% 3 1480 2100 1480 2190 4 0
% $S'$
\put(14.8000,-21.9000){\makebox(0,0)[rt]{$S'$}}%
% VECTOR 2 0 3 0
% 2 2351 1898 1631 2618
% 
\special{pn 8}%
\special{pa 2352 1898}%
\special{pa 1632 2618}%
\special{fp}%
\special{sh 1}%
\special{pa 1632 2618}%
\special{pa 1692 2586}%
\special{pa 1670 2580}%
\special{pa 1664 2558}%
\special{pa 1632 2618}%
\special{fp}%
% VECTOR 2 0 3 0
% 2 2596 2008 2596 3027
% 
\special{pn 8}%
\special{pa 2596 2008}%
\special{pa 2596 3028}%
\special{fp}%
\special{sh 1}%
\special{pa 2596 3028}%
\special{pa 2616 2960}%
\special{pa 2596 2974}%
\special{pa 2576 2960}%
\special{pa 2596 3028}%
\special{fp}%
% VECTOR 2 0 3 0
% 2 2596 1279 2596 260
% 
\special{pn 8}%
\special{pa 2596 1280}%
\special{pa 2596 260}%
\special{fp}%
\special{sh 1}%
\special{pa 2596 260}%
\special{pa 2576 328}%
\special{pa 2596 314}%
\special{pa 2616 328}%
\special{pa 2596 260}%
\special{fp}%
% VECTOR 2 0 3 0
% 2 2857 1396 3577 676
% 
\special{pn 8}%
\special{pa 2858 1396}%
\special{pa 3578 676}%
\special{fp}%
\special{sh 1}%
\special{pa 3578 676}%
\special{pa 3516 710}%
\special{pa 3540 714}%
\special{pa 3544 738}%
\special{pa 3578 676}%
\special{fp}%
% VECTOR 2 0 3 0
% 2 2857 1900 3577 2620
% 
\special{pn 8}%
\special{pa 2858 1900}%
\special{pa 3578 2620}%
\special{fp}%
\special{sh 1}%
\special{pa 3578 2620}%
\special{pa 3544 2560}%
\special{pa 3540 2582}%
\special{pa 3516 2588}%
\special{pa 3578 2620}%
\special{fp}%
% VECTOR 2 0 3 0
% 2 2972 1637 3991 1640
% 
\special{pn 8}%
\special{pa 2972 1638}%
\special{pa 3992 1640}%
\special{fp}%
\special{sh 1}%
\special{pa 3992 1640}%
\special{pa 3924 1620}%
\special{pa 3938 1640}%
\special{pa 3924 1660}%
\special{pa 3992 1640}%
\special{fp}%
% VECTOR 2 0 3 0
% 2 2252 1637 1234 1640
% 
\special{pn 8}%
\special{pa 2252 1638}%
\special{pa 1234 1640}%
\special{fp}%
\special{sh 1}%
\special{pa 1234 1640}%
\special{pa 1302 1660}%
\special{pa 1288 1640}%
\special{pa 1302 1620}%
\special{pa 1234 1640}%
\special{fp}%
% VECTOR 2 0 3 0
% 2 2351 1412 1631 692
% 
\special{pn 8}%
\special{pa 2352 1412}%
\special{pa 1632 692}%
\special{fp}%
\special{sh 1}%
\special{pa 1632 692}%
\special{pa 1664 754}%
\special{pa 1670 730}%
\special{pa 1692 726}%
\special{pa 1632 692}%
\special{fp}%
% STR 2 0 3 0
% 3 2920 2380 2920 2470 5 0
% $O$
\put(29.2000,-24.7000){\makebox(0,0){$O$}}%
% STR 2 0 3 0
% 3 2600 1550 2600 1640 5 0
% \textbf{BH}
\put(26.0000,-16.4000){\makebox(0,0){\textbf{BH}}}%
% CIRCLE 2 0 3 0
% 4 2600 1640 2600 2860 2600 2860 2600 2860
% 
\special{pn 8}%
\special{ar 2600 1640 1220 1220  0.0000000 6.2831853}%
% CIRCLE 2 0 3 0
% 4 2600 1650 2600 2150 2600 2150 2600 2150
% 
\special{pn 8}%
\special{ar 2600 1650 500 500  0.0000000 6.2831853}%
% STR 2 0 3 0
% 3 2800 1920 2800 2020 5 0
% $H$
\put(28.0000,-20.2000){\makebox(0,0){$H$}}%
% STR 2 0 3 0
% 3 2100 1740 2100 1840 4 0
% $S$
\put(21.0000,-18.4000){\makebox(0,0)[rt]{$S$}}%
\end{picture}%

%% file: tun01.tex
%WinTpicVersion3.08
\unitlength 0.1in
\begin{picture}( 21.0000, 16.1500)( 10.0000,-20.1500)
% BOX 2 5 1 0
% 2 2800 400 1000 2000
% 
\special{pn 8}%
\special{sh 0.300}%
\special{pa 2800 400}%
\special{pa 1000 400}%
\special{pa 1000 2000}%
\special{pa 2800 2000}%
\special{pa 2800 400}%
\special{ip}%
% STR 2 0 3 0
% 3 2800 2000 2800 2100 5 0
% Horizon
\put(28.0000,-21.0000){\makebox(0,0){Horizon}}%
% STR 2 0 3 0
% 3 1900 500 1900 600 5 0
% {\bf Black Hole}
\put(19.0000,-6.0000){\makebox(0,0){{\bf Black Hole}}}%
% CIRCLE 2 0 2 0
% 4 2450 1200 2450 1350 2450 1350 2450 1350
% 
\special{pn 8}%
\special{sh 0}%
\special{ar 2450 1200 150 150  0.0000000 6.2831853}%
% STR 2 0 3 0
% 3 2450 1100 2450 1200 5 0
% $P$
\put(24.5000,-12.0000){\makebox(0,0){$P$}}%
% POLYLINE 2 0 2 0
% 10 2650 1150 2950 1150 2950 1100 3100 1200 2950 1300 2950 1250 2650 1250 2650 1150 2650 1150 2650 1150
% 
\special{pn 8}%
\special{sh 0}%
\special{pa 2650 1150}%
\special{pa 2950 1150}%
\special{pa 2950 1100}%
\special{pa 3100 1200}%
\special{pa 2950 1300}%
\special{pa 2950 1250}%
\special{pa 2650 1250}%
\special{pa 2650 1150}%
\special{pa 2650 1150}%
\special{pa 2650 1150}%
\special{fp}%
% LINE 0 0 3 0
% 2 2800 400 2800 2000
% 
\special{pn 20}%
\special{pa 2800 400}%
\special{pa 2800 2000}%
\special{fp}%
% CIRCLE 2 0 2 0
% 4 2100 1200 2100 1350 2100 1350 2100 1350
% 
\special{pn 8}%
\special{sh 0}%
\special{ar 2100 1200 150 150  0.0000000 6.2831853}%
% POLYLINE 2 0 2 0
% 10 1900 1150 1600 1150 1600 1100 1450 1200 1600 1300 1600 1250 1900 1250 1900 1150 1900 1150 1900 1150
% 
\special{pn 8}%
\special{sh 0}%
\special{pa 1900 1150}%
\special{pa 1600 1150}%
\special{pa 1600 1100}%
\special{pa 1450 1200}%
\special{pa 1600 1300}%
\special{pa 1600 1250}%
\special{pa 1900 1250}%
\special{pa 1900 1150}%
\special{pa 1900 1150}%
\special{pa 1900 1150}%
\special{fp}%
% STR 2 0 3 0
% 3 2100 1100 2100 1200 5 0
% $\bar{P}$
\put(21.0000,-12.0000){\makebox(0,0){$\bar{P}$}}%
% STR 2 0 3 0
% 3 3500 500 3500 600 5 0
% {\bf Our Universe}
\put(35.0000,-6.0000){\makebox(0,0){{\bf Our Universe}}}%
% STR 2 0 3 0
% 3 3500 1100 3500 1200 5 0
% Radiation
\put(35.0000,-12.0000){\makebox(0,0){Radiation}}%
\end{picture}%

%% file: tun02.tex
%WinTpicVersion3.08
\unitlength 0.1in
\begin{picture}( 51.5000, 16.0000)(  6.0000,-18.0000)
% CIRCLE 2 0 1 0
% 4 1400 1000 1400 1800 1400 1800 1400 1800
% 
\special{pn 8}%
\special{sh 0.300}%
\special{ar 1400 1000 800 800  0.0000000 6.2831853}%
% CIRCLE 2 2 3 0
% 4 4000 1000 4000 1800 4000 1800 4000 1800
% 
\special{pn 8}%
\special{ar 4000 1000 800 800  0.0000000 0.0150000}%
\special{ar 4000 1000 800 800  0.0600000 0.0750000}%
\special{ar 4000 1000 800 800  0.1200000 0.1350000}%
\special{ar 4000 1000 800 800  0.1800000 0.1950000}%
\special{ar 4000 1000 800 800  0.2400000 0.2550000}%
\special{ar 4000 1000 800 800  0.3000000 0.3150000}%
\special{ar 4000 1000 800 800  0.3600000 0.3750000}%
\special{ar 4000 1000 800 800  0.4200000 0.4350000}%
\special{ar 4000 1000 800 800  0.4800000 0.4950000}%
\special{ar 4000 1000 800 800  0.5400000 0.5550000}%
\special{ar 4000 1000 800 800  0.6000000 0.6150000}%
\special{ar 4000 1000 800 800  0.6600000 0.6750000}%
\special{ar 4000 1000 800 800  0.7200000 0.7350000}%
\special{ar 4000 1000 800 800  0.7800000 0.7950000}%
\special{ar 4000 1000 800 800  0.8400000 0.8550000}%
\special{ar 4000 1000 800 800  0.9000000 0.9150000}%
\special{ar 4000 1000 800 800  0.9600000 0.9750000}%
\special{ar 4000 1000 800 800  1.0200000 1.0350000}%
\special{ar 4000 1000 800 800  1.0800000 1.0950000}%
\special{ar 4000 1000 800 800  1.1400000 1.1550000}%
\special{ar 4000 1000 800 800  1.2000000 1.2150000}%
\special{ar 4000 1000 800 800  1.2600000 1.2750000}%
\special{ar 4000 1000 800 800  1.3200000 1.3350000}%
\special{ar 4000 1000 800 800  1.3800000 1.3950000}%
\special{ar 4000 1000 800 800  1.4400000 1.4550000}%
\special{ar 4000 1000 800 800  1.5000000 1.5150000}%
\special{ar 4000 1000 800 800  1.5600000 1.5750000}%
\special{ar 4000 1000 800 800  1.6200000 1.6350000}%
\special{ar 4000 1000 800 800  1.6800000 1.6950000}%
\special{ar 4000 1000 800 800  1.7400000 1.7550000}%
\special{ar 4000 1000 800 800  1.8000000 1.8150000}%
\special{ar 4000 1000 800 800  1.8600000 1.8750000}%
\special{ar 4000 1000 800 800  1.9200000 1.9350000}%
\special{ar 4000 1000 800 800  1.9800000 1.9950000}%
\special{ar 4000 1000 800 800  2.0400000 2.0550000}%
\special{ar 4000 1000 800 800  2.1000000 2.1150000}%
\special{ar 4000 1000 800 800  2.1600000 2.1750000}%
\special{ar 4000 1000 800 800  2.2200000 2.2350000}%
\special{ar 4000 1000 800 800  2.2800000 2.2950000}%
\special{ar 4000 1000 800 800  2.3400000 2.3550000}%
\special{ar 4000 1000 800 800  2.4000000 2.4150000}%
\special{ar 4000 1000 800 800  2.4600000 2.4750000}%
\special{ar 4000 1000 800 800  2.5200000 2.5350000}%
\special{ar 4000 1000 800 800  2.5800000 2.5950000}%
\special{ar 4000 1000 800 800  2.6400000 2.6550000}%
\special{ar 4000 1000 800 800  2.7000000 2.7150000}%
\special{ar 4000 1000 800 800  2.7600000 2.7750000}%
\special{ar 4000 1000 800 800  2.8200000 2.8350000}%
\special{ar 4000 1000 800 800  2.8800000 2.8950000}%
\special{ar 4000 1000 800 800  2.9400000 2.9550000}%
\special{ar 4000 1000 800 800  3.0000000 3.0150000}%
\special{ar 4000 1000 800 800  3.0600000 3.0750000}%
\special{ar 4000 1000 800 800  3.1200000 3.1350000}%
\special{ar 4000 1000 800 800  3.1800000 3.1950000}%
\special{ar 4000 1000 800 800  3.2400000 3.2550000}%
\special{ar 4000 1000 800 800  3.3000000 3.3150000}%
\special{ar 4000 1000 800 800  3.3600000 3.3750000}%
\special{ar 4000 1000 800 800  3.4200000 3.4350000}%
\special{ar 4000 1000 800 800  3.4800000 3.4950000}%
\special{ar 4000 1000 800 800  3.5400000 3.5550000}%
\special{ar 4000 1000 800 800  3.6000000 3.6150000}%
\special{ar 4000 1000 800 800  3.6600000 3.6750000}%
\special{ar 4000 1000 800 800  3.7200000 3.7350000}%
\special{ar 4000 1000 800 800  3.7800000 3.7950000}%
\special{ar 4000 1000 800 800  3.8400000 3.8550000}%
\special{ar 4000 1000 800 800  3.9000000 3.9150000}%
\special{ar 4000 1000 800 800  3.9600000 3.9750000}%
\special{ar 4000 1000 800 800  4.0200000 4.0350000}%
\special{ar 4000 1000 800 800  4.0800000 4.0950000}%
\special{ar 4000 1000 800 800  4.1400000 4.1550000}%
\special{ar 4000 1000 800 800  4.2000000 4.2150000}%
\special{ar 4000 1000 800 800  4.2600000 4.2750000}%
\special{ar 4000 1000 800 800  4.3200000 4.3350000}%
\special{ar 4000 1000 800 800  4.3800000 4.3950000}%
\special{ar 4000 1000 800 800  4.4400000 4.4550000}%
\special{ar 4000 1000 800 800  4.5000000 4.5150000}%
\special{ar 4000 1000 800 800  4.5600000 4.5750000}%
\special{ar 4000 1000 800 800  4.6200000 4.6350000}%
\special{ar 4000 1000 800 800  4.6800000 4.6950000}%
\special{ar 4000 1000 800 800  4.7400000 4.7550000}%
\special{ar 4000 1000 800 800  4.8000000 4.8150000}%
\special{ar 4000 1000 800 800  4.8600000 4.8750000}%
\special{ar 4000 1000 800 800  4.9200000 4.9350000}%
\special{ar 4000 1000 800 800  4.9800000 4.9950000}%
\special{ar 4000 1000 800 800  5.0400000 5.0550000}%
\special{ar 4000 1000 800 800  5.1000000 5.1150000}%
\special{ar 4000 1000 800 800  5.1600000 5.1750000}%
\special{ar 4000 1000 800 800  5.2200000 5.2350000}%
\special{ar 4000 1000 800 800  5.2800000 5.2950000}%
\special{ar 4000 1000 800 800  5.3400000 5.3550000}%
\special{ar 4000 1000 800 800  5.4000000 5.4150000}%
\special{ar 4000 1000 800 800  5.4600000 5.4750000}%
\special{ar 4000 1000 800 800  5.5200000 5.5350000}%
\special{ar 4000 1000 800 800  5.5800000 5.5950000}%
\special{ar 4000 1000 800 800  5.6400000 5.6550000}%
\special{ar 4000 1000 800 800  5.7000000 5.7150000}%
\special{ar 4000 1000 800 800  5.7600000 5.7750000}%
\special{ar 4000 1000 800 800  5.8200000 5.8350000}%
\special{ar 4000 1000 800 800  5.8800000 5.8950000}%
\special{ar 4000 1000 800 800  5.9400000 5.9550000}%
\special{ar 4000 1000 800 800  6.0000000 6.0150000}%
\special{ar 4000 1000 800 800  6.0600000 6.0750000}%
\special{ar 4000 1000 800 800  6.1200000 6.1350000}%
\special{ar 4000 1000 800 800  6.1800000 6.1950000}%
\special{ar 4000 1000 800 800  6.2400000 6.2550000}%
% CIRCLE 2 0 1 0
% 4 4000 1000 4400 400 4400 400 4400 400
% 
\special{pn 8}%
\special{sh 0.300}%
\special{ar 4000 1000 722 722  0.0000000 6.2831853}%
% CIRCLE 2 0 3 0
% 4 5600 1000 5600 1150 5600 1150 5600 1150
% 
\special{pn 8}%
\special{ar 5600 1000 150 150  0.0000000 6.2831853}%
% STR 2 0 3 0
% 3 5200 900 5200 1000 5 0
% $+$
\put(52.0000,-10.0000){\makebox(0,0){$+$}}%
% POLYLINE 2 0 3 0
% 10 2500 900 2700 900 2700 800 2900 1000 2700 1200 2700 1100 2500 1100 2500 900 2500 900 2500 900
% 
\special{pn 8}%
\special{pa 2500 900}%
\special{pa 2700 900}%
\special{pa 2700 800}%
\special{pa 2900 1000}%
\special{pa 2700 1200}%
\special{pa 2700 1100}%
\special{pa 2500 1100}%
\special{pa 2500 900}%
\special{pa 2500 900}%
\special{pa 2500 900}%
\special{fp}%
% STR 2 0 3 0
% 3 1410 900 1410 1000 0 0
% 
\put(14.1000,-10.0000){\makebox(0,0)[lb]{}}%
% STR 2 0 3 0
% 3 1400 890 1400 990 0 0
% 
\put(14.0000,-9.9000){\makebox(0,0)[lb]{}}%
% STR 2 0 3 0
% 3 1400 900 1400 1000 5 0
% $M$
\put(14.0000,-10.0000){\makebox(0,0){$M$}}%
% STR 2 0 3 0
% 3 4000 900 4000 1000 5 0
% $M-\omega$
\put(40.0000,-10.0000){\makebox(0,0){$M-\omega$}}%
% STR 2 0 3 0
% 3 5600 900 5600 1000 5 0
% $\omega$
\put(56.0000,-10.0000){\makebox(0,0){$\omega$}}%
\end{picture}%

%% file: tun03.tex
%WinTpicVersion3.08
\unitlength 0.1in
\begin{picture}( 42.0000, 16.0000)(  6.0000,-18.0000)
% CIRCLE 2 0 1 0
% 4 1400 1000 1400 1800 1400 1800 1400 1800
% 
\special{pn 8}%
\special{sh 0.300}%
\special{ar 1400 1000 800 800  0.0000000 6.2831853}%
% CIRCLE 2 2 3 0
% 4 4000 1000 4000 1800 4000 1800 4000 1800
% 
\special{pn 8}%
\special{ar 4000 1000 800 800  0.0000000 0.0150000}%
\special{ar 4000 1000 800 800  0.0600000 0.0750000}%
\special{ar 4000 1000 800 800  0.1200000 0.1350000}%
\special{ar 4000 1000 800 800  0.1800000 0.1950000}%
\special{ar 4000 1000 800 800  0.2400000 0.2550000}%
\special{ar 4000 1000 800 800  0.3000000 0.3150000}%
\special{ar 4000 1000 800 800  0.3600000 0.3750000}%
\special{ar 4000 1000 800 800  0.4200000 0.4350000}%
\special{ar 4000 1000 800 800  0.4800000 0.4950000}%
\special{ar 4000 1000 800 800  0.5400000 0.5550000}%
\special{ar 4000 1000 800 800  0.6000000 0.6150000}%
\special{ar 4000 1000 800 800  0.6600000 0.6750000}%
\special{ar 4000 1000 800 800  0.7200000 0.7350000}%
\special{ar 4000 1000 800 800  0.7800000 0.7950000}%
\special{ar 4000 1000 800 800  0.8400000 0.8550000}%
\special{ar 4000 1000 800 800  0.9000000 0.9150000}%
\special{ar 4000 1000 800 800  0.9600000 0.9750000}%
\special{ar 4000 1000 800 800  1.0200000 1.0350000}%
\special{ar 4000 1000 800 800  1.0800000 1.0950000}%
\special{ar 4000 1000 800 800  1.1400000 1.1550000}%
\special{ar 4000 1000 800 800  1.2000000 1.2150000}%
\special{ar 4000 1000 800 800  1.2600000 1.2750000}%
\special{ar 4000 1000 800 800  1.3200000 1.3350000}%
\special{ar 4000 1000 800 800  1.3800000 1.3950000}%
\special{ar 4000 1000 800 800  1.4400000 1.4550000}%
\special{ar 4000 1000 800 800  1.5000000 1.5150000}%
\special{ar 4000 1000 800 800  1.5600000 1.5750000}%
\special{ar 4000 1000 800 800  1.6200000 1.6350000}%
\special{ar 4000 1000 800 800  1.6800000 1.6950000}%
\special{ar 4000 1000 800 800  1.7400000 1.7550000}%
\special{ar 4000 1000 800 800  1.8000000 1.8150000}%
\special{ar 4000 1000 800 800  1.8600000 1.8750000}%
\special{ar 4000 1000 800 800  1.9200000 1.9350000}%
\special{ar 4000 1000 800 800  1.9800000 1.9950000}%
\special{ar 4000 1000 800 800  2.0400000 2.0550000}%
\special{ar 4000 1000 800 800  2.1000000 2.1150000}%
\special{ar 4000 1000 800 800  2.1600000 2.1750000}%
\special{ar 4000 1000 800 800  2.2200000 2.2350000}%
\special{ar 4000 1000 800 800  2.2800000 2.2950000}%
\special{ar 4000 1000 800 800  2.3400000 2.3550000}%
\special{ar 4000 1000 800 800  2.4000000 2.4150000}%
\special{ar 4000 1000 800 800  2.4600000 2.4750000}%
\special{ar 4000 1000 800 800  2.5200000 2.5350000}%
\special{ar 4000 1000 800 800  2.5800000 2.5950000}%
\special{ar 4000 1000 800 800  2.6400000 2.6550000}%
\special{ar 4000 1000 800 800  2.7000000 2.7150000}%
\special{ar 4000 1000 800 800  2.7600000 2.7750000}%
\special{ar 4000 1000 800 800  2.8200000 2.8350000}%
\special{ar 4000 1000 800 800  2.8800000 2.8950000}%
\special{ar 4000 1000 800 800  2.9400000 2.9550000}%
\special{ar 4000 1000 800 800  3.0000000 3.0150000}%
\special{ar 4000 1000 800 800  3.0600000 3.0750000}%
\special{ar 4000 1000 800 800  3.1200000 3.1350000}%
\special{ar 4000 1000 800 800  3.1800000 3.1950000}%
\special{ar 4000 1000 800 800  3.2400000 3.2550000}%
\special{ar 4000 1000 800 800  3.3000000 3.3150000}%
\special{ar 4000 1000 800 800  3.3600000 3.3750000}%
\special{ar 4000 1000 800 800  3.4200000 3.4350000}%
\special{ar 4000 1000 800 800  3.4800000 3.4950000}%
\special{ar 4000 1000 800 800  3.5400000 3.5550000}%
\special{ar 4000 1000 800 800  3.6000000 3.6150000}%
\special{ar 4000 1000 800 800  3.6600000 3.6750000}%
\special{ar 4000 1000 800 800  3.7200000 3.7350000}%
\special{ar 4000 1000 800 800  3.7800000 3.7950000}%
\special{ar 4000 1000 800 800  3.8400000 3.8550000}%
\special{ar 4000 1000 800 800  3.9000000 3.9150000}%
\special{ar 4000 1000 800 800  3.9600000 3.9750000}%
\special{ar 4000 1000 800 800  4.0200000 4.0350000}%
\special{ar 4000 1000 800 800  4.0800000 4.0950000}%
\special{ar 4000 1000 800 800  4.1400000 4.1550000}%
\special{ar 4000 1000 800 800  4.2000000 4.2150000}%
\special{ar 4000 1000 800 800  4.2600000 4.2750000}%
\special{ar 4000 1000 800 800  4.3200000 4.3350000}%
\special{ar 4000 1000 800 800  4.3800000 4.3950000}%
\special{ar 4000 1000 800 800  4.4400000 4.4550000}%
\special{ar 4000 1000 800 800  4.5000000 4.5150000}%
\special{ar 4000 1000 800 800  4.5600000 4.5750000}%
\special{ar 4000 1000 800 800  4.6200000 4.6350000}%
\special{ar 4000 1000 800 800  4.6800000 4.6950000}%
\special{ar 4000 1000 800 800  4.7400000 4.7550000}%
\special{ar 4000 1000 800 800  4.8000000 4.8150000}%
\special{ar 4000 1000 800 800  4.8600000 4.8750000}%
\special{ar 4000 1000 800 800  4.9200000 4.9350000}%
\special{ar 4000 1000 800 800  4.9800000 4.9950000}%
\special{ar 4000 1000 800 800  5.0400000 5.0550000}%
\special{ar 4000 1000 800 800  5.1000000 5.1150000}%
\special{ar 4000 1000 800 800  5.1600000 5.1750000}%
\special{ar 4000 1000 800 800  5.2200000 5.2350000}%
\special{ar 4000 1000 800 800  5.2800000 5.2950000}%
\special{ar 4000 1000 800 800  5.3400000 5.3550000}%
\special{ar 4000 1000 800 800  5.4000000 5.4150000}%
\special{ar 4000 1000 800 800  5.4600000 5.4750000}%
\special{ar 4000 1000 800 800  5.5200000 5.5350000}%
\special{ar 4000 1000 800 800  5.5800000 5.5950000}%
\special{ar 4000 1000 800 800  5.6400000 5.6550000}%
\special{ar 4000 1000 800 800  5.7000000 5.7150000}%
\special{ar 4000 1000 800 800  5.7600000 5.7750000}%
\special{ar 4000 1000 800 800  5.8200000 5.8350000}%
\special{ar 4000 1000 800 800  5.8800000 5.8950000}%
\special{ar 4000 1000 800 800  5.9400000 5.9550000}%
\special{ar 4000 1000 800 800  6.0000000 6.0150000}%
\special{ar 4000 1000 800 800  6.0600000 6.0750000}%
\special{ar 4000 1000 800 800  6.1200000 6.1350000}%
\special{ar 4000 1000 800 800  6.1800000 6.1950000}%
\special{ar 4000 1000 800 800  6.2400000 6.2550000}%
% CIRCLE 2 0 3 0
% 4 4650 1000 4650 1150 4650 1150 4650 1150
% 
\special{pn 8}%
\special{ar 4650 1000 150 150  0.0000000 6.2831853}%
% POLYLINE 2 0 3 0
% 10 2500 900 2700 900 2700 800 2900 1000 2700 1200 2700 1100 2500 1100 2500 900 2500 900 2500 900
% 
\special{pn 8}%
\special{pa 2500 900}%
\special{pa 2700 900}%
\special{pa 2700 800}%
\special{pa 2900 1000}%
\special{pa 2700 1200}%
\special{pa 2700 1100}%
\special{pa 2500 1100}%
\special{pa 2500 900}%
\special{pa 2500 900}%
\special{pa 2500 900}%
\special{fp}%
% STR 2 0 3 0
% 3 1410 900 1410 1000 0 0
% 
\put(14.1000,-10.0000){\makebox(0,0)[lb]{}}%
% STR 2 0 3 0
% 3 1400 890 1400 990 0 0
% 
\put(14.0000,-9.9000){\makebox(0,0)[lb]{}}%
% CIRCLE 2 0 1 0
% 4 4000 1000 4500 1000 4500 1000 4500 1000
% 
\special{pn 8}%
\special{sh 0.300}%
\special{ar 4000 1000 500 500  0.0000000 6.2831853}%
% CIRCLE 2 0 1 0
% 4 2050 1000 2050 1150 2050 1150 2050 1150
% 
\special{pn 8}%
\special{sh 0.300}%
\special{ar 2050 1000 150 150  0.0000000 6.2831853}%
% VECTOR 2 0 3 0
% 4 2050 1000 2200 1000 2200 1000 2050 1000
% 
\special{pn 8}%
\special{pa 2050 1000}%
\special{pa 2200 1000}%
\special{fp}%
\special{sh 1}%
\special{pa 2200 1000}%
\special{pa 2134 980}%
\special{pa 2148 1000}%
\special{pa 2134 1020}%
\special{pa 2200 1000}%
\special{fp}%
\special{pa 2200 1000}%
\special{pa 2050 1000}%
\special{fp}%
\special{sh 1}%
\special{pa 2050 1000}%
\special{pa 2118 1020}%
\special{pa 2104 1000}%
\special{pa 2118 980}%
\special{pa 2050 1000}%
\special{fp}%
% STR 2 0 3 0
% 3 2120 970 2120 1070 5 0
% $\epsilon$
\put(21.2000,-10.7000){\makebox(0,0){$\epsilon$}}%
% STR 2 0 3 0
% 3 1680 610 1680 710 5 0
% $2M$
\put(16.8000,-7.1000){\makebox(0,0){$2M$}}%
% VECTOR 2 0 3 0
% 2 1770 640 1970 440
% 
\special{pn 8}%
\special{pa 1770 640}%
\special{pa 1970 440}%
\special{fp}%
\special{sh 1}%
\special{pa 1970 440}%
\special{pa 1910 474}%
\special{pa 1932 478}%
\special{pa 1938 502}%
\special{pa 1970 440}%
\special{fp}%
% VECTOR 2 0 3 0
% 2 1600 800 1400 1000
% 
\special{pn 8}%
\special{pa 1600 800}%
\special{pa 1400 1000}%
\special{fp}%
\special{sh 1}%
\special{pa 1400 1000}%
\special{pa 1462 968}%
\special{pa 1438 962}%
\special{pa 1434 940}%
\special{pa 1400 1000}%
\special{fp}%
% DOT 2 0 3 0
% 2 1400 1000 1400 1000
% 
\special{pn 8}%
\special{sh 1}%
\special{ar 1400 1000 10 10 0  6.28318530717959E+0000}%
\special{sh 1}%
\special{ar 1400 1000 10 10 0  6.28318530717959E+0000}%
% DOT 2 0 3 0
% 2 4000 1000 4000 990
% 
\special{pn 8}%
\special{sh 1}%
\special{ar 4000 1000 10 10 0  6.28318530717959E+0000}%
\special{sh 1}%
\special{ar 4000 990 10 10 0  6.28318530717959E+0000}%
% VECTOR 2 0 3 0
% 4 4500 1000 4650 1000 4650 1000 4500 1000
% 
\special{pn 8}%
\special{pa 4500 1000}%
\special{pa 4650 1000}%
\special{fp}%
\special{sh 1}%
\special{pa 4650 1000}%
\special{pa 4584 980}%
\special{pa 4598 1000}%
\special{pa 4584 1020}%
\special{pa 4650 1000}%
\special{fp}%
\special{pa 4650 1000}%
\special{pa 4500 1000}%
\special{fp}%
\special{sh 1}%
\special{pa 4500 1000}%
\special{pa 4568 1020}%
\special{pa 4554 1000}%
\special{pa 4568 980}%
\special{pa 4500 1000}%
\special{fp}%
% STR 2 0 3 0
% 3 4650 980 4650 1080 5 0
% $\epsilon$
\put(46.5000,-10.8000){\makebox(0,0){$\epsilon$}}%
% STR 2 0 3 0
% 3 4080 710 4080 810 5 0
% $2(M-\omega)$
\put(40.8000,-8.1000){\makebox(0,0){$2(M-\omega)$}}%
% STR 2 0 3 0
% 3 1400 1050 1400 1150 5 0
% {\bf BH}
\put(14.0000,-11.5000){\makebox(0,0){{\bf BH}}}%
% STR 2 0 3 0
% 3 3990 640 3990 740 0 0
% 
\put(39.9000,-7.4000){\makebox(0,0)[lb]{}}%
% VECTOR 2 0 3 0
% 2 4120 880 4020 980
% 
\special{pn 8}%
\special{pa 4120 880}%
\special{pa 4020 980}%
\special{fp}%
\special{sh 1}%
\special{pa 4020 980}%
\special{pa 4082 948}%
\special{pa 4058 942}%
\special{pa 4054 920}%
\special{pa 4020 980}%
\special{fp}%
% VECTOR 2 0 3 0
% 2 4250 750 4350 650
% 
\special{pn 8}%
\special{pa 4250 750}%
\special{pa 4350 650}%
\special{fp}%
\special{sh 1}%
\special{pa 4350 650}%
\special{pa 4290 684}%
\special{pa 4312 688}%
\special{pa 4318 712}%
\special{pa 4350 650}%
\special{fp}%
% VECTOR 2 0 3 0
% 2 4570 460 4590 440
% 
\special{pn 8}%
\special{pa 4570 460}%
\special{pa 4590 440}%
\special{fp}%
\special{sh 1}%
\special{pa 4590 440}%
\special{pa 4530 474}%
\special{pa 4552 478}%
\special{pa 4558 502}%
\special{pa 4590 440}%
\special{fp}%
% VECTOR 2 0 3 0
% 2 4380 620 4360 640
% 
\special{pn 8}%
\special{pa 4380 620}%
\special{pa 4360 640}%
\special{fp}%
\special{sh 1}%
\special{pa 4360 640}%
\special{pa 4422 608}%
\special{pa 4398 602}%
\special{pa 4394 580}%
\special{pa 4360 640}%
\special{fp}%
% STR 2 0 3 0
% 3 4490 430 4490 530 5 0
% $2\omega$
\put(44.9000,-5.3000){\makebox(0,0){$2\omega$}}%
% STR 2 0 3 0
% 3 4000 1050 4000 1150 5 0
% {\bf BH}
\put(40.0000,-11.5000){\makebox(0,0){{\bf BH}}}%
\end{picture}%

%% file: tun04.tex
%WinTpicVersion3.08
\unitlength 0.1in
\begin{picture}( 29.3000, 23.2500)( 12.7000,-27.0500)
% BOX 2 5 1 0
% 2 3200 2600 1270 380
% 
\special{pn 8}%
\special{sh 0.300}%
\special{pa 3200 2600}%
\special{pa 1270 2600}%
\special{pa 1270 380}%
\special{pa 3200 380}%
\special{pa 3200 2600}%
\special{ip}%
% STR 2 0 3 0
% 3 1740 1480 1740 1580 5 0
% {\bf BH}
\put(17.4000,-15.8000){\makebox(0,0){{\bf BH}}}%
% VECTOR 2 0 3 0
% 2 3792 1275 4150 1275
% 
\special{pn 8}%
\special{pa 3792 1276}%
\special{pa 4150 1276}%
\special{fp}%
\special{sh 1}%
\special{pa 4150 1276}%
\special{pa 4084 1256}%
\special{pa 4098 1276}%
\special{pa 4084 1296}%
\special{pa 4150 1276}%
\special{fp}%
% VECTOR 2 0 3 0
% 2 3670 1275 3330 1275
% 
\special{pn 8}%
\special{pa 3670 1276}%
\special{pa 3330 1276}%
\special{fp}%
\special{sh 1}%
\special{pa 3330 1276}%
\special{pa 3398 1296}%
\special{pa 3384 1276}%
\special{pa 3398 1256}%
\special{pa 3330 1276}%
\special{fp}%
% STR 2 0 3 0
% 3 3730 1460 3730 1560 5 0
% outside
\put(37.3000,-15.6000){\makebox(0,0){outside}}%
% STR 2 0 3 0
% 3 3520 1020 3520 1120 5 0
% $\phi^{L} _{\rm{out}}$
\put(35.2000,-11.2000){\makebox(0,0){$\phi^{L} _{\rm{out}}$}}%
% STR 2 0 3 0
% 3 3940 1020 3940 1120 5 0
% $\phi^{R} _{\rm{out}}$
\put(39.4000,-11.2000){\makebox(0,0){$\phi^{R} _{\rm{out}}$}}%
% LINE 0 0 3 0
% 2 3200 400 3200 2600
% 
\special{pn 20}%
\special{pa 3200 400}%
\special{pa 3200 2600}%
\special{fp}%
% VECTOR 2 0 3 0
% 2 2730 1280 3088 1280
% 
\special{pn 8}%
\special{pa 2730 1280}%
\special{pa 3088 1280}%
\special{fp}%
\special{sh 1}%
\special{pa 3088 1280}%
\special{pa 3022 1260}%
\special{pa 3036 1280}%
\special{pa 3022 1300}%
\special{pa 3088 1280}%
\special{fp}%
% VECTOR 2 0 3 0
% 2 2600 1280 2260 1280
% 
\special{pn 8}%
\special{pa 2600 1280}%
\special{pa 2260 1280}%
\special{fp}%
\special{sh 1}%
\special{pa 2260 1280}%
\special{pa 2328 1300}%
\special{pa 2314 1280}%
\special{pa 2328 1260}%
\special{pa 2260 1280}%
\special{fp}%
% STR 2 0 3 0
% 3 2670 1470 2670 1570 5 0
% inside
\put(26.7000,-15.7000){\makebox(0,0){inside}}%
% STR 2 0 3 0
% 3 2470 1020 2470 1120 5 0
% $\phi^{L} _{\rm{in}}$
\put(24.7000,-11.2000){\makebox(0,0){$\phi^{L} _{\rm{in}}$}}%
% STR 2 0 3 0
% 3 2900 1030 2900 1130 5 0
% $\phi^{R} _{\rm{in}}$
\put(29.0000,-11.3000){\makebox(0,0){$\phi^{R} _{\rm{in}}$}}%
% LINE 2 2 3 0
% 2 2190 390 2190 2590
% 
\special{pn 8}%
\special{pa 2190 390}%
\special{pa 2190 2590}%
\special{dt 0.045}%
% LINE 2 2 3 0
% 2 4200 390 4200 2590
% 
\special{pn 8}%
\special{pa 4200 390}%
\special{pa 4200 2590}%
\special{dt 0.045}%
% VECTOR 2 0 3 0
% 4 3200 2190 4200 2190 4190 2190 3200 2190
% 
\special{pn 8}%
\special{pa 3200 2190}%
\special{pa 4200 2190}%
\special{fp}%
\special{sh 1}%
\special{pa 4200 2190}%
\special{pa 4134 2170}%
\special{pa 4148 2190}%
\special{pa 4134 2210}%
\special{pa 4200 2190}%
\special{fp}%
\special{pa 4190 2190}%
\special{pa 3200 2190}%
\special{fp}%
\special{sh 1}%
\special{pa 3200 2190}%
\special{pa 3268 2210}%
\special{pa 3254 2190}%
\special{pa 3268 2170}%
\special{pa 3200 2190}%
\special{fp}%
% STR 2 0 3 0
% 3 3710 2230 3710 2330 5 0
% $\varepsilon$
\put(37.1000,-23.3000){\makebox(0,0){$\varepsilon$}}%
% VECTOR 2 0 3 0
% 4 2175 2195 3175 2195 3165 2195 2175 2195
% 
\special{pn 8}%
\special{pa 2176 2196}%
\special{pa 3176 2196}%
\special{fp}%
\special{sh 1}%
\special{pa 3176 2196}%
\special{pa 3108 2176}%
\special{pa 3122 2196}%
\special{pa 3108 2216}%
\special{pa 3176 2196}%
\special{fp}%
\special{pa 3166 2196}%
\special{pa 2176 2196}%
\special{fp}%
\special{sh 1}%
\special{pa 2176 2196}%
\special{pa 2242 2216}%
\special{pa 2228 2196}%
\special{pa 2242 2176}%
\special{pa 2176 2196}%
\special{fp}%
% STR 2 0 3 0
% 3 2685 2235 2685 2335 5 0
% $\varepsilon$
\put(26.8500,-23.3500){\makebox(0,0){$\varepsilon$}}%
% STR 2 0 3 0
% 3 3200 2690 3200 2790 5 0
% ${\rm Horizon}$
\put(32.0000,-27.9000){\makebox(0,0){${\rm Horizon}$}}%
% STR 2 0 3 0
% 3 4190 2690 4190 2790 5 0
% $r_++\varepsilon$
\put(41.9000,-27.9000){\makebox(0,0){$r_++\varepsilon$}}%
% STR 2 0 3 0
% 3 2180 2690 2180 2790 5 0
% $r_+-\varepsilon$
\put(21.8000,-27.9000){\makebox(0,0){$r_+-\varepsilon$}}%
\end{picture}%